\documentclass[10pt]{article}

\usepackage[dvips]{epsfig} % this package is for manipulating postscript graphics
\usepackage{amssymb,amsmath,amsthm,mathrsfs} %this package contains all the special ams fonts etc.
\usepackage{graphicx}
\usepackage{lineno}

\usepackage[authoryear,sort&compress]{natbib}
\usepackage{url}
\usepackage{enumerate}
\usepackage{colordvi,color,pspicture}
\usepackage{psfrag}
\usepackage{rotating}
\newcommand{\X}{\mathcal{X}}
\newcommand{\U}{\mathcal{U}}

\newcommand{\E}{\mathbf{E}}
\newcommand{\Var}{\mathbf{Var}}
\newcommand{\Cov}{\mathbf{Cov}}

\newcommand{\ah}{\widehat{\alpha}}

\newcommand{\pasy}{p_{\text{asy}}}
\newcommand{\pmc}{p_{\text{mc}}}
\newcommand{\prand}{p_{\text{rand}}}

\theoremstyle{plain}

\theoremstyle{definition}

\theoremstyle{remark}

\begin{document}

%\begin{singlespace}

\title{Technical Report \# KU-EC-13-1:\\
New Cell-Specific and Overall Tests of Spatial Interaction
Based on Nearest Neighbor Contingency Tables}
\author{
Elvan Ceyhan
\thanks{Department of Mathematics, Ko\c{c} University, Sar{\i}yer, 34450, Istanbul, Turkey.}
%\and
%Carey E. Priebe\thanks{Department of Applied Mathematics and Statistics,
%The Johns Hopkins University,
%Baltimore, Md. 21218}
}
\date{\today}
\maketitle

%\linenumbers
%\modulolinenumbers[2]

\begin{abstract}
\noindent
Spatial interaction patterns such as segregation and association can be tested
using nearest neighbor contingency tables (NNCTs).
We introduce new cell-specific (or pairwise) and overall segregation tests
and
determine their asymptotic distributions.
In particular, we demonstrate that cell-specific tests enjoy asymptotic normality,
while overall tests have chi-square distributions asymptotically.
We also perform an extensive Monte Carlo simulation study to compare
the finite sample performance of the tests in terms of empirical size and power.
In addition to the cell-specific tests as post-hoc tests for overall tests,
we discuss one-class-versus-rest type of NNCT-tests after
an overall test yields significant interaction.
We also introduce the concepts of total, strong, and partial segregation/association
to label levels of these patterns.
We compare these new tests with the existing NNCT-tests in literature with simulations as well
and illustrate the NNCT-tests on an ecological data set.
\end{abstract}

\noindent
%{\scriptsize
{\small {\it Keywords:} Association; completely mapped data;
complete spatial randomness; post-hoc tests; random labeling; segregation; sparse sampling

\vspace{.25 in}

$^*$corresponding author.\\
\indent {\it e-mail:} elceyhan@ku.edu.tr (E.~Ceyhan) }

%\end{singlespace}

%\newpage

%\linenumbers
%\modulolinenumbers[2]

\section{Introduction}
\label{sec:intro}
Multivariate clustering patterns such as segregation and association
result from multivariate interaction between two or more classes (or species).
For convenience,
categories of the points or units are referred to as ``classes",
e.g., a class can stand for species, sex, or some other characteristic of the unit/subject.
Segregation is the spatial pattern in which points from the same class are closer to each other,
while association is the pattern in which points from different classes are closer to each other.
These patterns may have important implications in ecology, plant biology, or epidemiology.
See, for example, \cite{whipple:1980}, \cite{diggle:2003}, and \cite{hamill:1986}.
In particular, in ecology,
two tree species could be highly dependent on each other (as a result of, say,
symbiosis or mutualism),
and thus, coexist in a close vicinity (i.e., they are associated),
or they could be enjoying the company of conspecifics and thus form one-class clumps or groups
(i.e., they are segregated).
In epidemiology, cases might be clustered compared to controls,
due to infectious nature of a disease or closeness to a source of the disease
(i.e., cases and controls are segregated).
In a social context,
segregation of residences due to the socioeconomic status or ethnicity can be investigated by generative models
(\cite{fossett:2011}).
In literature,
spatial segregation is also used to refer to a univariate pattern of spatial clustering (\cite{robertson:2011}),
which is referred to as aggregation (\cite{ceyhan:overall}).
In a social network,
segregation of individuals are also modeled via random graph theoretical tools (\cite{henry:2011}).
In veterinary epidemiology,
a nonparametric method for detecting spatial segregation
according to the genotype and year of occurrence of bovine tuberculosis
is employed by \cite{diggle:2005}.

Many univariate (i.e., one-class) or multivariate (multi-class) spatial clustering tests are proposed in literature
(see \cite{kulldorff:2006} for an extensive review).
%Essentially these methods can be classified as first order or second order methods.
These methods include Ripley's $K$-function (\cite{ripley:2004}),
or $J$-function (\cite{lieshout:1999}),
nearest neighbor (NN) methods (\cite{diggle:2003}) and so on.
Among NN methods,
this article concerns the nearest neighbor contingency tables (NNCTs).
\cite{pielou:1961} introduced various tests based on NNCTs,
however, \cite{dixon:1994} extended these tests in various directions,
and also determined the correct asymptotic distribution of the proposed tests.
\cite{ceyhan:class2009,ceyhan:overall} compared NNCT-tests in literature,
and also proposed various tests based on NNCTs.

In this article, we introduce various new cell-specific segregation tests
and overall tests based on the cell-specific tests.
We compare these tests with the existing NNCT-tests in literature
(\cite{dixon:1994, dixon:NNCTEco2002} and \cite{ceyhan:corrected}).
We demonstrate that cell-specific tests are asymptotically normal,
and overall tests tend to chi-square distribution with the corresponding degrees of freedom.
In practice, cell-specific tests serve as post-hoc tests to be performed
when an overall test yields a significant result.
As an alternative post-hoc test after a significant overall test,
we discuss one-class-versus-rest (or one-vs-rest) type of NNCT-tests.
By extensive Monte Carlo simulations,
we compare the newly proposed tests to the ones in literature in terms of empirical size and power,
and determine which tests perform better for the segregation or association alternative
and which ones are more robust to differences in relative abundances (of the classes).

We describe the NNCTs and provide the null and alternative patterns in Section \ref{sec:null-alt-nncts},
provide the cell-specific tests in Section \ref{sec:cell-spec},
overall tests in Section \ref{sec:overall},
empirical size analysis in the two- and three-class cases in Sections \ref{sec:monte-carlo-2Cl}
and \ref{sec:monte-carlo-3Cl}, respectively,
and
empirical power analysis under segregation and association in the two- and three-class cases
in Sections \ref{sec:emp-power-2Cl} and \ref{sec:emp-power-3Cl}, respectively.
We present the empirical size and power analysis for the one-vs-rest type testing in
the three-class case in Section \ref{sec:one-vs-rest-size-power},
the illustration on the example data set in Section \ref{sec:example},
and our conclusions and guidelines for using the tests in Section \ref{sec:disc-conc}.

\section{Null and Alternative Spatial Patterns and NNCTs}
\label{sec:null-alt-nncts}
We describe the spatial point patterns for two classes only; the extension to multi-class case is straightforward.
Our null hypothesis is
$$H_o: \text{randomness in the NN structure with NN probabilities being proportional to class frequencies}$$
which may result from \emph{random labeling} (RL) or \emph{independence} of points from two classes.
Under independence,
the two classes result independently from the same stochastic process,
so that their spatial distribution is identical.
In this article,
among independence patterns we will only consider
complete spatial randomness (CSR)
of points from two classes.
Roughly, under \emph{CSR independence},
two classes are independently uniformly distributed in a region of interest,
while RL is the pattern in which,
given a fixed set of points in a region,
class labels are assigned to these fixed
points randomly so that the labels are independent of the locations.

As alternatives, we consider two major types of deviations from $H_o$:
segregation and association.
{\em Segregation} occurs if the
NN of an individual is more likely to be of the same
class as the individual than to be from a different class.
That is,
the probability that this individual having a NN from the same class
is larger than the relative frequency of the same class
(see, e.g., \cite{pielou:1961}).
{\em Association} occurs if the NN of an individual is more
likely to be from another class than to be of the same class as the individual.
That is,
the probability that this individual having a NN from another class
is larger than the relative frequency of the other class in question.
These patterns are not symmetric,
e.g., for two classes,
one class might be more associated with another class.
For example, plant species $X$ could be more dependent on species $Y$,
hence $X$ plants occur in close vicinity of $Y$ plants,
while the reverse relation may not be in the same level or type.
Also, class $X$ points might exhibit a stronger clustering,
compared to class $Y$ points,
and so might be more segregated compared to class $Y$ points.
See \cite{ceyhan:corrected} for more detail on the null and alternative patterns.

NNCTs are constructed using the NN frequencies of classes.
The construction of NNCTs for two classes is described, e.g., in \cite{ceyhan:overall},
here we provide a brief description for $m \ge 2$ classes.
Suppose there are $m$ classes labeled as $\{1,2,\ldots,m\}$.
NNCTs are constructed using NN frequencies for each class.
Let $N_i$ be the number of points from class $i$ for $i \in \{1,2,\ldots,m\}$
and $n=\sum_{i=1}^m N_i$.
If we record the class of each point and its nearest
neighbor, the NN relationships fall into $m^2$ categories:
$$(1,1),\,(1,2),\ldots,(1,m);\,(2,1),\,(2,2),\ldots,(2,m);\ldots,(m,m)$$
where in category $(i,j)$, class $i$ is the base class,
while class $j$ is the class of the NN.
Denoting $N_{ij}$ as the observed frequency of category $(i,j)$ for $i,j \in \{1,2,\ldots,m\}$,
we obtain the NNCT in Table \ref{tab:NNCT-mxm}
where $C_j$ is the sum of column $j$; i.e., number of times class
$j$ points serve as NNs
for $j\in\{1,2,\ldots,m\}$.
Note also that $n=\sum_{i,j}N_{ij}$, $n_i=\sum_{j=1}^m\, N_{ij}$, and $C_j=\sum_{i=1}^m\, N_{ij}$.
In cell $(i,j)$,
class $i$ is called the \emph{base class}, and class $j$ is called the \emph{NN class}.
Here we adopt the
convention that variables denoted by upper case letters are random
quantities, while variables denoted by lower case letters are fixed quantities.
Thus, in our NNCT-analysis, row sums are assumed to be fixed
(i.e., class sizes are given), while column sums are assumed to be
random and depend on the NN relationships between the classes.

\begin{table} [ht]
\centering
\begin{tabular}{cc|ccc|c}
\multicolumn{2}{c}{}& \multicolumn{3}{c}{NN class}& \\
\multicolumn{2}{c}{}&    class 1 &  $\ldots$ & class $m$   &   total  \\
\hline
&class 1 &    $N_{11}$    &  $\ldots$ &   $N_{1m}$    &   $n_1$  \\
\raisebox{1.5ex}[0pt]{base class}& $\vdots$ & $\vdots$ & $\ddots$ & $\vdots$ & $\vdots$\\
&class $m$ &    $N_{m1}$    & $\ldots$ &    $N_{mm}$    &   $n_m$  \\
\hline
&total     &    $C_1$             &  $\ldots$ &    $C_m$             &   $n$  \\
\end{tabular}
\caption{ \label{tab:NNCT-mxm}
The NNCT for $m$ classes.}
\end{table}

Under CSR independence or RL,
cell counts $N_{ij}$ would be close to their expected values,
while under segregation the diagonal counts $N_{ii}$ would be larger,
while under association the off-diagonal counts $N_{ij}$ would be larger than expected under $H_o$.
When \cite{pielou:1961} developed NNCT-tests, she used Pearson's $\chi^2$ test of
independence for testing segregation
which is not appropriate
due to the dependence structure in a NNCT.
\cite{dixon:1994} derived the correct asymptotic distribution of cell counts and
hence the appropriate test which also has a $\chi^2$-distribution asymptotically.
\cite{ceyhan:overall} determines the conditions when Pielou's test is appropriate,
and when Dixon's test is appropriate, and discusses their use in practice.

\subsection{Total, Strong, and Partial Segregation and Association}
\label{sec:part-seg}
When $H_o$ is rejected,
if the diagonal entries
($N_{ii}$ values) tend to be higher than expected,
there is {\em segregation};
if the off-diagonal entries are larger than expected,
there is {\em association}.
These types of patterns  are easy to detect for $m=2$ classes,
but for $m > 2$,
rejecting $H_o$ only indicates that there is some sort of deviation
from the null case, but with many possible directions,
since rejecting $H_o$ only implies that for some class $i$,
there exists classes that are more likely to serve as NN to class $i$ or less
likely to serve as NN to class $i$ than expected under $H_o$.
Let $\pi_{ij}$ be the probability that a point is from class $i$
and its NN is from class $j$.
For example, for a fixed class $i$,
if $\pi_{ii} \ge \sum_{j \not= i}\pi_{ij}$,
then we have {\em total segregation} of class $i$ from other classes;
that is,
class $i$ is more likely to have a same class NN than all other classes combined.
If $\pi_{ii} \ge \pi_{ij}$ for all $i \not= j$, then
we have {\em strong segregation}, which implies that class $i$ is
more likely to have a con-specific NN compared to all other classes one at a time.
Notice that total segregation implies strong segregation.
The strict inequalities in the above definitions yield
strict versions of total and strong segregation patterns.

For fixed classes $i$ and $j$, with $i \ne j$,
if $\pi_{ij} \ge \sum_{k \not= j}\pi_{ik}$ then we have {\em total association} of class $j$ with class $i$;
that is, class $j$ is more likely to be a NN of class $i$ than all other classes combined.
If $\pi_{ij} \ge \pi_{ik}$ for all $k \not= j$,
then we have {\em strong association} of class $j$ with class $i$,
which implies that class $j$ is more likely to be a NN of class $i$
compared to all other classes one at a time.
Notice that total association implies strong association.
Furthermore,
the strict inequalities in the above definitions yield
strict versions of total and strong association patterns.

On the other hand, if $\pi_{ii} \ge \pi_{ij}$ for all
$j \in S_1 \subsetneq \{1,2, \ldots,m\}\setminus \{i\}$ and
$\pi_{ii} \le \pi_{ij}$
for all $j \in S_2=\{1,2,\ldots,m\} \setminus (S_1\cup \{i\})$,
then we say that class $i$ is {\em more segregated} from the classes in $S_1$
and {\em more associated} with the classes in $S_2$.
Such cases are called {\em partial segregation} of class $i$ with respect to
classes in $S_1$ and {\em partial association} of class $i$ with classes in $S_2$.

\section{Cell-Specific Segregation Tests}
\label{sec:cell-spec}

We describe cell-specific segregation tests of Dixon and introduce new cell-specific tests labeled as
\emph{type I-IV} cell-specific tests, henceforth.

\subsection{Dixon's Cell-Specific Segregation Tests}
\label{sec:dix-cell-spec}
Dixon's cell-specific tests are used to measure the deviation of observed count in cell $(i,j)$ in a NNCT
from its expected value under $H_o$
described in detail in, e.g., \cite{dixon:1994, dixon:NNCTEco2002}.
The test statistic suggested by Dixon for cell $(i,j)$ is given by
\begin{equation}
\label{eqn:dixon-Zij}
Z^D_{ij}=\frac{N_{ij}-\E[N_{ij}]}{\sqrt{\Var[N_{ij}]}},
\end{equation}
where
$\E[N_{ij}]$ is the expected cell count and $\Var[N_{ij}]$ is the variance of cell count $N_{ij}$.

For $m \ge 2$ classes, under RL or CSR independence, the expected cell count for cell $(i,j)$ is
\begin{equation}
\label{eqn:Exp[Nij]}
\E[N_{ij}]=
\begin{cases}
n_i(n_i-1)/(n-1) & \text{if $i=j$,}\\
n_i\,n_j/(n-1)     & \text{if $i \not= j$,}
\end{cases}
\end{equation}
where $n_i$ is the fixed sample size for class $i$ for $i=1,2,\ldots,m$.
Observe that the expected cell counts depend only on the size of
each class (i.e., row sums), but not on column sums.
And the variance is
{\small
\begin{equation}
\label{eqn:VarNij}
\Var[N_{ij}]=
\begin{cases}
(n+R)\,p_{ii}+(2\,n-2\,R+Q)\,p_{iii}+(n^2-3\,n-Q+R)\,p_{iiii}-(n\,p_{ii})^2 & \text{if $i=j$,}\\
n\,p_{ij}+Q\,p_{iij}+(n^2-3\,n-Q+R)\,p_{iijj} -(n\,p_{ij})^2             & \text{if $i \not= j$,}
\end{cases}
\end{equation}
}
\noindent
with $p_{xx}$, $p_{xxx}$, and $p_{xxxx}$
are the probabilities that a randomly picked pair,
triplet, or quartet of points, respectively, are the indicated classes and
are given by
\begin{align}
\label{eqn:probs}
p_{ii}&=\frac{n_i\,(n_i-1)}{n\,(n-1)},  & p_{ij}&=\frac{n_i\,n_j}{n\,(n-1)},\nonumber\\
p_{iii}&=\frac{n_i\,(n_i-1)\,(n_i-2)}{n\,(n-1)\,(n-2)}, &
p_{iij}&=\frac{n_i\,(n_i-1)\,n_j}{n\,(n-1)\,(n-2)},\\
p_{iiii}&=\frac{n_i\,(n_i-1)\,(n_i-2)\,(n_i-3)}{n\,(n-1)\,(n-2)\,(n-3)}, &
p_{iijj}&=\frac{n_i\,(n_i-1)\,n_j\,(n_j-1)}{n\,(n-1)\,(n-2)\,(n-3)}.\nonumber
\end{align}
Furthermore, $R$ is twice the number of reflexive pairs
and $Q$ is the number of points with shared  NNs,
which occurs when two or more
points share a NN.
Then $Q=2\,(Q_2+3\,Q_3+6\,Q_4+10\,Q_5+15\,Q_6)$
where $Q_k$ is the number of points that serve
as a NN to other points $k$ times.

\subsection{Type I Cell-Specific Segregation Tests}
\label{sec:type-I-cell-spec}
In standard cases like multinomial sampling for contingency tables with fixed
row totals and conditioning on the column totals, $C_j=c_j$,
the expected cell count for cell $(i,j)$ in contingency tables
is $\E[N_{ij}]=\frac{n_i\,c_j}{n}$.
We first consider the difference $N_{ij}-\frac{n_i\,c_j}{n}$ for cell $(i,j)$.
However under RL, $N_i=n_i$ are fixed,
but $C_j$ are random quantities and $C_j=\sum_{i=1}^m N_{ij}$,
hence we suggest as the first type of cell-specific segregation test as
$$T^I_{ij}=N_{ij}-\frac{n_i\,C_j}{n}.$$
Then under RL,
\begin{equation}
%\label{eqn:Exp[Deltaij]}
\E\left[T^I_{ij}\right]=
\begin{cases}
\frac{n_i(n_i-1)}{(n-1)}-\frac{n_i}{n}\,\E[C_i] & \text{if $i=j$,}\\
\frac{n_i\,n_j}{(n-1)}-\frac{n_i}{n}\,\E[C_j]  & \text{if $i \not= j$.}
\end{cases}
\end{equation}

\noindent
For all $j$, $\E[C_j]=n_j$, since
\begin{multline*}
\E[C_j]=\sum_{i=1}^m \E[N_{ij}]=
\frac{n_j(n_j-1)}{(n-1)}+\sum_{i\neq j} \frac{n_i n_j}{(n-1)}
=\frac{n_j(n_j-1)}{(n-1)}+\frac{n_j}{(n-1)}\sum_{i\neq j} n_i\\
=\frac{n_j(n_j-1)}{(n-1)}+\frac{n_j}{(n-1)}(n-n_j)=n_j.
\end{multline*}

\noindent
Therefore,
\begin{equation}
\label{eqn:Exp[TIij]}
\E\left[T^I_{ij}\right]=
\begin{cases}
\frac{n_i(n_i-n)}{n(n-1)} & \text{if $i=j$,}\\
\frac{n_i\,n_j}{n(n-1)}  & \text{if $i \not= j$.}
\end{cases}
\end{equation}

For the variance of $T^I_{ij}$,
we have
\begin{equation}
\label{eqn:Var[TIij]} \Var\left[T^I_{ij}\right]=
\Var[N_{ij}]+\left(\frac{n_i^2}{n^2}\right)\Var[C_j]-2\left(\frac{n_i}{n}\right)\Cov[N_{ij},C_j]
\end{equation}
where
$\Var[N_{ij}]$ are as in Equation \eqref{eqn:VarNij},
$\Var[C_j]=\sum_{i=1}^m \Var[N_{ij}]+\sum_{k \neq i}\sum_i \Cov[N_{ij},N_{kj}]$
and
$\Cov[N_{ij},C_j]=\sum_{k=1}^m \Cov[N_{ij},N_{kj}]$
with $\Cov[N_{ij},N_{kl}]$ are as in Equations (4)-(12) of \cite{dixon:NNCTEco2002}.

As a new cell-specific test, we propose
\begin{equation}
\label{eqn:new-ZIij}
Z_{ij}^I=\frac{T^I_{ij}-\E\left[T^I_{ij}\right]}{\sqrt{\Var\left[T^I_{ij}\right]}}.
\end{equation}

\subsection{Type II Cell-Specific Segregation Tests}
\label{sec:type-II-cell-spec}
In Section \ref{sec:type-I-cell-spec},
we suggested $N_{ij}-\frac{n_i\,C_j}{n}$ as the test statistic for cell $(i,j)$.
However, under RL, $\E[C_j]=n_j$,
so we suggest as the second type of segregation test as
$$T^{II}_{ij}=N_{ij}-\frac{n_i\,n_j}{n}.$$
Then under RL,
$\E\left[T^{II}_{ij}\right]=\E\left[T^I_{ij}\right]$ which is provided in Equation \eqref{eqn:Exp[TIij]}.
Moreover, the variance of $T^I_{ij}$ is
$\Var\left[T^{II}_{ij}\right]=\Var[N_{ij}]$,
since $n_i$, $n_j$ and $n$ are fixed.

As a cell-specific test, we propose
\begin{equation}
\label{eqn:new-ZIij}
Z^{II}_{ij}=\frac{T^{II}_{ij}-\E\left[T^{II}_{ij}\right]}{\sqrt{\Var\left[T^{II}_{ij}\right]}}.
\end{equation}

\subsection{Type III Cell-Specific Segregation Tests}
\label{sec:type-III-cell-spec}
In the previous sections,
$\E\left[T^{I}_{ij}\right]=\E\left[T^{II}_{ij}\right] \not=0$
under RL.
Hence, instead of these test statistics,
in order to obtain 0 expected value for our test statistic,
we suggest the following:
\begin{equation}
\label{eqn:TIIIij}
T^{III}_{ij}=
\begin{cases}
N_{ii}-\frac{(n_i-1)}{(n-1)}C_i & \text{if $i=j$,}\\
N_{ij}-\frac{n_i}{(n-1)}C_j     & \text{if $i \not= j$.}
\end{cases}
\end{equation}
Then $\E\left[T^{III}_{ij}\right]=0$, since,
for $i=j$,
$$
\E\left[T^{III}_{ii}\right]=\E[N_{ii}]-\frac{(n_i-1)}{(n-1)}\E[C_i]=
\frac{n_i(n_i-1)}{(n-1)}-\frac{(n_i-1)}{(n-1)}n_i=0,
$$
and
for $i \neq j$,
$$
\E\left[T^{III}_{ij}\right]=\E[N_{ij}]-\frac{(n_i-1)}{(n-1)}\E[C_j]=
\frac{n_i\,n_j}{(n-1)}-\frac{(n_i-1)}{(n-1)}n_j=0.
$$

As for the variance of $T^{III}_{ij}$,
we have
\begin{equation}
\label{eqn:Var[TIIIij]}
\Var\left[T^{III}_{ij}\right]=
\begin{cases}
\Var[N_{ii}]+\frac{(n_i-1)^2}{(n-1)^2}\Var[C_i]-2\frac{(n_i-1)}{(n-1)}\Cov[N_{ii},C_i]
& \text{if $i=j$,} \vspace{.05 in}\\
\Var[N_{ij}]+\frac{n_i^2}{(n-1)^2}\Var[C_j]-2\frac{n_i}{(n-1)}\Cov[N_{ij},C_j] & \text{if $i \not= j$.}
\end{cases}
\end{equation}

As a new cell-specific test, we propose
\begin{equation}
\label{eqn:new-ZIIIij}
Z^{III}_{ij}=\frac{T^{III}_{ij}}{\sqrt{\Var\left[T^{III}_{ij}\right]}}.
\end{equation}

Notice that this is same as the new cell-specific test introduced in (\cite{ceyhan:corrected})
and details of this test are provided for the sake of completeness.

\subsection{Type IV Cell-Specific Segregation Tests}
\label{sec:type-IV-cell-spec}
For $T^{III}_{ij}$,
we introduced a coefficient in front of the second term,
i.e., $n_i\,C_j/n$,
to obtain a zero expected value for our statistic under RL.
In this section,
we modify the first term and obtain the following test statistic:
\begin{equation}
\label{eqn:TIVij}
T^{IV}_{ij}=
\begin{cases}
\frac{n_i(n-1)}{n(n_i-1)}N_{ii}-\frac{n_i}{n}C_i =\frac{n_i}{n}\left(\frac{n-1}{n_i-1}N_{ii}-C_i\right) & \text{if $i=j$,}\\
\frac{n-1}{n}N_{ij}-\frac{n_i}{n}C_j =\frac{1}{n} \left( (n-1)\,N_{ij}-n_i \, C_j \right)    & \text{if $i \not= j$.}
\end{cases}
\end{equation}
Then $\E\left[T^{IV}_{ij}\right]=0$, since,
for $i=j$,
$$
\E\left[T^{IV}_{ii}\right]=
\frac{n_i}{n}\left(\frac{n-1}{n_i-1}\E[N_{ii}]-\E[C_i]\right)=
\frac{n_i}{n}\left(\frac{n-1}{n_i-1}\frac{n_i(n_i-1)}{n-1}-n_i\right)=0,
$$
and
for $i \neq j$,
$$
\E\left[T^{IV}_{ij}\right]=
\frac{1}{n}\left((n-1)\E[N_{ij}]-n_i\E[C_j]\right)=
\frac{1}{n}\left((n-1)\frac{n_i\,n_j}{n-1}-n_i\, n_j\right)=0.
$$

As for the variance of $T^{IV}_{ij}$,
we have
\begin{equation}
\label{eqn:Var[TIVij]}
\Var\left[T^{IV}_{ij}\right]=
\begin{cases}
\frac{n_i^2}{n^2}\left(\frac{(n-1)^2}{(n_i-1)^2}\Var[N_{ii}]+\Var[C_i]-2\frac{(n-1)}{(n_i-1)}\Cov[N_{ii},C_i] \right)
& \text{if $i=j$,} \vspace{.05 in}\\
\frac{1}{n^2}\left((n-1)^2\Var[N_{ij}]+n_i^2\Var[C_j]-2(n-1)n_i\Cov[N_{ij},C_j]\right) & \text{if $i \not= j$.}
\end{cases}
\end{equation}

As a new cell-specific test, we propose
\begin{equation}
\label{eqn:new-ZIVij}
Z^{IV}_{ij}=\frac{T^{IV}_{ij}}{\sqrt{\Var\left[T^{IV}_{ij}\right]}}.
\end{equation}

\section{Overall Segregation Tests}
\label{sec:overall}
In this section, we describe the overall segregation tests in literature
and introduce new overall tests based on
cell-specific tests in Section \ref{sec:cell-spec}.

\subsection{Dixon's Overall Segregation Test}
\label{sec:dix-overall}
In the multi-class case with $m$ classes,
combining the $m^2$ cell-specific tests in Section \ref{sec:dix-cell-spec},
\cite{dixon:NNCTEco2002} suggests the quadratic
form to obtain the overall segregation test as follows:
\begin{equation}
\label{eqn:dix-chisq-mxm}
\X_D=(\mathbf{N}-\E[\mathbf{N}])'\Sigma_D^-(\mathbf{N}-\E[\mathbf{N}])
\end{equation}
where $\mathbf{N}$ is the $m^2\times1$ vector of $m$ rows
of the NNCT concatenated row-wise,
$\E[\mathbf{N}]$ is the vector of $\E[N_{ij}]$ which are as in Equation (\ref{eqn:Exp[Nij]}),
$\Sigma_D$ is the $m^2 \times m^2$ variance-covariance matrix for
the cell count vector $\mathbf{N}$ with diagonal entries equal to
$\Var[N_{ij}]$ and off-diagonal entries being $\Cov[N_{ij},\,N_{kl}]$
for $(i,j) \neq (k,l)$.
The explicit forms of the variance and covariance terms are provided in (\cite{dixon:NNCTEco2002}).
Also, $\Sigma_D^-$ is a generalized inverse of $\Sigma_D$ (\cite{searle:2006})
and $'$ stands for the transpose of a vector or matrix.
Then under RL,
$\X_D$ has a $\chi^2_{m(m-1)}$ distribution asymptotically.
%Furthermore, the test statistics $Z_{ij}^D$ are dependent,
%hence their squares do not sum to $C_N$.
%Under CSR independence, the distribution of $\X_D$
%is conditional on $Q$ and $R$.

\subsection{Type I Overall Segregation Test}
\label{sec:typeIoverall}
We can also combine the type I cell-specific tests of Section \ref{sec:type-I-cell-spec}.
Let $\mathbf {T_I}$ be the vector of $m^2$ $T^I_{ij}$ values, i.e.,
$$\mathbf {T_I}=\left(T^I_{11},T^I_{12},\ldots,T^I_{1m},T^I_{21},T^I_{22},\ldots,T^I_{2m},\ldots,T^I_{mm}\right)',$$
and let $\E\left[\mathbf {T_I}\right]$ be the vector of $\E\left[T^I_{ij}\right]$ values.
Note that $\E\left[\mathbf {T_I}\right]=
\Big(\E\left[T^I_{11}\right],\E\left[T^I_{12}\right],\ldots,\E\left[T^I_{1m}\right],\E\left[T^I_{21}\right],\E\left[T^I_{22}\right],\ldots,$
$\E\left[T^I_{2m}\right],\ldots,\E\left[T^I_{mm}\right]\Big)'$.
Hence to obtain a new overall segregation test, referred to as type I overall test, we use the following quadratic form:
\begin{equation}
\label{eqn:new-overall]}
\X_I=\left(\mathbf {T_I}-\E\left[\mathbf {T_I}\right]\right)'\Sigma_I^-\left(\mathbf {T_I}-\E\left[\mathbf {T_I}\right]\right)
\end{equation}
where $\Sigma_I$ is the $m^2 \times m^2$ variance-covariance matrix of $\mathbf {T_I}$.

Under RL, the diagonal entries in the variance-covariance matrix $\Sigma_I$ are
$\Var\left[T^I_{ij}\right]$ which are provided in Equation \eqref{eqn:Var[TIij]}.
For the off-diagonal entries in $\Sigma_I$, i.e., $\Cov\left[T^I_{ij},T^I_{kl}\right]$ with $(i,j \not= (k,l)$,
we have
\begin{multline*}
\Cov\left[T^I_{ij},T^I_{kl}\right]=\Cov \left[ N_{ij}-\frac{n_i}{n}C_j,N_{kl}-\frac{n_k}{n}C_l \right]=\\
\Cov[N_{ij},N_{kl}]-\frac{n_k}{n}\Cov[N_{ij},C_l]-\frac{n_i}{n}\Cov[N_{kl},C_j]
+\frac{n_i n_k}{n^2}\Cov[C_j,C_l].
\end{multline*}

\subsection{Type II Overall Segregation Test}
\label{sec:typeIIoverall}
We also combine the type II cell-specific tests of Section \ref{sec:type-II-cell-spec}.
Let $\mathbf {T_{II}}$ be the vector of $m^2$ $T^{II}_{ij}$ values, i.e.,
$$\mathbf {T_{II}}=\left(T^{II}_{11},T^{II}_{12},\ldots,T^{II}_{1m},T^{II}_{21},T^{II}_{22},\ldots,T^{II}_{2m},\ldots,T^{II}_{mm}\right)',$$
and let $\E\left[\mathbf {T_{II}}\right]$ be the vector of $\E\left[T^{II}_{ij}\right]$ values.
%Note that $\E\left[\mathbf {T_{II}}\right]=
%[\E[T^{II}_{11}],\E[T^{II}_{12}],\ldots,\E[T^{II}_{1m}],\E[T^{II}_{21}],\E[T^{II}_{22}],\ldots,\E[T^{II}_{2m}],\ldots,\E[T^{II}_{mm}]]'$.
As the type II overall segregation test, we use the following quadratic form:
\begin{equation}
\label{eqn:typeIIoverall]}
\X_{II}=\left(\mathbf {T_{II}}-\E\left[\mathbf {T_{II}}\right]\right)'\Sigma_{II}^-\left(\mathbf {T_{II}}-\E\left[\mathbf {T_{II}}\right]\right)
\end{equation}
where $\Sigma_{II}$ is the $m^2 \times m^2$ variance-covariance matrix of $\mathbf {T_{II}}$.

Under RL, the diagonal entries in the variance-covariance matrix $\Sigma_N$ are
$\Var\left[T^{II}_{ij}\right]$ which are same as $\Var[N_{ij}]$.
For the off-diagonal entries in $\Sigma_{II}$,
i.e., $\Cov\left[T^{II}_{ij},T^{II}_{kl}\right]$ with $(i,j) \not= (k,l)$,
we have $\Cov\left[T^{II}_{ij},T^{II}_{kl}\right]=
\Cov[N_{ij}-\frac{n_i n_j}{n},N_{kl}-\frac{n_k n_l}{n}]= \Cov[N_{ij},N_{kl}]$.

\subsection{Type III Overall Segregation Test}
\label{sec:typeIIIoverall}
When we combine the type III cell-specific tests of Section \ref{sec:type-III-cell-spec},
we obtain type III overall test as follows.
Let $\mathbf {T_{III}}$ be the vector of $m^2$ $T^{III}_{ij}$ values, i.e.,
$$\mathbf {T_{III}}=\left(T^{III}_{11},T^{III}_{12},\ldots,T^{III}_{1m},T^{III}_{21},T^{III}_{22},\ldots,T^{III}_{2m},\ldots,T^{III}_{mm}\right)',$$
and let $\E\left[\mathbf {T_{III}}\right]$ be the vector of $\E\left[T^{III}_{ij}\right]$ values.
Note that $\E\left[\mathbf {T_{III}}\right]=\mathbf 0$.
As the type III overall segregation test, we use the following quadratic form:
\begin{equation}
\label{eqn:typeIIIoverall]}
\X_{III}=\left(\mathbf {T_{III}}\right)'\Sigma_{III}^-\left(\mathbf {T_{III}}\right)
\end{equation}
where $\Sigma_{III}$ is the $m^2 \times m^2$ variance-covariance matrix of $\mathbf {T_{III}}$.

Under RL, the diagonal entries in the variance-covariance matrix $\Sigma_{III}$ are
$\Var\left[T^{III}_{ij}\right]$ which are provided in Equation \eqref{eqn:Var[TIIIij]}.
For the off-diagonal entries in $\Sigma_{III}$, i.e., $\Cov\left[T^{III}_{ij},T^{III}_{kl}\right]$ with $(i,j) \not= (k,l)$,
there are four cases to consider:\\
\noindent
\textbf{case 1:} $i=j$ and $k=l$,
then
\begin{multline}
\Cov\left[T^{III}_{ii},T^{III}_{kk}\right]=\Cov \left[ N_{ii}-\frac{(n_i-1)}{(n-1)}C_i,N_{kk}-\frac{(n_k-1)}{(n-1)}C_k \right]=\\
\Cov[N_{ii},N_{kk}]-\frac{(n_k-1)}{(n-1)}\Cov[N_{ii},C_k]-\frac{(n_i-1)}{(n-1)}\Cov[N_{kk},C_i]
+\frac{(n_i-1)(n_k-1)}{(n-1)^2}\Cov[C_i,C_k].
\end{multline}

\noindent
\textbf{case 2:} $i=j$ and $k \neq l$,
then
\begin{multline}
\Cov\left[T^{III}_{ii},T^{III}_{kl}\right]=\Cov \left[N_{ii}-\frac{(n_i-1)}{(n-1)}C_i,N_{kl}-\frac{n_k}{(n-1)}C_l \right]=\\
\Cov[N_{ii},N_{kl}]-\frac{n_k}{(n-1)}\Cov[N_{ii},C_l]-\frac{(n_i-1)}{(n-1)}\Cov[N_{kl},C_i]
+\frac{(n_i-1)n_k}{(n-1)^2}\Cov[C_i,C_l].
\end{multline}

\noindent
\textbf{case 3:} $i \neq j$ and $k = l$,
then
$
\Cov\left[T^{III}_{ij},T^{III}_{kk}\right]=\Cov[T^{III}_{kk},T^{III}_{ij}],
$
which is essentially \textbf{case 2} above.

\noindent
\textbf{case 4:} $i \neq j$ and $k \neq l$,
then
\begin{multline}
\Cov\left[T^{III}_{ij},T^{III}_{kl}\right]=\Cov \left[ N_{ij}-\frac{n_i}{(n-1)}C_j,N_{kl}-\frac{n_k}{(n-1)}C_l \right]=\\
\Cov[N_{ij},N_{kl}]-\frac{n_k}{(n-1)}\Cov[N_{ij},C_l]-\frac{n_i}{(n-1)}\Cov[N_{kl},C_j]
+\frac{n_i n_k}{(n-1)^2}\Cov[C_j,C_l].
\end{multline}

Note that type III overall segregation is same as the new overall test provided in (\cite{ceyhan:corrected}).

\subsection{Type IV Overall Segregation Test}
\label{sec:typeIVoverall}
When we combine the type IV cell-specific tests of Section \ref{sec:type-IV-cell-spec},
we obtain type IV overall test as follows.
Let $\mathbf {T_{IV}}$ be the vector of $m^2$ $T^{IV}_{ij}$ values, i.e.,
$$\mathbf {T_{IV}}=\left[T^{IV}_{11},T^{IV}_{12},\ldots,T^{IV}_{1m},T^{IV}_{21},T^{IV}_{22},\ldots,T^{IV}_{2m},\ldots,T^{IV}_{mm}\right]',$$
and let $\E\left[\mathbf {T_{IV}}\right]$ be the vector of $\E\left[T^{IV}_{ij}\right]$ values.
Note that $\E\left[\mathbf {T_{IV}}\right]=\mathbf 0$.
As the type IV overall segregation test, we use the following quadratic form:
\begin{equation}
\label{eqn:typeIVoverall]}
\X_{IV}=\left(\mathbf {T_{IV}}\right)'\Sigma_{IV}^-\left(\mathbf {T_{IV}}\right)
\end{equation}
where $\Sigma_{IV}$ is the $m^2 \times m^2$ variance-covariance matrix of $\mathbf {T_{IV}}$.

Under RL, the diagonal entries in the variance-covariance matrix $\Sigma_{IV}$ are
$\Var\left[T^{IV}_{ij}\right]$ which are provided in Equation \eqref{eqn:Var[TIVij]}.
For the off-diagonal entries in $\Sigma_{IV}$, i.e., $\Cov\left[T^{IV}_{ij},T^{IV}_{kl}\right]$ with $(i,j) \not= (k,l)$,
there are four cases to consider:\\
\noindent
\textbf{case 1:} $i=j$ and $k=l$,
then
\begin{multline}
\Cov\left[T^{IV}_{ii},T^{IV}_{kk}\right]=
\Cov \left[ \frac{n_i}{n}\left( \frac{n-1}{n_i-1} N_{ii}-C_i\right),
\frac{n_k}{n}\left( \frac{n-1}{n_k-1} N_{kk}-\frac{(n_k-1)}{(n-1)}C_k\right) \right]=\\
\frac{n_i n_k}{n^2}\,\Cov\left[ \frac{n-1}{n_i-1} N_{ii}-C_i, \frac{n-1}{n_k-1} N_{kk}-\frac{(n_k-1)}{(n-1)}C_k \right]=\\
\frac{n_i n_k}{n^2}\,\left(\frac{(n-1)^2}{(n_i-1)}\Cov[N_{ii},N_{kk}]-
\frac{(n-1)}{(n_i-1)}\Cov[N_{ii},C_k]-\frac{(n-1)}{(n_k-1)}\Cov[N_{kk},C_i]+\Cov[C_i,C_k] \right).
\end{multline}

\noindent
\textbf{case 2:} $i=j$ and $k \neq l$,
then
\begin{multline}
\Cov\left[T^{IV}_{ii},T^{IV}_{kl}\right]=
\Cov \left[ \frac{n_i}{n}\left( \frac{n-1}{n_i-1} N_{ii}-C_i\right),
\frac{1}{n}\left( (n-1) N_{kl}-n_k C_l\right) \right]=\\
\frac{n_i}{n^2}\,\Cov\left[ \frac{n-1}{n_i-1} N_{ii}-C_i, (n-1) N_{kl}-n_k C_l \right]=\\
\frac{n_i}{n^2}\,\left(\frac{(n-1)^2}{(n_i-1)(n_k-1)}\Cov[N_{ii},N_{kl}]-
\frac{(n-1)n_k}{(n_i-1)}\Cov[N_{ii},C_l]-(n-1)\Cov[N_{kl},C_i]+n_k\Cov[C_i,C_l] \right).
\end{multline}

\noindent
\textbf{case 3:} $i \neq j$ and $k = l$,
then
$
\Cov\left[T^{IV}_{ij},T^{IV}_{kk}\right]=\Cov[T^{IV}_{kk},T^{IV}_{ij}],
$
which is essentially \textbf{case 2} above.

\noindent
\textbf{case 4:} $i \neq j$ and $k \neq l$,
then
\begin{multline}
\Cov\left[T^{IV}_{ij},T^{IV}_{kl}\right]=
\Cov \left[ \frac{1}{n}\left( (n-1) N_{ij}-n_i C_j\right),
\frac{1}{n}\left( (n-1) N_{kl}-n_k C_l\right) \right]=\\
\frac{1}{n^2}\,\Cov\left[ (n-1) N_{ij}-n_i C_j, (n-1) N_{kl}-n_k C_l \right]=\\
\frac{1}{n^2}\,\left((n-1)^2 \Cov[N_{ij},N_{kl}]-
(n-1)n_k \Cov[N_{ij},C_l]-(n-1)n_i \Cov[N_{kl},C_j]+n_i n_k\Cov[C_j,C_l] \right).
\end{multline}

\subsection{Remarks on NNCT-Tests}
\label{sec:further-remarks}
Under RL,
$Z^D_{ii}$ is shown to have $N(0,1)$ distribution asymptotically,
while for $m>2$ the asymptotic normality of the off-diagonal
cells in NNCTs is not rigorously established,
although extensive Monte Carlo simulations indicate approximate
normality for large samples (\cite{dixon:NNCTEco2002}).
Furthermore,
Dixon's cell specific test and type II cell-specific test are equivalent,
and so are types III and IV cell-specific tests.
The same holds for the corresponding overall tests,
since they are constructed based on the cell-specific tests.
However, the cell-specific test statistics are dependent,
hence their squares do not sum to the corresponding overall segregation tests.

Asymptotically, under RL,
$\X_D$ and $\X_{II}$ has a $\chi^2_{m(m-1)}$ distribution
since rank of $\Sigma_D$ and $\Sigma_{II}$ is $m(m-1)$,
in fact,
$\Sigma_D=\Sigma_{II}$.
$\X_{III}$ has a $\chi^2_{(m-1)^2}$ distribution asymptotically
since rank of $\Sigma_{III}$ is $(m-1)^2$.
Similarly,
$\X_I$ and $\X_{IV}$ also have $\chi^2_{(m-1)^2}$ distribution asymptotically.
The asymptotic distributions of the overall tests provide a natural classification of these
NNCT-tests.
More specifically,
Dixon's and type II overall tests only use the cell counts and row sums (i.e., class sizes) in the
corresponding cell-specific tests and hence asymptotically have $\chi^2$ distribution with $m(m-1)$ df,
while type I, III and IV overall tests use the column sums in addition to cell counts and row sums
and hence have asymptotic $\chi^2$ distribution with $(m-1)^2$ df.
That is,
if only the row sums are incorporated,
then one df is lost in each row as the sums of the row cells yield the fixed class size.
On the other hand,
if both row and column sums are incorporated,
one row and column can be obtained given the row and column sums,
hence leaving only $(m-1)^2$ df for the overall tests.
Dixon's and type II tests being identical can be easily established.
Type III and IV cell-specific (after standardization) and overall tests are also identical,
although the $T_{ij}$ values and the variance-covariance matrices are different.
Additionally,
type I tests and type III tests are similar (although not identical),
and hence give similar results.

In all the above cases, $\Cov[N_{ij},N_{kl}]$ are as in \cite{dixon:NNCTEco2002},
$\Cov[N_{ij},C_l]=\sum_{k=1}^m \Cov[N_{ij},N_{kl}]$ and
$\Cov[C_i,C_j]=\sum_{k=1}^m \sum_{l=1}^m \Cov[N_{ki},N_{lj}]$.

Under CSR independence,
the cell-specific and overall tests are as in RL case.
However, under RL,
$Q$ and $R$ are fixed quantities,
as they depend only on the location of the points,
but not the types of NNs,
while under CSR independence,
they are random.
Under CSR independence, the distributions of the test statistics above
are similar to the RL case.
The only difference is that the new cell-specific tests asymptotically have
$N(0,1)$ distribution conditional on $Q$ and $R$.
Hence,
under CSR independence,
$\Var[N_{ij}]$,
$\Cov[N_{ij},N_{kl}]$,
$\Cov[N_{ij},C_k]$,
$\Cov[C_i,C_j]$,
and all other quantities depending on $Q$ and $R$
are conditional on $Q$ and $R$.
The unconditional variances can be obtained by replacing $Q$ and $R$
with their expectations (see \cite{ceyhan:corrected} for more detail).
Since $Q$ and $R$ are random under CSR independence,
the variances of the cell-specific test statistics tend to be larger compared to the ones under RL.

Each of the cell-specific tests measures the deviation
of the test statistic from its expected value under $H_o$.
Dixon's and type II cell-specific tests depend on $N_{ij}$ (i.e., cell counts) and row sums only,
and types I, III, and IV cell-specific tests incorporate column sums as well.
For the cell-specific tests,
the $z$-score for cell $(i,j)$ indicates the level and direction of spatial interaction
between classes $i$ and $j$.
If the $z$-score for cell $(i,i)$ is significantly larger (less) than zero,
then class $i$ exhibits (lack of) segregation from other classes.
If the $z$-score for cell $(i,j)$ with $i \not= j$ is significantly larger (less) than zero,
then class $j$ exhibits (lack of) association with class $i$.
Moreover,
for cells $(i,j)$ with $i \not= j$,
the cell-specific tests are not symmetric.
For example,
the cell-specific test for cell $(i,j)$ may exhibit a different level of interaction
compared to the cell $(j,i)$.
The overall tests combine cell-specific tests in one compound summary statistic.
The performance of cell-specific tests are expected to
carry over to the overall tests,
provided the correct degrees of freedom is employed.

Recall that in the two-class case,
each cell count $N_{ij}$ has asymptotic normal distribution (\cite{cuzick:1990}).
Hence, the new cell-specific tests $Z^I_{ij}$, $Z^{II}_{ij}$, $Z^{III}_{ij}$
and $Z^{IV}_{ij}$ also converges in law to $N(0,1)$ as $n \rightarrow \infty$
(with $n_i \rightarrow \infty$ for all $i$).
Moreover, one and two-sided versions of these tests are also possible.
For the diagonal cells,
the right-sided (left-sided) version of these tests are for (lack of) segregation and
for the off-diagonal cells,
the right-sided (left-sided) version of these tests are for (lack of) association.
In the two-class case,
at most two cells contain all the information provided by the NNCT.
In particular,
for $i \not= j$,
segregation of class $i$ from class $j$ implies lack of association
between classes $i$ and $j$
and lack of segregation of class $i$ from class $j$ implies association between
classes $i$ and $j$.
For Dixon's cell-specific test,
we have $Z^D_{i1}=-Z^D_{i2}$ for $i=1,2$.
For type I cell-specific test,
$Z^I_{11}=Z^I_{22}=-Z^I_{12}=-Z^I_{21}$;
and for type II cell-specific test,
we have $Z^D_{ij}=Z^{II}_{ij}$;
for type III cell-specific test,
we have $Z^{III}_{1j}=-Z^{III}_{2j}$ for $j=1,2$;
and
for type IV cell-specific test,
we have
$Z^{III}_{ij}=Z^{IV}_{ij}$ for $i,j=1,2$.

In the multi-class case with $m>2$,
a positive $z$-score for the diagonal cell $(i,i)$ indicates segregation,
but it does not necessarily mean lack of association between class $i$
and class $j$ ($i\not=j$), since it could be the case that class $i$
could be associated with one class, yet not associated with another one.
%Likewise for the new cell-specific tests.
See also Section \ref{sec:part-seg}.

The cell-specific and overall tests are all consistent under both segregation and association alternative,
which can be shown with the same mechanism as in \cite{ceyhan:overall}.

\subsection{Post-hoc Tests after the Overall Tests: Class-Specific, Pairwise, and One-vs-rest Type Tests}
\label{post-hoc-tests}
In our construction of the NNCT-tests,
although we first introduce the cell-specific tests and then develop overall tests based on
the cell-specific tests,
in practice,
it is more natural to conduct the tests in reverse order.
That is,
first an overall NNCT-test could be performed,
and if significant,
then one can perform cell-specific tests to determine
the types and levels of the spatial interaction patterns between the classes.
This procedure is somewhat analogous to ANOVA $F$-test to compare multiple groups,
in the sense that
if the $F$-test yields a significant result,
then one performs pairwise tests to determine which pairs are different.
However,
NNCT-tests provide more alternatives (compared to the ANOVA $F$-test)
as post-hoc tests after an overall test is significant.
In the multi-class case,
when an overall test is rejected;
i.e., there is evidence in favor of some sort of deviation from
randomness of the spatial pattern,
the next natural question is
what type of deviation occurs for each class (or species).
To this end, one can conduct several post-hoc tests.
One type of post-hoc tests is the class-specific tests discussed in \cite{dixon:NNCTEco2002}
and \cite{ceyhan:class2009}.
For pairwise comparison of the interaction between classes,
one can resort to two options:
(i) in an $m \times m$ NNCT,
one can consider cell-specific tests for each cell (which also provides interaction of the class with itself on the diagonal cells)
and
(ii) one can restrict attention to the pair of classes $i,j$ with $i \not= j$
one at a time and conduct the tests as in the two-class case with a $2 \times 2$ NNCT.
We recommend the approach in (i),
since it incorporates all the classes in question and provides the types of interaction in the presence of all classes,
while the approach in (ii) ignores the possible effects of classes different from the pair in question.
This in practice might not give the exact picture of the mixed relationships
between all the classes.

As another alternative post-hoc procedure,
\cite{dixon:NNCTEco2002} suggested the following.
For class $i$, we pool the remaining classes and treat them as
the other class in a two-class setting.
Then we apply the two-class tests to the resulting NNCT.
To emphasize the difference,
this version of the class-specific test
is called {\em one-vs-rest type test}.
For $m>2$ classes,
let $\mathcal N$ be the $m \times m$ NNCT with cell counts being $N_{ij}$
and let $\widetilde{\mathcal N}$ be the $2 \times 2$ NNCT for the
one-versus-rest type procedure with cell counts being $\widetilde N_{ij}$.
When we are performing a one-versus-rest type testing for class $i$,
without loss of generality,
we can assign the first row in $\widetilde{\mathcal N}$ to class $i$
and the second row to the rest (of the classes).
Then $\widetilde N_{11}=N_{ii}$,
$\widetilde N_{12}=\sum_{j \not= i} N_{ij}$,
$\widetilde N_{21}=\sum_{j \not= i} N_{ji}$,
and
$\widetilde N_{22}=\sum_{j \not= i, k \not= i} N_{jk}$.
Hence in the one-versus-rest type testing,
the cell-specific test for cell $(1,1)$ in $\widetilde{\mathcal N}$
would be same as the cell-specific test for cell $(i,i)$ in $\mathcal N$.
Therefore,
to extract information from $\widetilde{\mathcal N}$ that is not provided by $\mathcal N$,
we consider the cell-specific tests for cell $(2,2)$ in $\widetilde{\mathcal N}$.
The overall test statistics for $\widetilde{\mathcal N}$
are also different than the ones for $\mathcal N$.

In a multi-class case with $m \ge 2$ classes,
there are $m$ class-specific and one-vs-rest
types of tests and ${m \choose 2}=m(m-1)/2$ pairwise tests
and $m^2$ cell-specific tests.
As $m$ increases the class-specific tests are less
intensive computationally and easier to interpret,
whereas the pairwise tests might yield conflicting results.

\section{Empirical Size Analysis in the Two-Class Case}
\label{sec:monte-carlo-2Cl}
We provide the empirical significance levels
for Dixon's and the new cell-specific and overall segregation tests
in the two-class case under CSR independence and RL patterns.
Our Monte Carlo simulation set-up is same as in \cite{ceyhan:corrected}.

\subsection{Empirical Size Analysis under CSR Independence of Two Classes}
\label{sec:CSR-emp-sign-2Cl}
For the CSR independence pattern,
in the two-class case, we label the classes as $X$ and $Y$,
or
class 1 and class 2, interchangeably.
We generate $n_1$ points from class $X$ and $n_2$ points from class $Y$
both of which are independent of each other and
independently uniformly distributed on the unit square, $(0,1) \times (0,1)$.
We use the class size combinations $(n_1,n_2) \in \{(10,10),(10,30), (10,50), (30,30), (30,50), (50,50),(50,100), (100,100)\}$
and perform $N_{mc}=10000$ replications.
The empirical sizes are calculated as
the ratio of number of significant results to the number of Monte
Carlo replications, $N_{mc}$.
We use .05 as our nominal significance level.

\begin{figure} [hbp]
\centering
%\psfrag{Density}{ \Huge{\bf{Density}}}
Empirical Size Plots for the NNCT-Tests under CSR Independence of Two Classes\\
\rotatebox{-90}{ \resizebox{2.1 in}{!}{\includegraphics{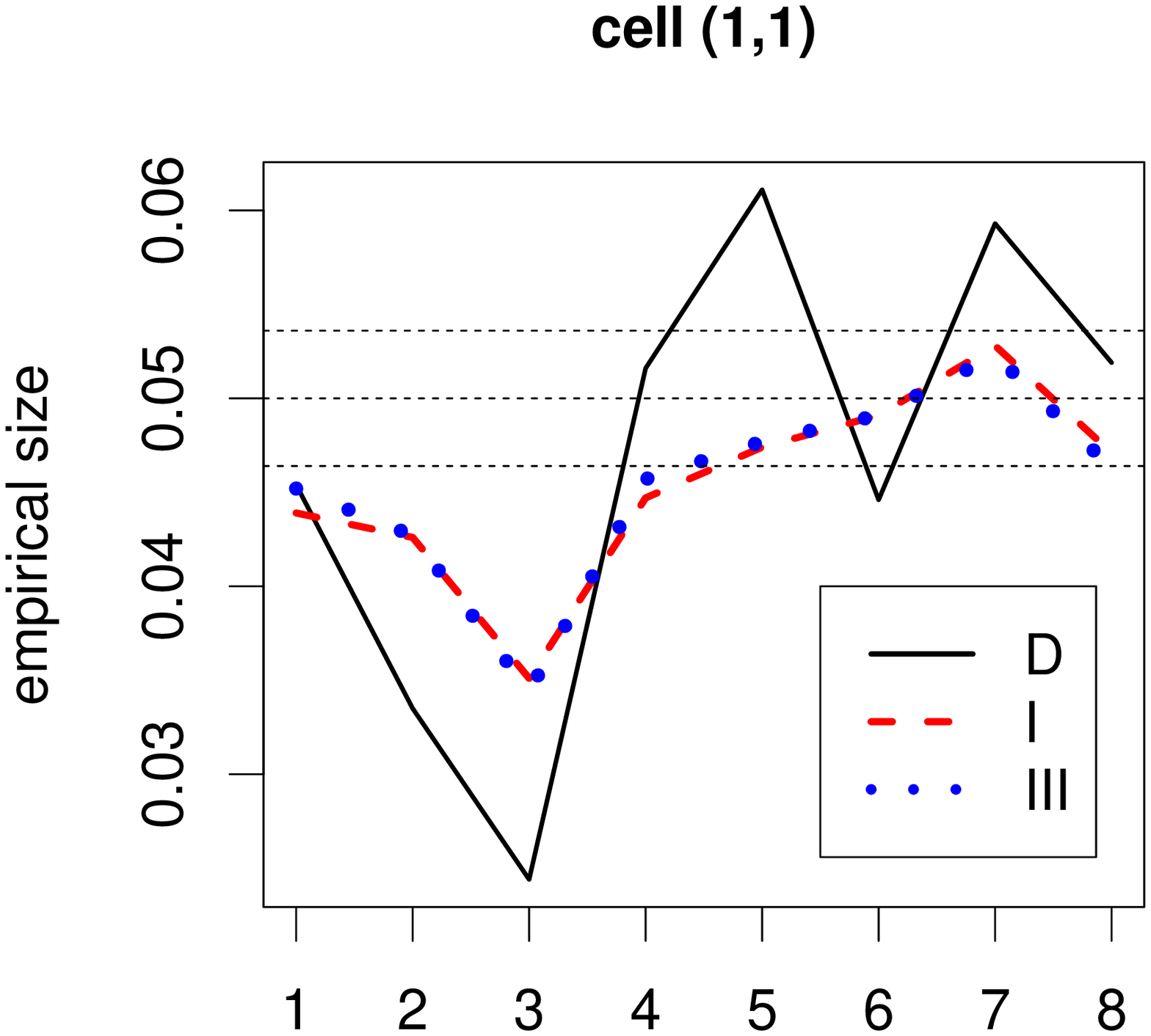} }}
\rotatebox{-90}{ \resizebox{2.1 in}{!}{\includegraphics{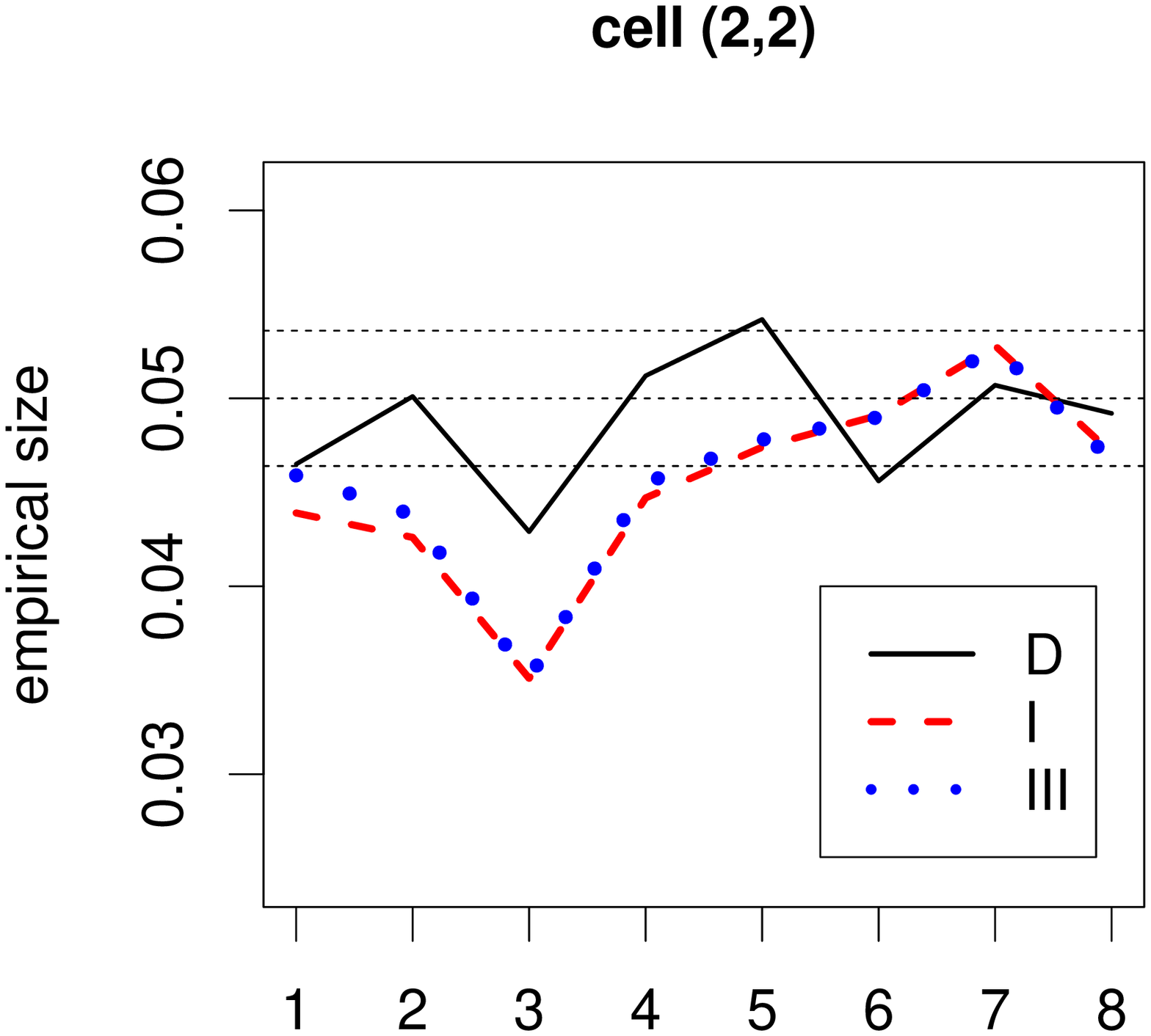} }}
\rotatebox{-90}{ \resizebox{2.1 in}{!}{\includegraphics{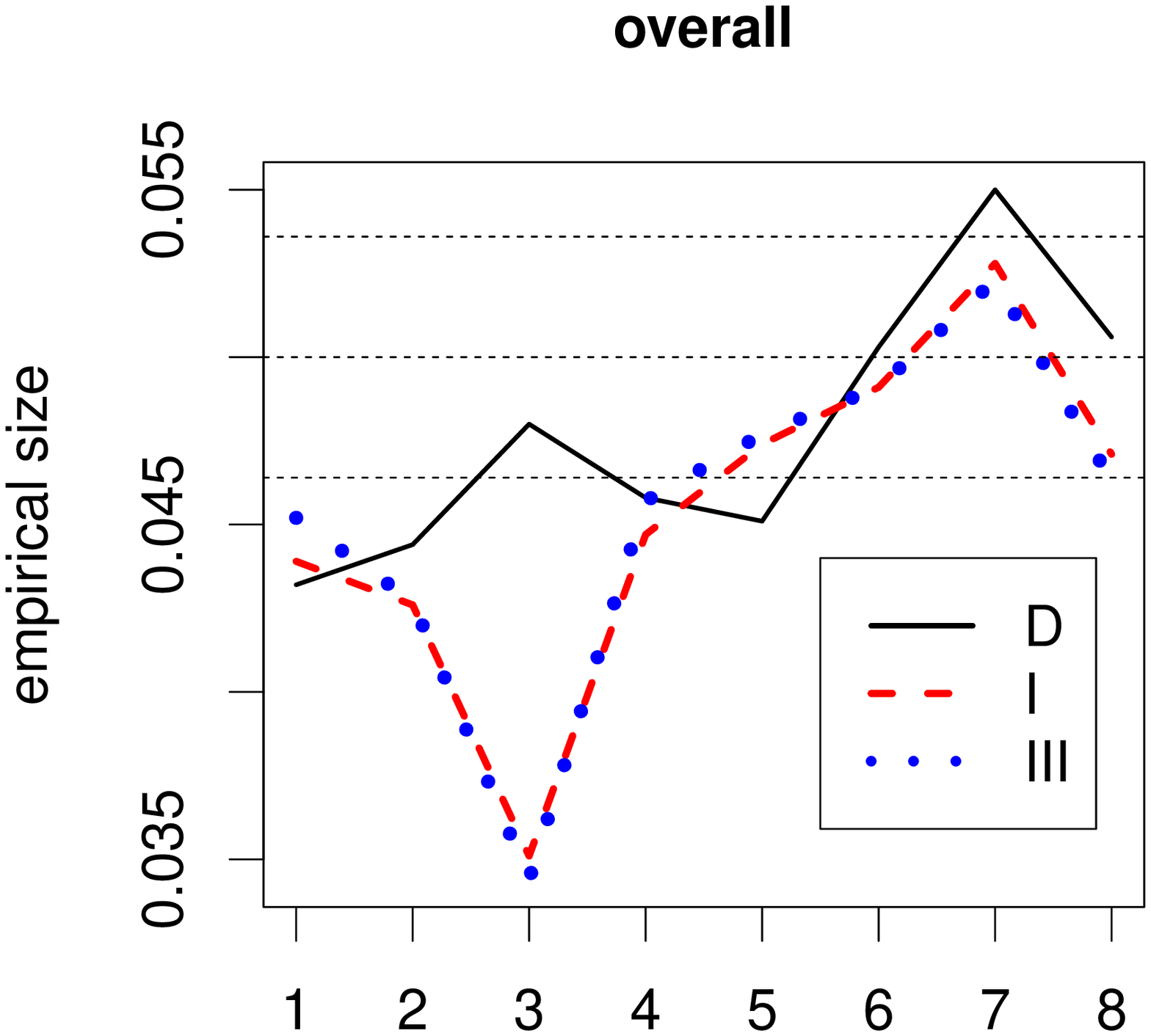} }}
\caption{
\label{fig:emp-size-CSR-2cl}
The empirical size estimates of the cell-specific tests for cells (1,1) (left),
cell (2,2) (middle), and overall segregation tests (right) under the CSR independence pattern
in the two-class case.
The horizontal lines are located at .0464 (upper threshold for conservativeness),
.0500 (nominal level), and .0536 (lower threshold for liberalness).
Notice that $y$-axis for overall size plot is differently scaled.
The horizontal axis labels:
1=(10,10), 2=(10,30), 3=(10,50), 4=(30,30), 5=(30,50),
6=(50,50), 7=(50,100), 8=(100,100).
The legend labeling:
D= Dixon's,
I= type I,
and
III= type III cell-specific or overall tests.
}
\end{figure}

We present the empirical significance
levels for the NNCT-tests in Figure \ref{fig:emp-size-CSR-2cl}.
The empirical sizes significantly smaller (larger) than .05 are deemed as conservative (liberal).
The asymptotic normal approximation to proportions are used in determining the significance of
the deviations of the empirical sizes from the nominal level of .05.
For these proportion  tests,
we also use $\alpha=.05$ to test against empirical size being equal to .05.
With $N_{mc}=10000$, empirical sizes less than .0464 are deemed conservative,
greater than .0536 are deemed liberal at $\alpha=.05$ level.
These thresholds are indicated as the dashed horizontal lines in Figure \ref{fig:emp-size-CSR-2cl}.
Note also that the class sizes are arranged in the increasing order for the first and then the second entries.
The size values for discrete class size combinations are joined
by piecewise straight lines for better visualization.
Let $\ah_{i,j}^D$, $\ah_{i,j}^I$-$\ah_{i,j}^{IV}$ be the empirical significance
levels of Dixon's and the cell-specific tests of types I-IV, respectively,
$\ah_D$ be for Dixon's and $\ah_I$-$\ah_{IV}$ be for the overall segregation tests.
Notice that in the two-class case
$\ah_{1,1}^D=\ah_{1,2}^D$ and $\ah_{2,1}^D=\ah_{2,2}^D$ for the two-sided alternative,
since $N_{12}=n_1-N_{11}$ and $N_{21}=n_2-N_{22}$.
The same holds for $\ah_{i,j}^{II}$.
Furthermore,
$\ah_{1,1}^{III}=\ah_{2,1}^{III}$ and $\ah_{1,2}^{III}=\ah_{2,2}^{III}$ for the two-sided alternative,
and the same holds for $\ah_{i,j}^{IV}$.
On the other hand $\ah_{i,j}^I$ are equal for all $i,j$ for the two-sided alternative.
So we only present cell-specific tests for cells $(1,1)$ and $(2,2)$.
Furthermore,
since Dixon's cell specific test and type II cell-specific test are equivalent,
and so are types III and IV cell-specific tests,
we only present Dixon's, type I and III cell-specific tests.
Since the same holds for the overall test,
we present only Dixon's, type I and III overall tests as well.

For cell $(1,1)$, Dixon's cell-specific test has empirical
size close to the nominal level of 0.05 for balanced class sizes
(i.e., for $n_1 \approx n_2$ or when relative abundance of classes are similar),
while for unbalanced class sizes, it tends to be liberal or conservative.
On the other hand,
types I and III cell-specific tests are less severely affected
by the differences in relative abundances of the classes,
i.e., they are closer to the nominal level for all class size combinations.
For cell $(2,2)$, Dixon's cell-specific test is
much closer to 0.05 for all class size combinations,
while type I and III cell-specific tests have similar performance as in cell $(1,1)$.
Thus, Dixon's cell-specific test has much better empirical size performance
for the diagonal cell corresponding to the class with larger size,
while types I and III cell-specific tests have better size performance
for the diagonal cell corresponding to the class with smaller size.

For the overall tests,
Dixon's test has better size performance for smaller classes.
Type I and III overall tests are conservative for smaller classes,
while they have the desired level for larger classes.
%Dixon's, types I and III are closer to the nominal level,
%while they are slightly conservative for smaller classes.
%In general Dixon's, type I, or type III overall tests are recommended in the two class case.

\subsection{Empirical Size Analysis under RL of Two Classes}
\label{sec:RL-emp-sign-2Cl}
For the RL pattern, we consider three cases,
in each of which, we first determine the locations of points
and then assign labels to them randomly.
See \cite{ceyhan:corrected} for more detail.
We generate $n_1$ points iid $\U(S_1)$ and
$n_2$ points iid $\U(S_2)$
for the same combinations of $n_1,n_2$ as in CSR independence case.
The locations of these points are taken to be the fixed locations
for which we assign the labels randomly.
For each class size combination $(n_1,n_2)$,
we randomly choose $n_1$ points (without replacement) and label them
as $X$ points and the remaining $n_2$ points as $Y$ points.
We repeat the RL procedure $N_{mc}=10000$ times for each class size combination.
Empirical sizes are estimated as in the CSR independence case.

In RL case (1), we have $S_1=S_2=(0,1) \times (0,1)$ (i.e., the unit square),
in RL case (2), $S_1=(0,2/3) \times (0,2/3)$ and $S_2=(1/3,1) \times (1/3,1)$,
and
in RL case (3), $S_1 =(0,1) \times (0,1)$ and $S_2=(2,3) \times (0,1)$.

\begin{figure} [hbp]
\centering
%\psfrag{Density}{ \Huge{\bf{Density}}}
\rotatebox{-90}{ \resizebox{2.1 in}{!}{\includegraphics{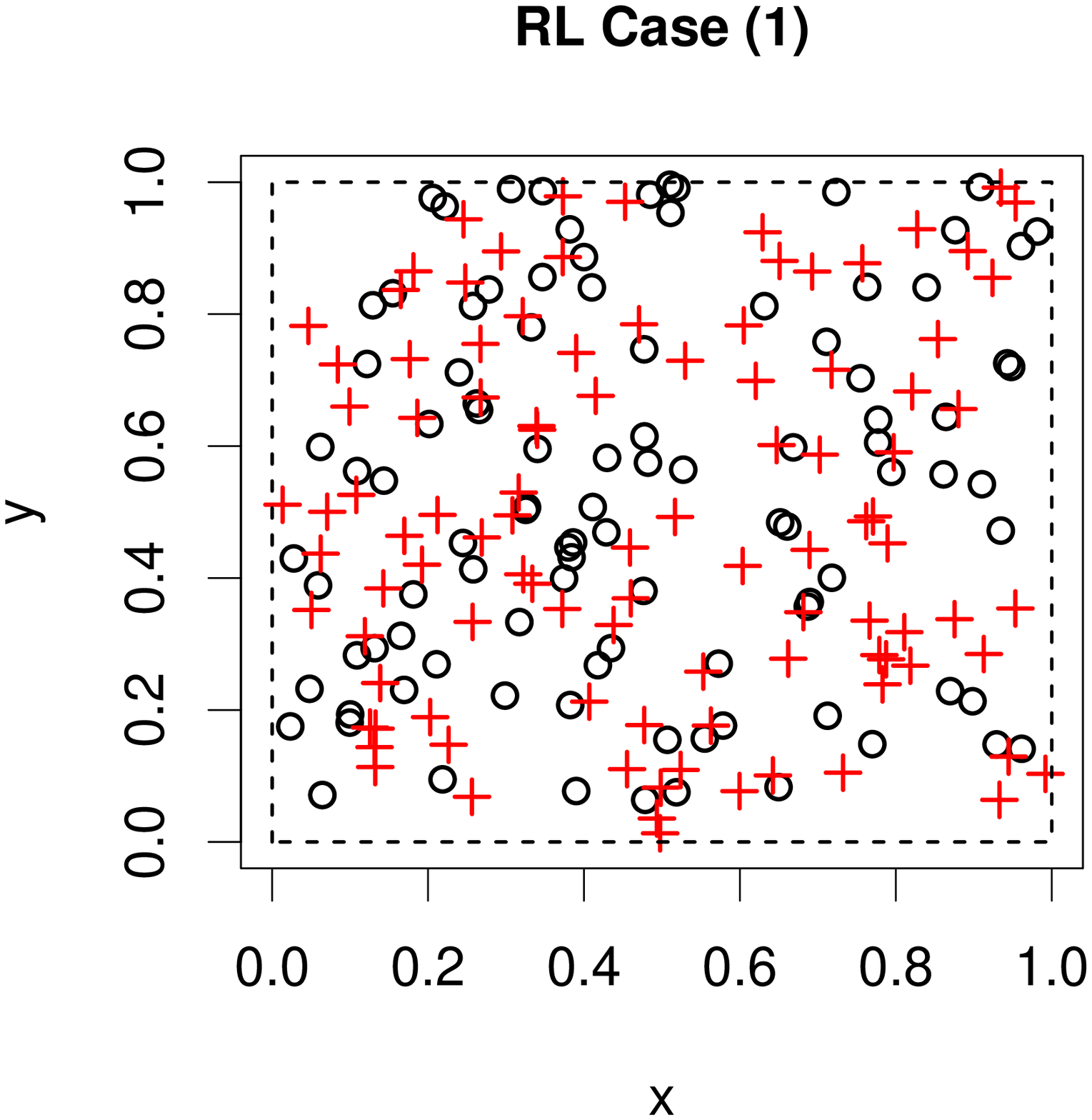} }}
\rotatebox{-90}{ \resizebox{2.1 in}{!}{\includegraphics{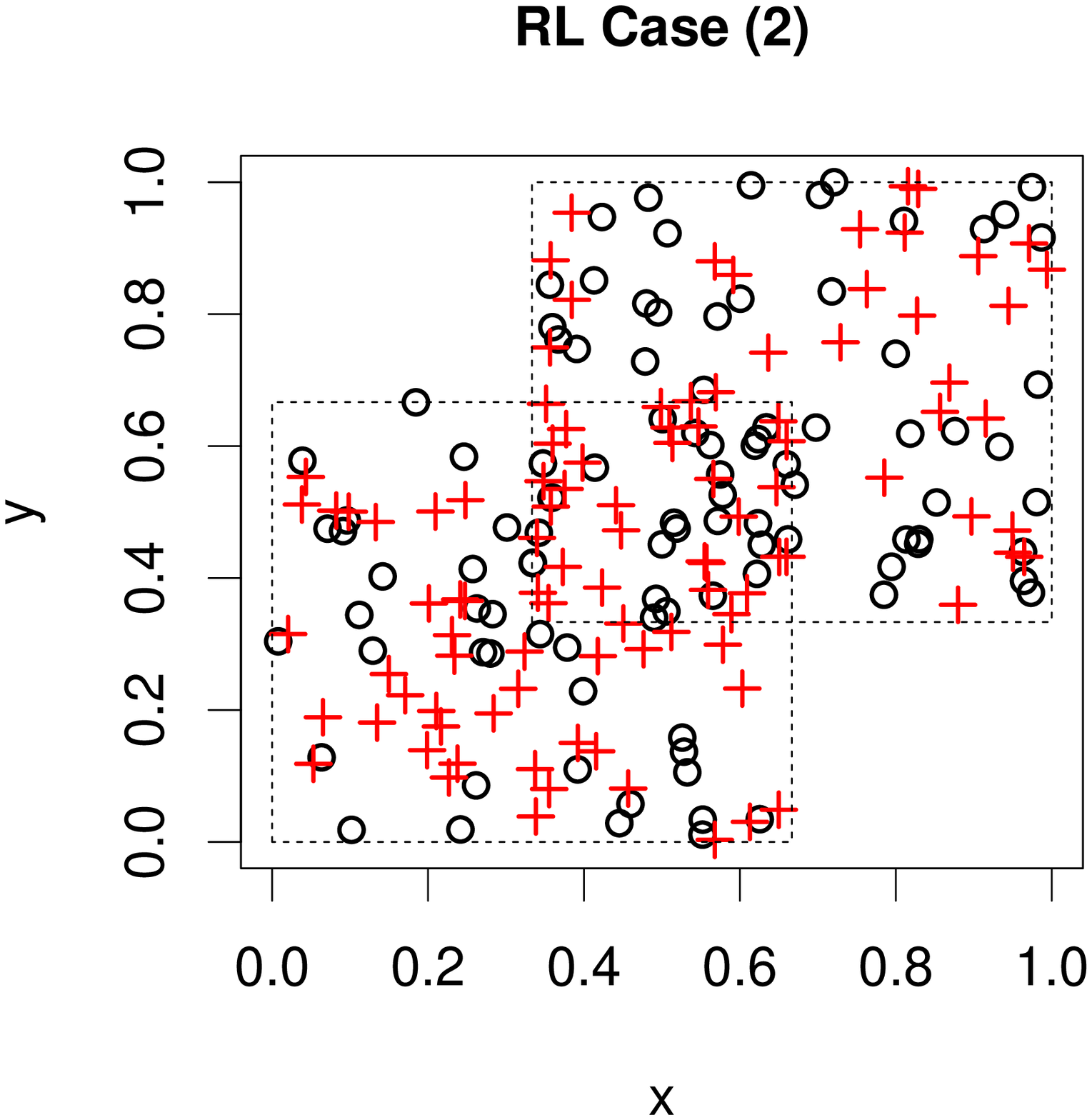} }}
\rotatebox{-90}{ \resizebox{2.1 in}{!}{\includegraphics{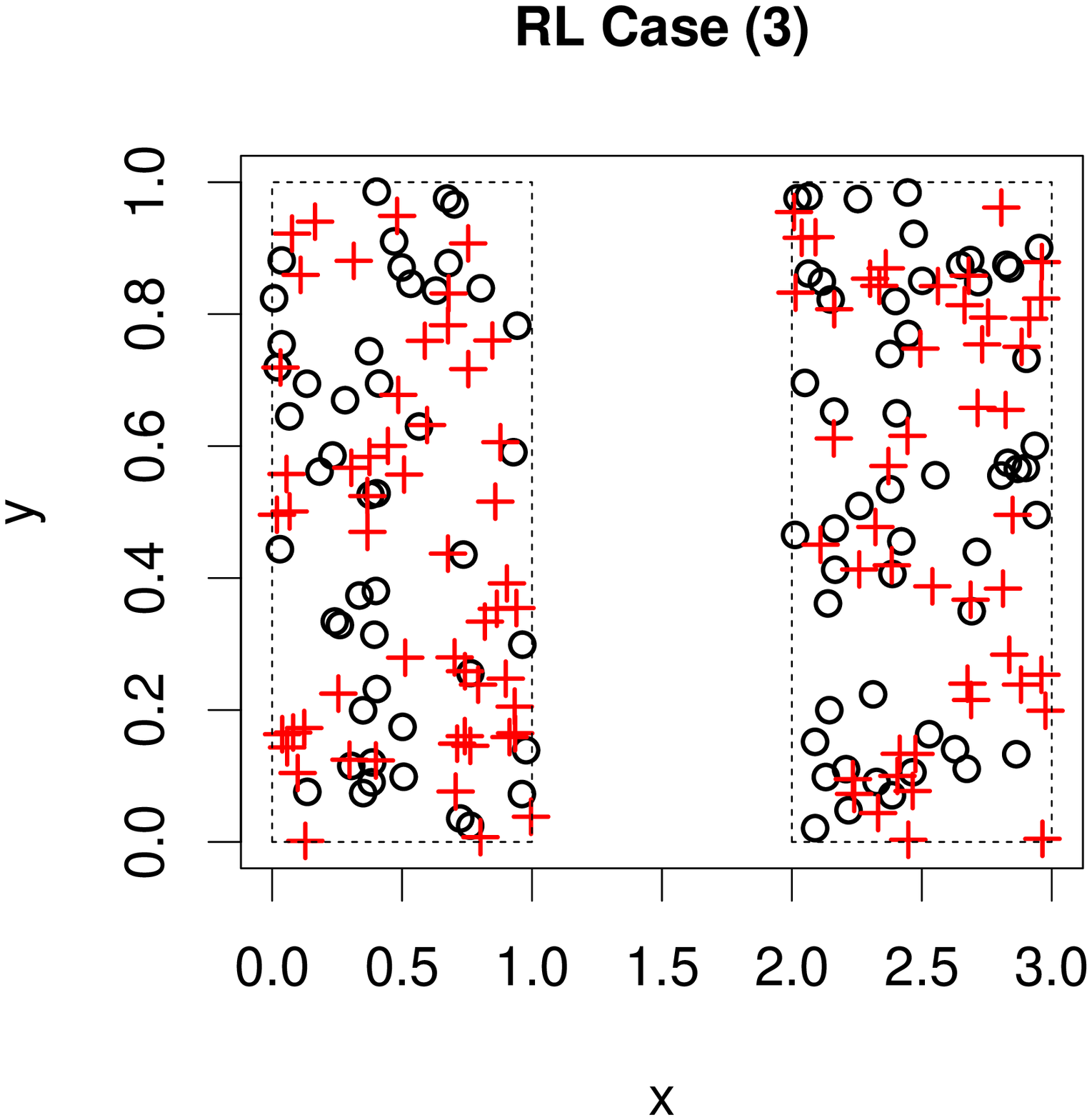} }}
 \caption{
\label{fig:RLcases}
The fixed locations for which RL procedure is applied for RL cases (1)-(3) with $n_1=n_2=100$
in the two-class case.
A realization of the RL of the two classes are indicated with circles ($\circ$) and triangles ({\footnotesize $\triangle$}).
Notice that $x$-axis for RL case (3) is differently scaled.
}
\end{figure}

The locations for which the RL procedure is applied in RL cases (1)-(3) are plotted
in Figure \ref{fig:RLcases} for $n_1=n_2=100$.
Observe that in RL case (1), the set of points are iid $\U((0,1) \times (0,1))$,
i.e., it can be assumed to be from a Poison process in the unit square.
The set of locations are from two overlapping clusters in RL case (2), and
from two disjoint clusters in RL case (3).

\begin{figure} [hbp]
\centering
%\psfrag{Density}{ \Huge{\bf{Density}}}
Empirical Size Plots for the NNCT-Tests for Two Classes under RL Case (1) \\
\rotatebox{-90}{ \resizebox{2.1 in}{!}{\includegraphics{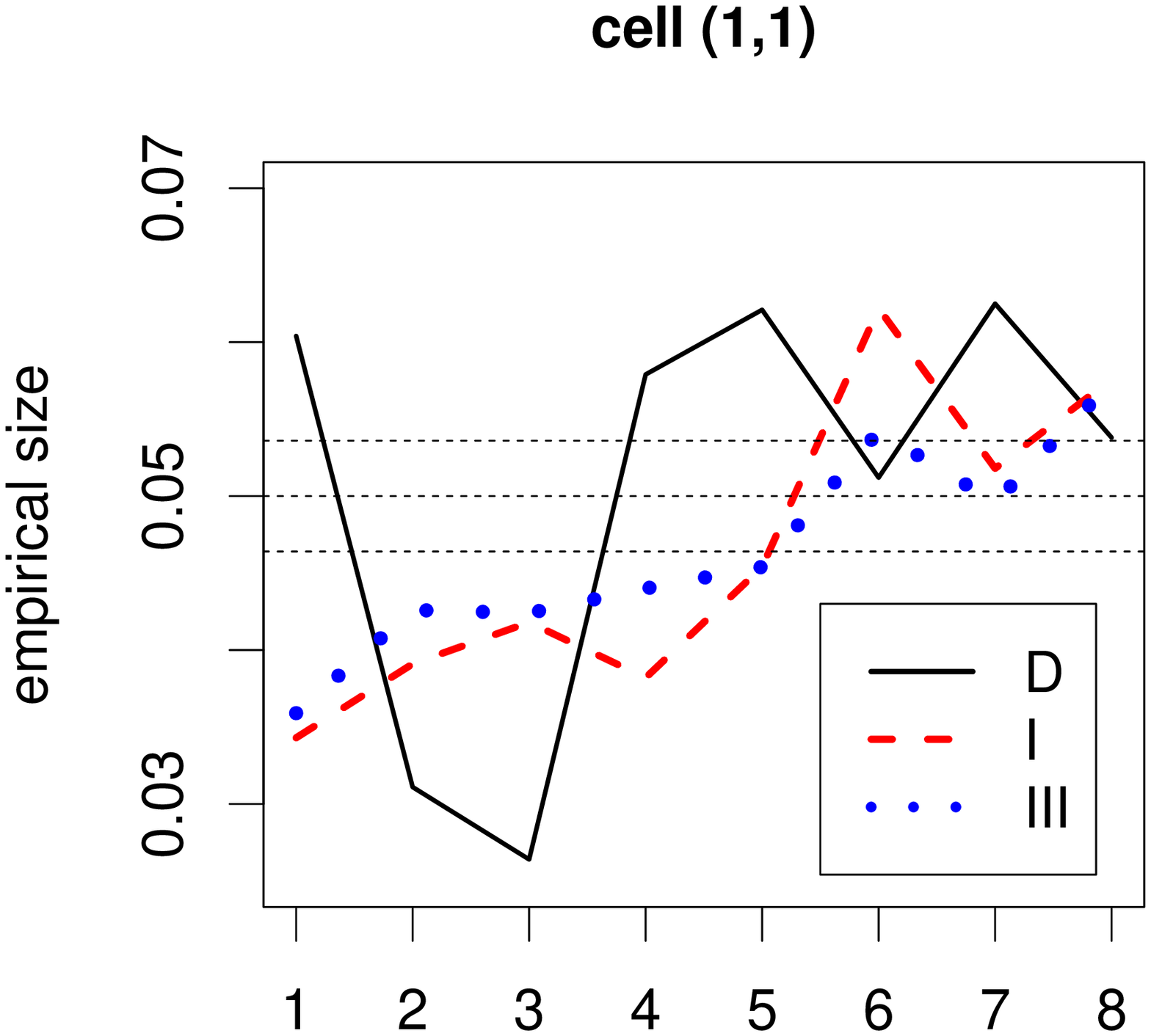} }}
\rotatebox{-90}{ \resizebox{2.1 in}{!}{\includegraphics{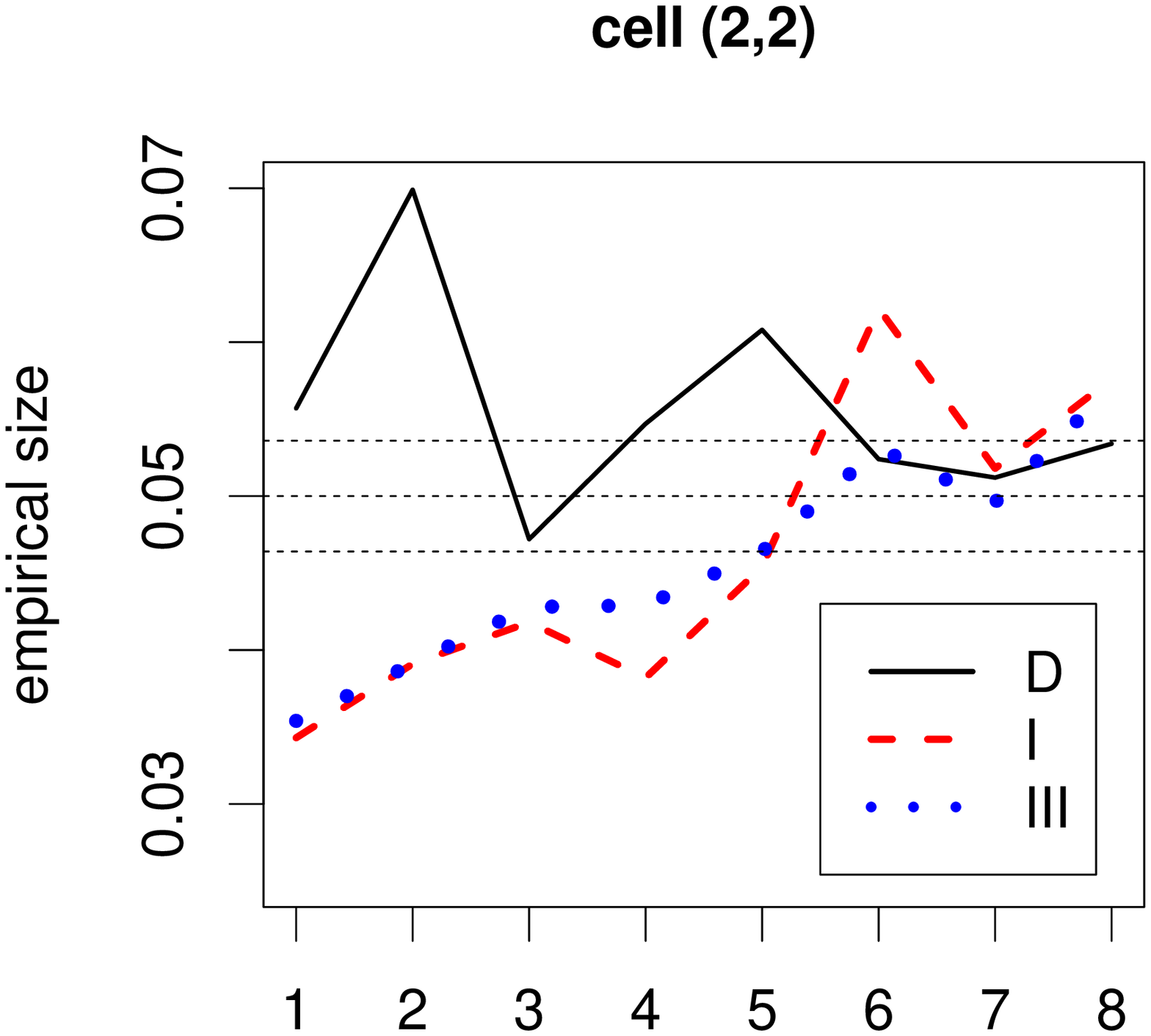} }}
\rotatebox{-90}{ \resizebox{2.1 in}{!}{\includegraphics{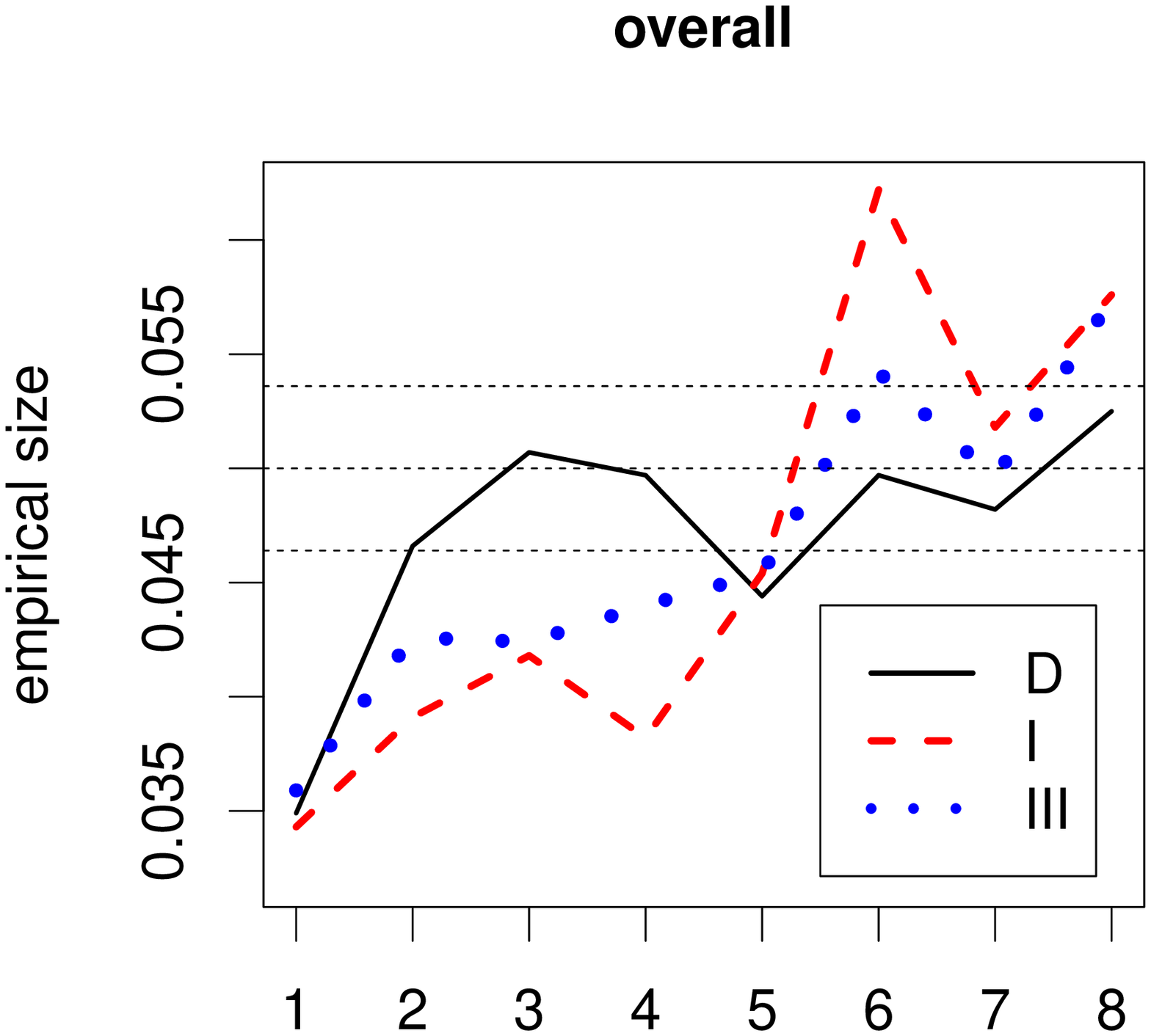} }}
RL Case (2) \\
\rotatebox{-90}{ \resizebox{2.1 in}{!}{\includegraphics{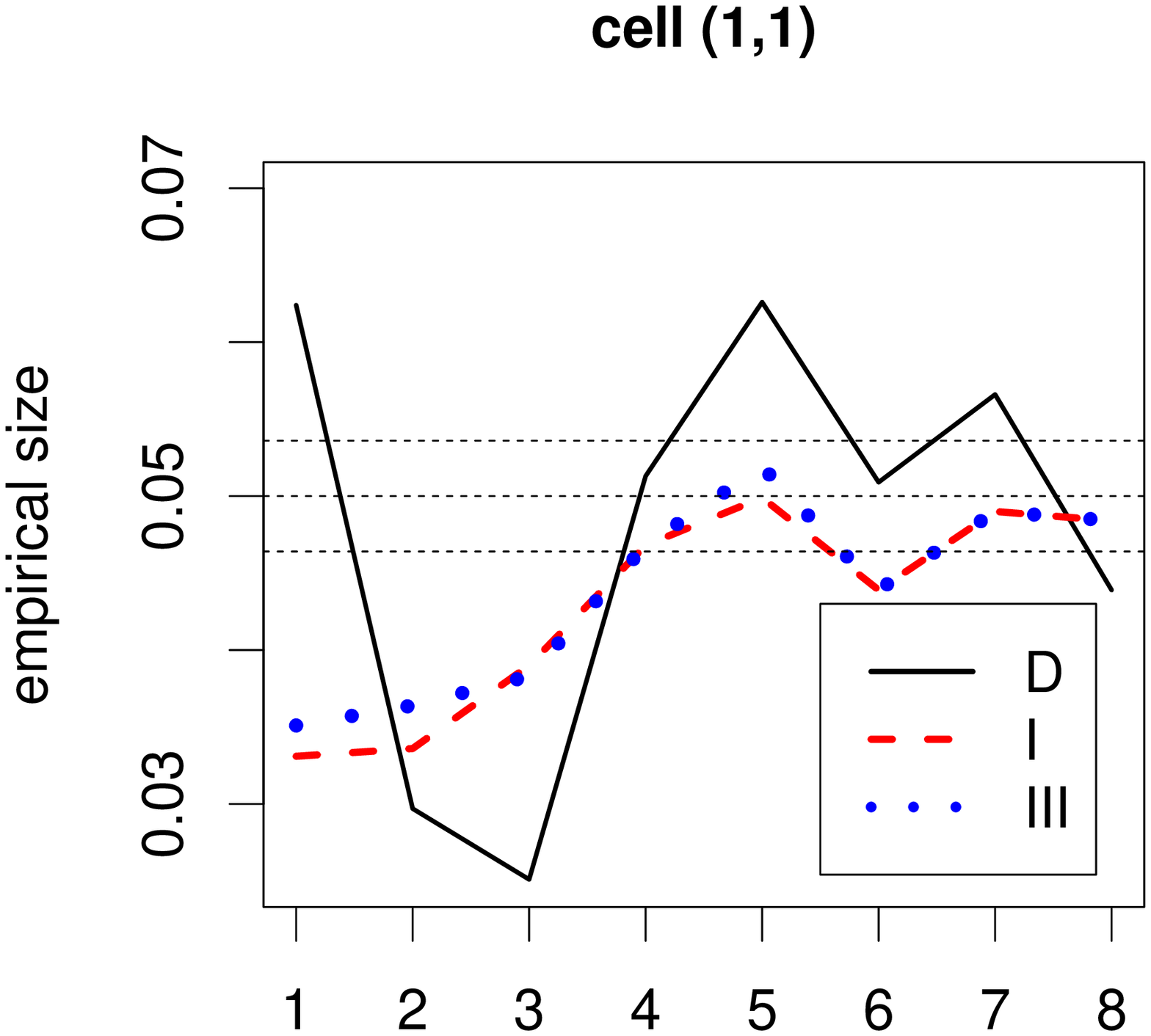} }}
\rotatebox{-90}{ \resizebox{2.1 in}{!}{\includegraphics{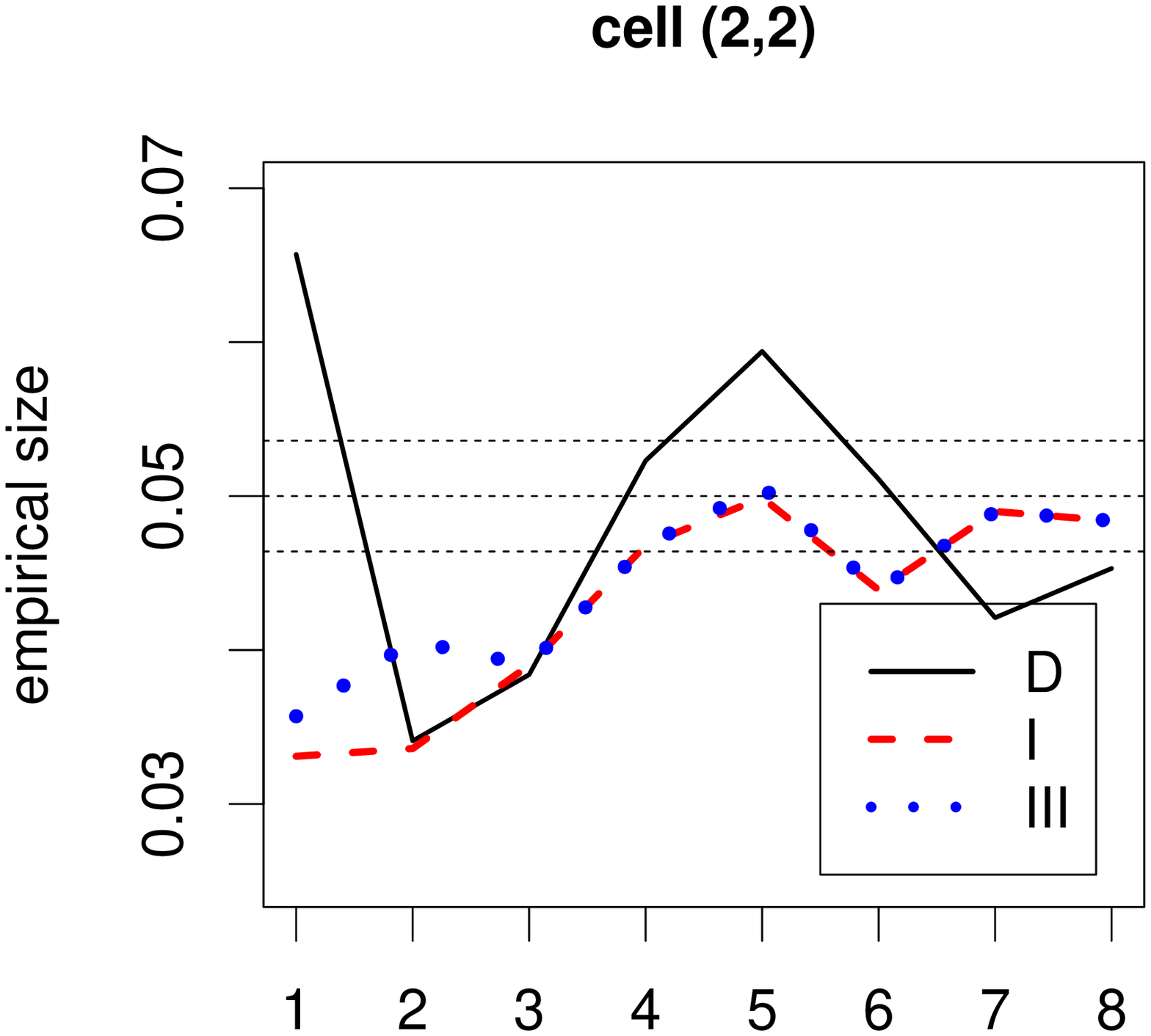} }}
\rotatebox{-90}{ \resizebox{2.1 in}{!}{\includegraphics{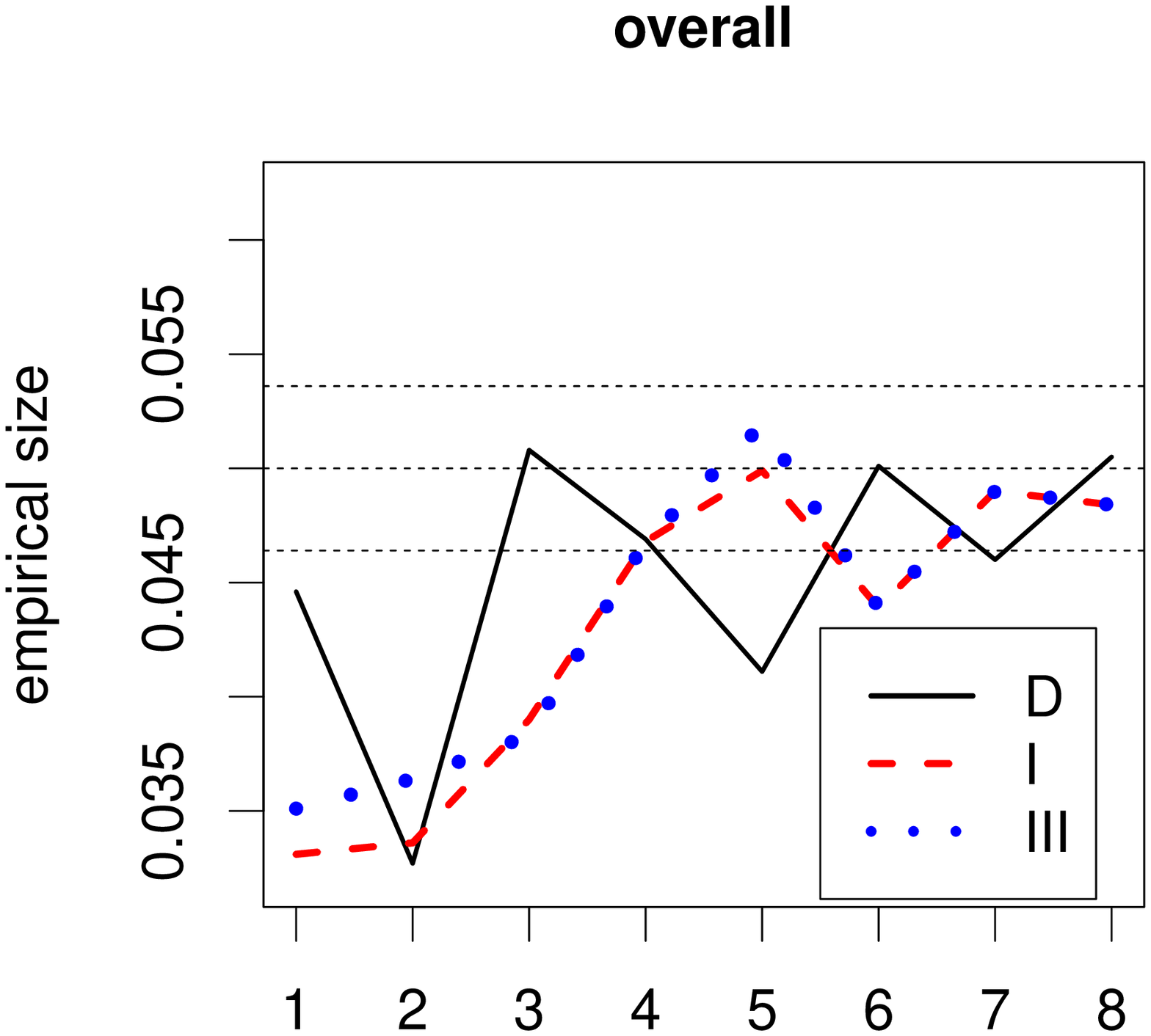} }}
RL Case (3) \\
\rotatebox{-90}{ \resizebox{2.1 in}{!}{\includegraphics{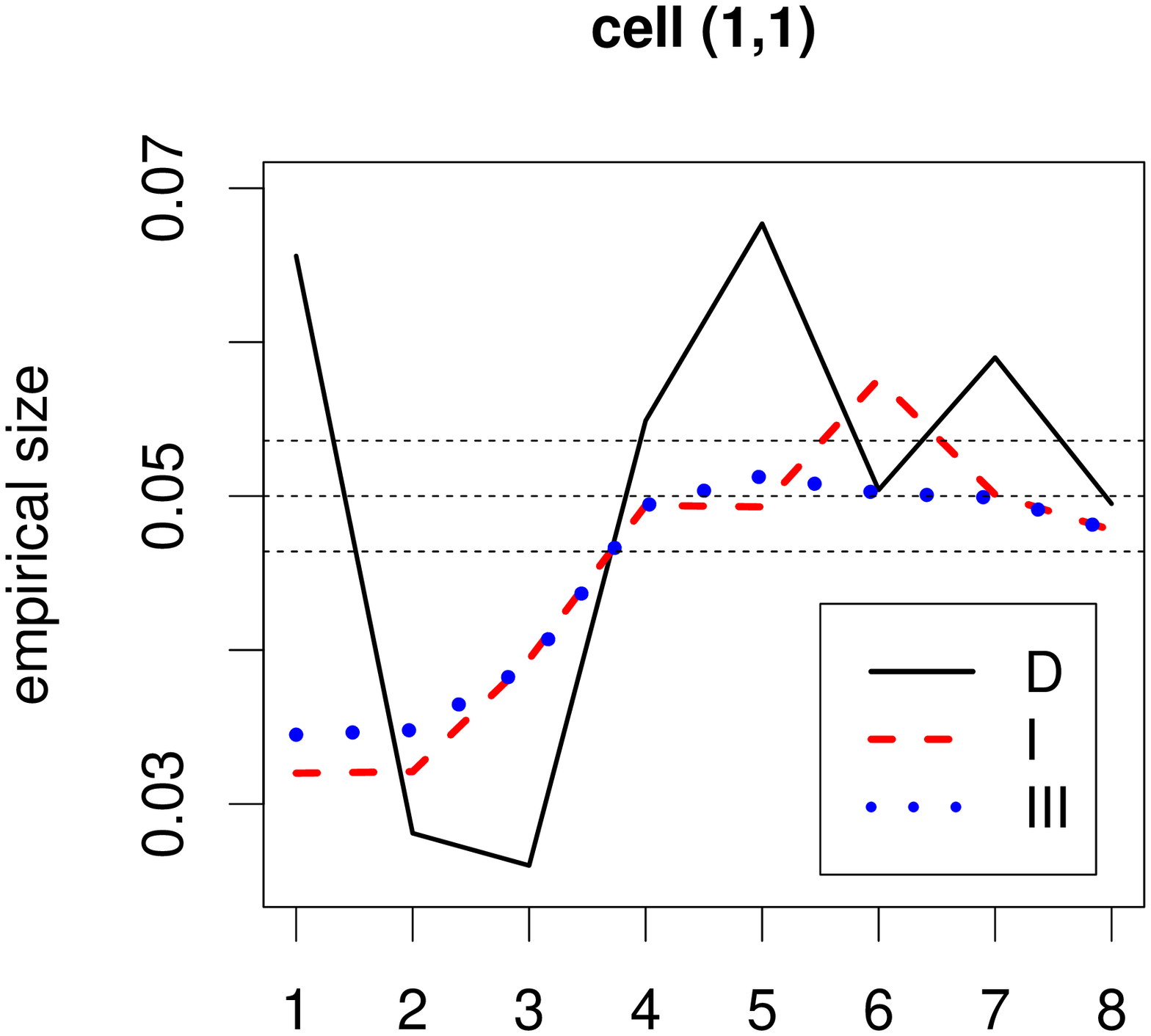} }}
\rotatebox{-90}{ \resizebox{2.1 in}{!}{\includegraphics{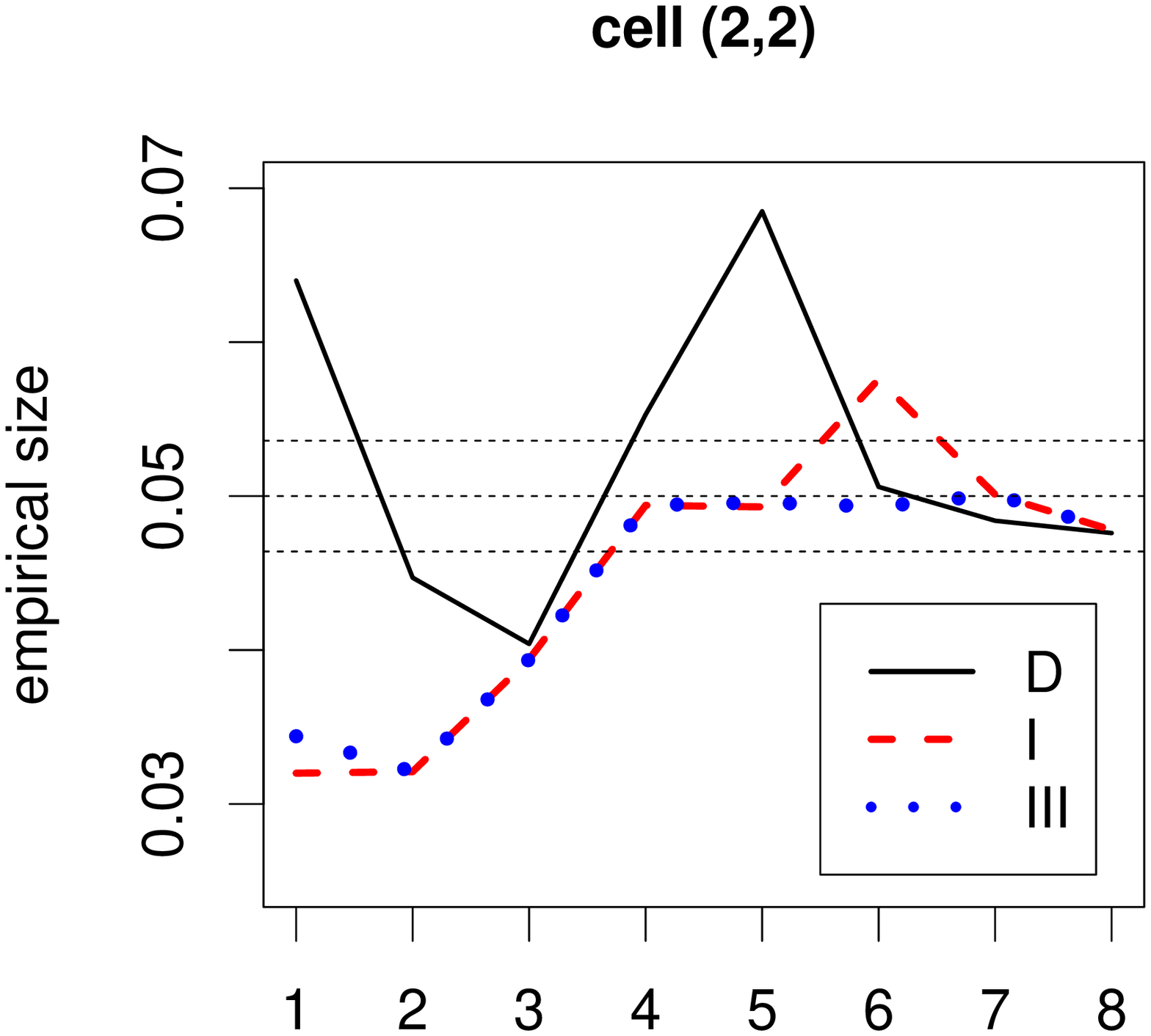} }}
\rotatebox{-90}{ \resizebox{2.1 in}{!}{\includegraphics{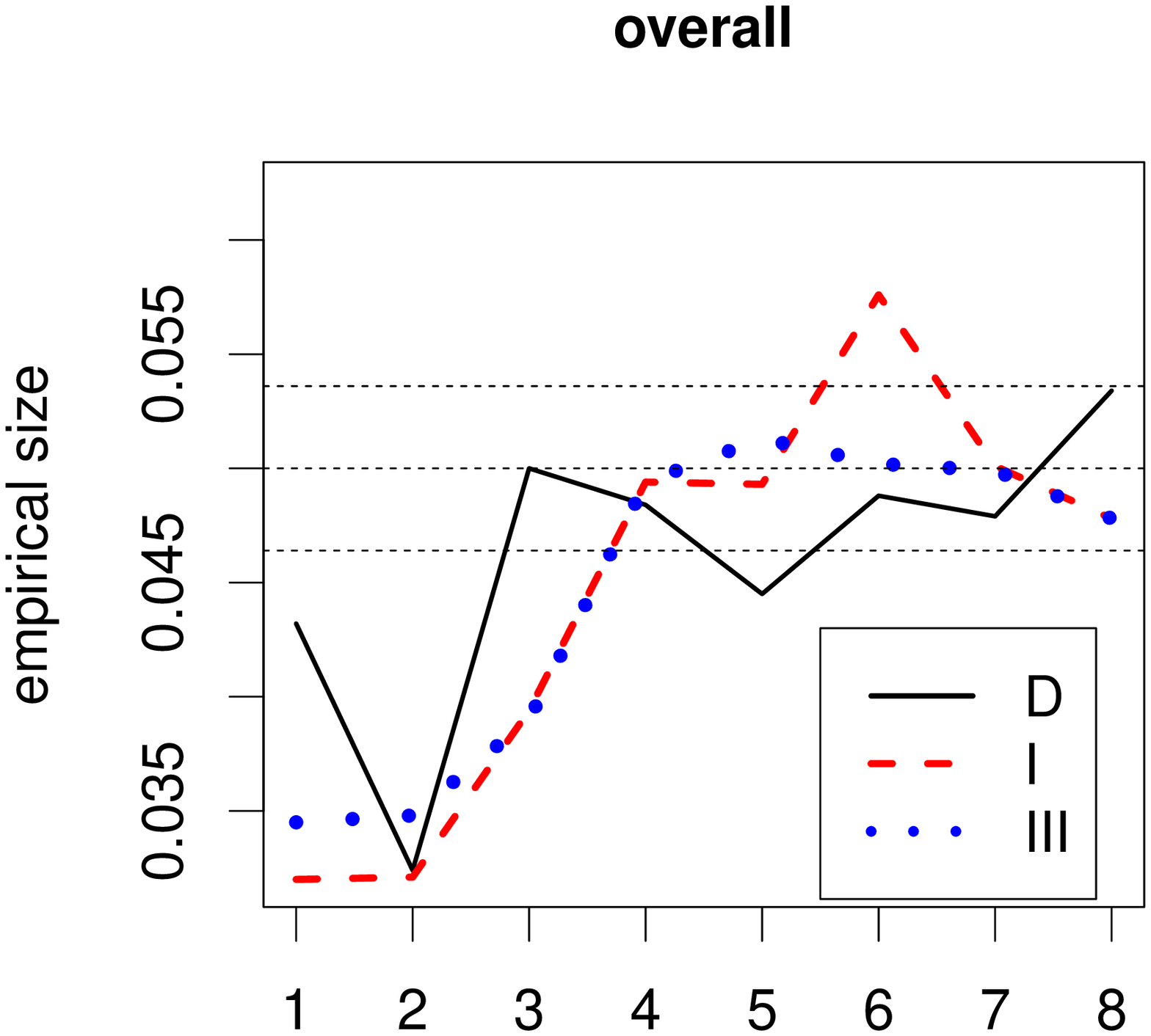} }}
\caption{
\label{fig:emp-size-RL-2cl}
The empirical size estimates of the cell-specific tests for cells (1,1) (left)
and (2,2) (middle) and overall segregation test (right) under the RL cases (1)-(3)
in the two-class case.
The horizontal lines, axis labels, and legend labeling are as in Figure \ref{fig:emp-size-CSR-2cl}.
}
\end{figure}

We present the empirical significance
levels for the NNCT-tests under the RL cases (1)-(3) in Figure \ref{fig:emp-size-RL-2cl}.
Under RL cases (1)-(3),
for cell $(1,1)$, type I and III cell-specific tests are closer (type III is closest) to the desired size
compared to Dixon's cell-specific tests, which is severely conservative when the cell size is small.
For cell $(2,2)$,
type I and III have similar performance as in cell $(1,1)$,
while due to the increase in expected cell counts,
Dixon's test gets closer to desired size, although still fluctuates around conservativeness and liberalness.
%In both cells, the tests are about the desired level for larger class sizes.
For the overall test,
type III overall test has the best performance.

\section{Empirical Size Analysis in the Three-Class Case}
\label{sec:monte-carlo-3Cl}
In this section, we provide the empirical significance levels
for Dixon's and the new overall and cell-specific segregation tests
in the three-class case under RL and CSR independence patterns.

\subsection{Empirical Size Analysis under CSR Independence of Three Classes}
\label{sec:CSR-emp-sign-3Cl}
The symmetry in cell counts for
rows in Dixon's cell-specific tests
and columns in the new type cell-specific tests occur only in the two-class case.
To better evaluate the performance of the cell-specific and overall tests,
we also consider the three-class case.
In the three-class case, we label the classes as class 1, 2, and 3 or $X$, $Y$, and $Z$ interchangeably.
We generate $n_1,\,n_2,\,n_3$ points distributed independently uniformly
on the unit square $(0,1) \times (0,1)$
from these classes.
We use
\begin{multline*}
(n_1,n_2,n_3) \in \{(10,10,10), (10,10,30), (10,10,50), (10,30,30), (10,30,50),
(30,30,30), (10,50,50),\\
(30,30,50),(30,50,50), (50,50,50), (50,50,100), (50,100,100), (100,100,100)\};
\end{multline*}
and $N_{mc}=10000$.
The empirical sizes and the significance of their deviation from .05 are calculated
as in Section \ref{sec:CSR-emp-sign-2Cl}.

\begin{figure} [hbp]
\centering
%\psfrag{Density}{ \Huge{\bf{Density}}}
Empirical Size Plots for the Cell-Specific Tests under CSR Independence\\
\rotatebox{-90}{ \resizebox{2.1 in}{!}{\includegraphics{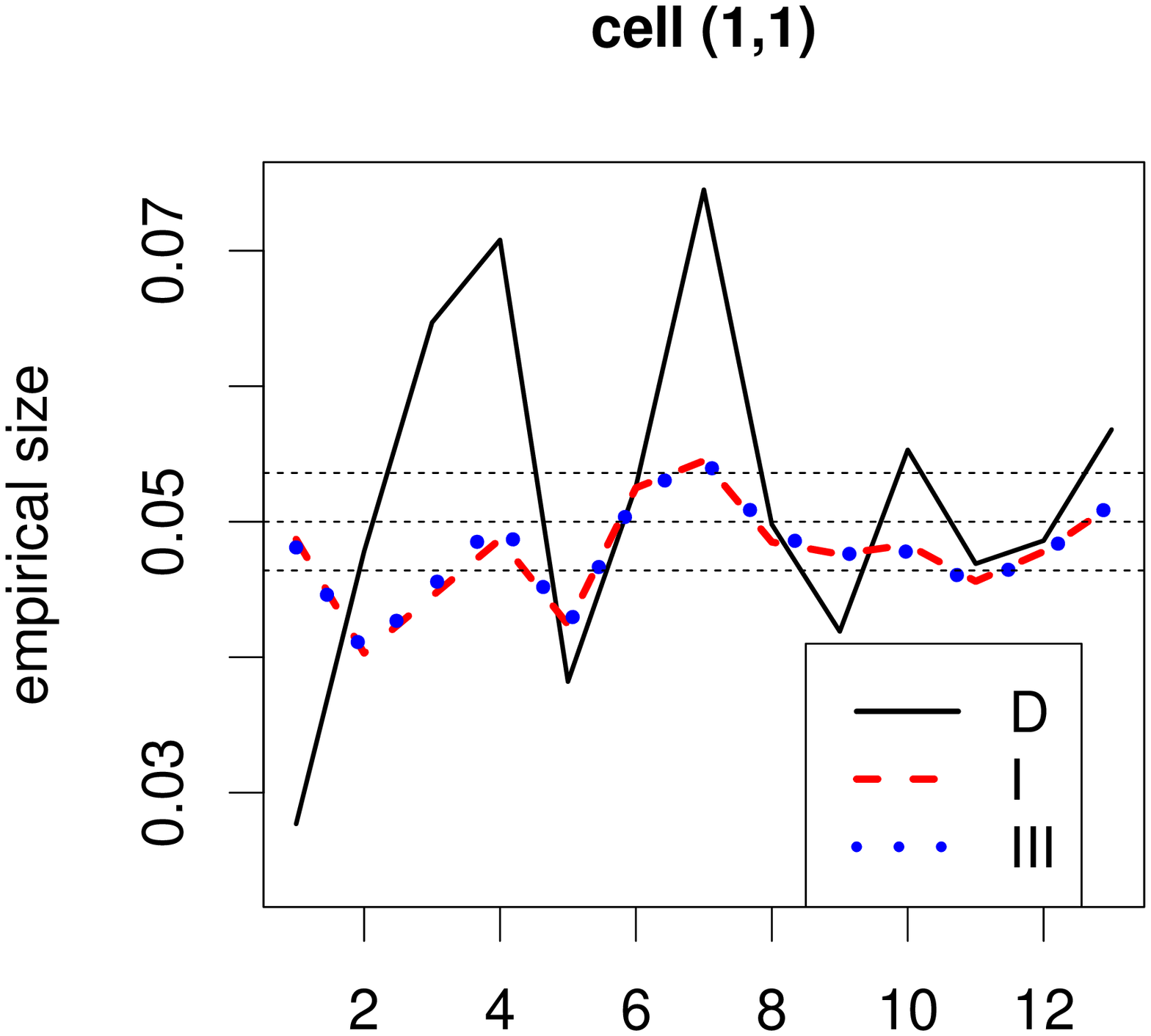} }}
\rotatebox{-90}{ \resizebox{2.1 in}{!}{\includegraphics{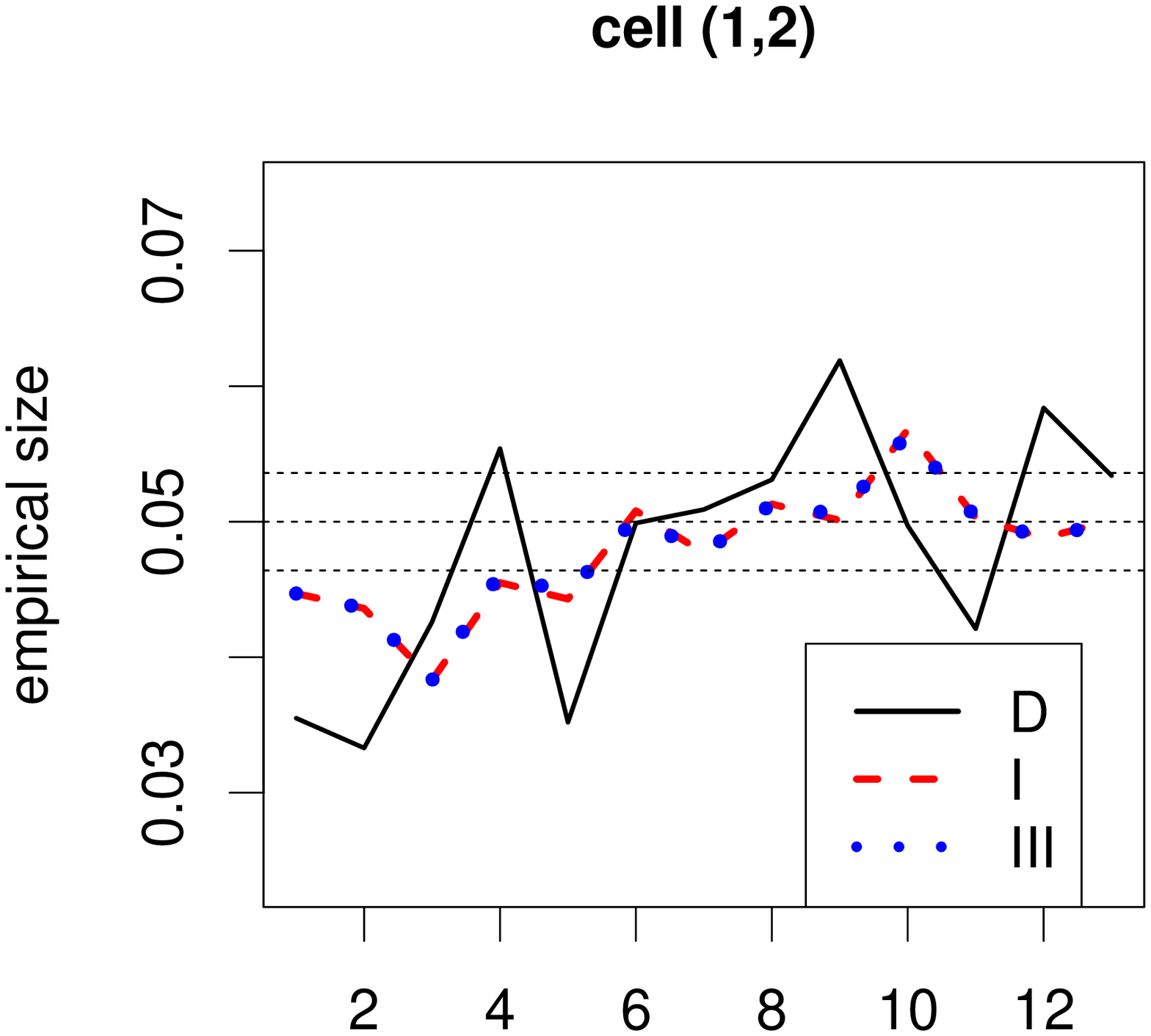} }}
\rotatebox{-90}{ \resizebox{2.1 in}{!}{\includegraphics{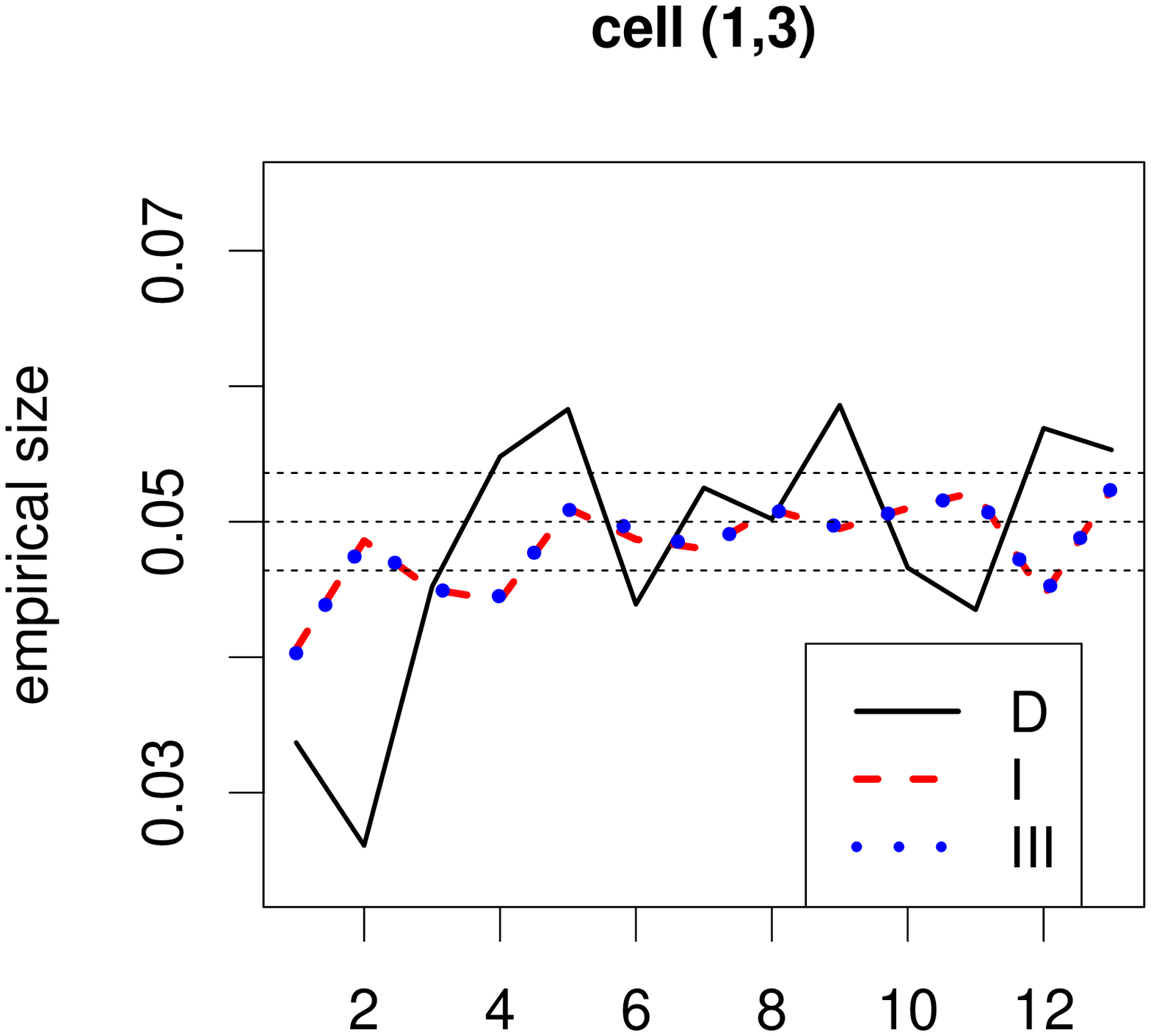} }}
\rotatebox{-90}{ \resizebox{2.1 in}{!}{\includegraphics{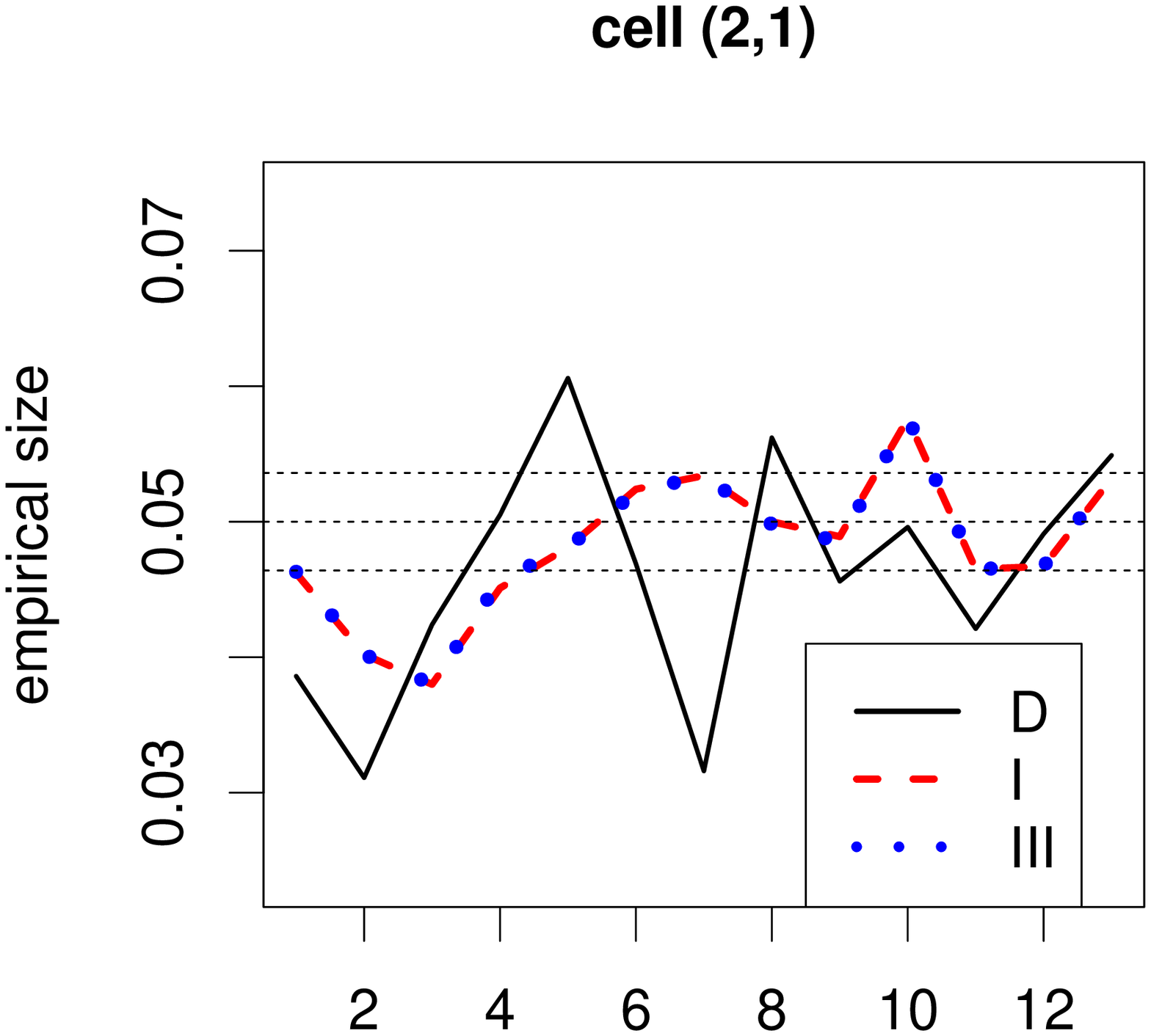} }}
\rotatebox{-90}{ \resizebox{2.1 in}{!}{\includegraphics{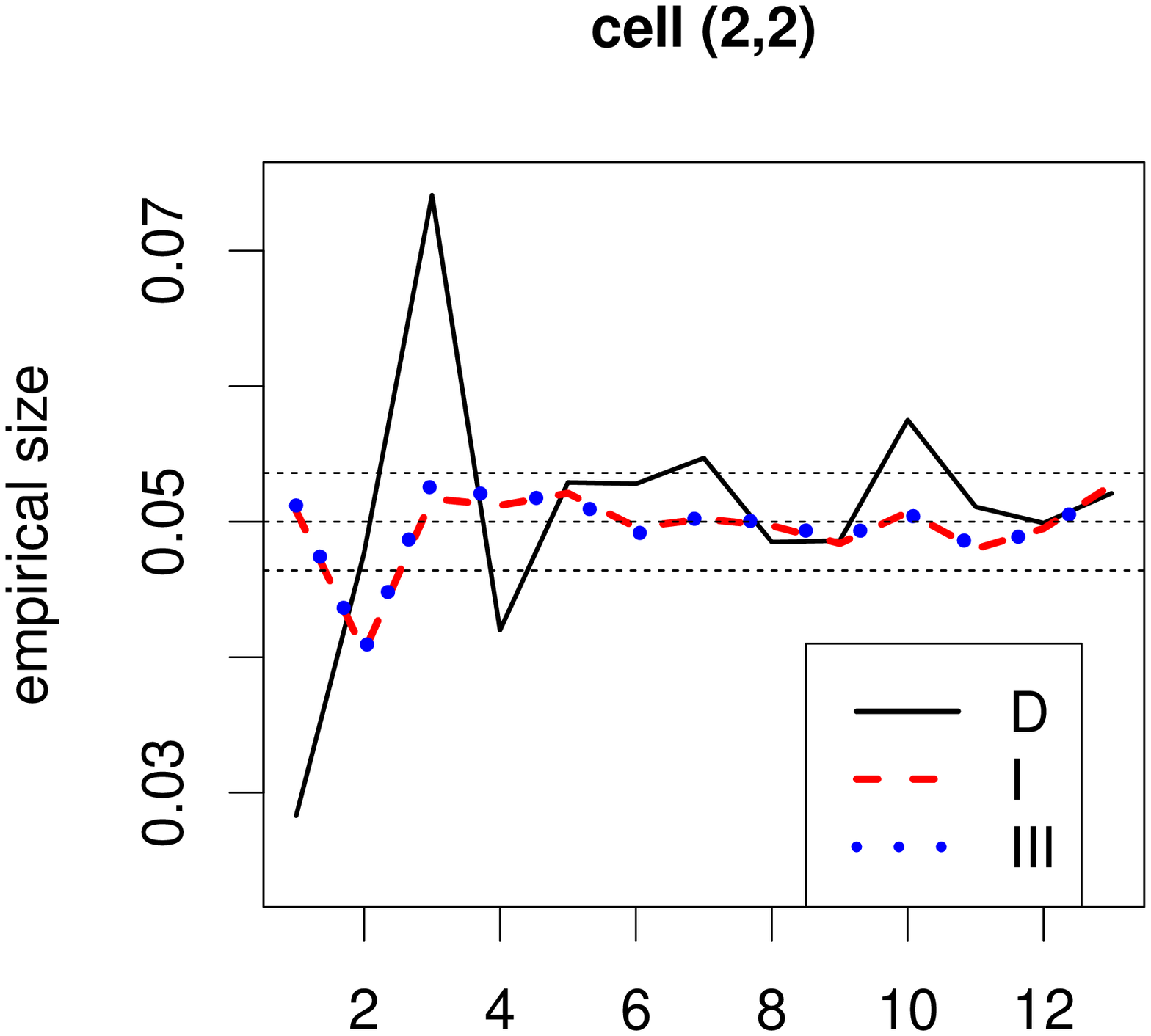} }}
\rotatebox{-90}{ \resizebox{2.1 in}{!}{\includegraphics{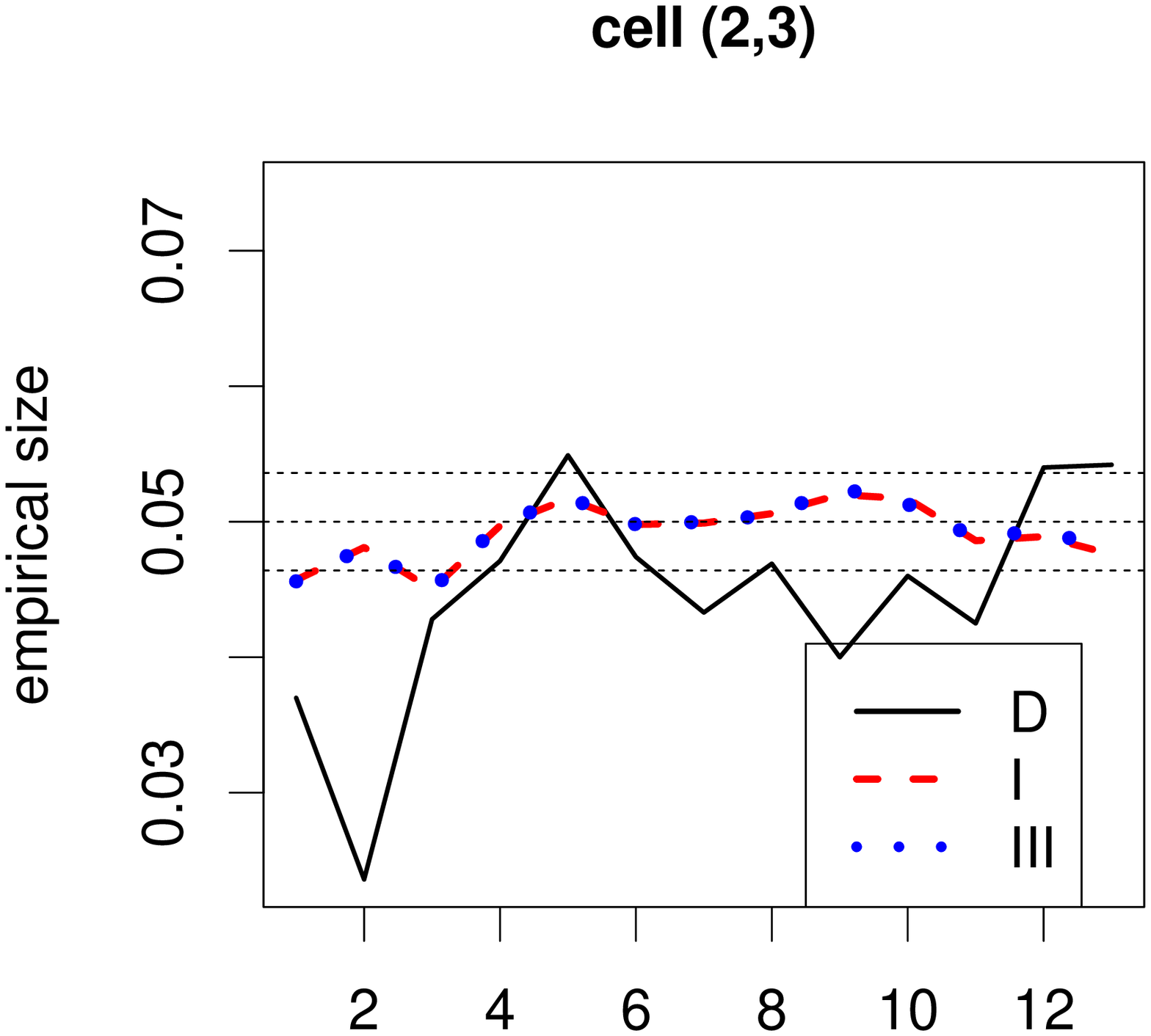} }}
\rotatebox{-90}{ \resizebox{2.1 in}{!}{\includegraphics{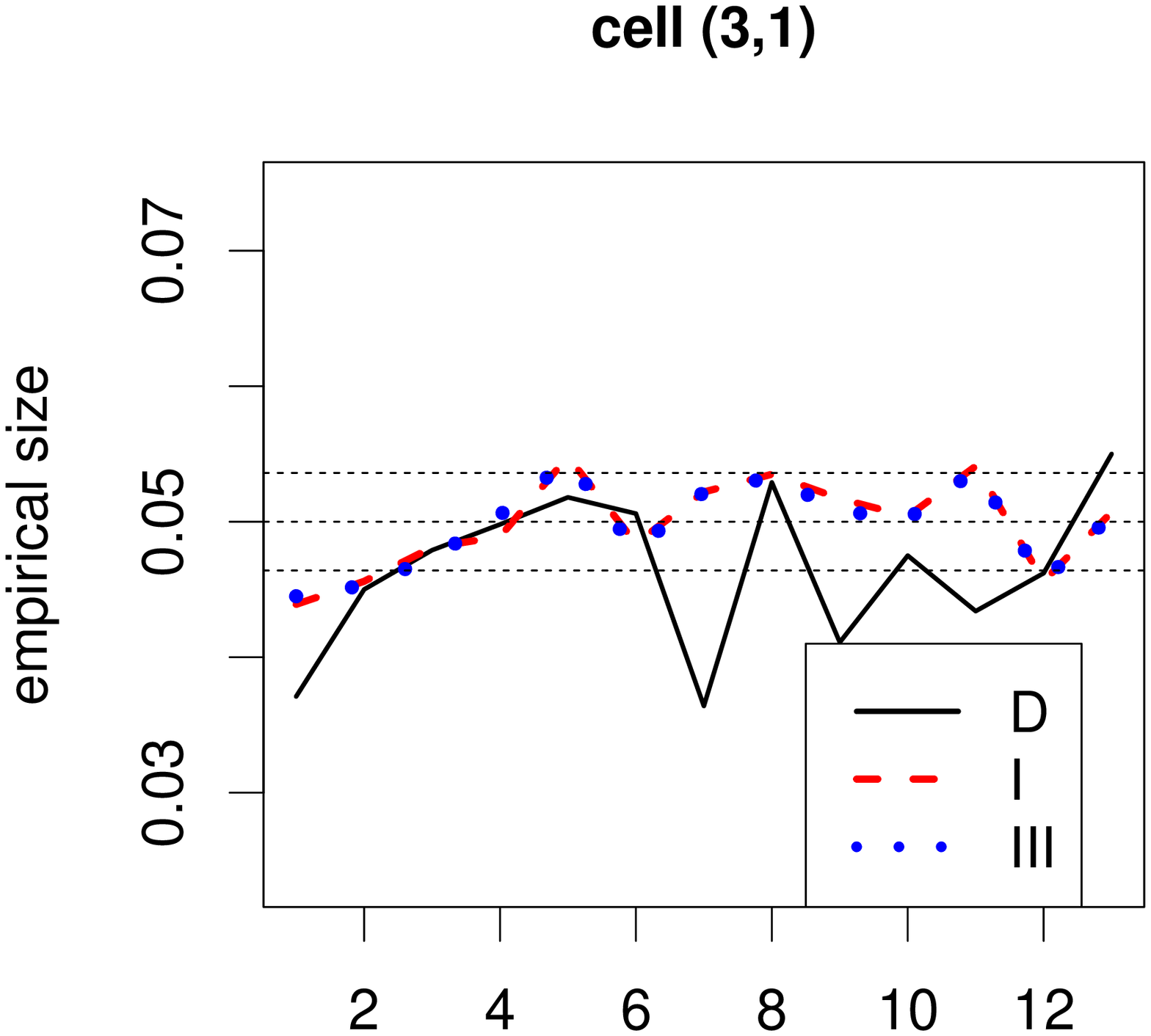} }}
\rotatebox{-90}{ \resizebox{2.1 in}{!}{\includegraphics{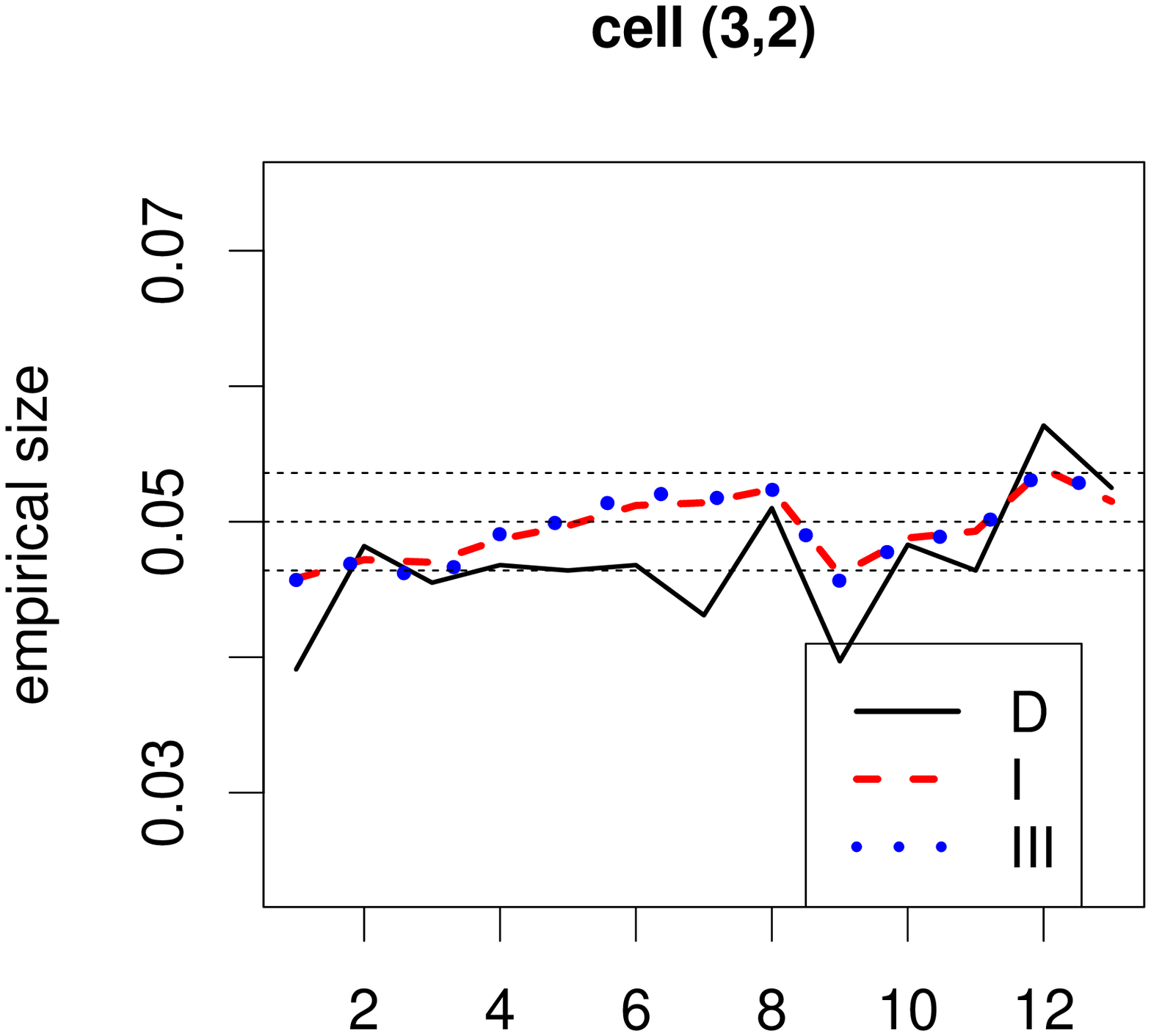} }}
\rotatebox{-90}{ \resizebox{2.1 in}{!}{\includegraphics{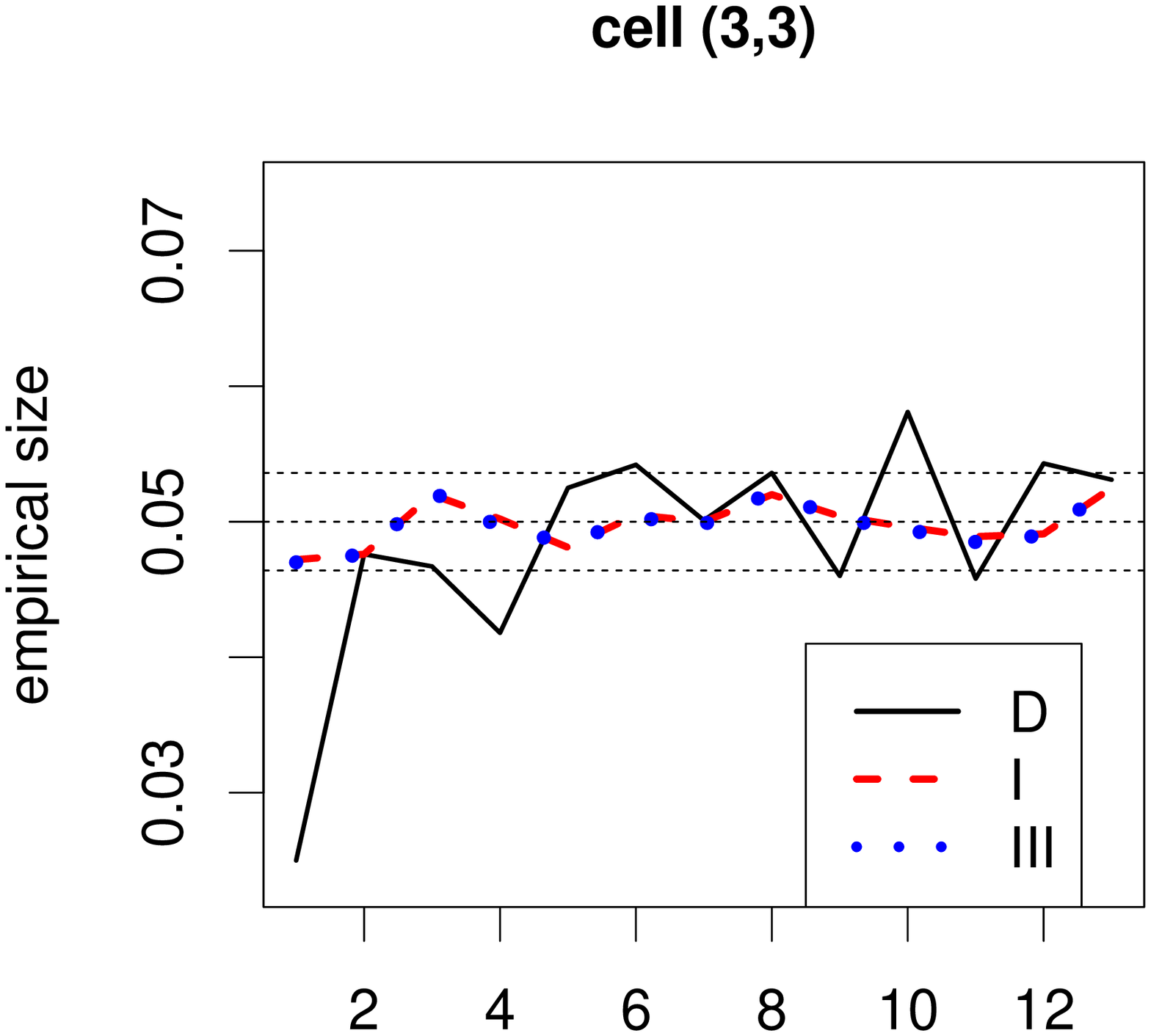} }}
\caption{
\label{fig:emp-size-CSR-cell-3cl}
The empirical size estimates of the cell-specific tests for cells $(1,1)-(3,3)$
under the CSR independence pattern in the three-class case.
The horizontal lines and legend labeling are as in Figure \ref{fig:emp-size-CSR-2cl}.
The horizontal axis labels are:
1=(10,10,10), 2=(10,10,30), 3=(10,10,50), 4=(10,30,30), 5=(10,30,50),
6=(30,30,30), 7=(10,50,50), 8=(30,30,50), 9=(30,50,50), 10=(50,50,50),
11=(50,50,100), 12=(50,100,100), 13=(100,100,100).
}
\end{figure}

\begin{figure} [hbp]
\centering
%\psfrag{Density}{ \Huge{\bf{Density}}}
Empirical Size Plots for the Overall Tests under CSR Independence\\
\rotatebox{-90}{ \resizebox{2.5 in}{!}{\includegraphics{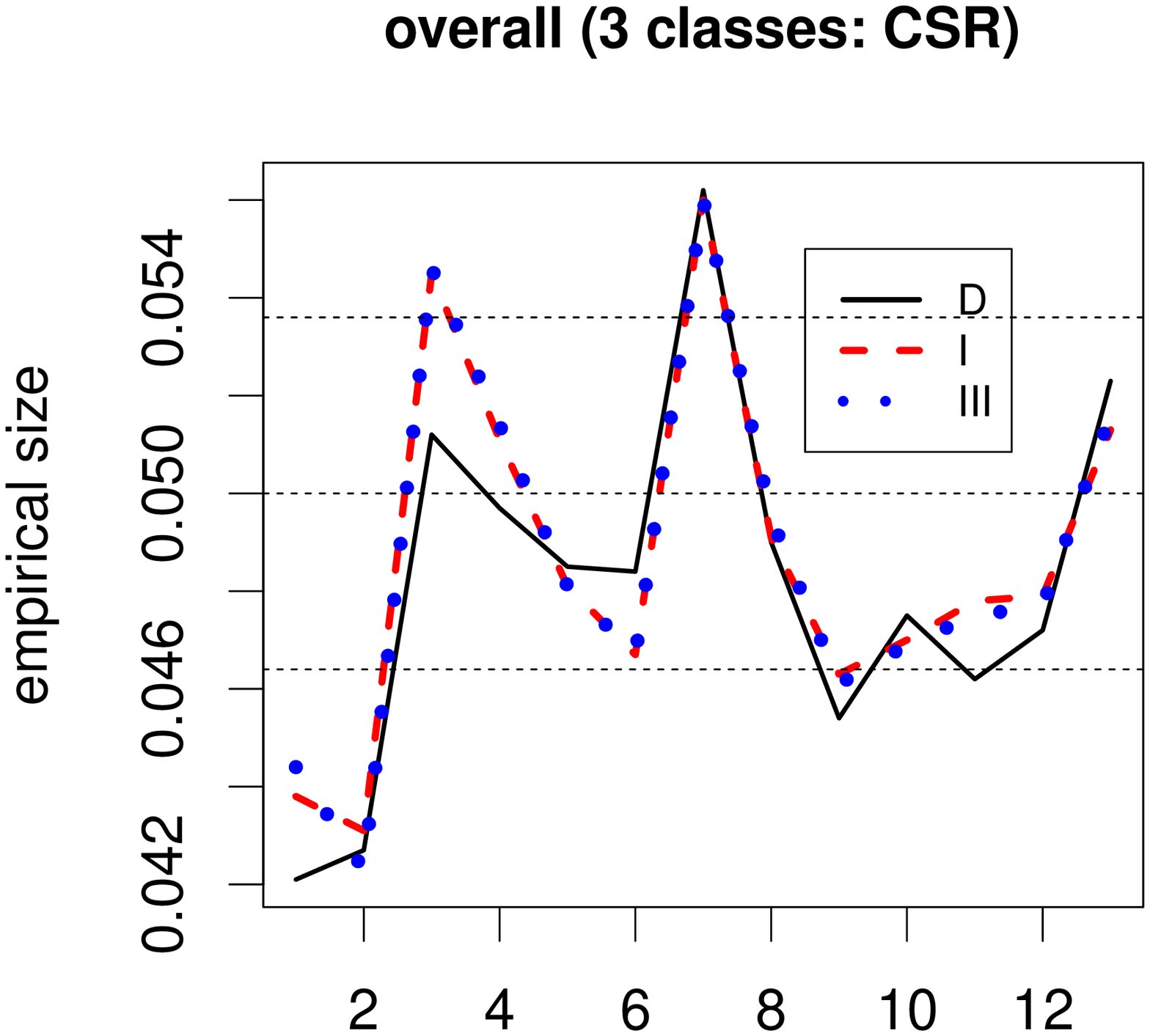} }}
\caption{
\label{fig:emp-size-CSR-overall-3cl}
The empirical size estimates of the
overall tests under the CSR independence pattern
in the three-class case.
The horizontal lines, and legend labeling are as in Figure \ref{fig:emp-size-CSR-2cl}
 and axis labels are as in Figure \ref{fig:emp-size-CSR-cell-3cl}.
}
\end{figure}

We present the empirical significance
levels for the cell-specific tests in Figure \ref{fig:emp-size-CSR-cell-3cl} and
for the overall tests in Figure \ref{fig:emp-size-CSR-overall-3cl}.
For the cell-specific tests,
clearly, type I and III tests are closer to the desired level,
and are less affected by the differences in class sizes.
On the other hand,
Dixon's test is extremely liberal or conservative,
when class sizes are very different (which may result in smaller
expected cell counts).
The overall tests have similar size performance
with Dixon's test being slightly better for smaller classes,
while type I and III slightly better for larger classes.

\subsection{Empirical Size Analysis under RL of Three Classes}
\label{sec:RL-emp-sign-3Cl}
We also perform Monte Carlo simulations under RL for the three class case to compare the tests
without conditioning on $Q$ and $R$.
Under RL,
we consider two cases, in each of which we first determine the locations of the points,
and then assign the labels randomly.
We generate $n_1$ points iid $\U(S_1)$,
$n_2$ points iid $\U(S_2)$,
and
$n_3$ points iid $\U(S_3)$
for each combination of $n_1,n_2,n_3$ as in CSR independence.
The locations of these points are taken to be fixed
and we assign the labels randomly.
For each class size combination $(n_1,n_2,n_3)$
we pick $n_1$ points (without replacement) and label them as $X$,
pick $n_2$ points from the remaining points (without replacement) and label them as $Y$ points,
and label the remaining $n_3$ points as $Z$ points.
We estimate the empirical size estimates based on $N_{mc}=10000$ replications for each class size combination
as in the CSR independence case.

\begin{figure} [hbp]
\centering
%\psfrag{Density}{ \Huge{\bf{Density}}}
\rotatebox{-90}{ \resizebox{2.5 in}{!}{\includegraphics{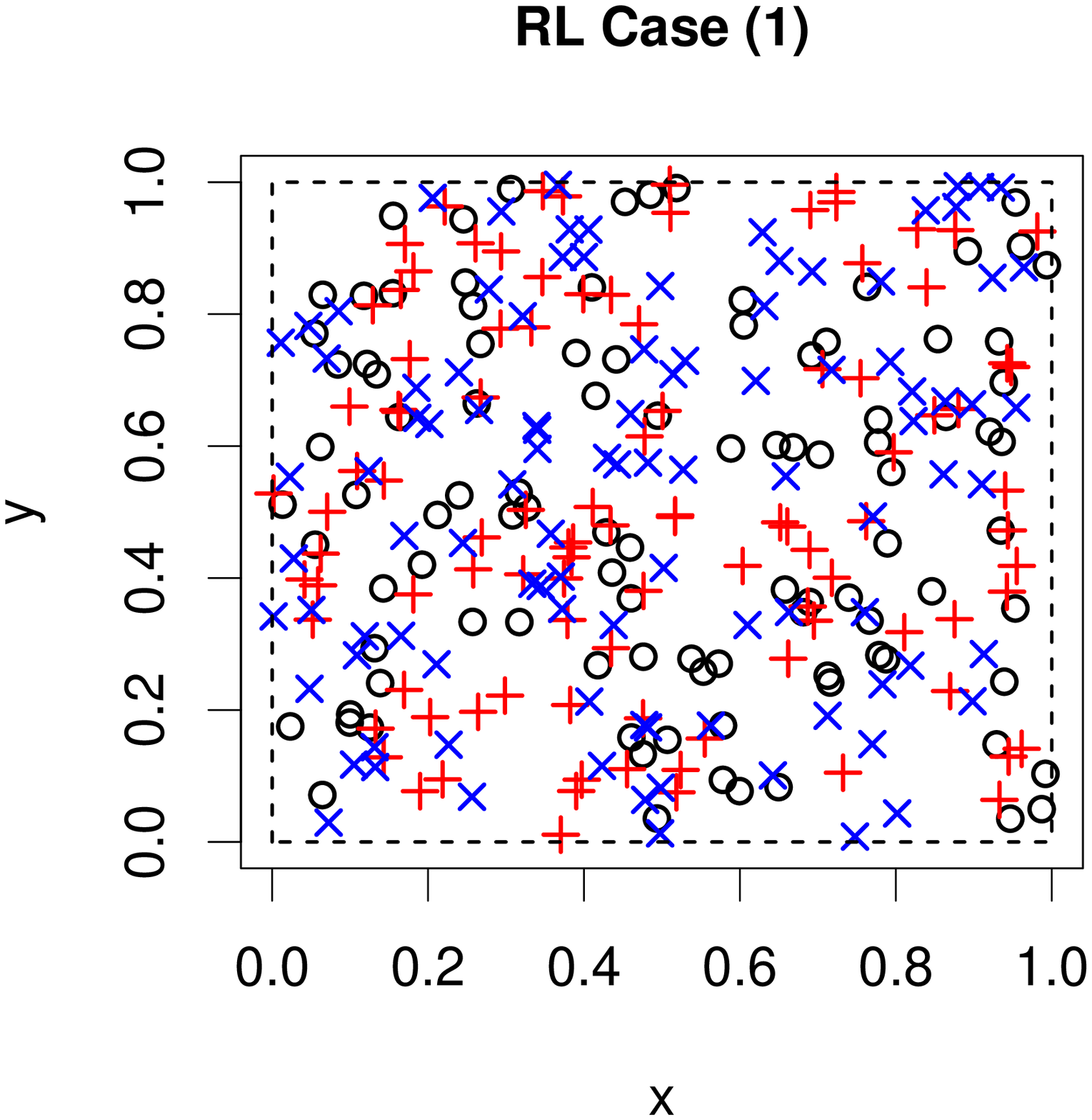} }}
\rotatebox{-90}{ \resizebox{2.5 in}{!}{\includegraphics{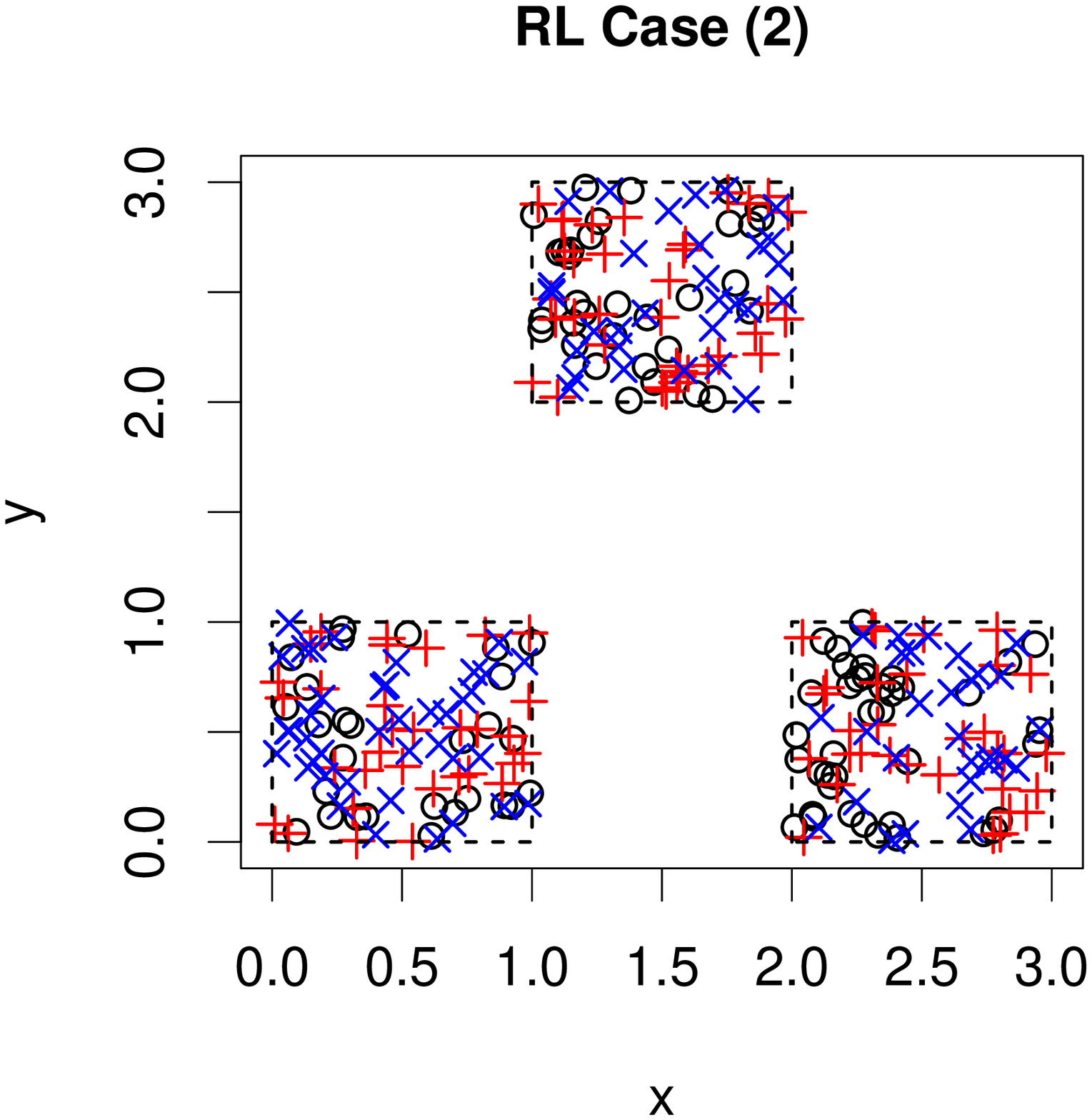} }}
 \caption{
\label{fig:RL3Clcases}
The fixed locations for which RL procedure is applied for RL cases (1) and (2) with $n_1=n_2=n_3=100$
in the three-class case.
A realization of the RL of the three classes are indicated with circles ($\circ$), pluses ($+$), and crosses ($\times$).
Notice that $x$-axis for RL case (2) is differently scaled.
}
\end{figure}

In RL case (1), we take $S_1=S_2=S_3=(0,1) \times (0,1)$,
and
in RL case (2), $S_1=(0,1) \times (0,1)$,
$S_2=(2,3) \times (0,1)$,
and
$S_3=(1,2) \times (2,3)$.
The locations for which the RL procedure is applied in RL cases (1) and (2) are plotted
in Figure \ref{fig:RL3Clcases} for $n_1=n_2=n_3=100$.
In RL case (1), the locations of the points can be assumed to
be from a Poisson process in the unit square.
In RL case (2), the locations of the points are from three disjoint clusters.

\begin{figure} [hbp]
\centering
%\psfrag{Density}{ \Huge{\bf{Density}}}
Empirical Size Plots for the Cell-Specific Tests under RL case (1)\\
\rotatebox{-90}{ \resizebox{2.1 in}{!}{\includegraphics{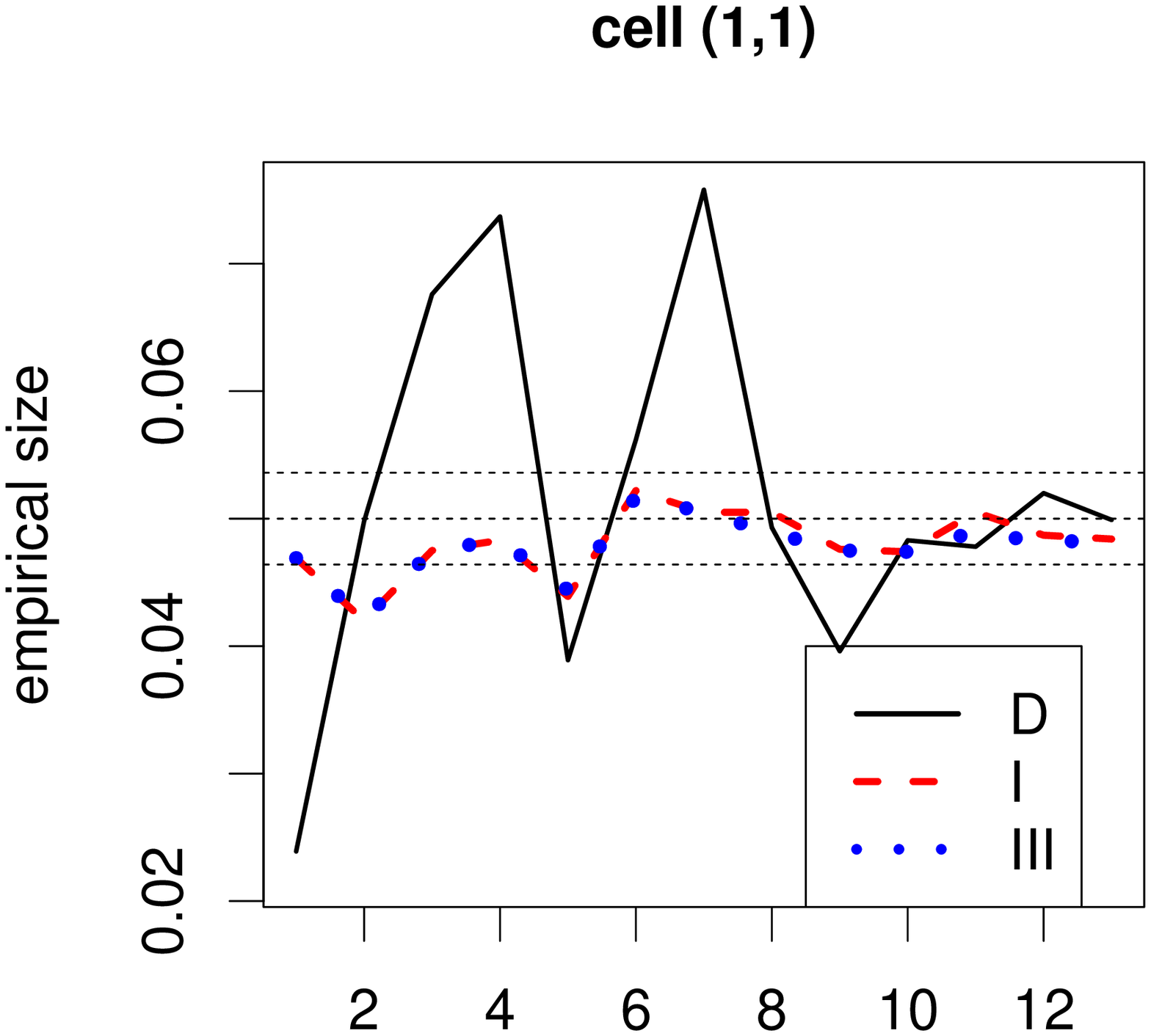} }}
\rotatebox{-90}{ \resizebox{2.1 in}{!}{\includegraphics{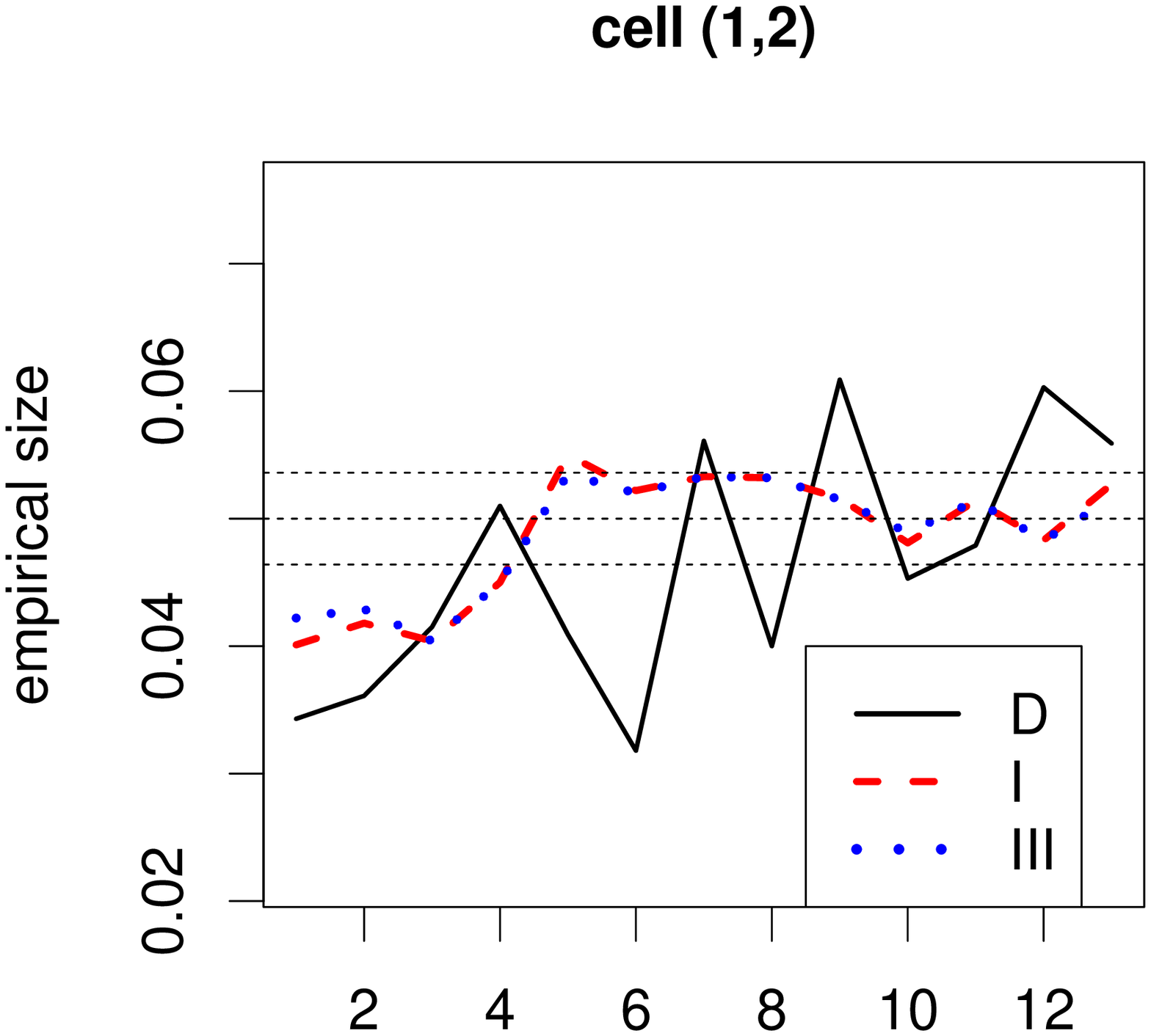} }}
\rotatebox{-90}{ \resizebox{2.1 in}{!}{\includegraphics{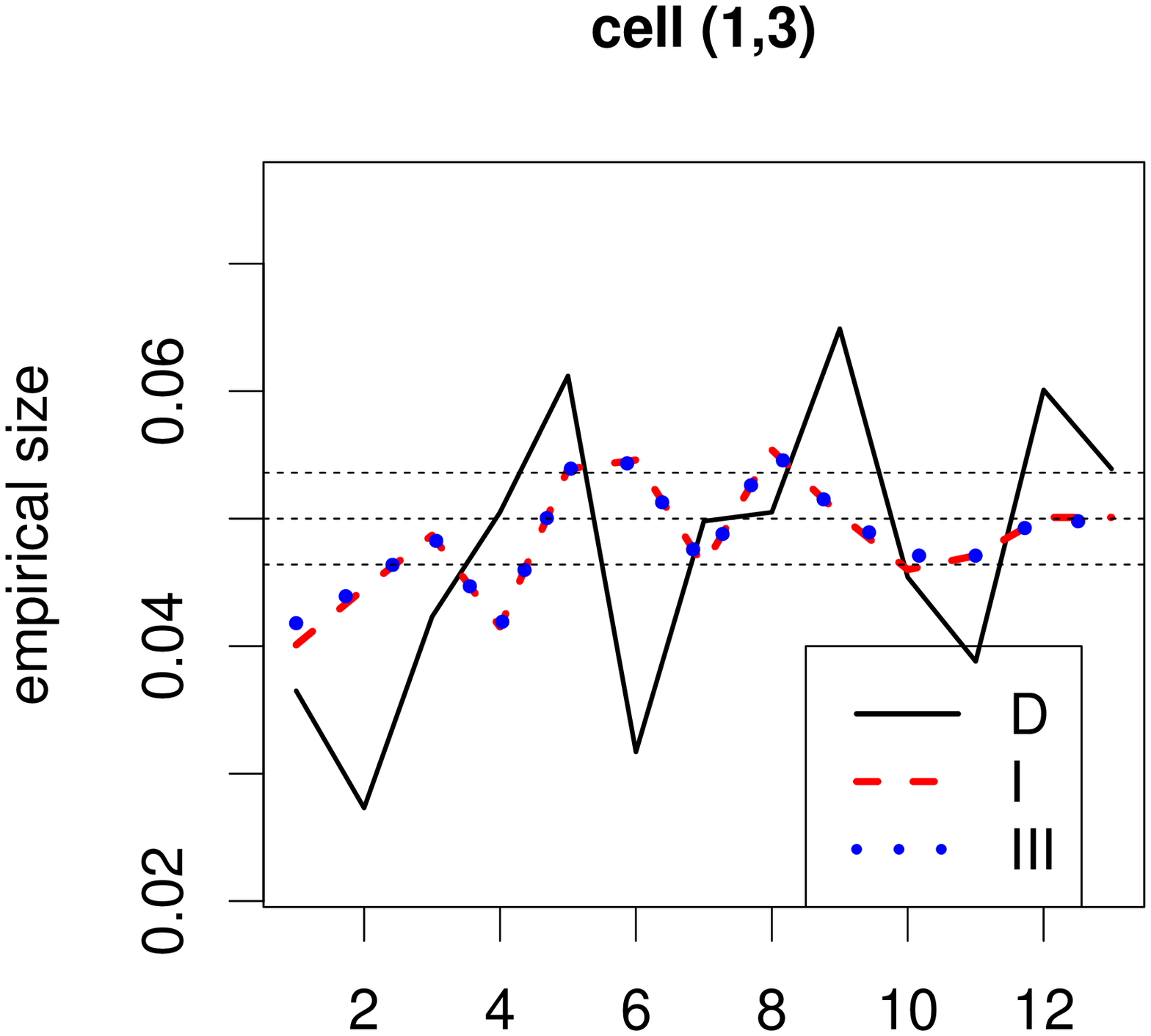} }}
\rotatebox{-90}{ \resizebox{2.1 in}{!}{\includegraphics{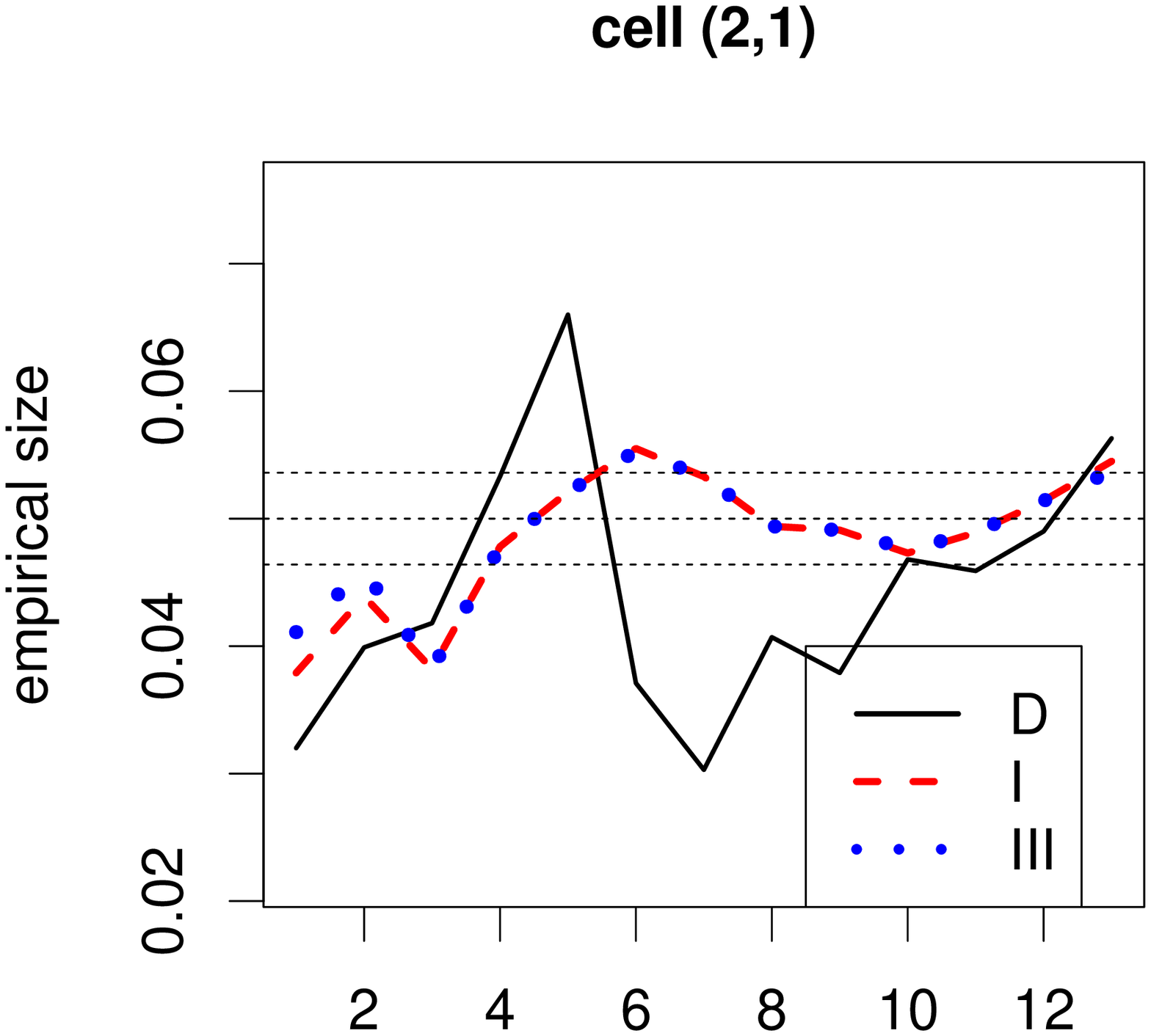} }}
\rotatebox{-90}{ \resizebox{2.1 in}{!}{\includegraphics{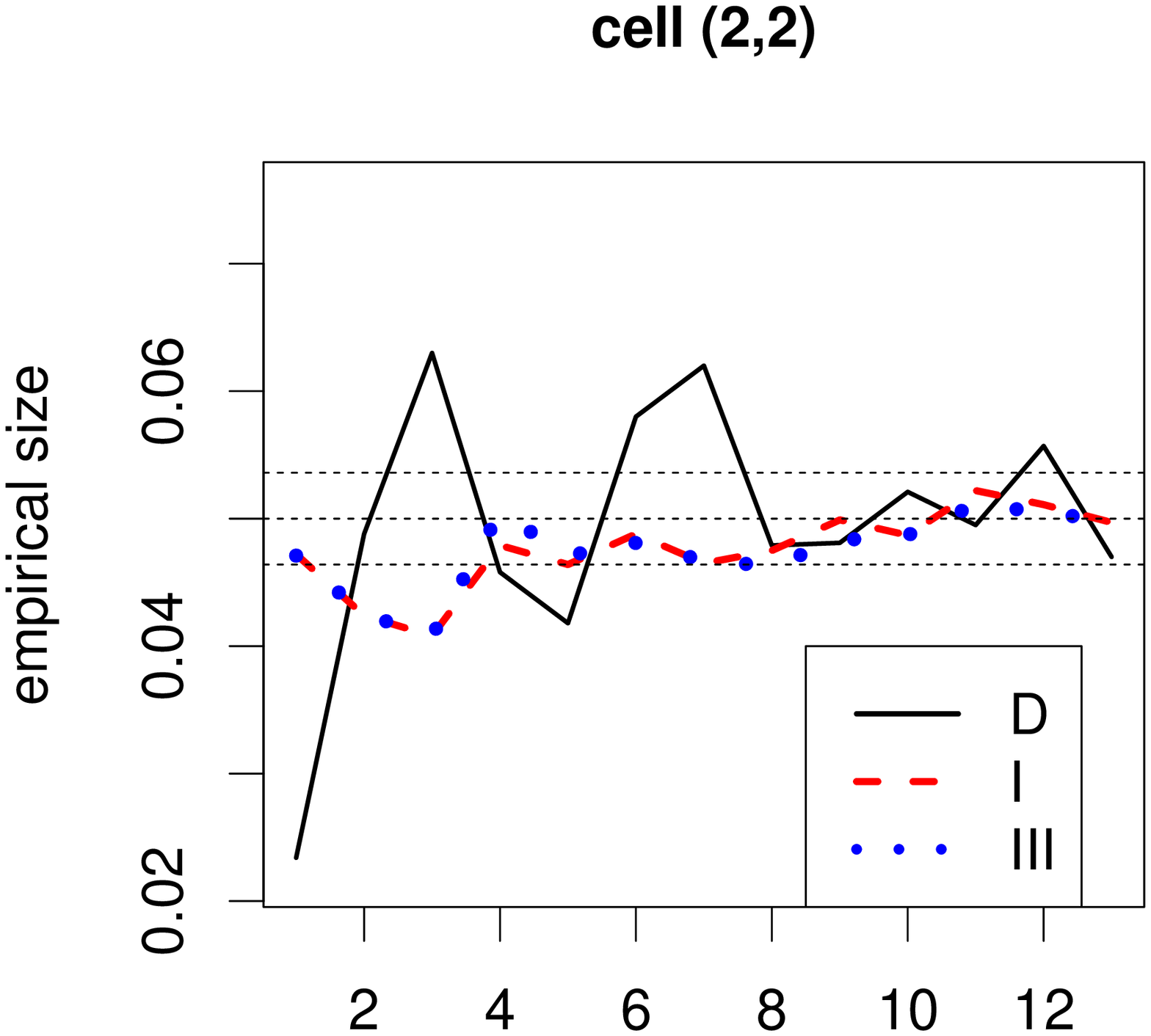} }}
\rotatebox{-90}{ \resizebox{2.1 in}{!}{\includegraphics{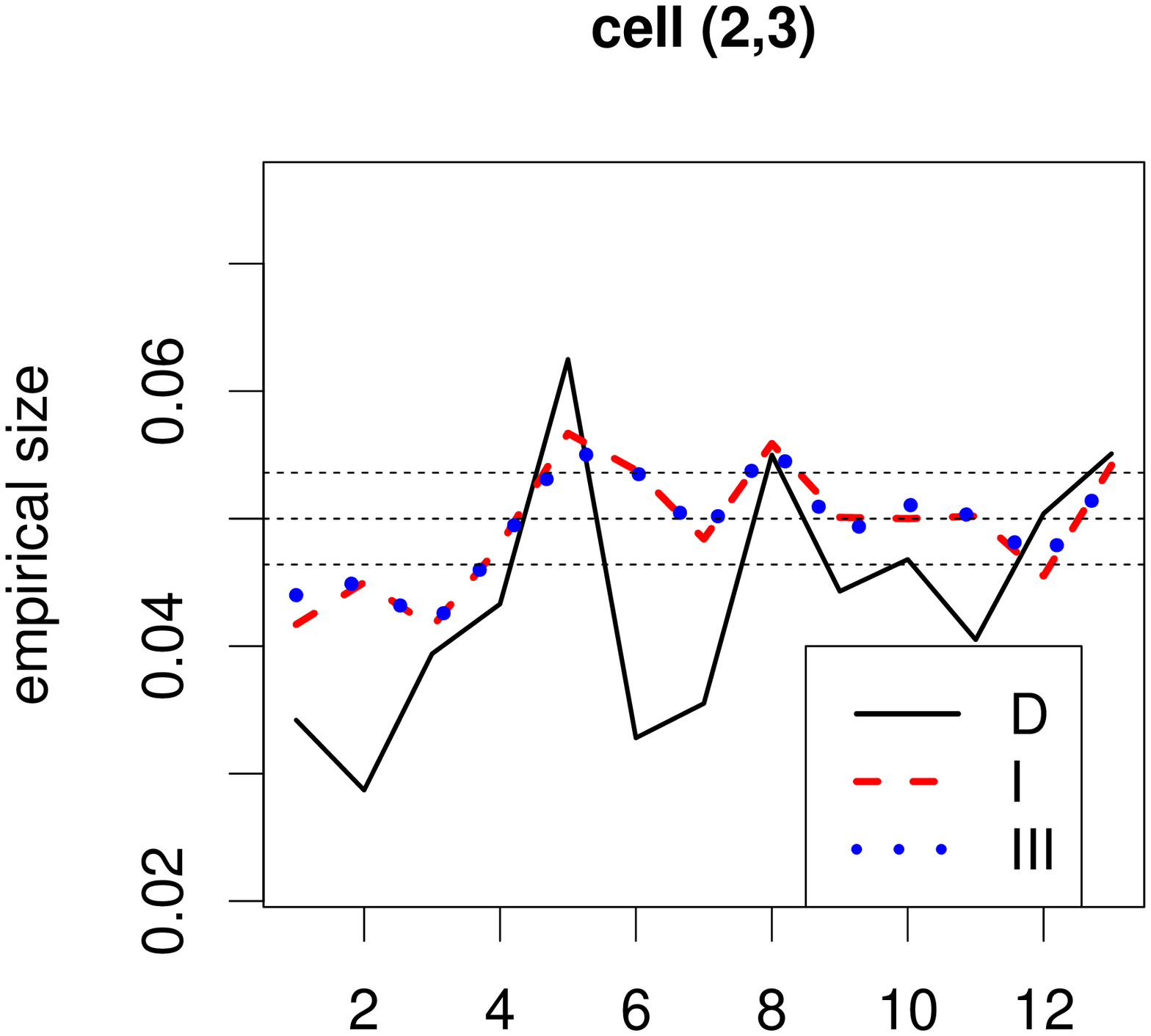} }}
\rotatebox{-90}{ \resizebox{2.1 in}{!}{\includegraphics{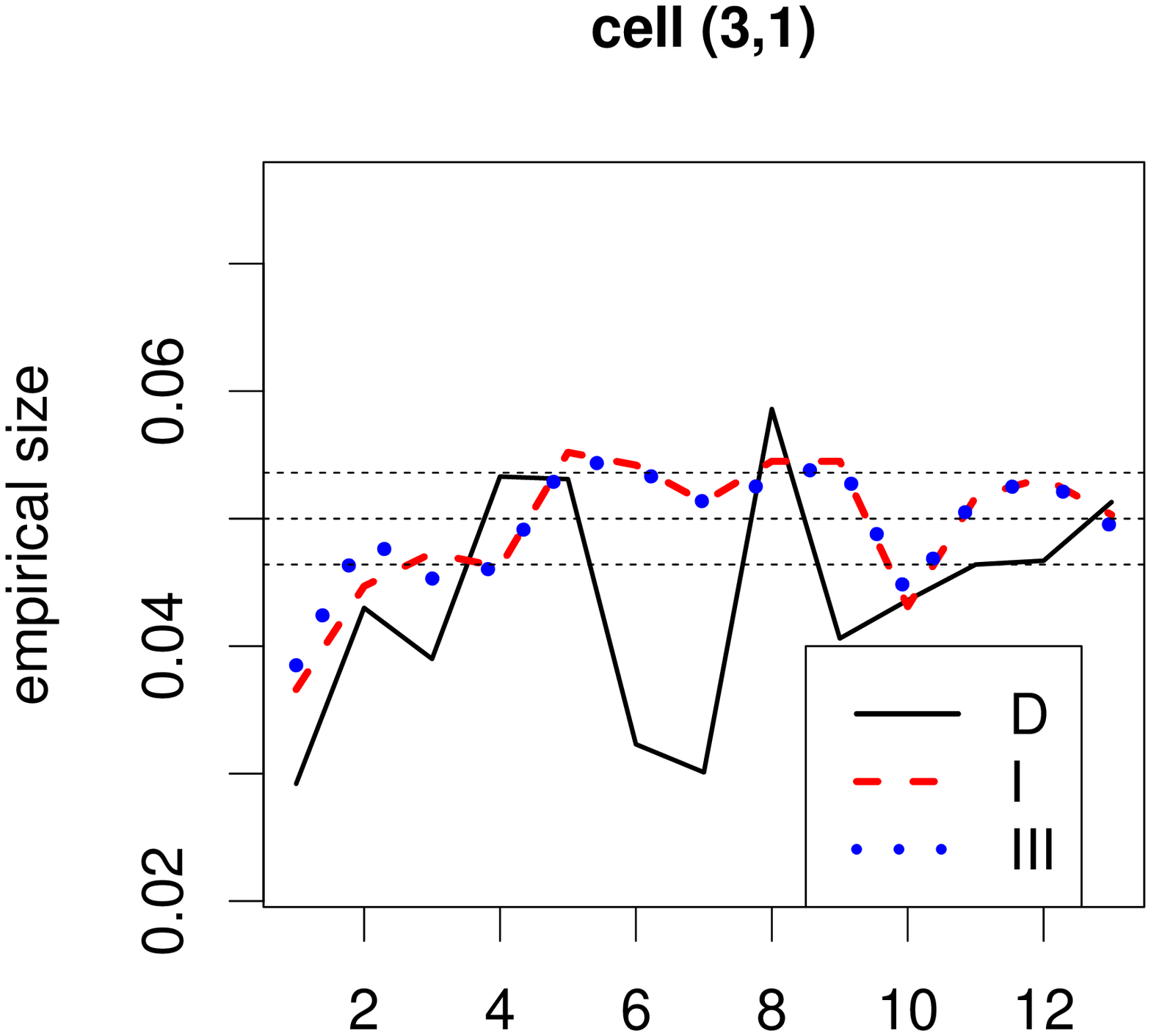} }}
\rotatebox{-90}{ \resizebox{2.1 in}{!}{\includegraphics{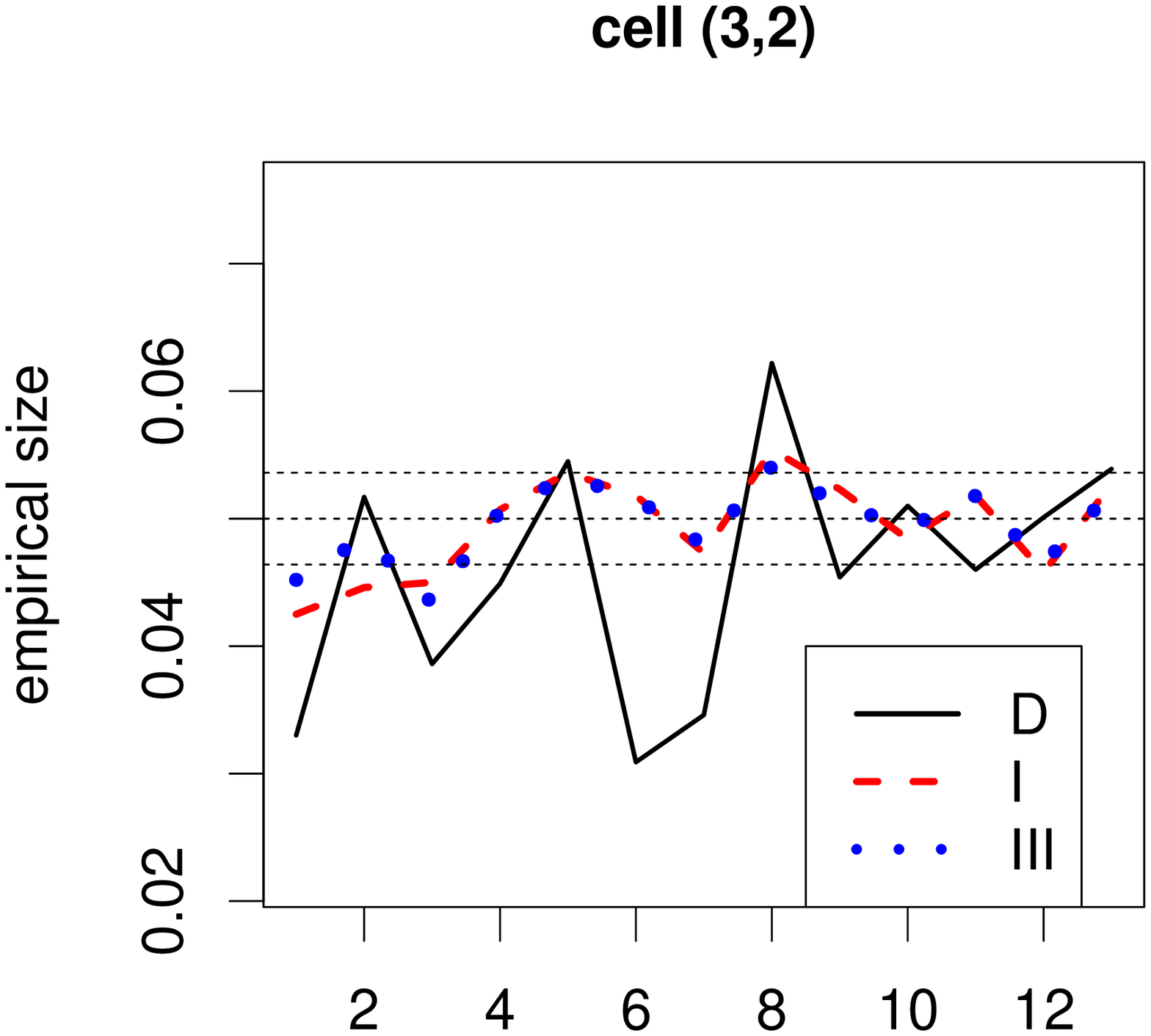} }}
\rotatebox{-90}{ \resizebox{2.1 in}{!}{\includegraphics{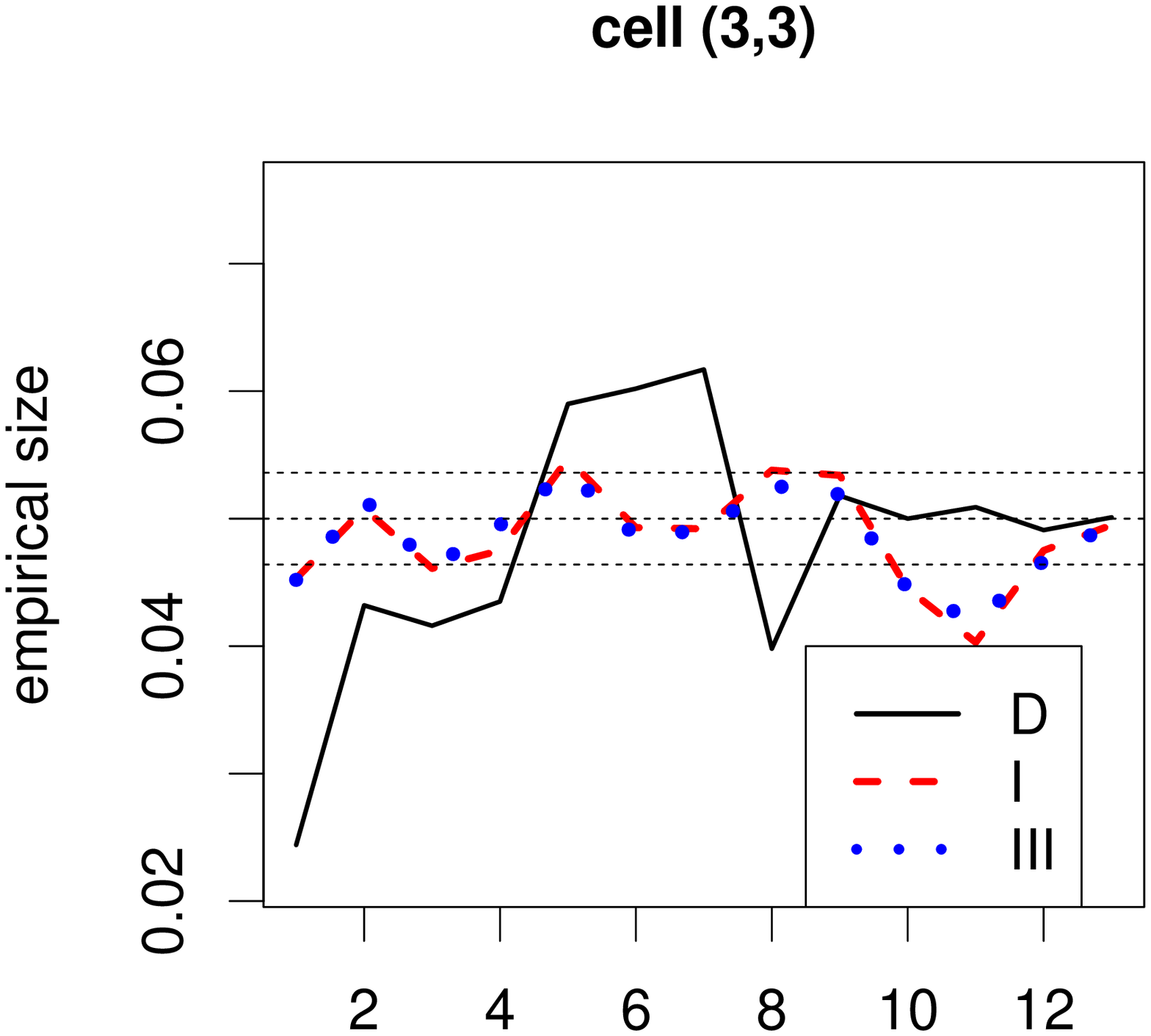} }}
\caption{
\label{fig:emp-size-RL1-cell-3cl}
The empirical size estimates of the cell-specific tests for cells $(1,1)-(3,3)$
under the RL case 1 in the three-class case.
The horizontal lines and legend labeling are as in Figure \ref{fig:emp-size-CSR-2cl}
and axis labeling are as in Figure \ref{fig:emp-size-CSR-cell-3cl}.
}
\end{figure}

\begin{figure} [hbp]
\centering
%\psfrag{Density}{ \Huge{\bf{Density}}}
Empirical Size Plots for the Cell-Specific Tests under RL case (2)\\
\rotatebox{-90}{ \resizebox{2.1 in}{!}{\includegraphics{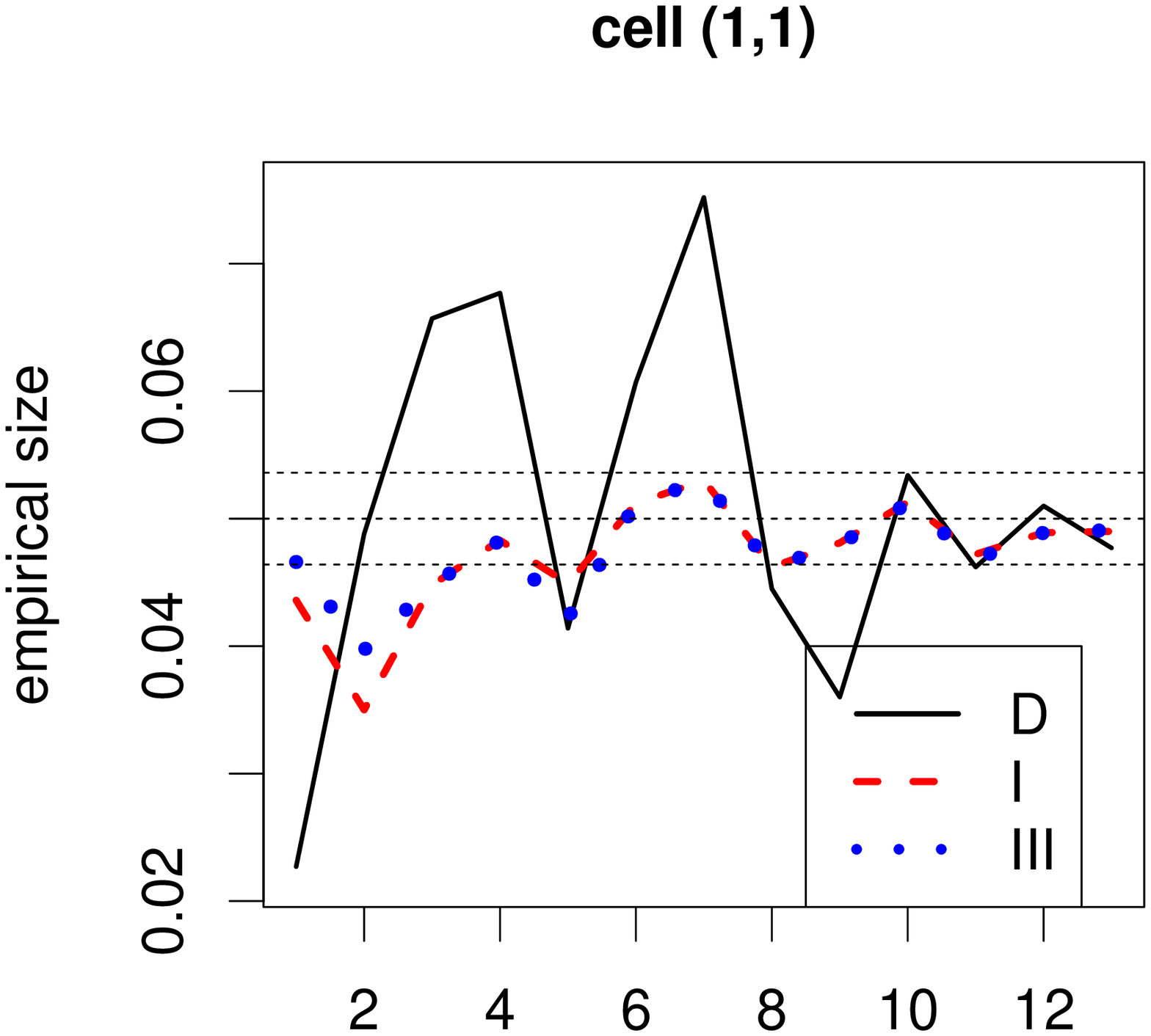} }}
\rotatebox{-90}{ \resizebox{2.1 in}{!}{\includegraphics{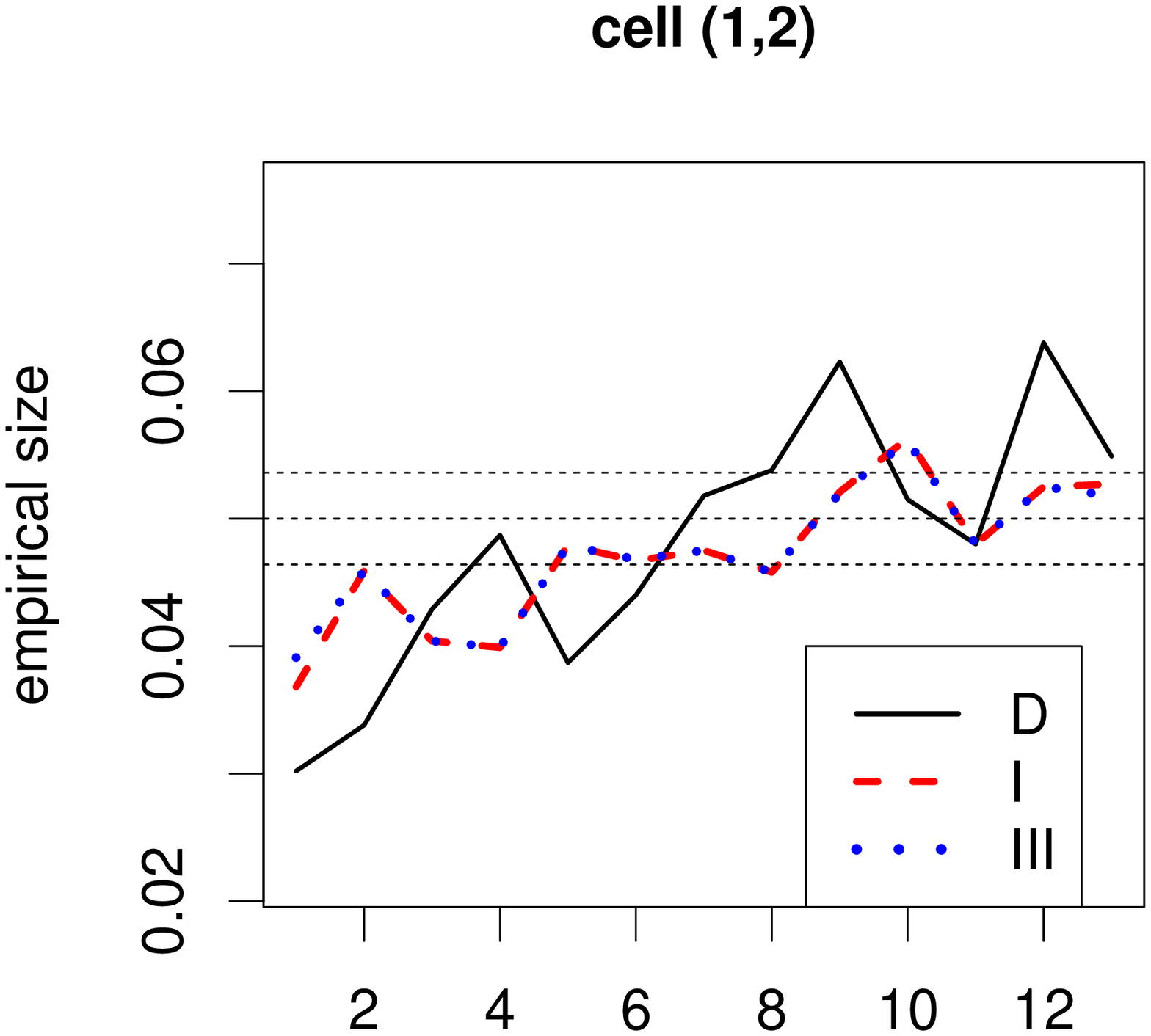} }}
\rotatebox{-90}{ \resizebox{2.1 in}{!}{\includegraphics{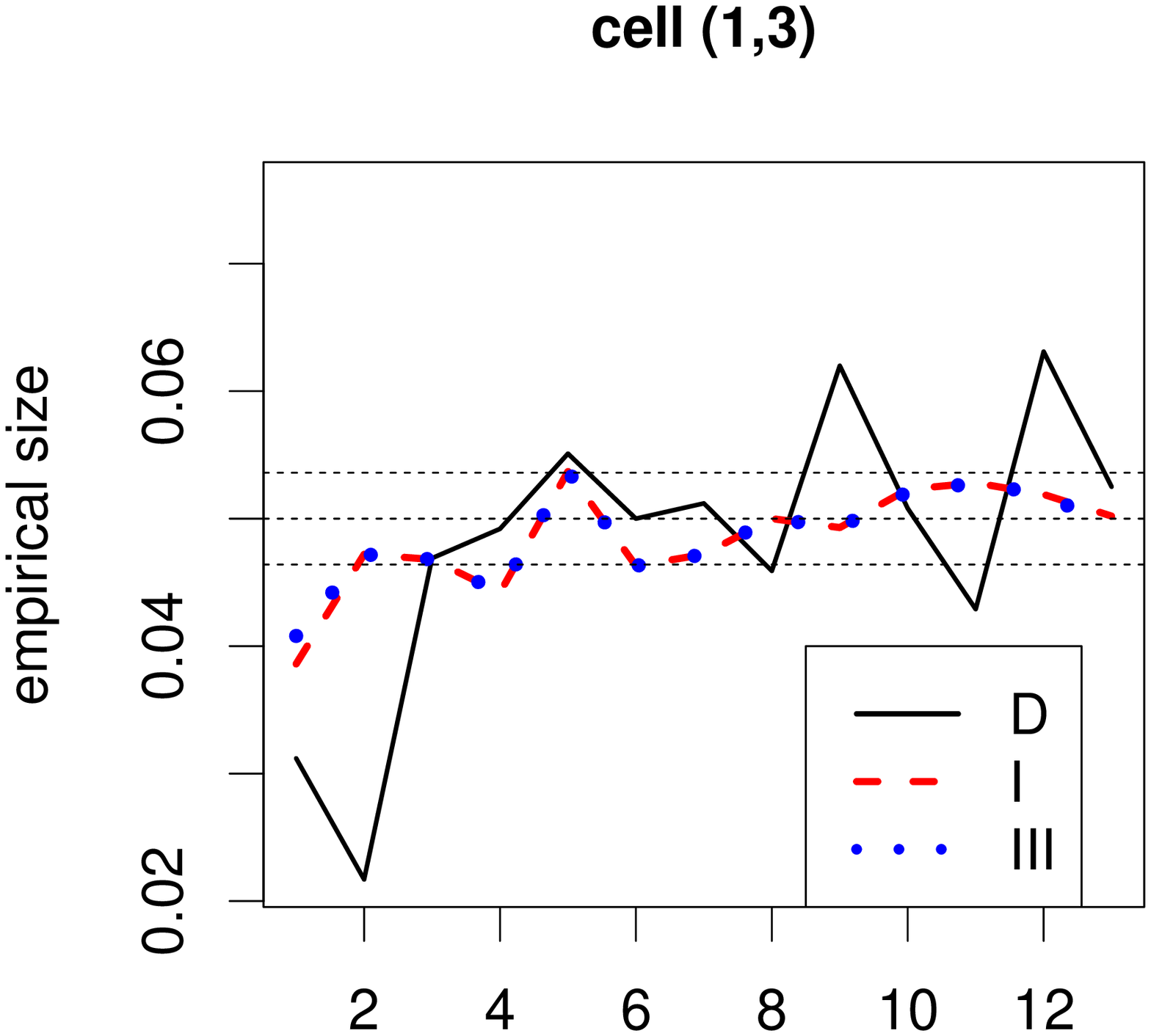} }}
\rotatebox{-90}{ \resizebox{2.1 in}{!}{\includegraphics{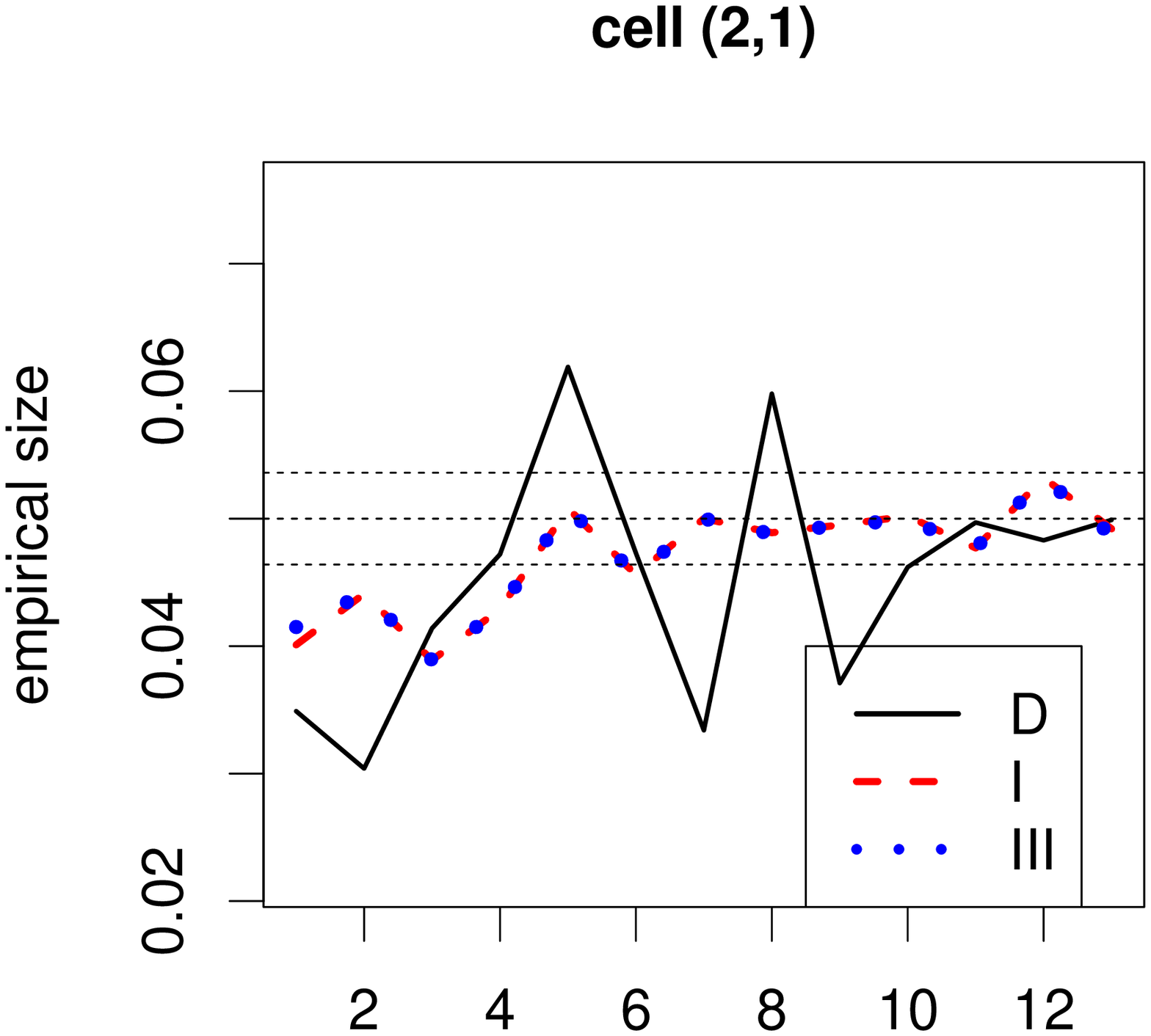} }}
\rotatebox{-90}{ \resizebox{2.1 in}{!}{\includegraphics{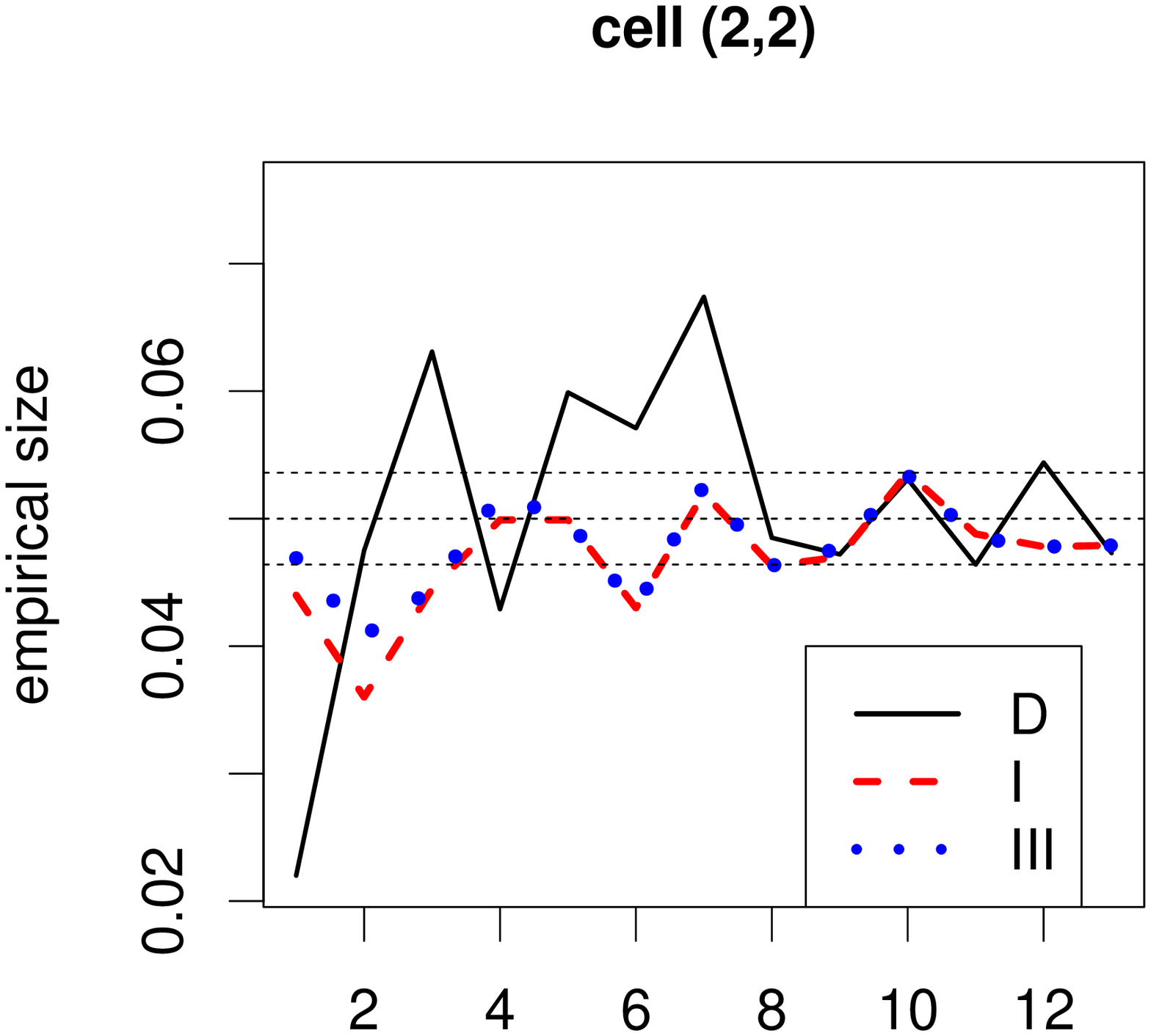} }}
\rotatebox{-90}{ \resizebox{2.1 in}{!}{\includegraphics{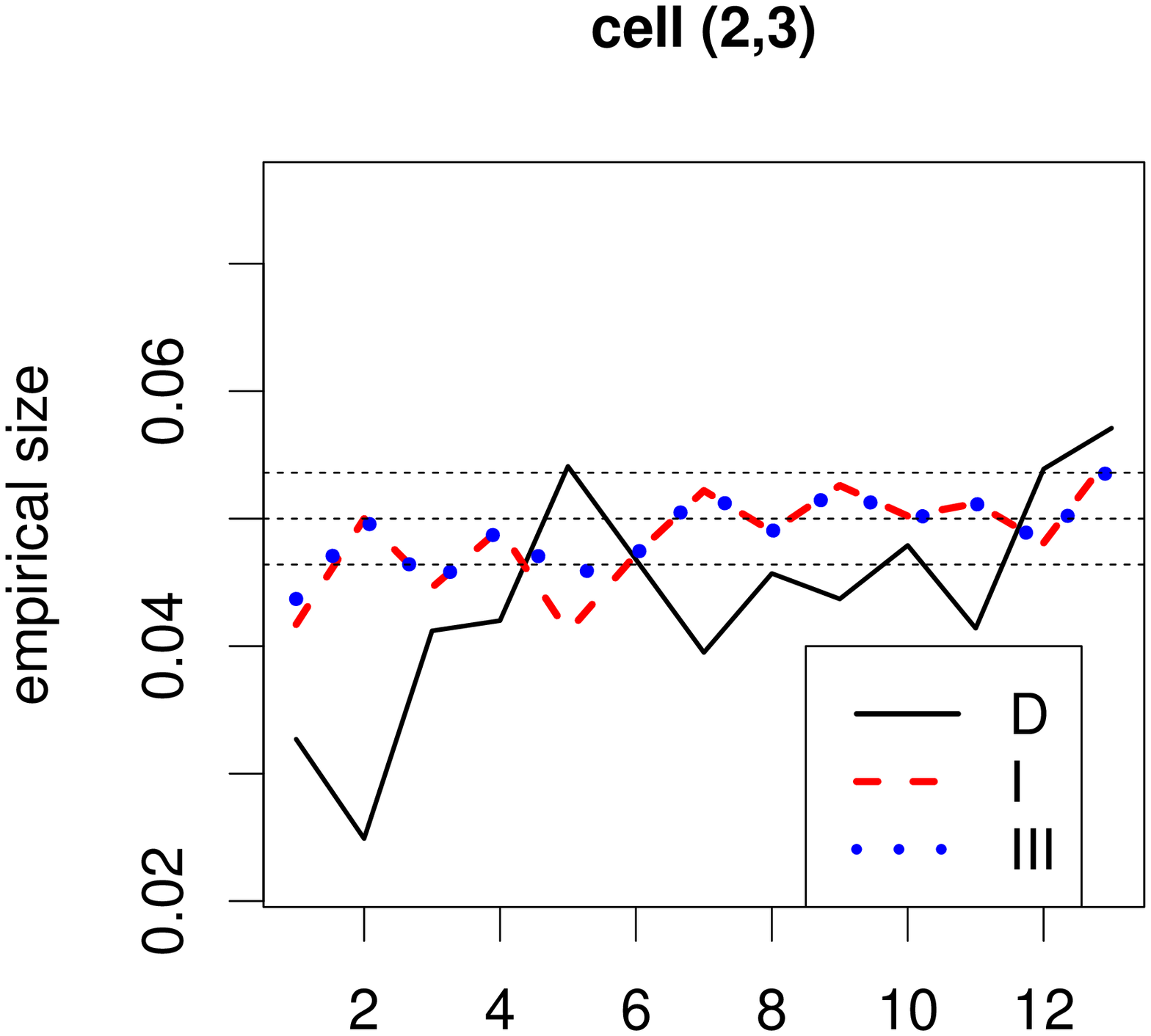} }}
\rotatebox{-90}{ \resizebox{2.1 in}{!}{\includegraphics{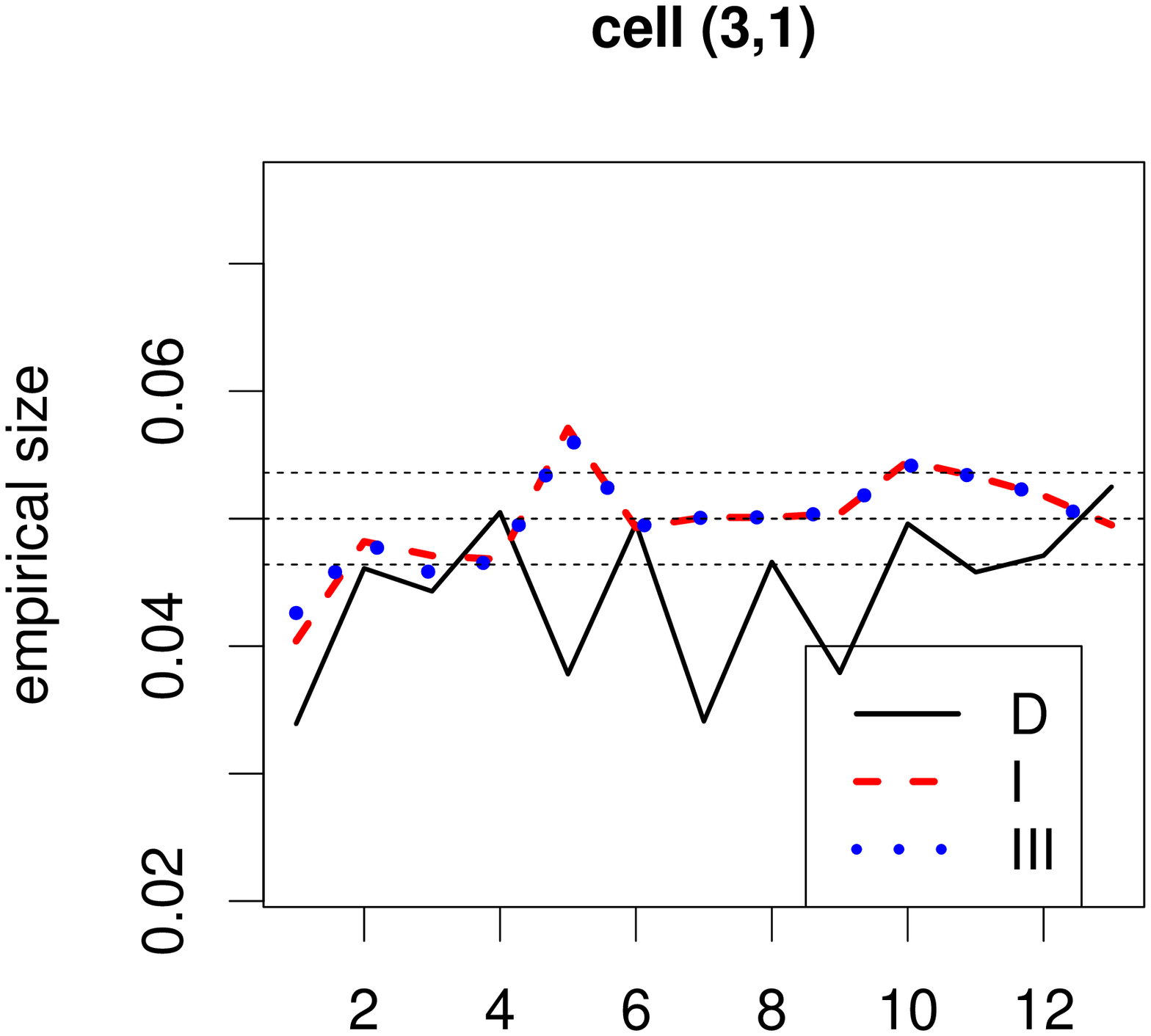} }}
\rotatebox{-90}{ \resizebox{2.1 in}{!}{\includegraphics{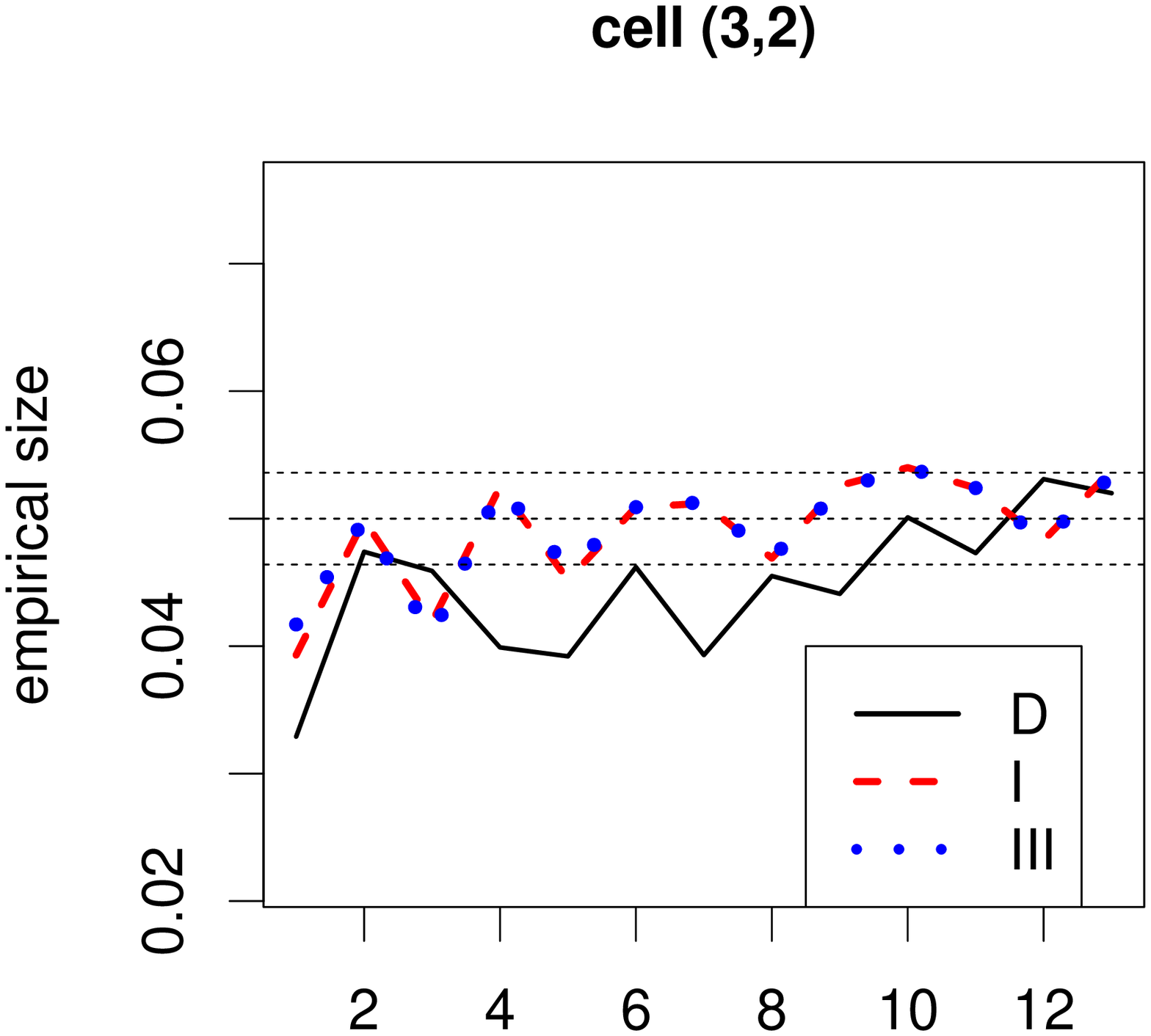} }}
\rotatebox{-90}{ \resizebox{2.1 in}{!}{\includegraphics{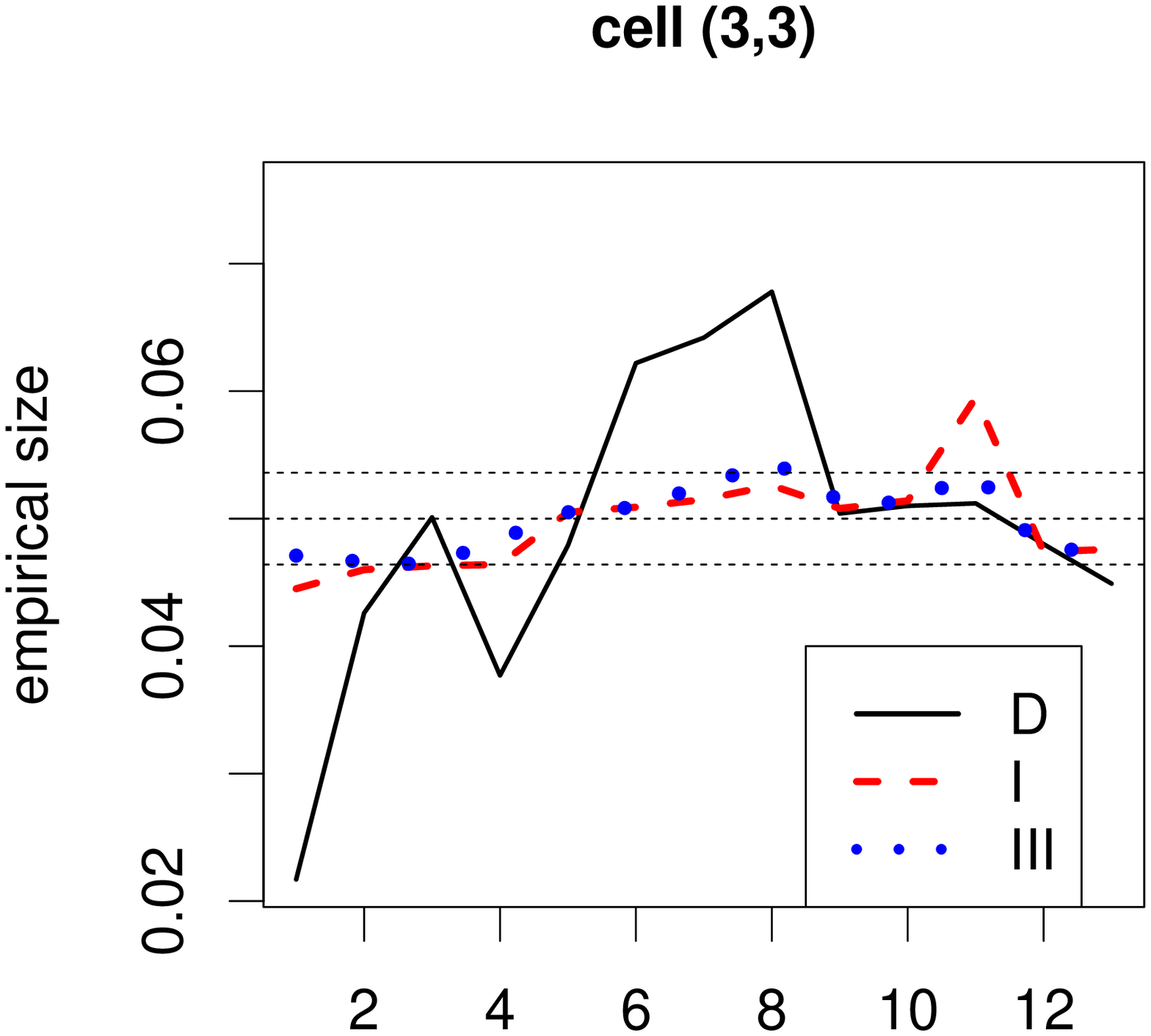} }}
\caption{
\label{fig:emp-size-RL2-cell-3cl}
The empirical size estimates of the cell-specific tests for cells $(1,1)-(3,3)$
under the RL case 2 in the three-class case.
The horizontal lines and legend labeling are as in Figure \ref{fig:emp-size-CSR-2cl}
and axis labeling are as in Figure \ref{fig:emp-size-CSR-cell-3cl}.
}
\end{figure}

\begin{figure} [hbp]
\centering
%\psfrag{Density}{ \Huge{\bf{Density}}}
Empirical Size Plots for the Overall Tests under RL cases 1 and 2\\
\rotatebox{-90}{ \resizebox{2.5 in}{!}{\includegraphics{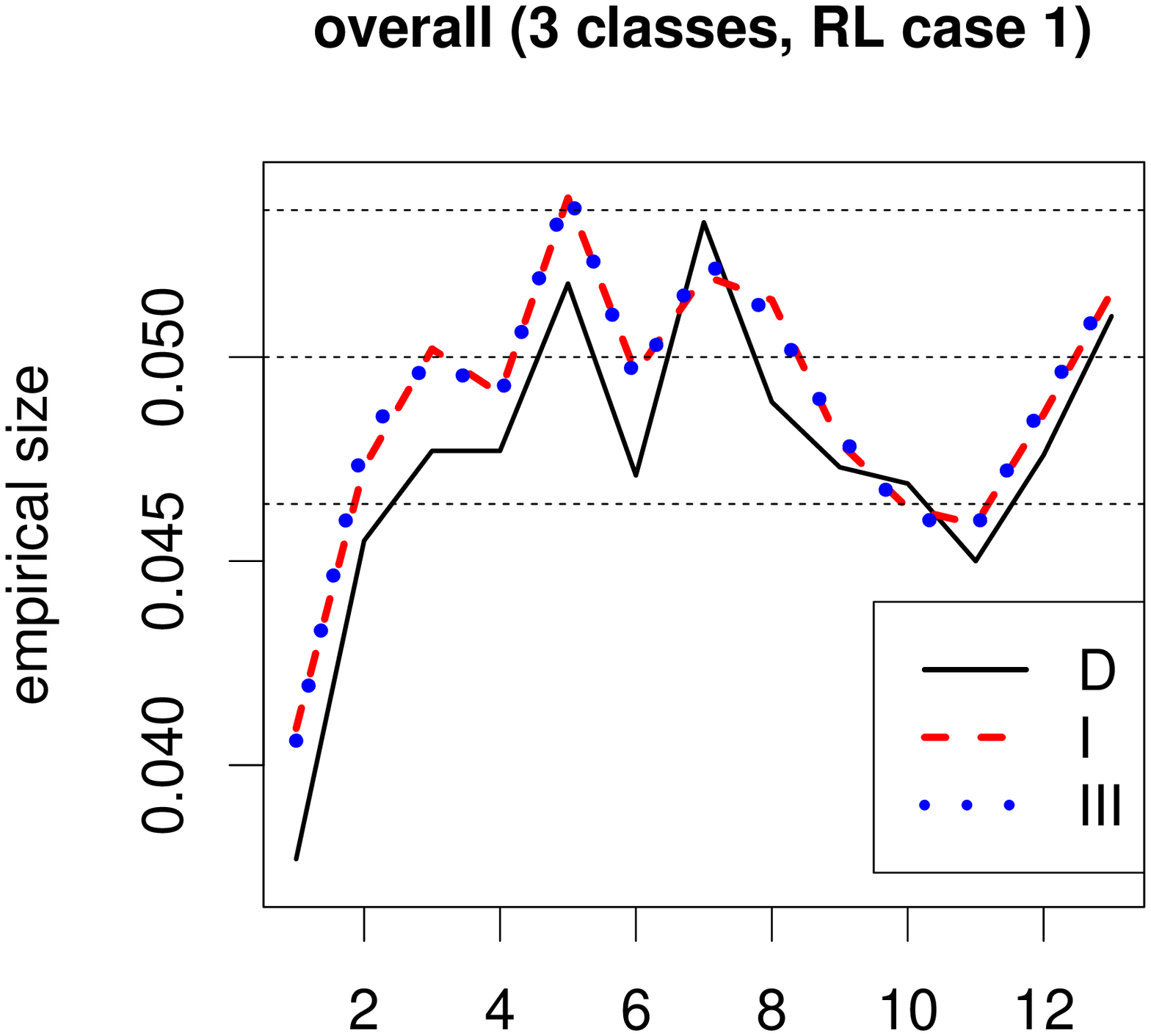} }}
\rotatebox{-90}{ \resizebox{2.5 in}{!}{\includegraphics{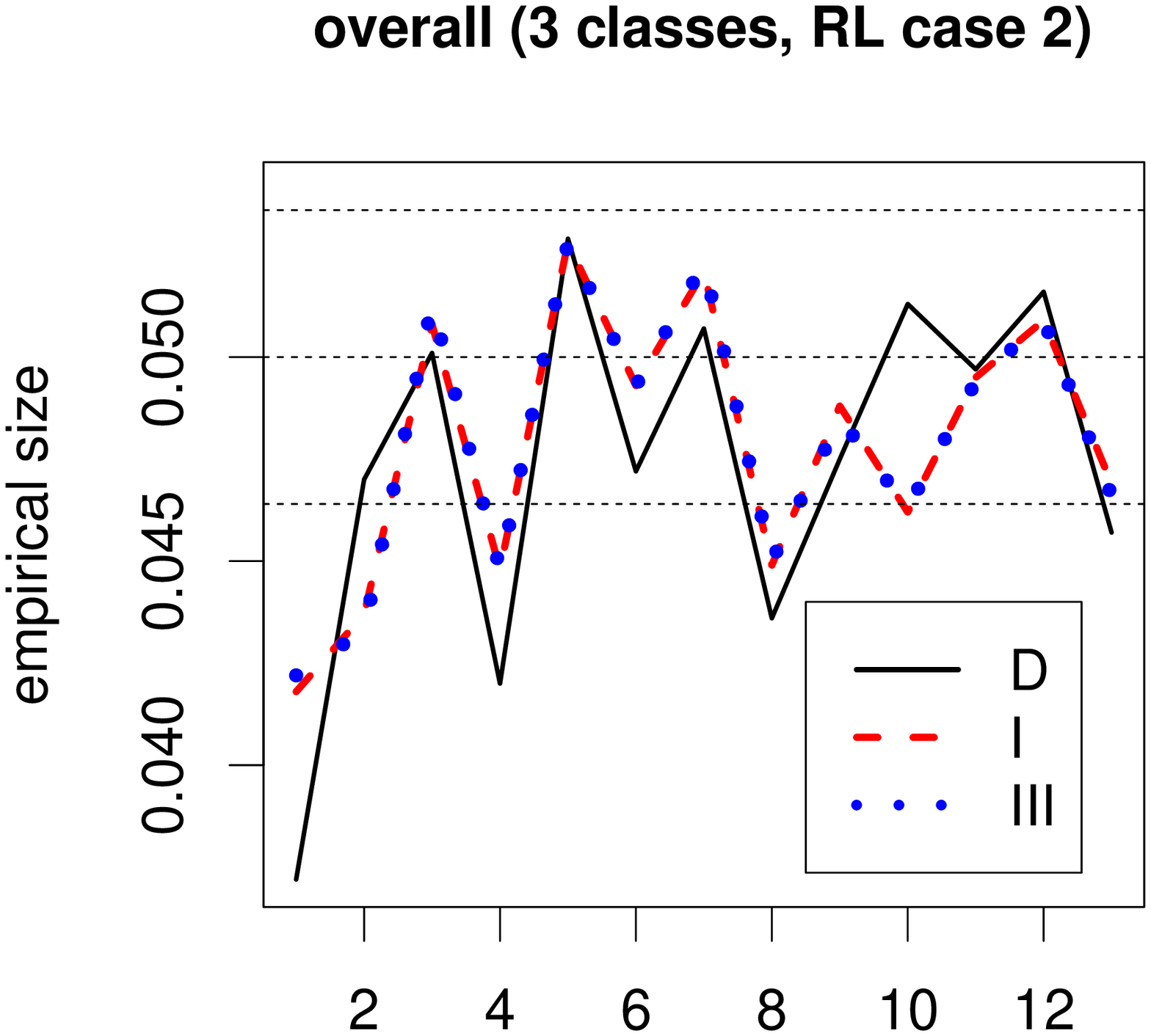} }}
\caption{
\label{fig:emp-size-RL1-overall-3cl}
The empirical size estimates of the overall tests under the RL cases 1 and 2
in the three-class case.
The horizontal lines and legend labeling are as in Figure \ref{fig:emp-size-CSR-2cl}
and axis labeling are as in Figure \ref{fig:emp-size-CSR-cell-3cl}.
}
\end{figure}

We present the empirical significance
levels under RL case 1 in Figure \ref{fig:emp-size-RL1-cell-3cl}
and under RL case 2 in Figure \ref{fig:emp-size-RL2-cell-3cl}.
Under both RL cases,
type I and III cell-specific tests perform better in terms of empirical size
(i.e., their empirical sizes are closer to the desired level)
and less affected by smaller cell counts and unbalanced class sizes,
compared to Dixon's cell-specific tests.
Empirical sizes for the overall tests under RL cases 1 and 2 are presented in Figure \ref{fig:emp-size-RL1-overall-3cl}.
Type I and III cell-specific tests are closer to the nominal level compared to Dixon's test.

\section{Empirical Power Analysis in the Two-Class Case}
\label{sec:emp-power-2Cl}
We consider three cases for each of segregation and association alternatives
in the two-class case.

\subsection{Empirical Power Analysis under Segregation of Two Classes}
\label{sec:power-comp-seg-2Cl}
For the segregation alternatives, we generate
$X_i \stackrel{iid}{\sim} \U(S_1)$ and
$Y_j \stackrel{iid}{\sim} \U(S_2)$
where $S_1=(0,1-s)\times(0,1-s)$ and $S_2=(s,1)$
for $i=1,\ldots,n_1$ and $j=1,\ldots,n_2$ and $s \in (0,1)$.
We consider the following three segregation alternatives:
\begin{equation}
\label{eqn:seg-alt-2Cl}
H_S^I: s=1/6, \;\;\; H_S^{II}: s=1/4, \text{ and } H_S^{III}: s=1/3.
\end{equation}

Notice that, the level of segregation increases as $s$ increases;
that is, segregation gets stronger from $H_S^I$ to $H_S^{III}$.
We calculate the power estimates using the asymptotic critical values
based on the standard normal distribution
for the cell-specific tests
and the corresponding $\chi^2$-distributions for the overall tests.

\begin{figure} [hbp]
\centering
%\psfrag{Density}{ \Huge{\bf{Density}}}
Empirical Power Estimates of the NNCT-Tests under $H^I_S$ \\
\rotatebox{-90}{ \resizebox{2.1 in}{!}{\includegraphics{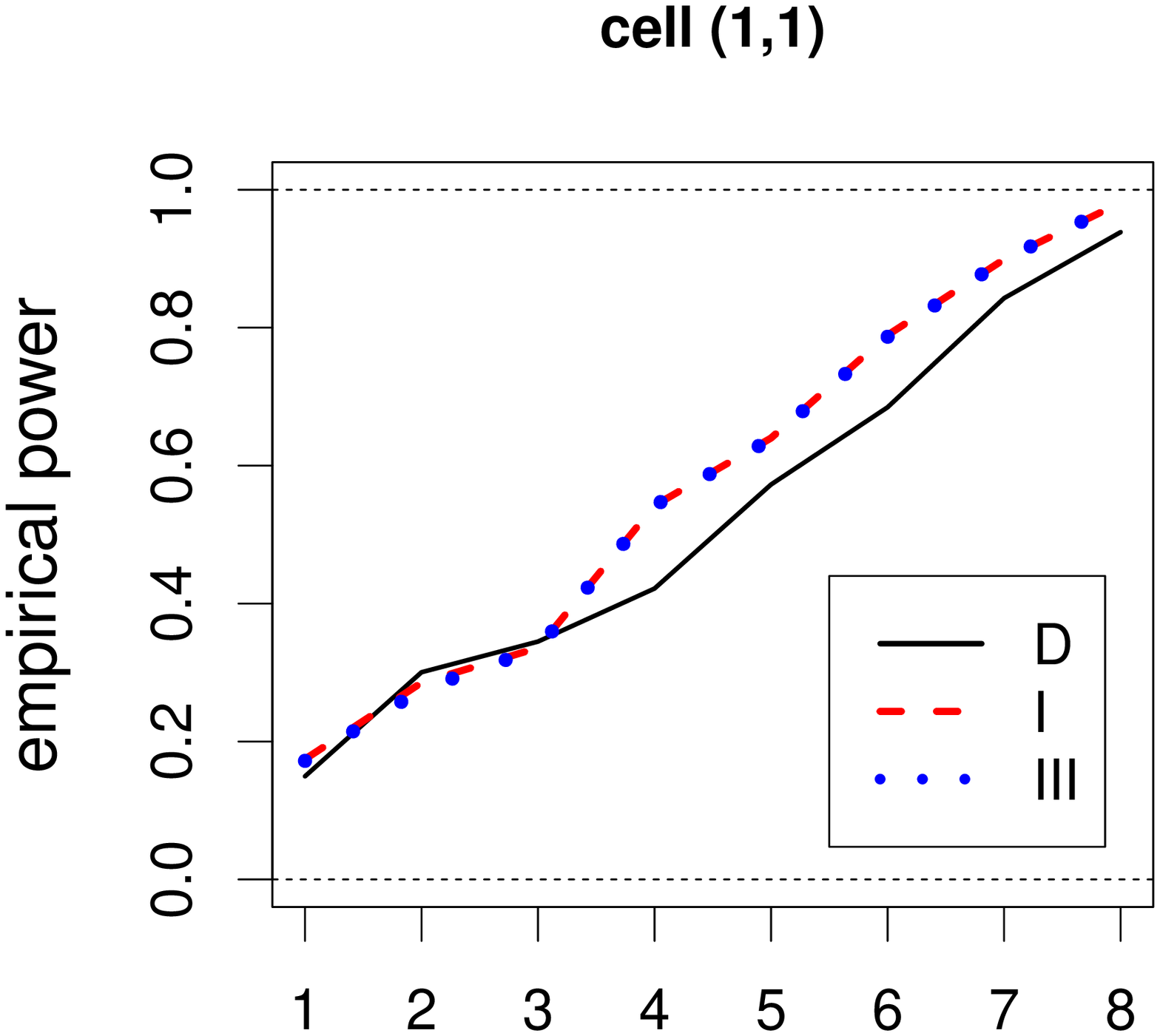} }}
\rotatebox{-90}{ \resizebox{2.1 in}{!}{\includegraphics{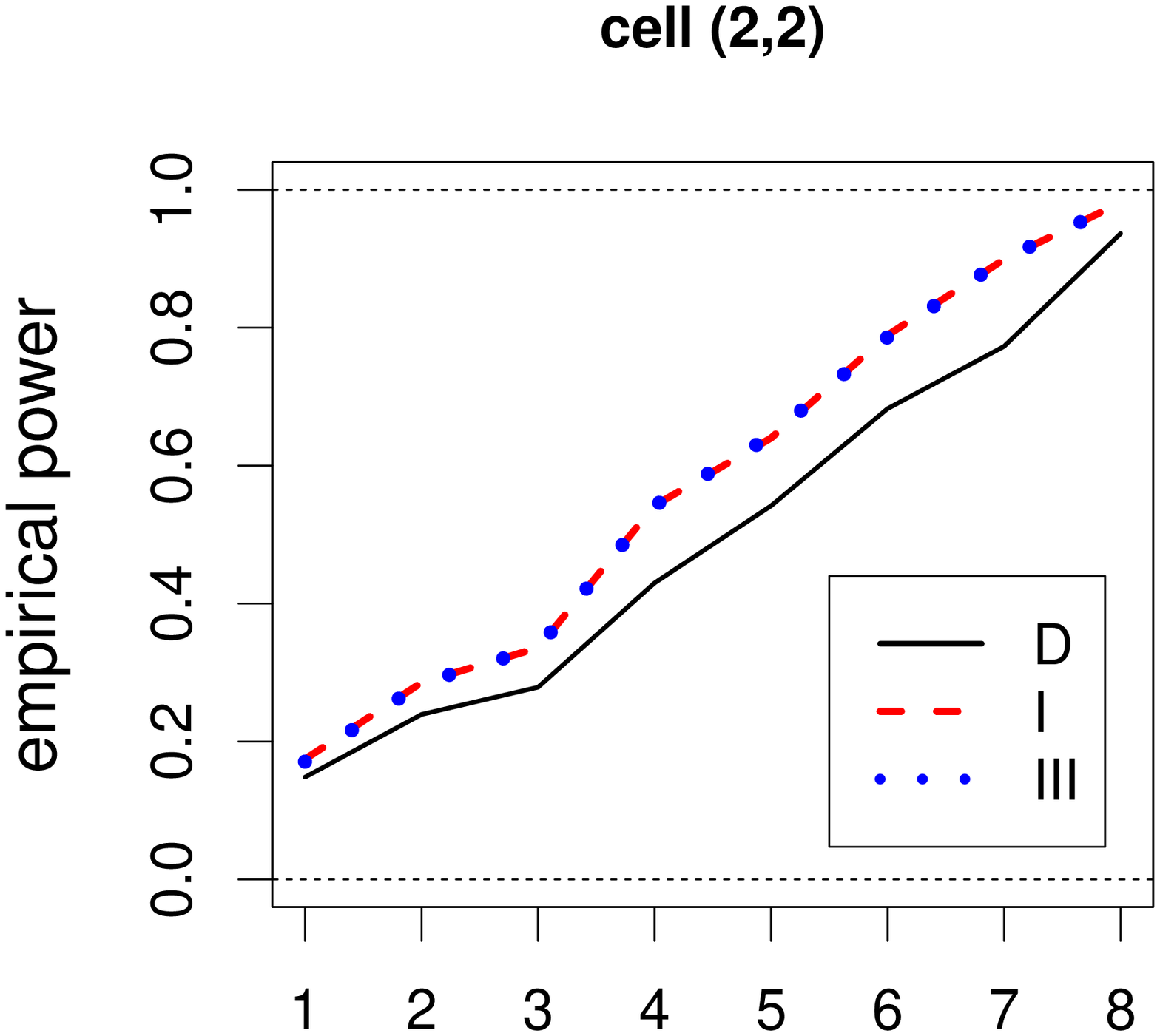} }}
\rotatebox{-90}{ \resizebox{2.1 in}{!}{\includegraphics{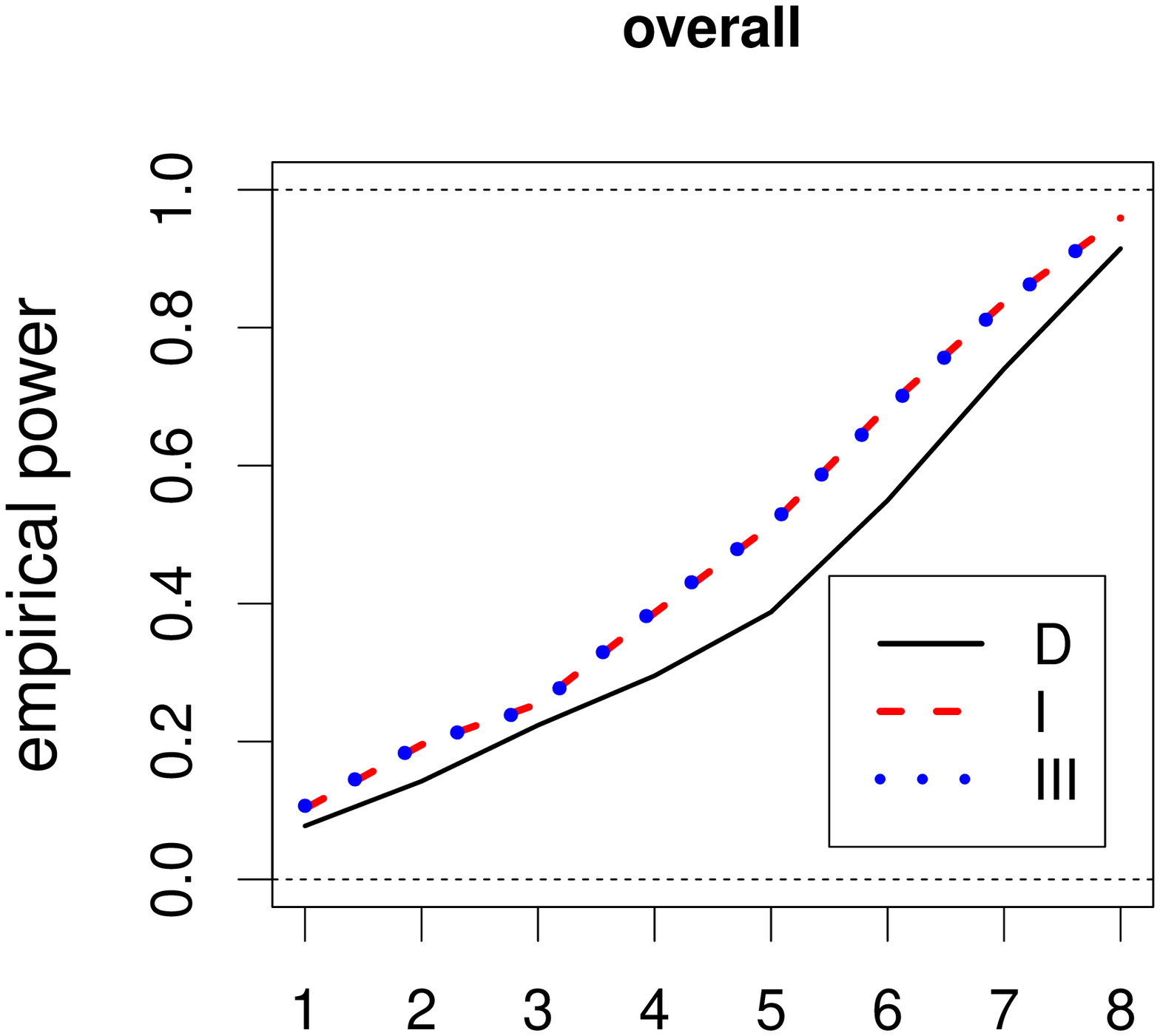} }}
Power Estimates under $H^{II}_S$\\
\rotatebox{-90}{ \resizebox{2.1 in}{!}{\includegraphics{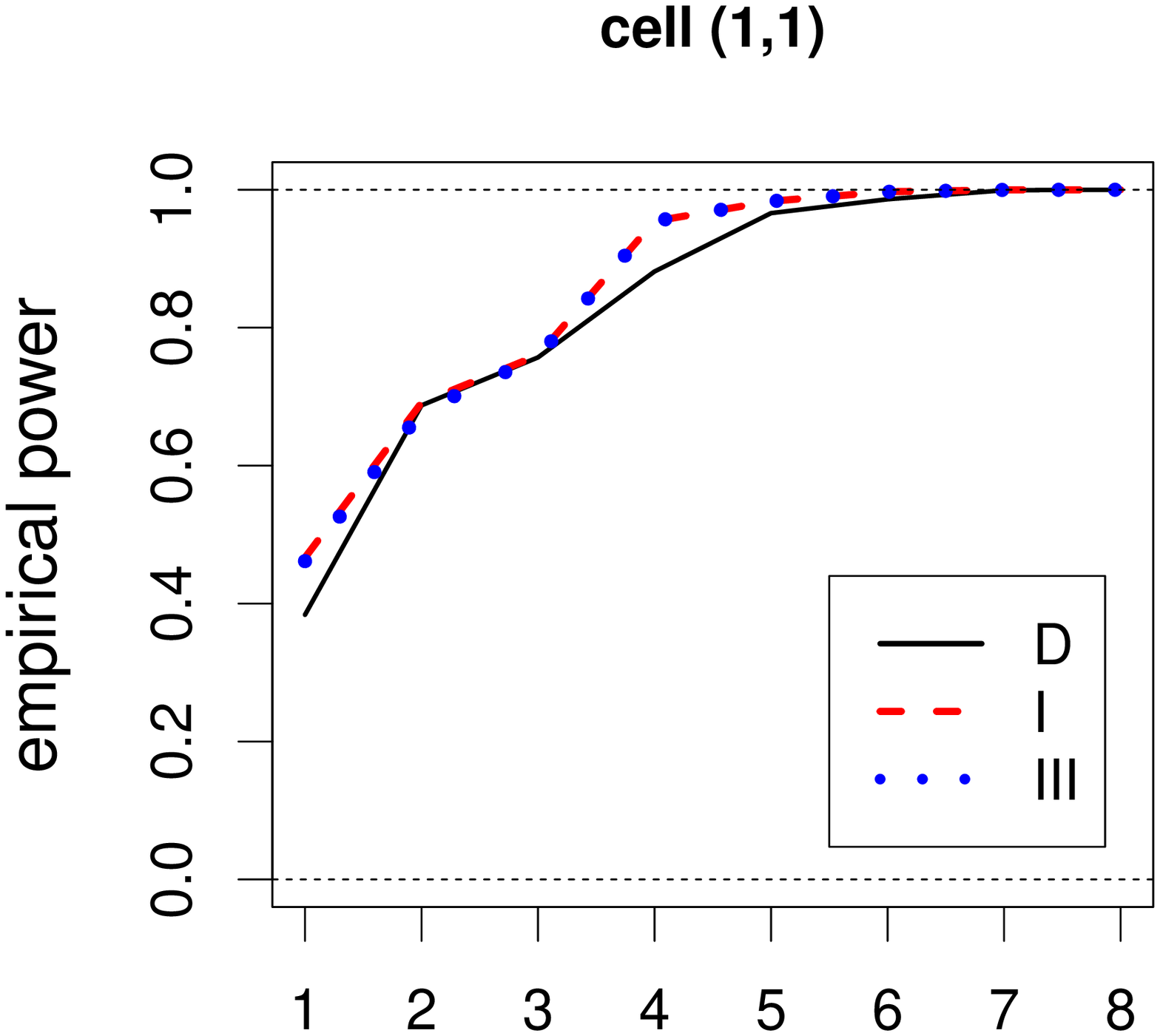} }}
\rotatebox{-90}{ \resizebox{2.1 in}{!}{\includegraphics{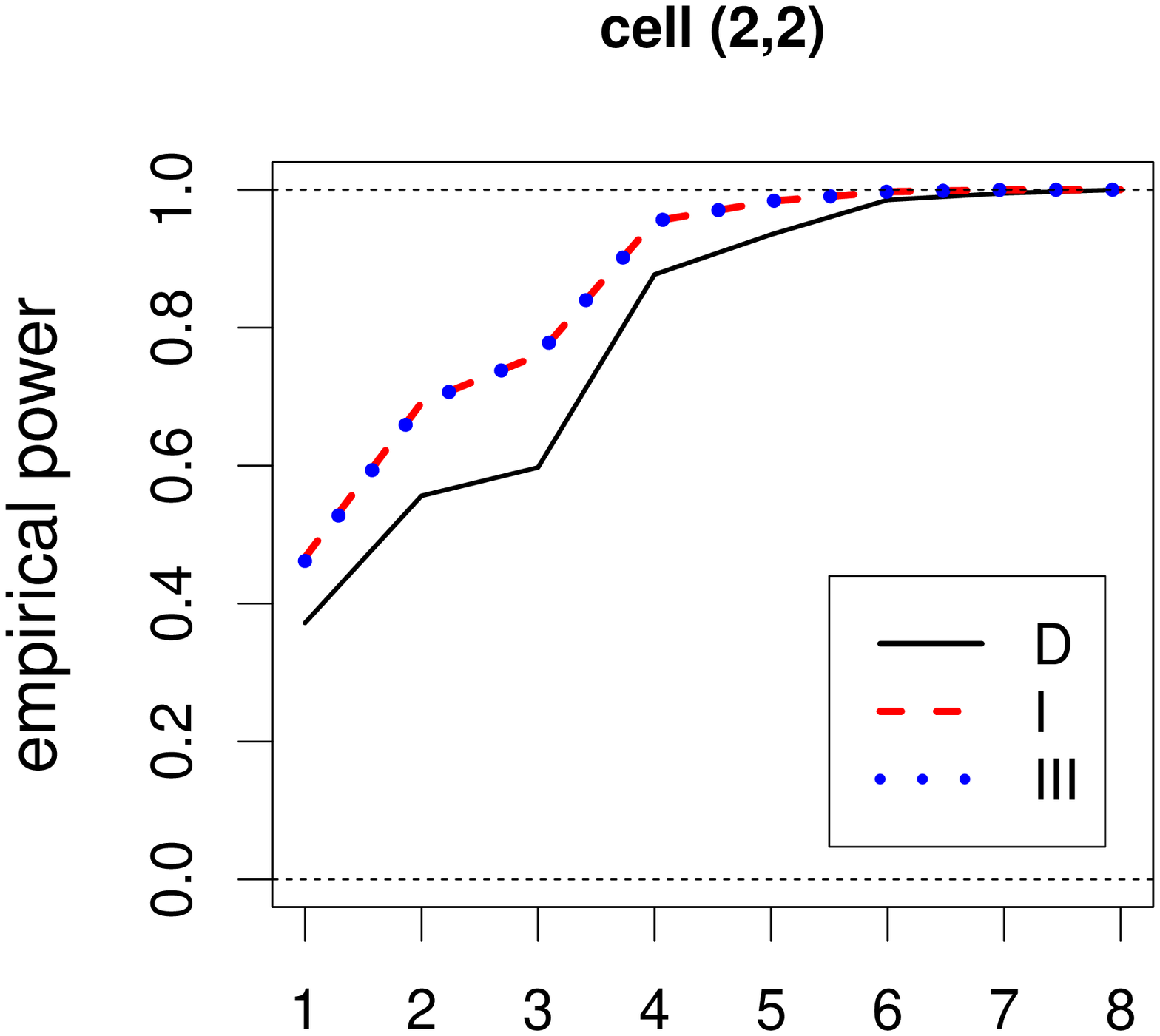} }}
\rotatebox{-90}{ \resizebox{2.1 in}{!}{\includegraphics{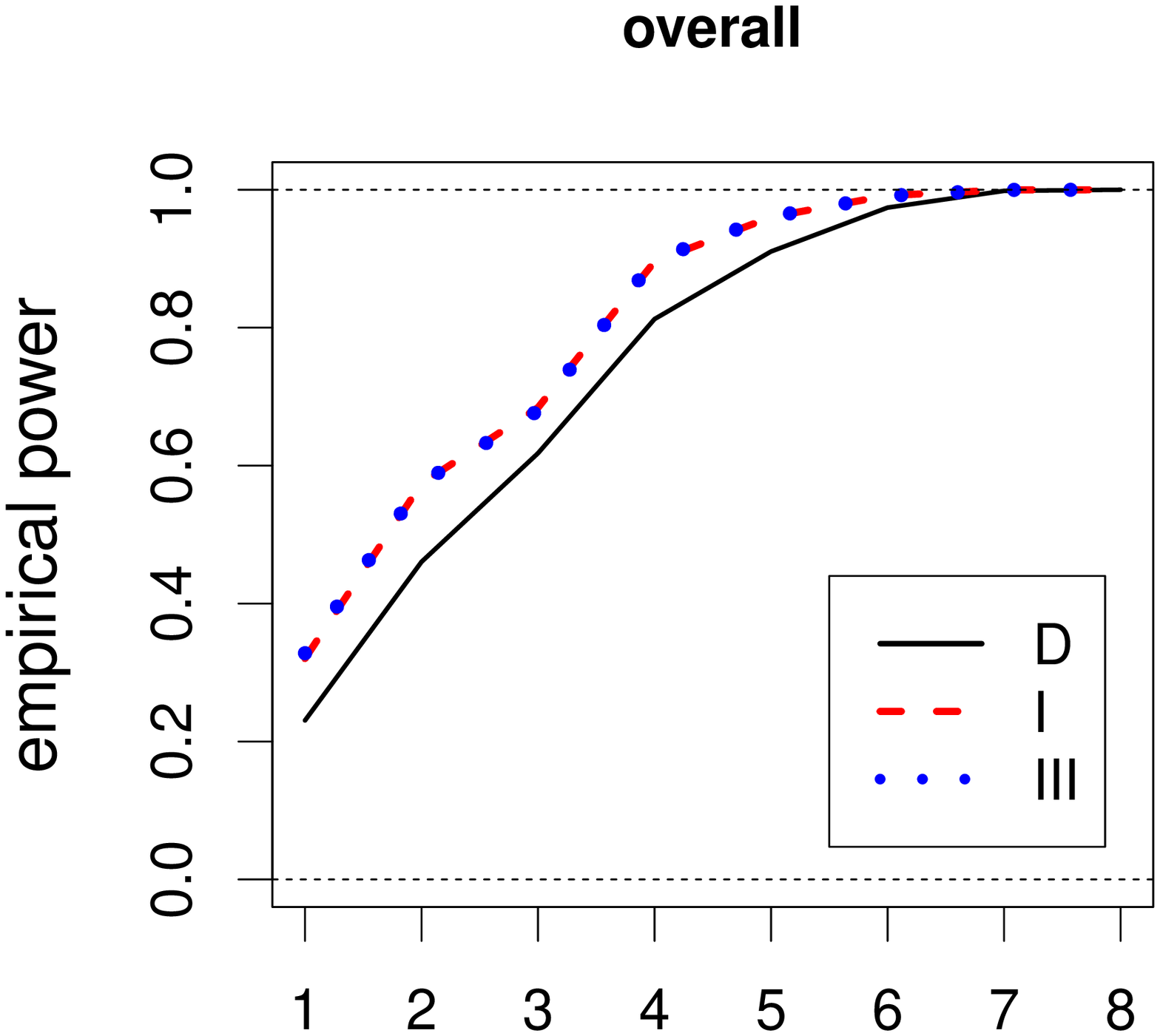} }}
Power Estimates under $H^{III}_S$\\
\rotatebox{-90}{ \resizebox{2.1 in}{!}{\includegraphics{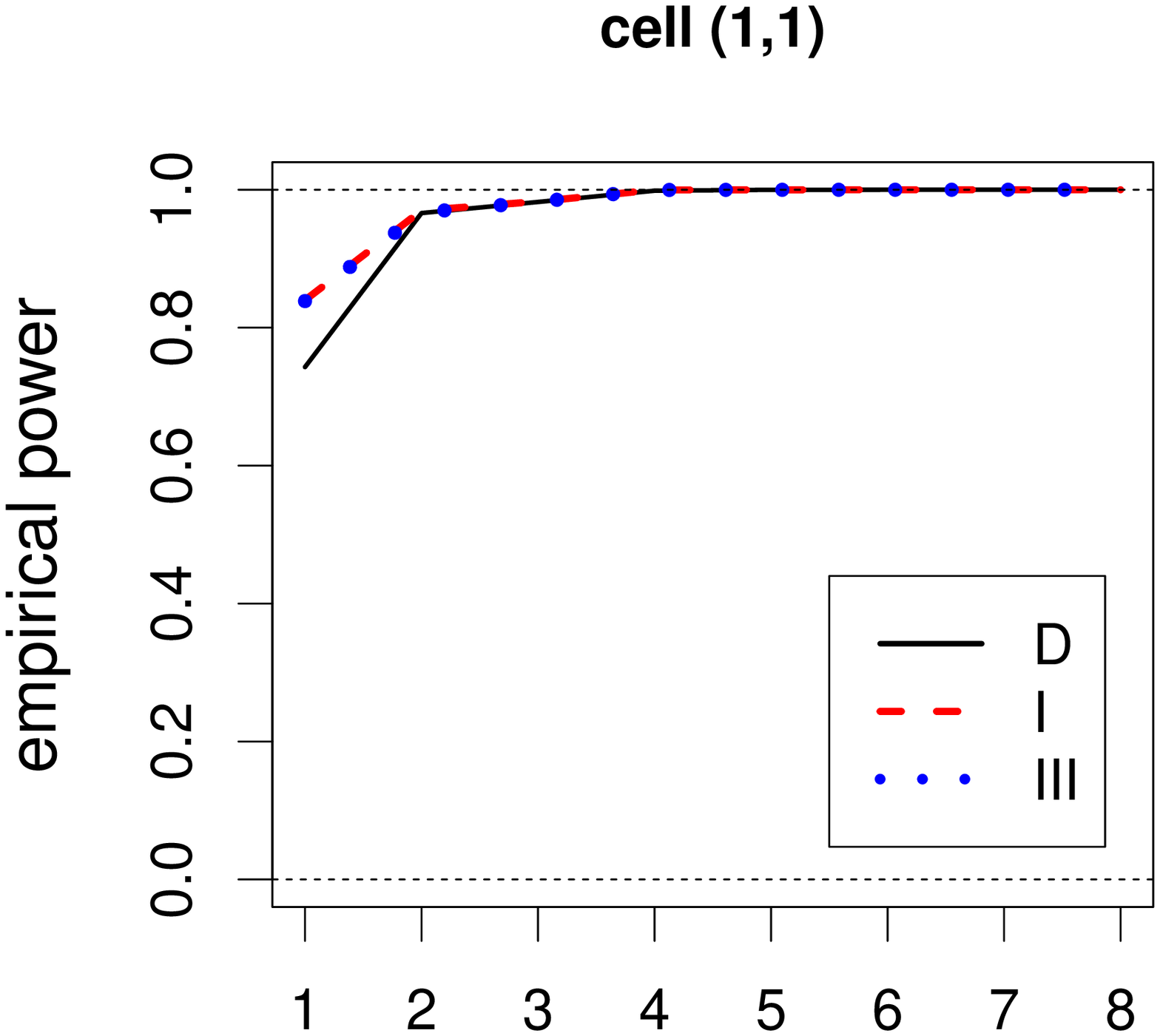} }}
\rotatebox{-90}{ \resizebox{2.1 in}{!}{\includegraphics{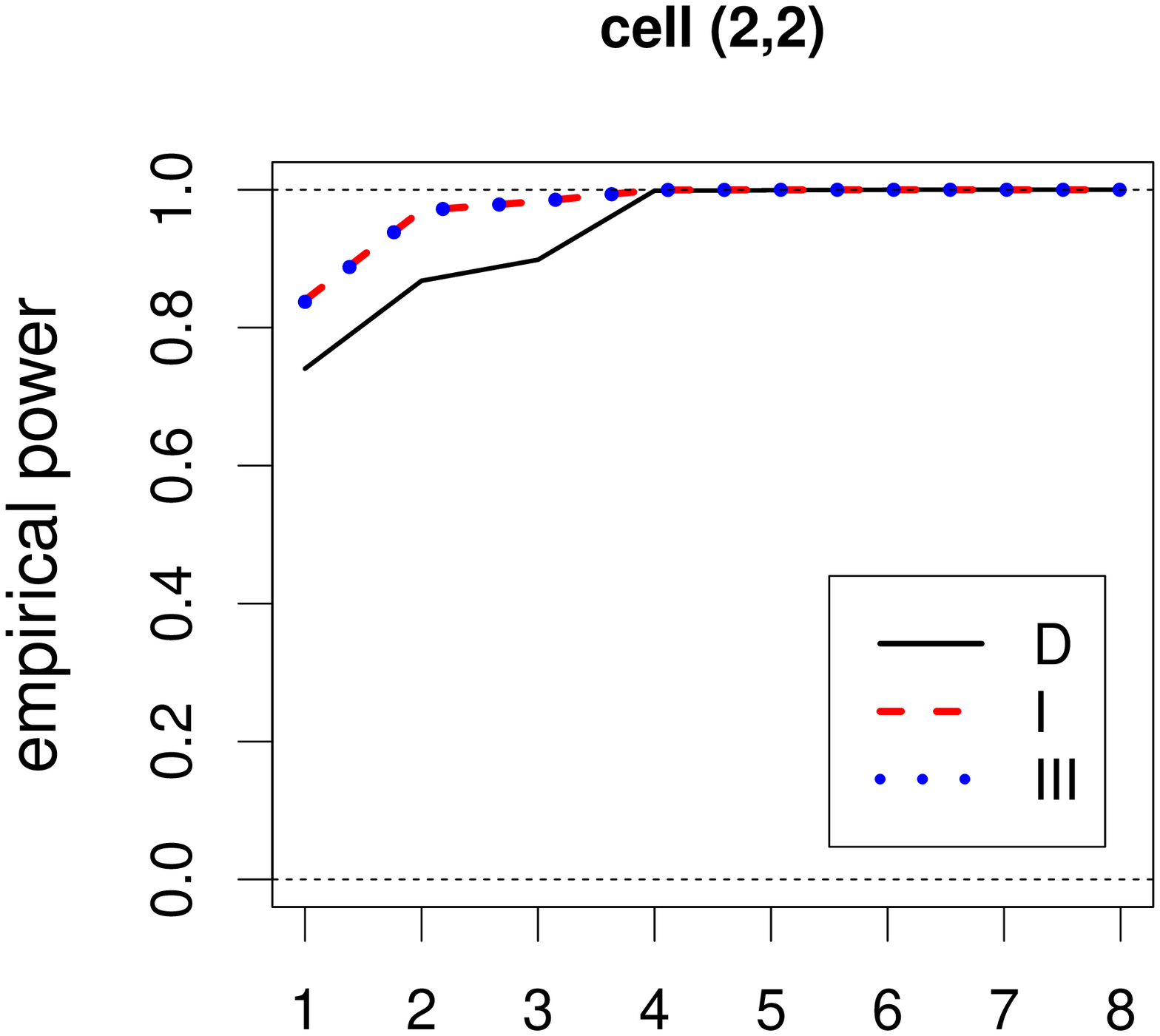} }}
\rotatebox{-90}{ \resizebox{2.1 in}{!}{\includegraphics{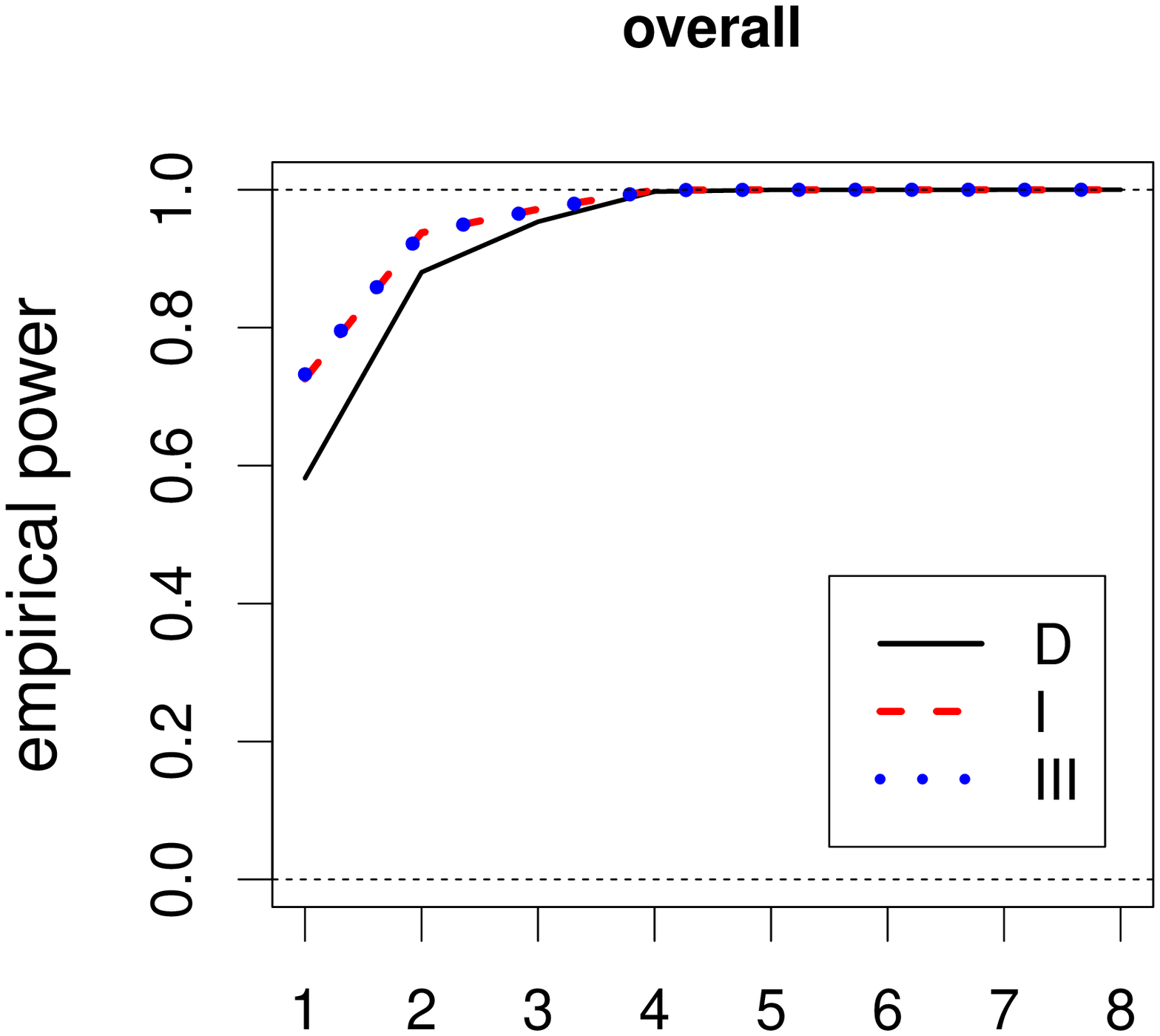} }}
 \caption{
\label{fig:power-seg-2cl}
The empirical power estimates for the cell-specific (left and middle columns)
and the overall tests (right column) under the segregation alternatives, $H_S^I - H_S^{III}$
in the two-class case.
The horizontal axis labels and legend labeling are as in Figure \ref{fig:emp-size-CSR-2cl}.
}
\end{figure}

The power estimates based on the asymptotic critical values
are presented in Figure \ref{fig:power-seg-2cl}.
We omit the power estimates of the cell-specific tests for cells $(1,2)$ and $(2,1)$,
since they would be same as cells $(1,1)$ and $(2,2)$
(but for the left-sided alternative).
As expected,
the power estimates increase as segregation gets stronger
and also as class size increases.
For the cell-specific and overall tests,
type I and III tests have higher power estimates.

\subsection{Empirical Power Analysis under Association of Two Classes}
\label{sec:power-comp-assoc-2Cl}
Under the association alternatives, we consider three cases also.
In each case,
we generate $X_i \stackrel{iid}{\sim} \U((0,1)\times(0,1))$ for $i=1,2,\ldots,n_1$.
Then we generate $Y_j$ associated with $X$'s for $j=1,2,\ldots,n_2$ as follows.
For each $j$, select an $i$ randomly, and
set $Y_j =X_i+R_j\,\left( \cos T_j, \sin T_j \right)'$ where
$R_j \stackrel{iid}{\sim} \U(0,r)$ with $r \in (0,1)$ and
$T_j \stackrel{iid}{\sim} \U(0,2\,\pi)$.
We consider the following association alternatives:
\begin{equation}
\label{eqn:assoc-alt-2Cl} H_A^{I}: r=1/(2\sqrt{n_t}),\;\;\; H_A^{II}: r=1/(3\sqrt{n_t}),
\text{ and } H_A^{III}: r=1/(4\sqrt{n_t})
%H_A^{I}: r=1/7,\;\;\; H_A^{II}: r=1/10, H_A^{III}: r=1/20,
%\text{ and } H_A^{IV}: r=1/(2\sqrt{n_1}).
\end{equation}
where $n_t=n_1+n_2$.
Notice that association gets stronger as $r$ decreases;
that is, association gets stronger from $H_A^I$ to $H_A^{III}$.
Furthermore,
by construction,
the association of $Y$ points with $X$ points is stronger,
compared to the association of $X$ points with $Y$ points.
These association alternatives are motivated from the expected distance between points from
homogeneous Poisson Process (HPP).
Letting $D$ be the distance from a randomly chosen point to the nearest other point in a HPP with intensity $\rho$,
we have
$\E[D]=1/(2\,\sqrt{\rho})$ and $\Var[D]=(4-\pi)/(4\,\pi\,\rho)$ (\cite{dixon:EncycEnv2002}).
In our case, under CSR independence, intensity of $n_t$ points would be $\widehat \rho=n_t$,
since area of the unit square is 1.
Hence we have set $r=1/(2\sqrt{n_t})$, $r=1/(3\sqrt{n_t})$, and $r=1/(4\sqrt{n_t})$ for $H_A^I$ to $H_A^{III}$.
For example,
under $H_A^I$,
the displacements of $Y_j$ around $X_i$ would be limited by the average distance between $n_t$ points under $H_o$.

\begin{figure} [hbp]
\centering
%\psfrag{Density}{ \Huge{\bf{Density}}}
Empirical Power Estimates of the NNCT-Tests under $H^I_A$ \\
\rotatebox{-90}{ \resizebox{2.1 in}{!}{\includegraphics{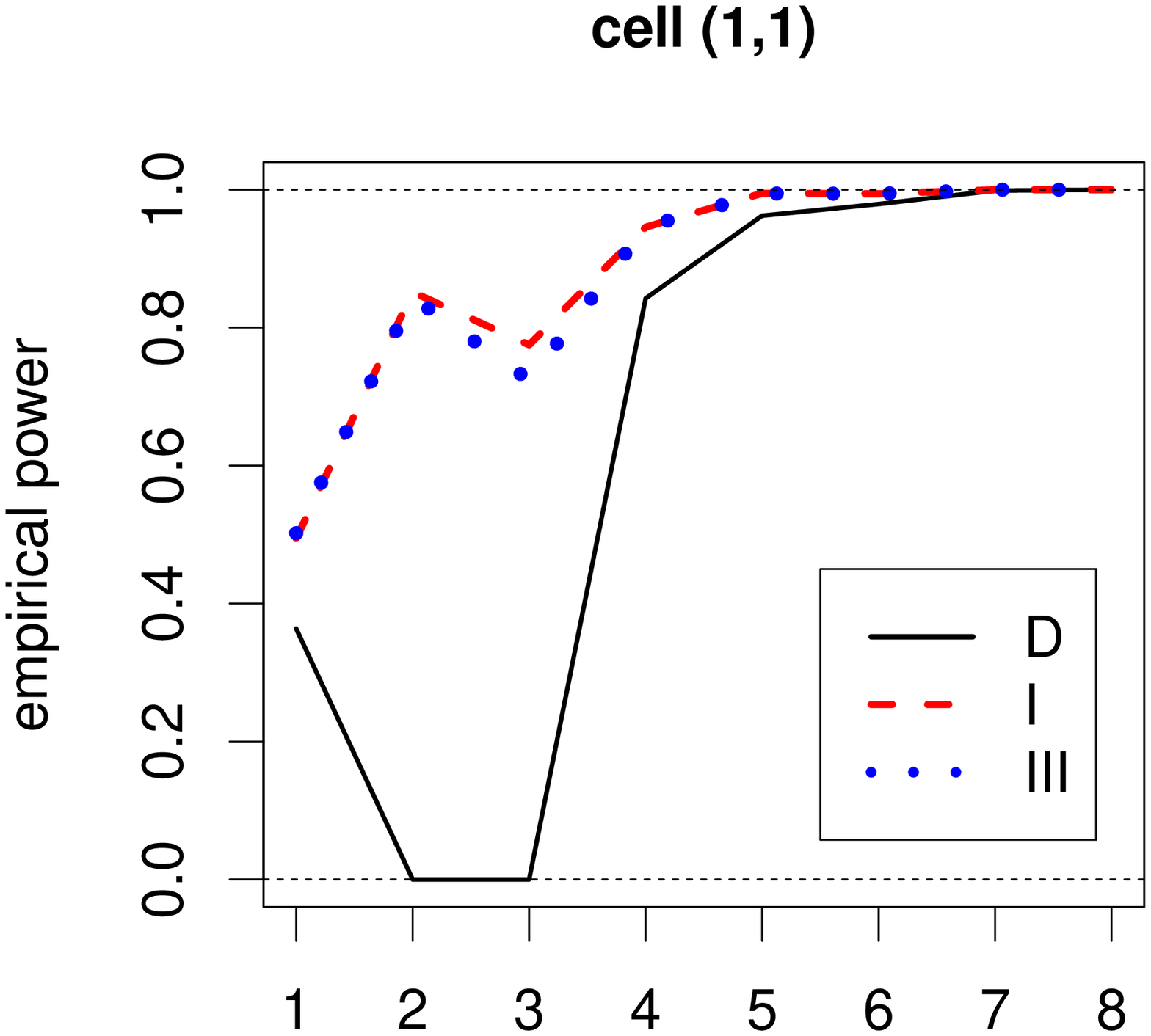} }}
\rotatebox{-90}{ \resizebox{2.1 in}{!}{\includegraphics{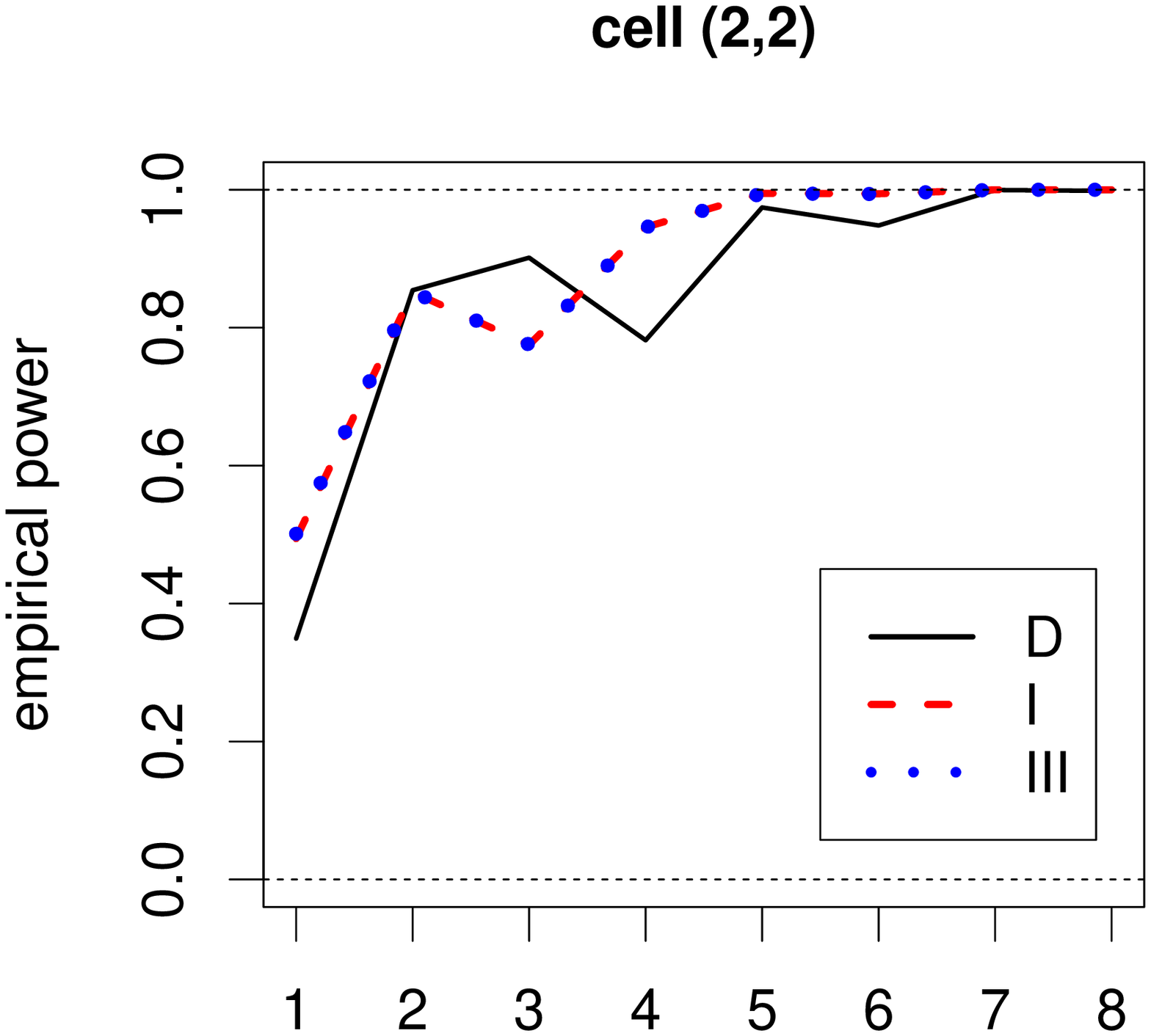} }}
\rotatebox{-90}{ \resizebox{2.1 in}{!}{\includegraphics{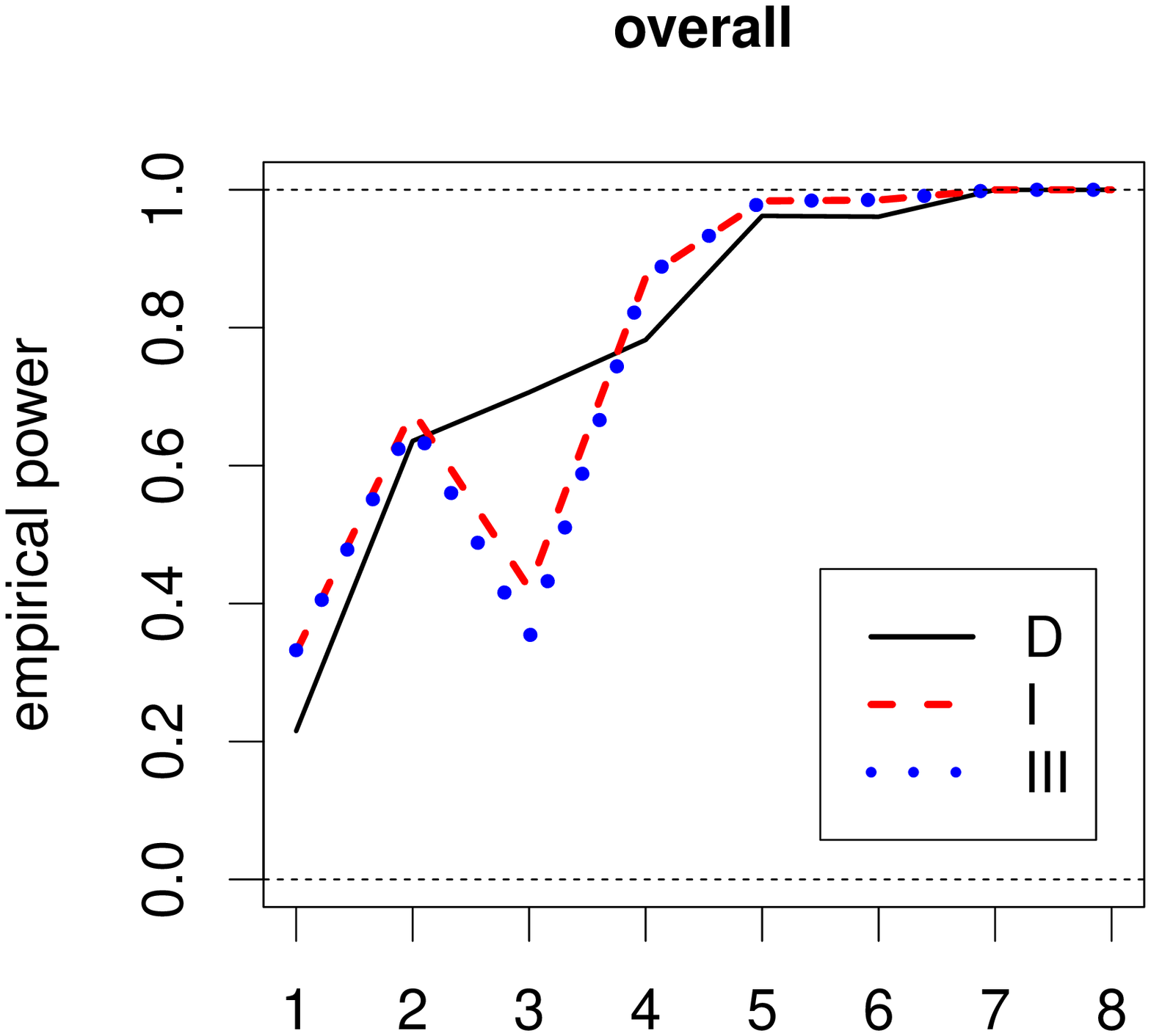} }}
Power Estimates under $H^{II}_A$\\
\rotatebox{-90}{ \resizebox{2.1 in}{!}{\includegraphics{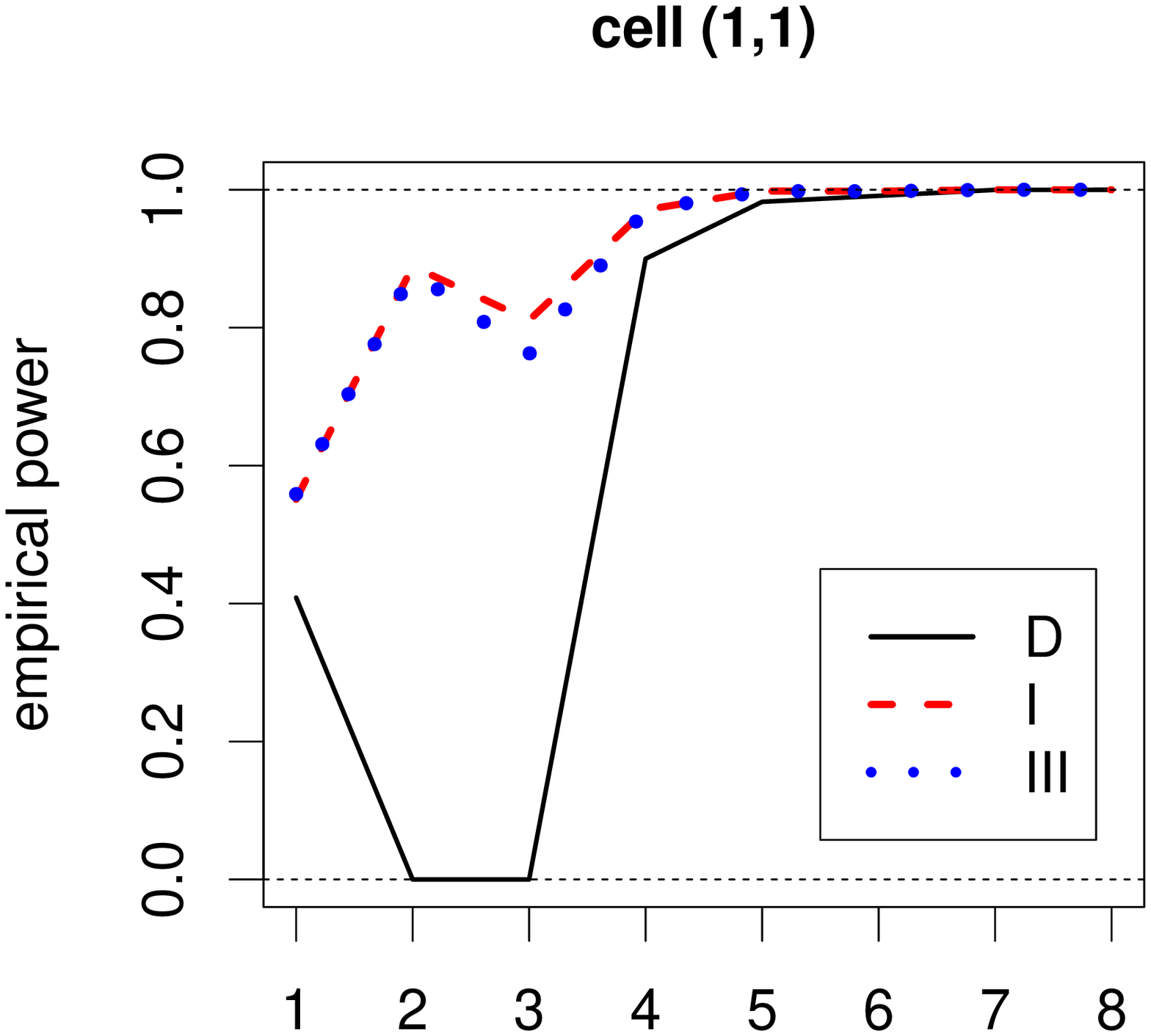} }}
\rotatebox{-90}{ \resizebox{2.1 in}{!}{\includegraphics{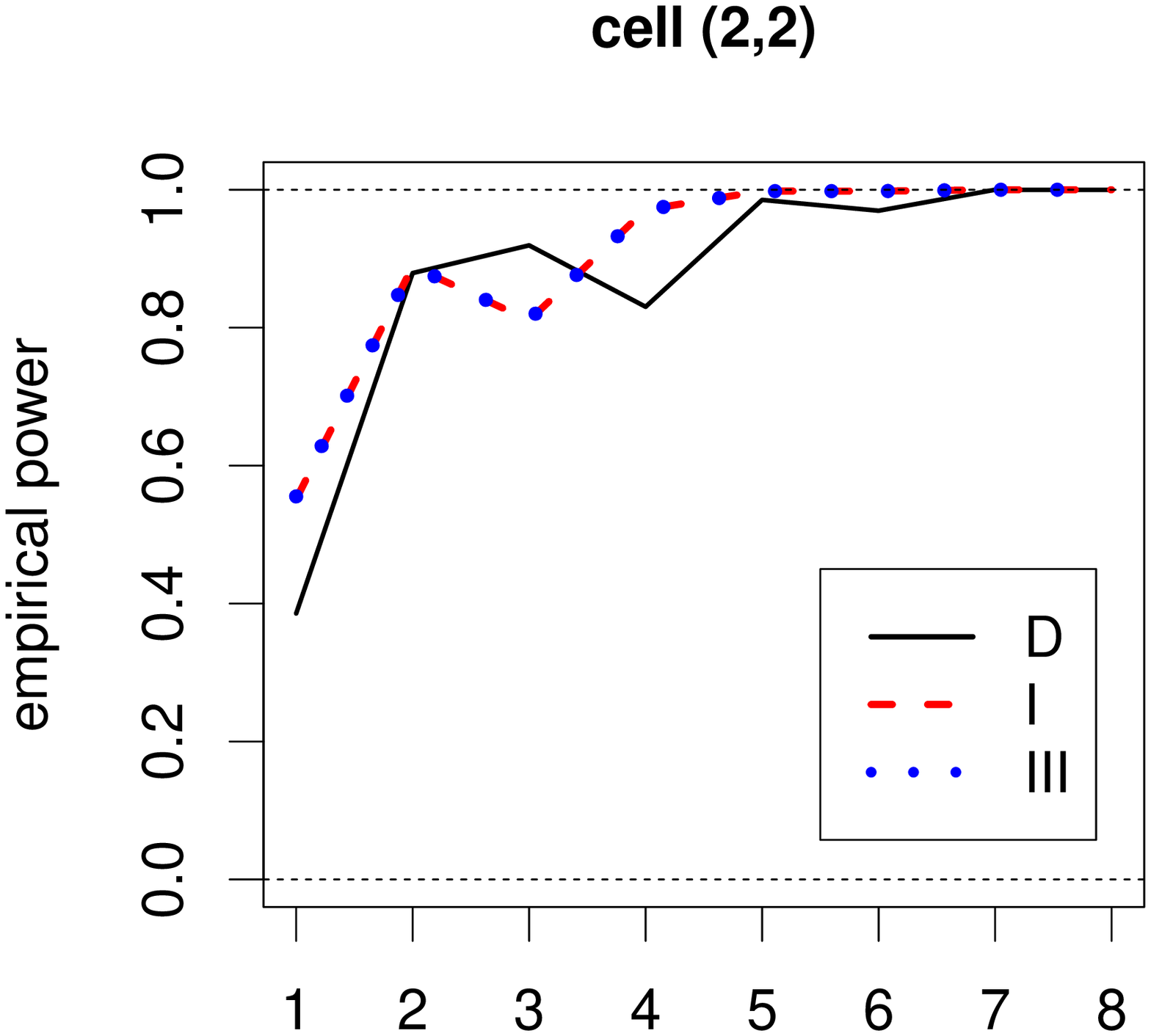} }}
\rotatebox{-90}{ \resizebox{2.1 in}{!}{\includegraphics{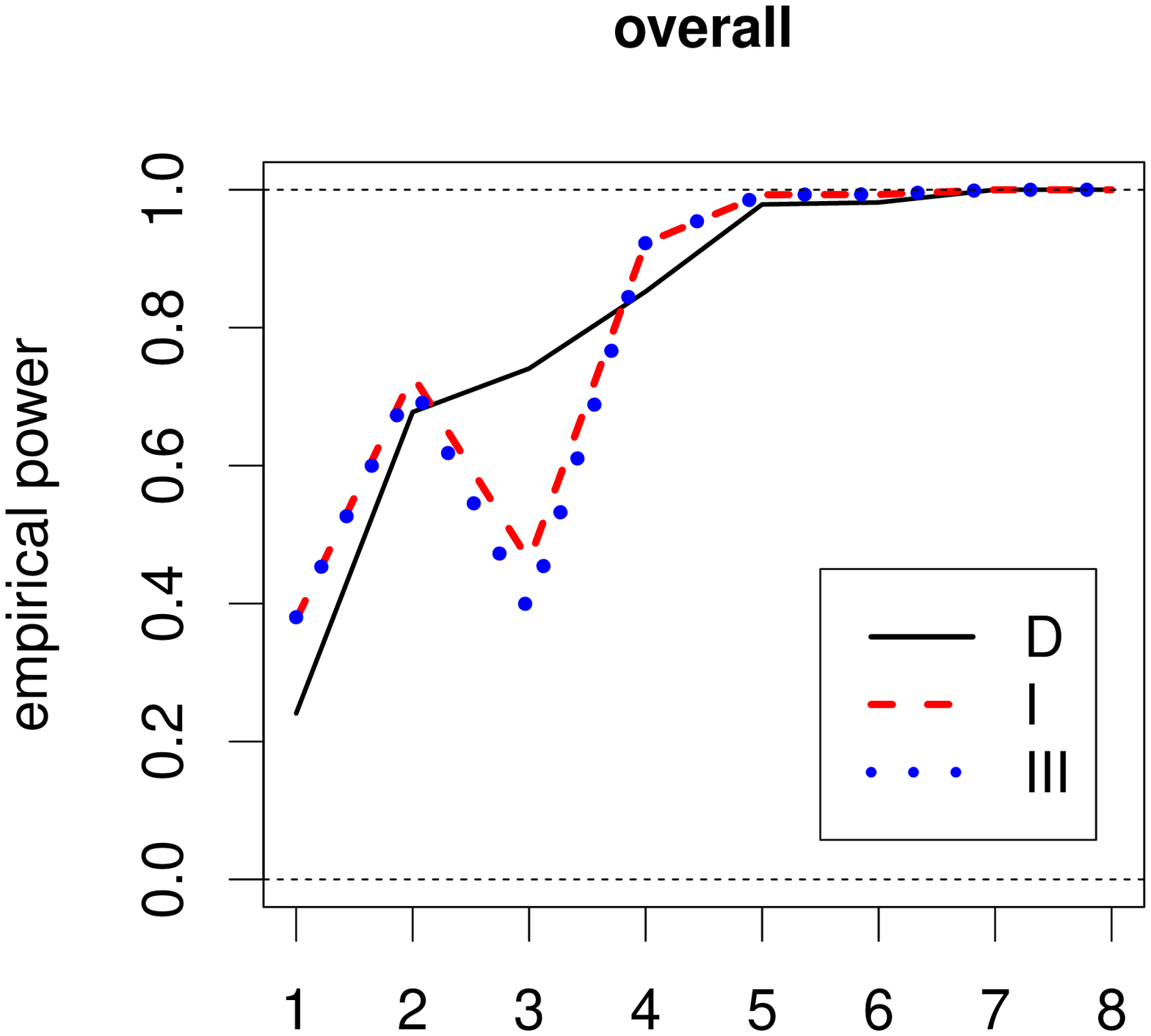} }}
Power Estimates under $H^{III}_A$\\
\rotatebox{-90}{ \resizebox{2.1 in}{!}{\includegraphics{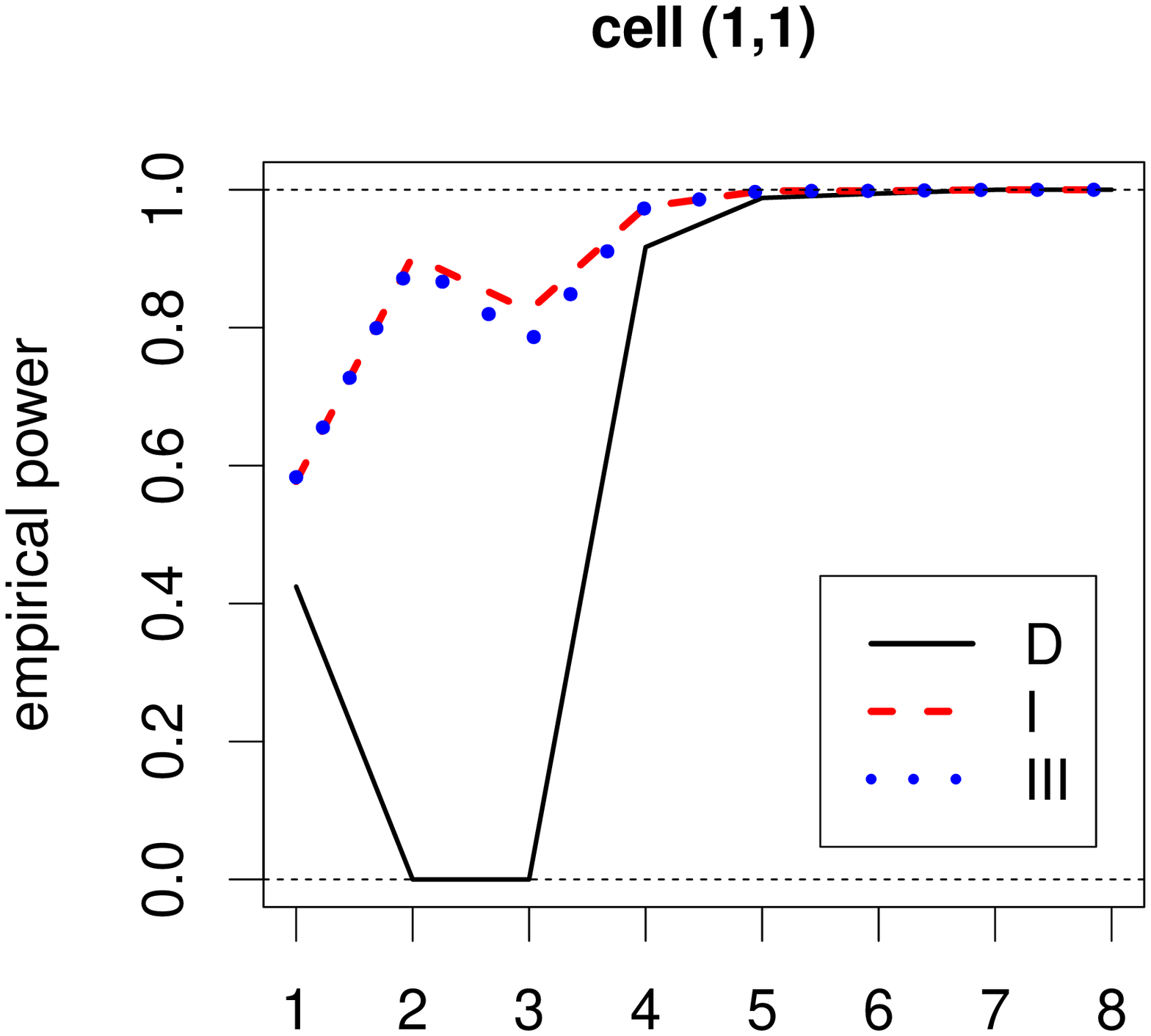} }}
\rotatebox{-90}{ \resizebox{2.1 in}{!}{\includegraphics{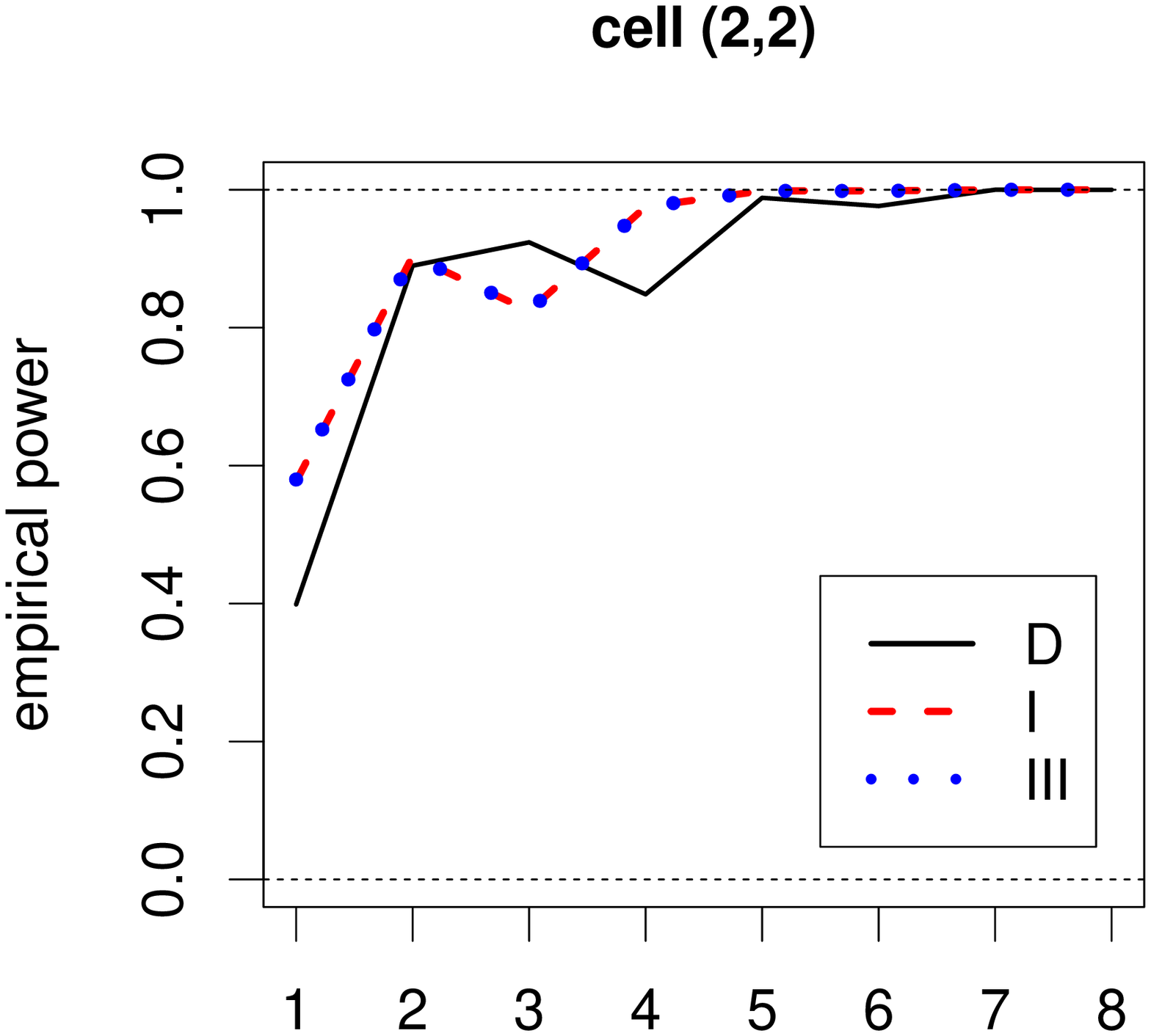} }}
\rotatebox{-90}{ \resizebox{2.1 in}{!}{\includegraphics{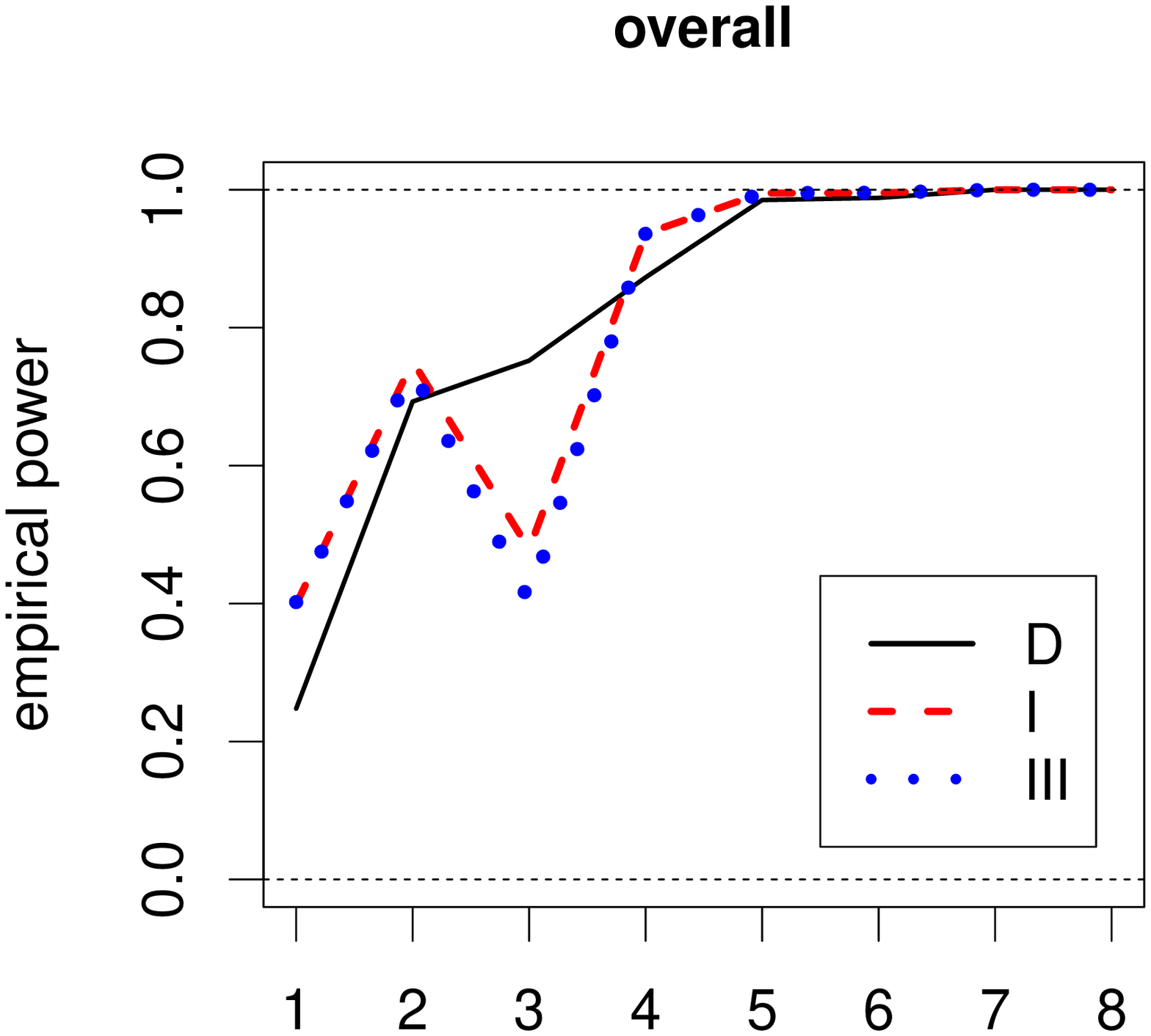} }}
\caption{
\label{fig:power-assoc-2cl}
The empirical power estimates for the NNCT-tests under the association alternatives in the two-class case.
The horizontal axis labels and legend labeling are as in Figure \ref{fig:emp-size-CSR-2cl}.
}
\end{figure}

The empirical power estimates under association are presented in Figure \ref{fig:power-assoc-2cl}.
As association gets stronger,
the power estimates increase.
However, there is a decline in power from $(n_1,n_2)=(10,10)$ to $(10,30)$ and $(10,50)$,
and this decline is more drastic for Dixon's cell $(1,1)$ test.
For balanced class sizes,
the power tends to increase as $n_t$ increases.
Furthermore,
type I and III have higher power for all class size combinations for cell $(1,1)$,
and for most class size combinations for cell $(2,2)$ and the overall test.

\section{Empirical Power Analysis in the Three-Class Case}
\label{sec:emp-power-3Cl}
We also consider three cases for each of segregation and association alternatives in the
three-class case.

\subsection{Empirical Power Analysis under Segregation of Three Classes}
\label{sec:power-comp-seg-3Cl}
Under the segregation alternatives, we generate
$X_i \stackrel{iid}{\sim} \U(S_1)$,
$Y_j \stackrel{iid}{\sim} \U(S_2)$, and
$Z_k \stackrel{iid}{\sim} \U(S_3)$
for $i=1,\ldots,n_1$, $j=1,\ldots,n_2$, and $k=1,\ldots,n_3$
where $S_1=(0,1-2s)\times(0,1-2s)$, $S_2=(2s,1)\times(2s,1)$, and $S_3=(s,1-s)\times(s,1-s)$
with $s \in (0,1/2)$.
We consider the following segregation alternatives:
\begin{equation}
\label{eqn:seg-alt-3Cl}
H_{S_1}: s=1/12, \;\;\; H_{S_2}: s=1/8, \text{ and } H_{S_3}: s=1/6.
\end{equation}

Notice that,
as $s$ increases,
segregation between the classes gets stronger;
that is,
segregation gets stronger from $H_{S_1}$ to $H_{S_3}$.
Furthermore, by construction,
classes $X$ and $Y$ are more segregated
compared to $Z$ and $X$ or $Z$ and $Y$.
In fact, the segregation between $X$ and $Z$ and segregation between $Y$ and $Z$ are identical (as a stochastic process).

\begin{figure} [hbp]
\centering
%\psfrag{Density}{ \Huge{\bf{Density}}}
Empirical Power Estimates of Cell-Specific Tests under $H_{S_1}$\\
\rotatebox{-90}{ \resizebox{2.1 in}{!}{\includegraphics{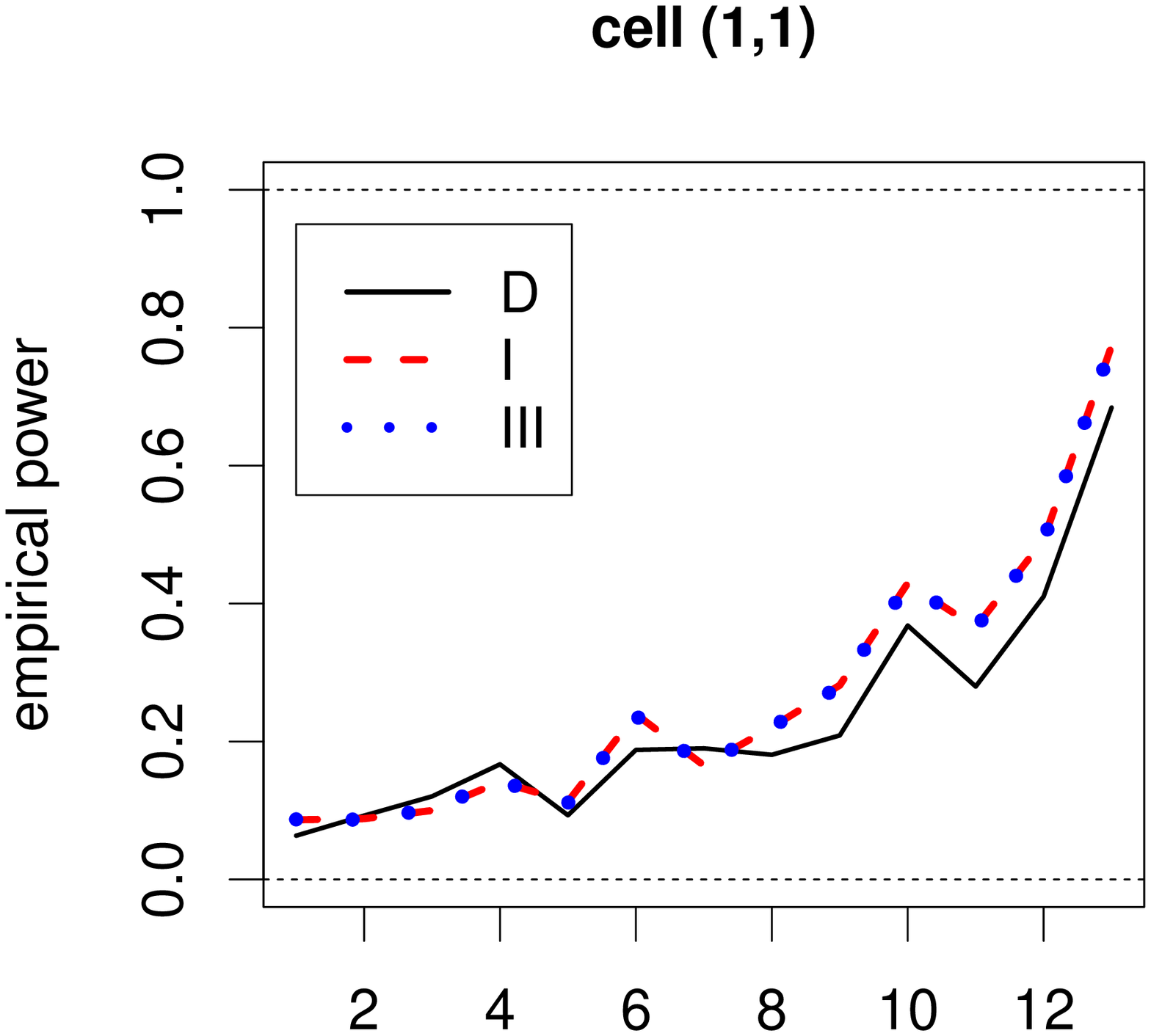} }}
\rotatebox{-90}{ \resizebox{2.1 in}{!}{\includegraphics{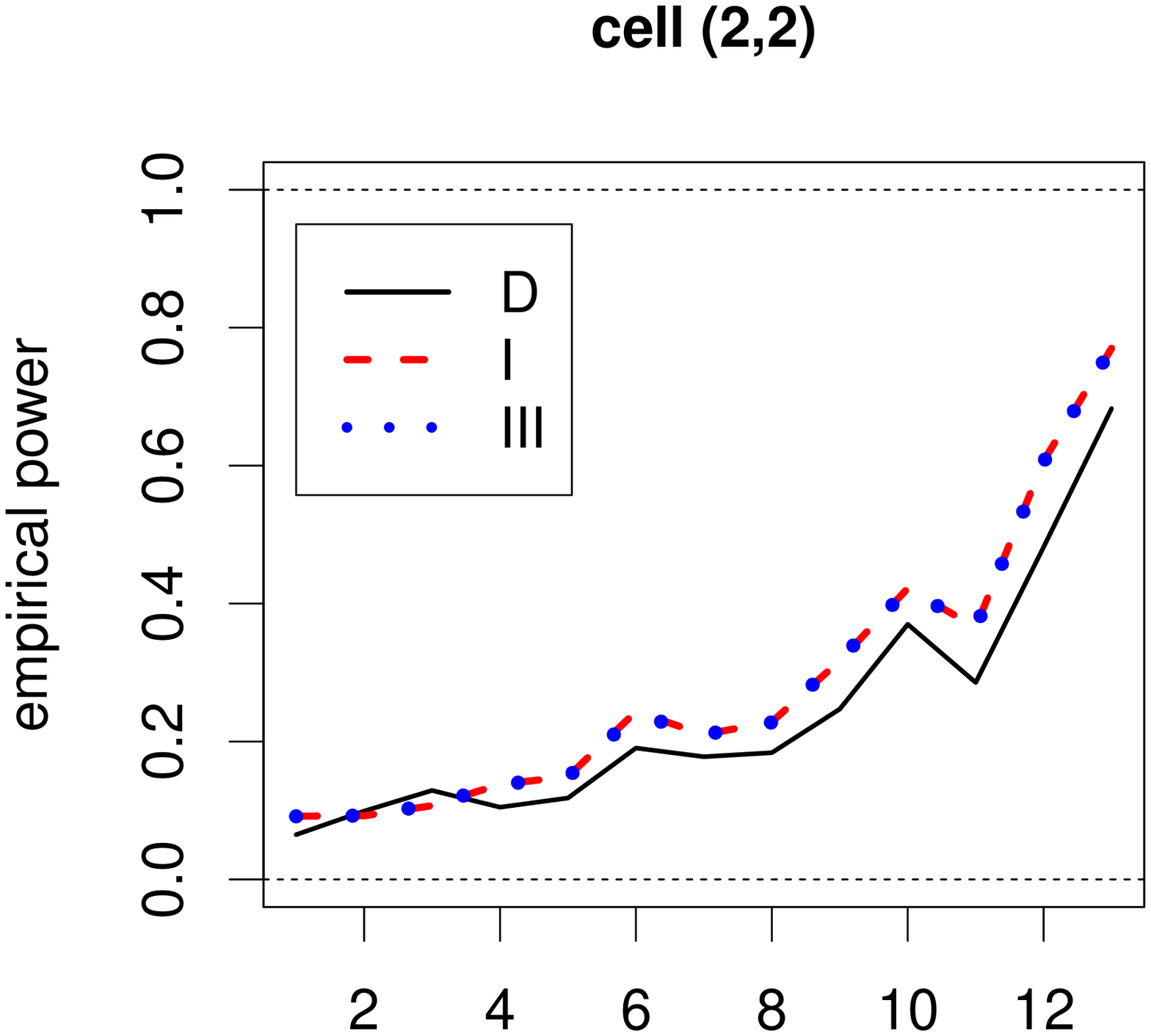} }}
\rotatebox{-90}{ \resizebox{2.1 in}{!}{\includegraphics{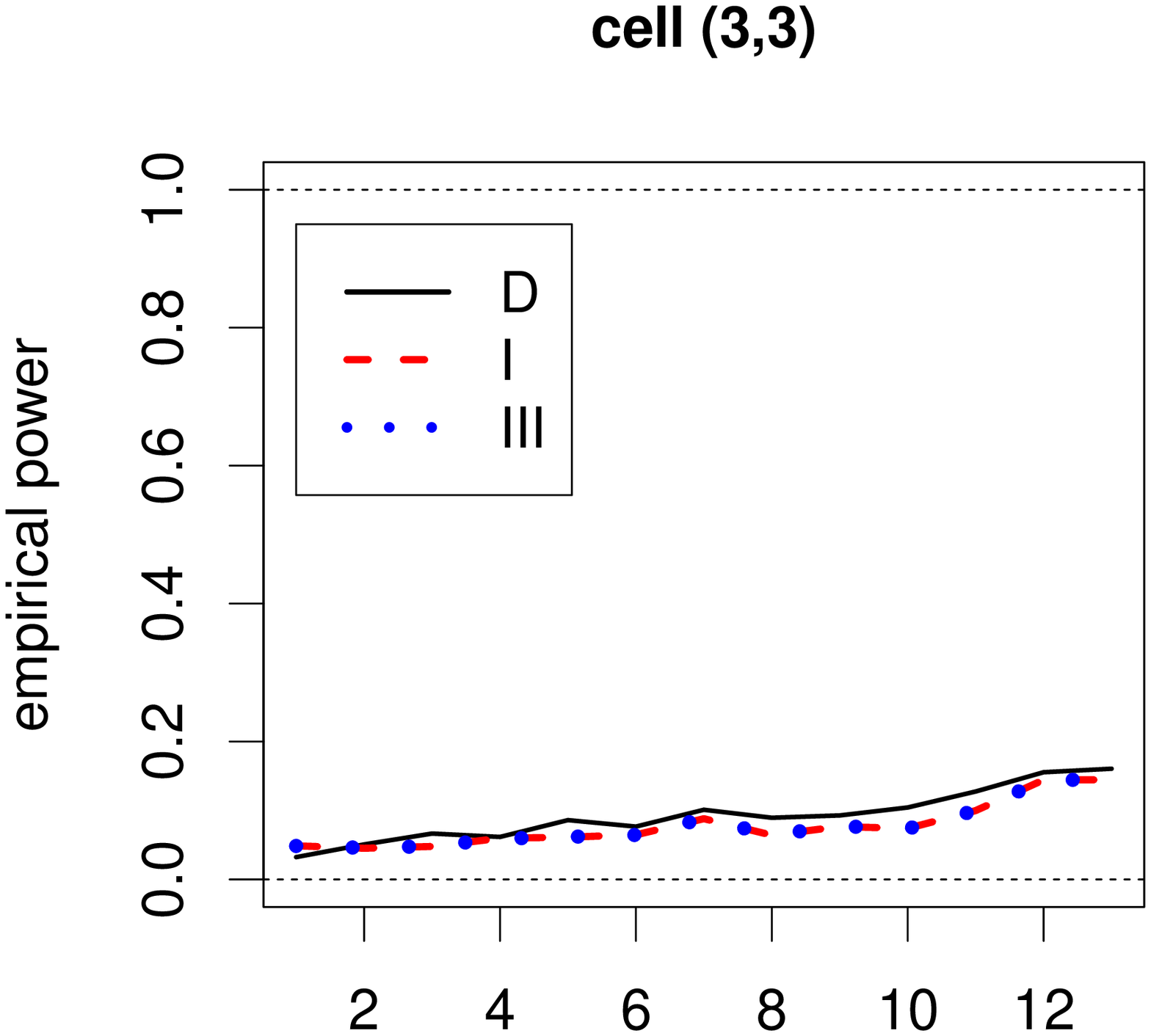} }}
Power Estimates under $H_{S_2}$\\
\rotatebox{-90}{ \resizebox{2.1 in}{!}{\includegraphics{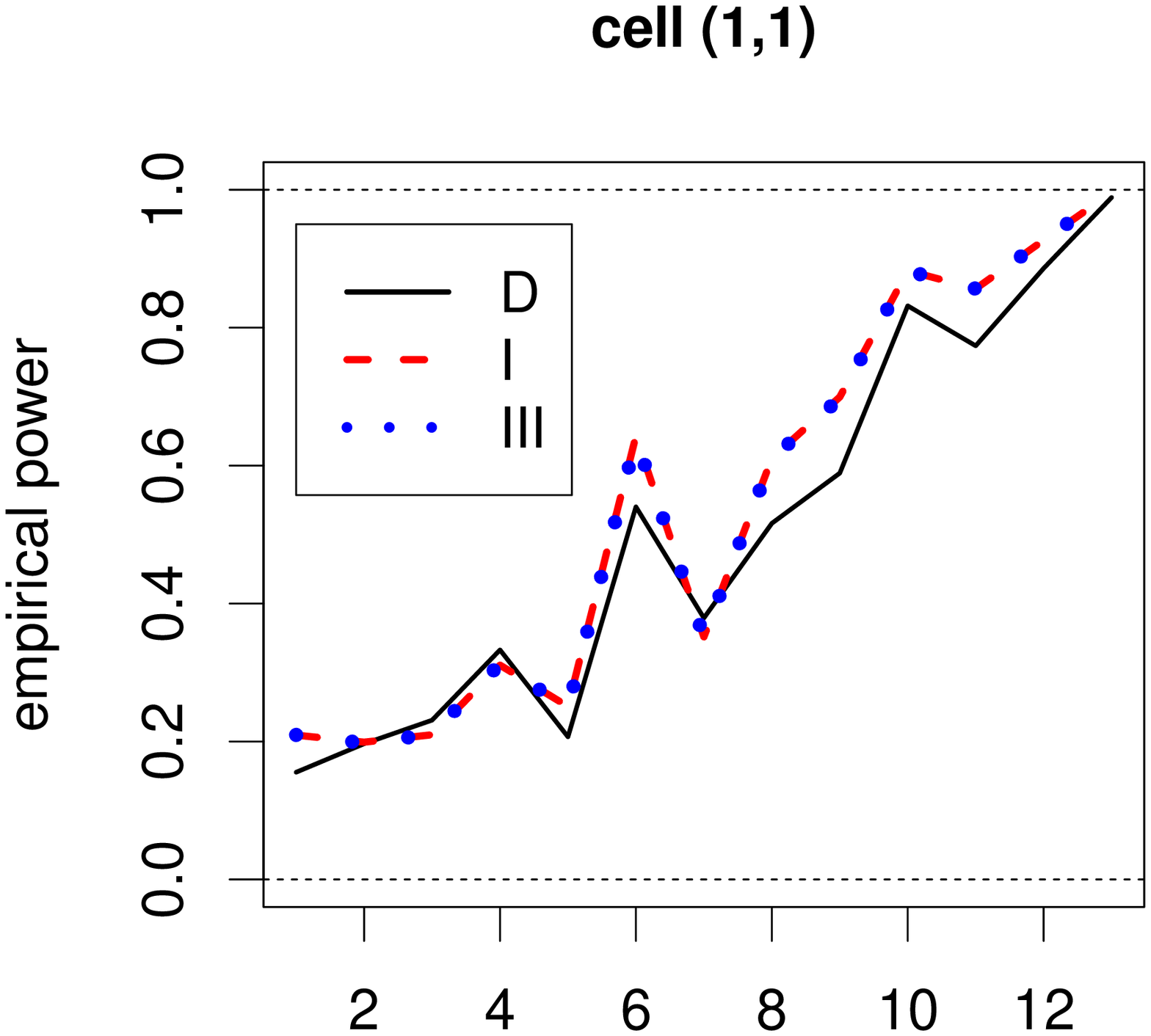} }}
\rotatebox{-90}{ \resizebox{2.1 in}{!}{\includegraphics{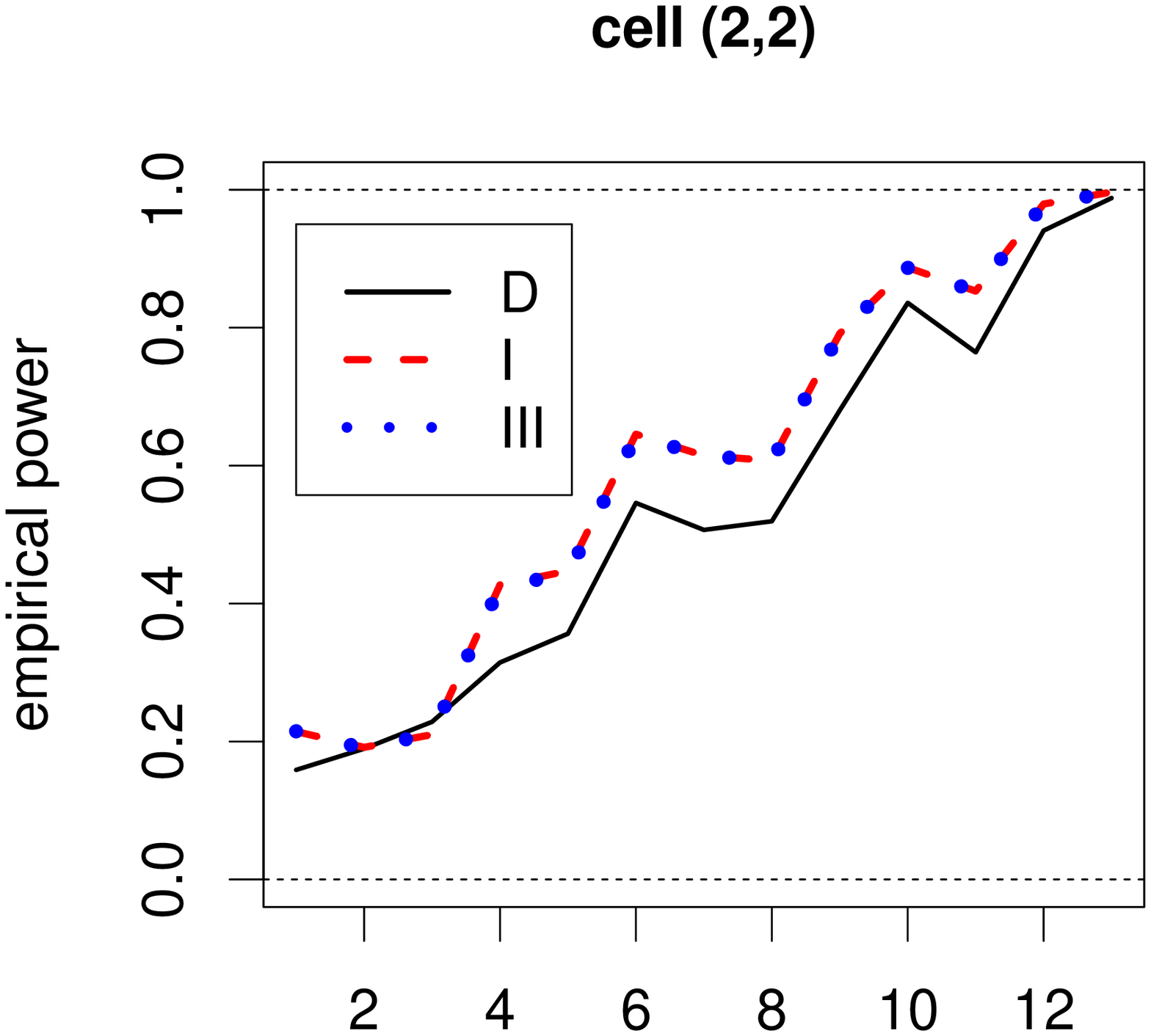} }}
\rotatebox{-90}{ \resizebox{2.1 in}{!}{\includegraphics{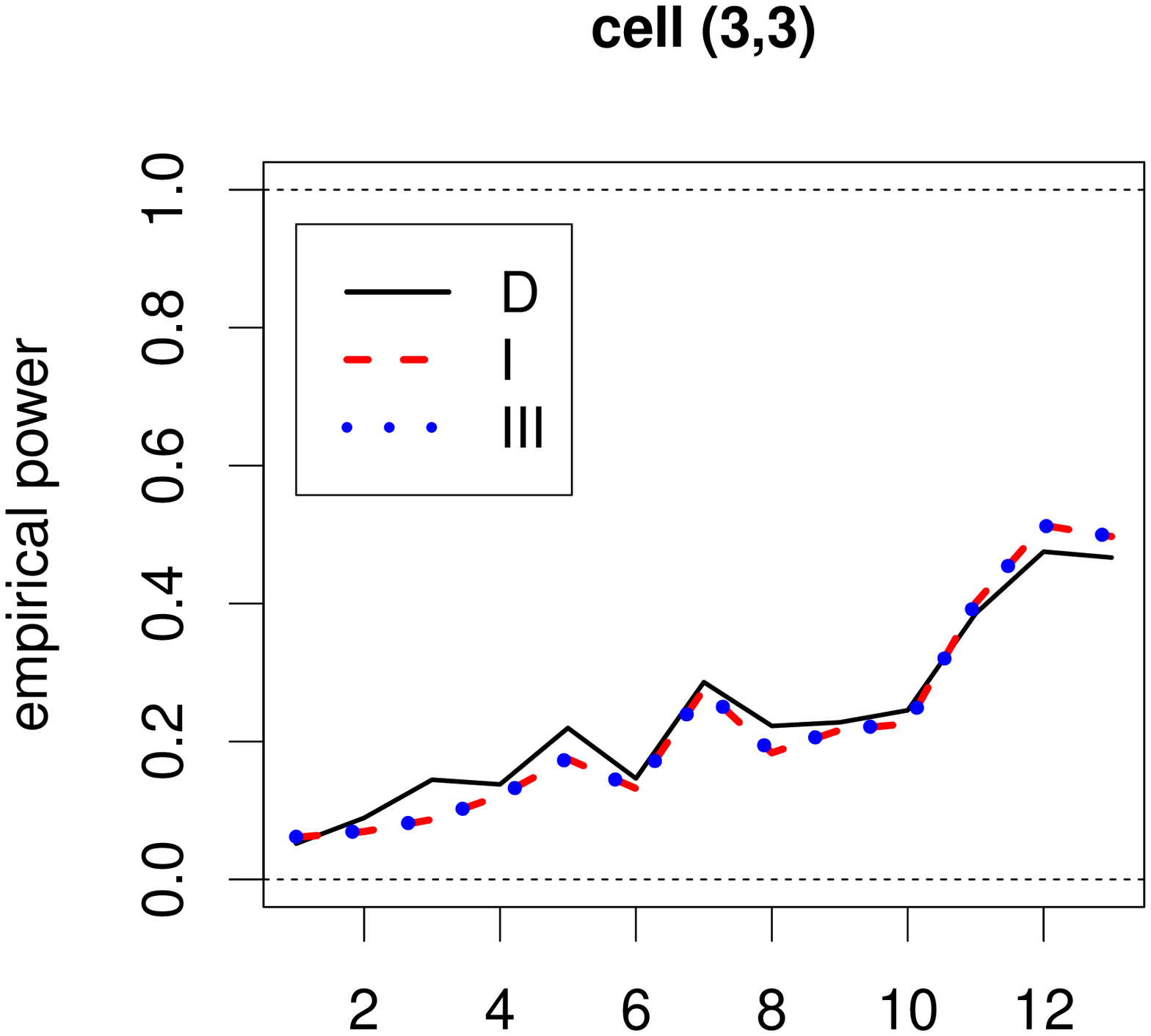} }}
Power Estimates under $H_{S_3}$\\
\rotatebox{-90}{ \resizebox{2.1 in}{!}{\includegraphics{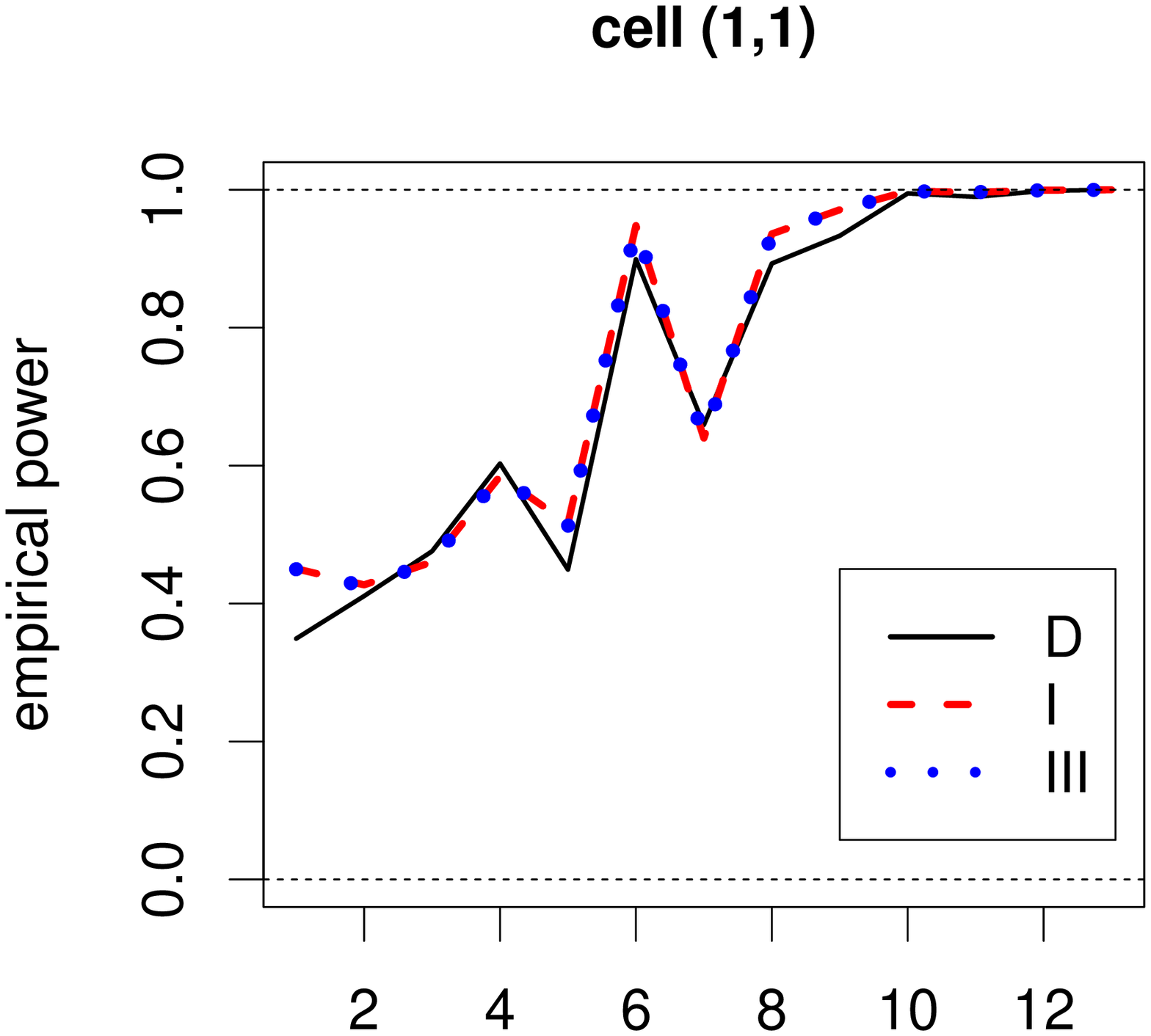} }}
\rotatebox{-90}{ \resizebox{2.1 in}{!}{\includegraphics{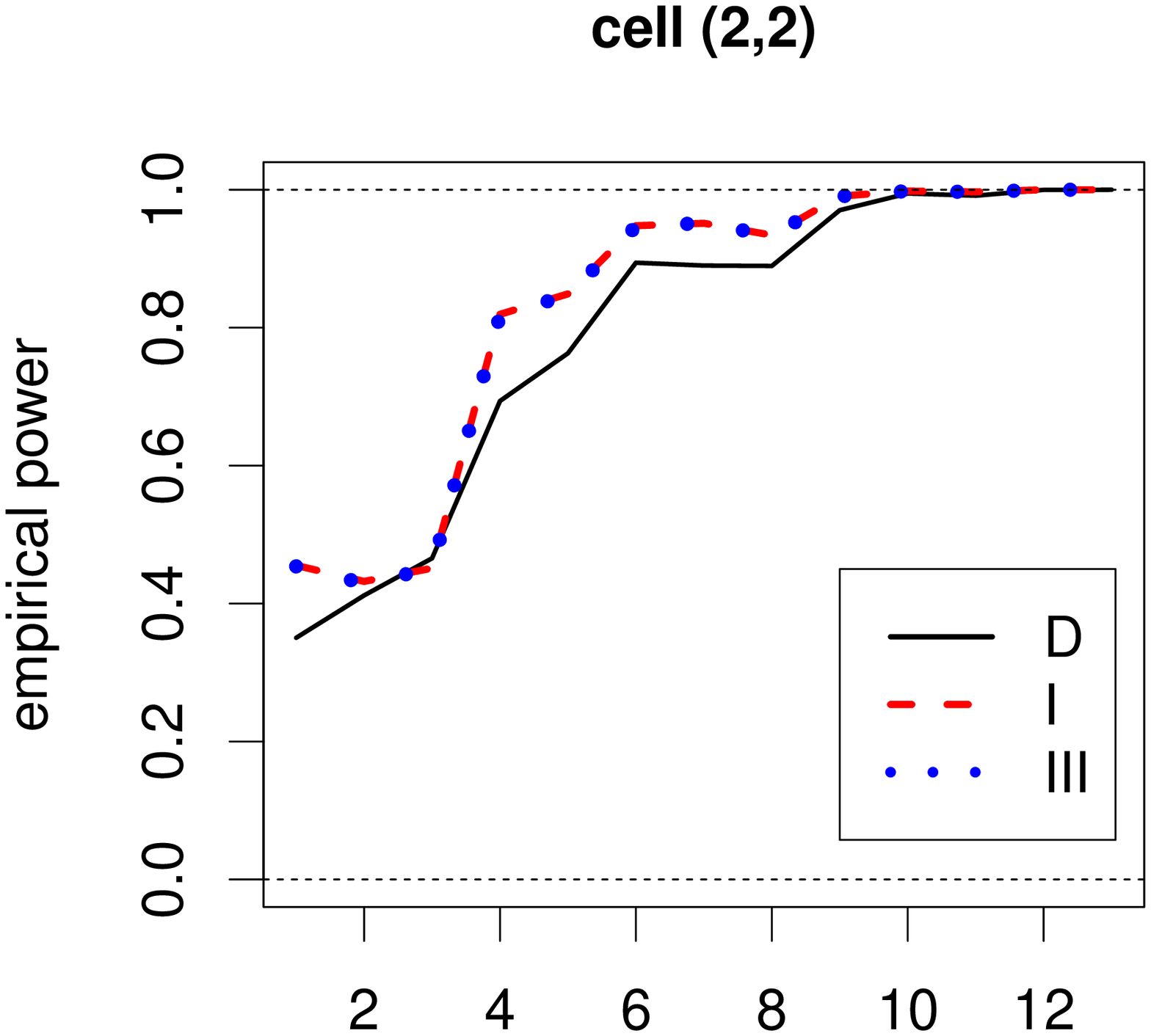} }}
\rotatebox{-90}{ \resizebox{2.1 in}{!}{\includegraphics{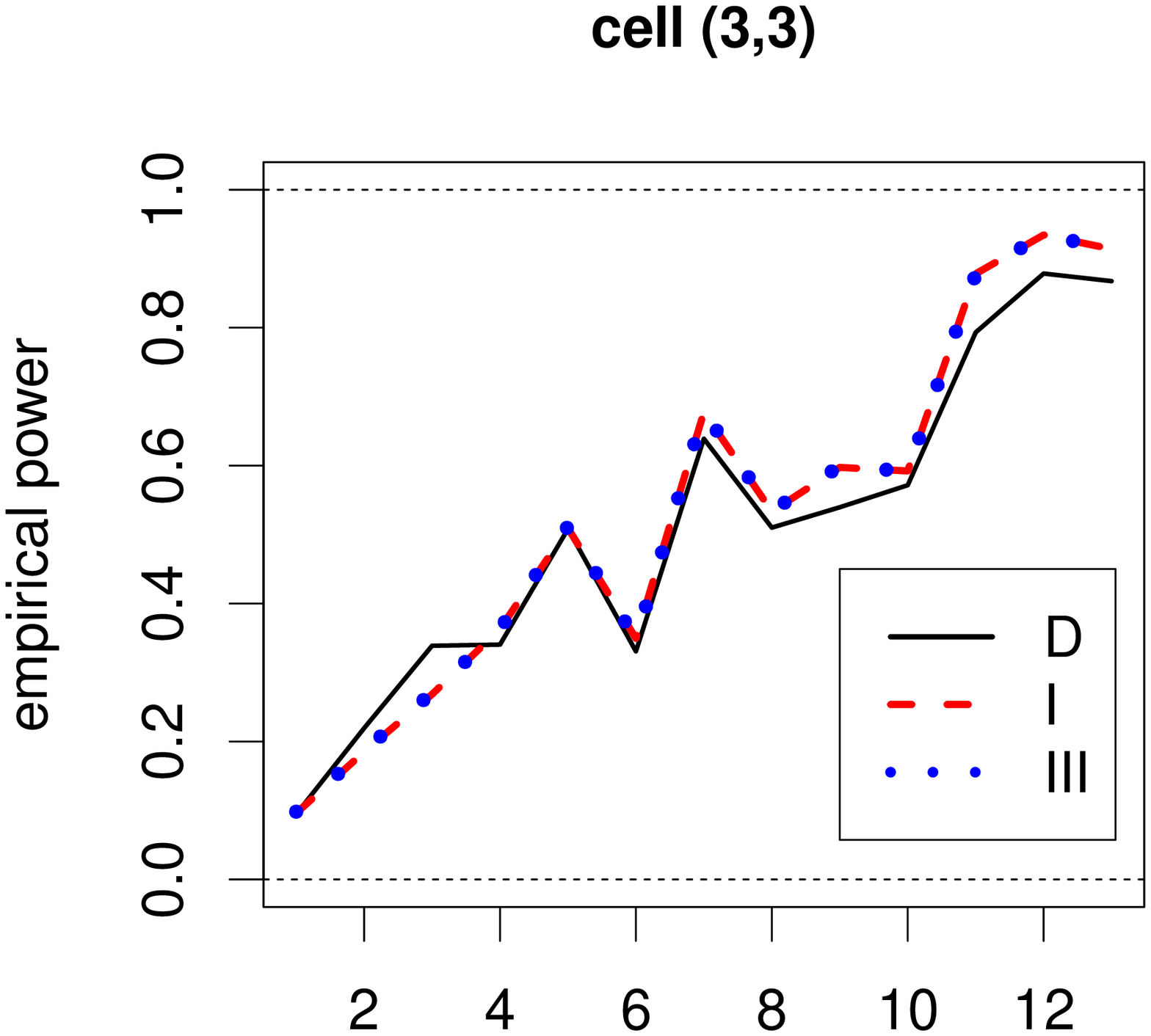} }}
\caption{
\label{fig:power-seg-cell-3cl-diag}
The empirical power estimates of the cell-specific tests
for cells $(1,1)$, $(2,2)$, and $(3,3)$ under the segregation alternatives
$H_{S_1}$ (top), $H_{S_2}$ (middle), and $H_{S_3}$ (bottom) in the three-class case.
The legend labeling is as in Figure \ref{fig:emp-size-CSR-2cl}
and
horizontal axis labels are as in Figure \ref{fig:emp-size-CSR-cell-3cl}.
}
\end{figure}

\begin{figure} [hbp]
\centering
%\psfrag{Density}{ \Huge{\bf{Density}}}
Empirical Power Estimates of Cell-Specific Tests under $H_{S_1}$\\
\rotatebox{-90}{ \resizebox{2.1 in}{!}{\includegraphics{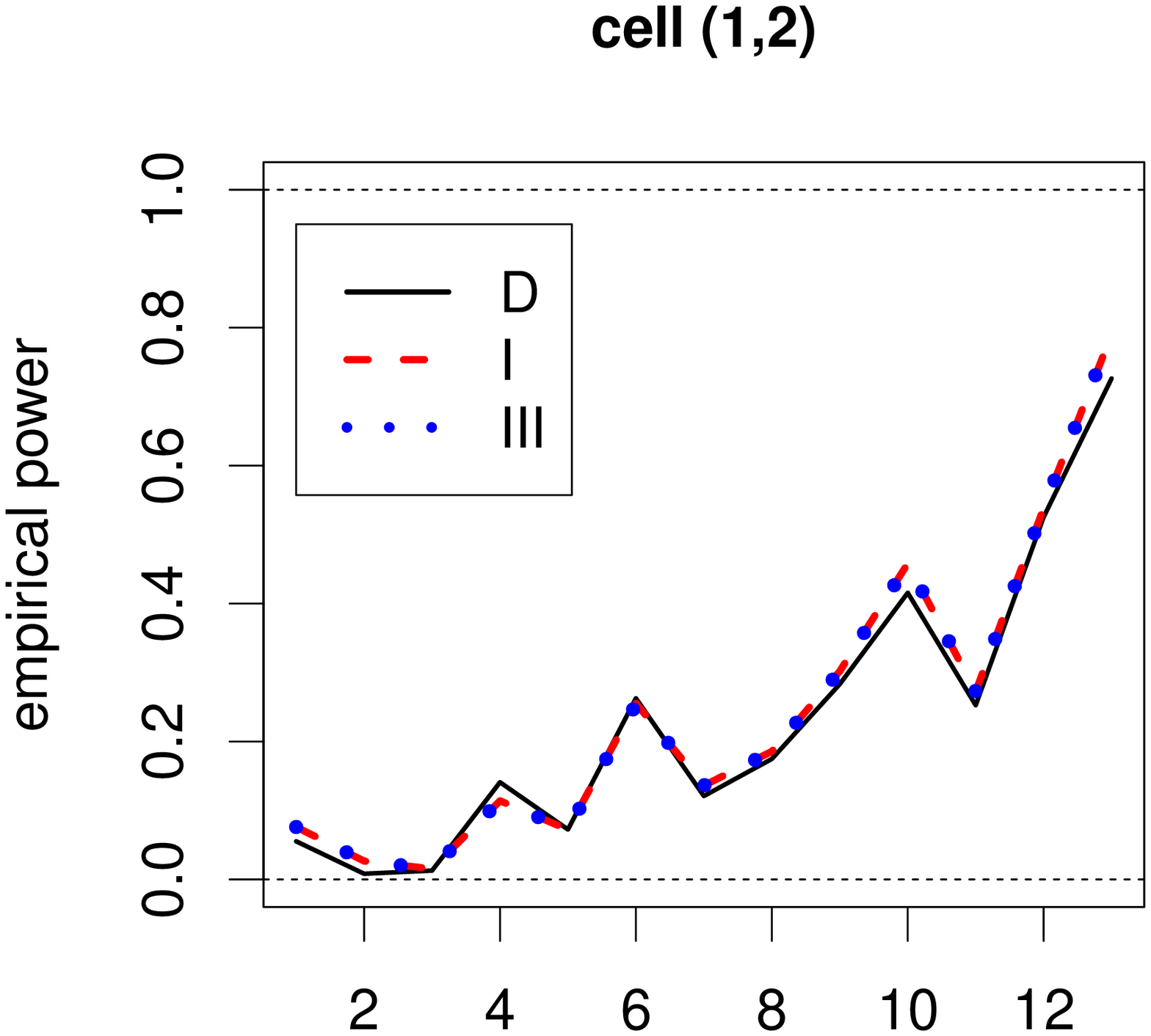} }}
\rotatebox{-90}{ \resizebox{2.1 in}{!}{\includegraphics{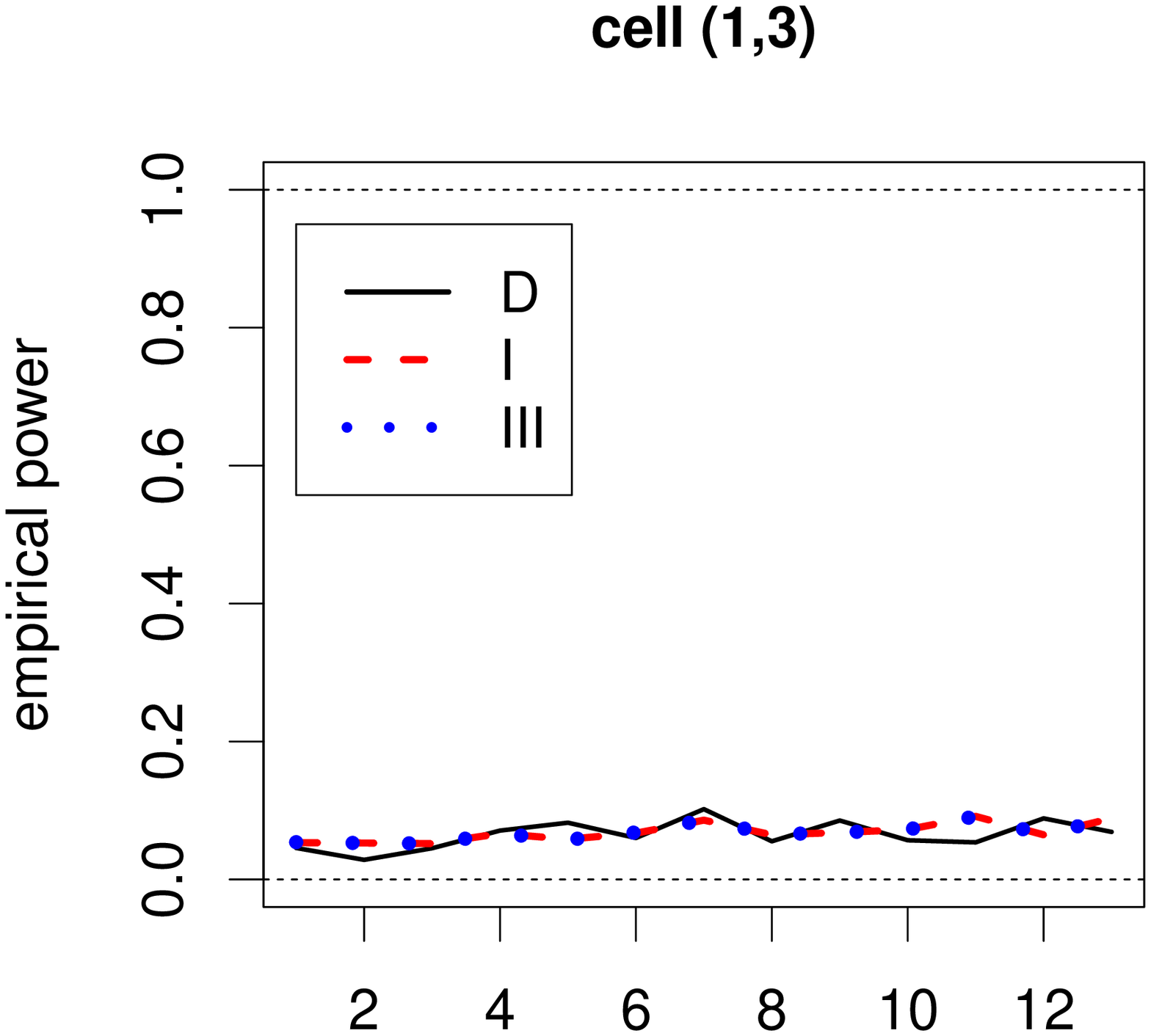} }}
\rotatebox{-90}{ \resizebox{2.1 in}{!}{\includegraphics{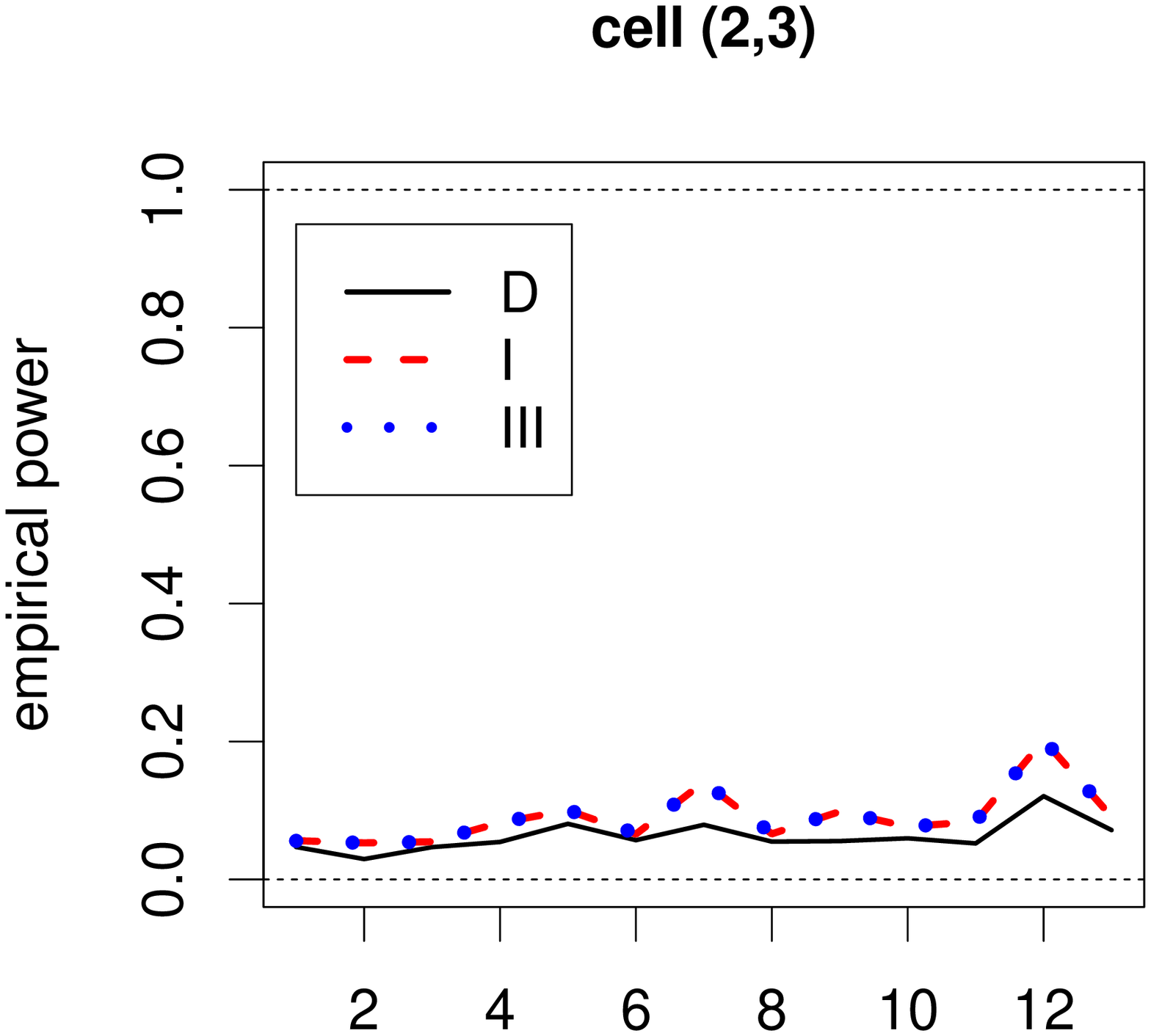} }}
Power Estimates under $H_{S_2}$\\
\rotatebox{-90}{ \resizebox{2.1 in}{!}{\includegraphics{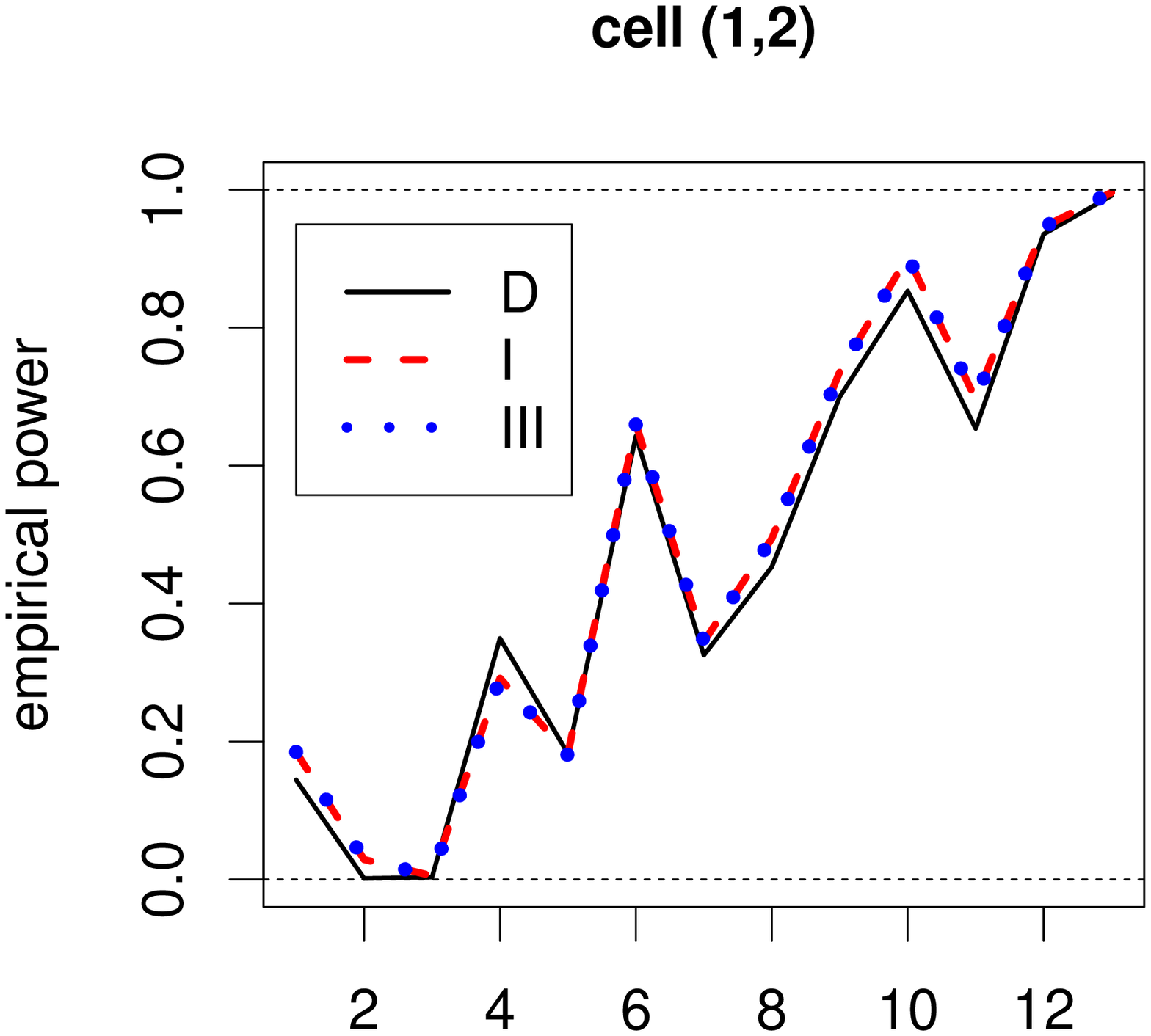} }}
\rotatebox{-90}{ \resizebox{2.1 in}{!}{\includegraphics{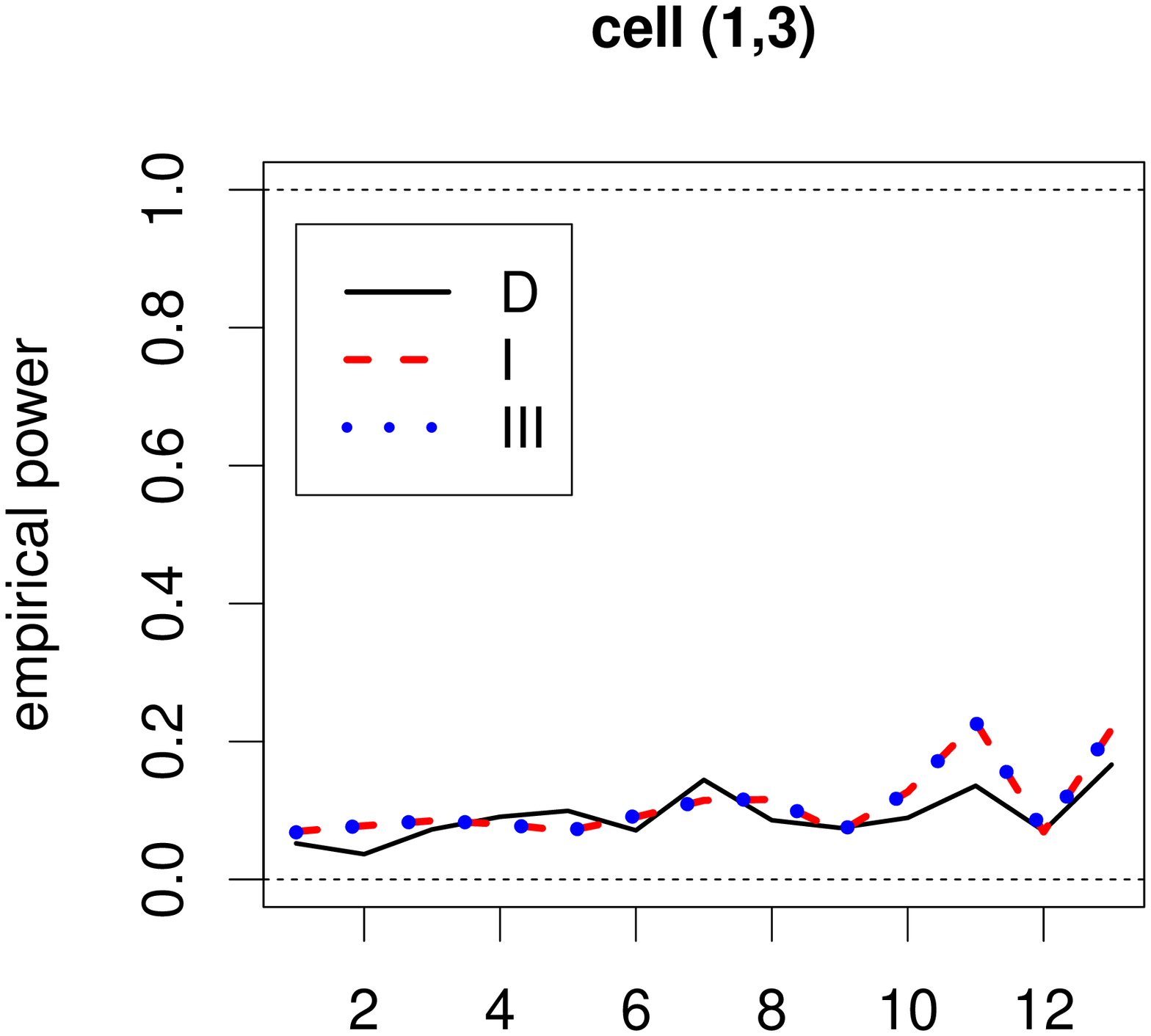} }}
\rotatebox{-90}{ \resizebox{2.1 in}{!}{\includegraphics{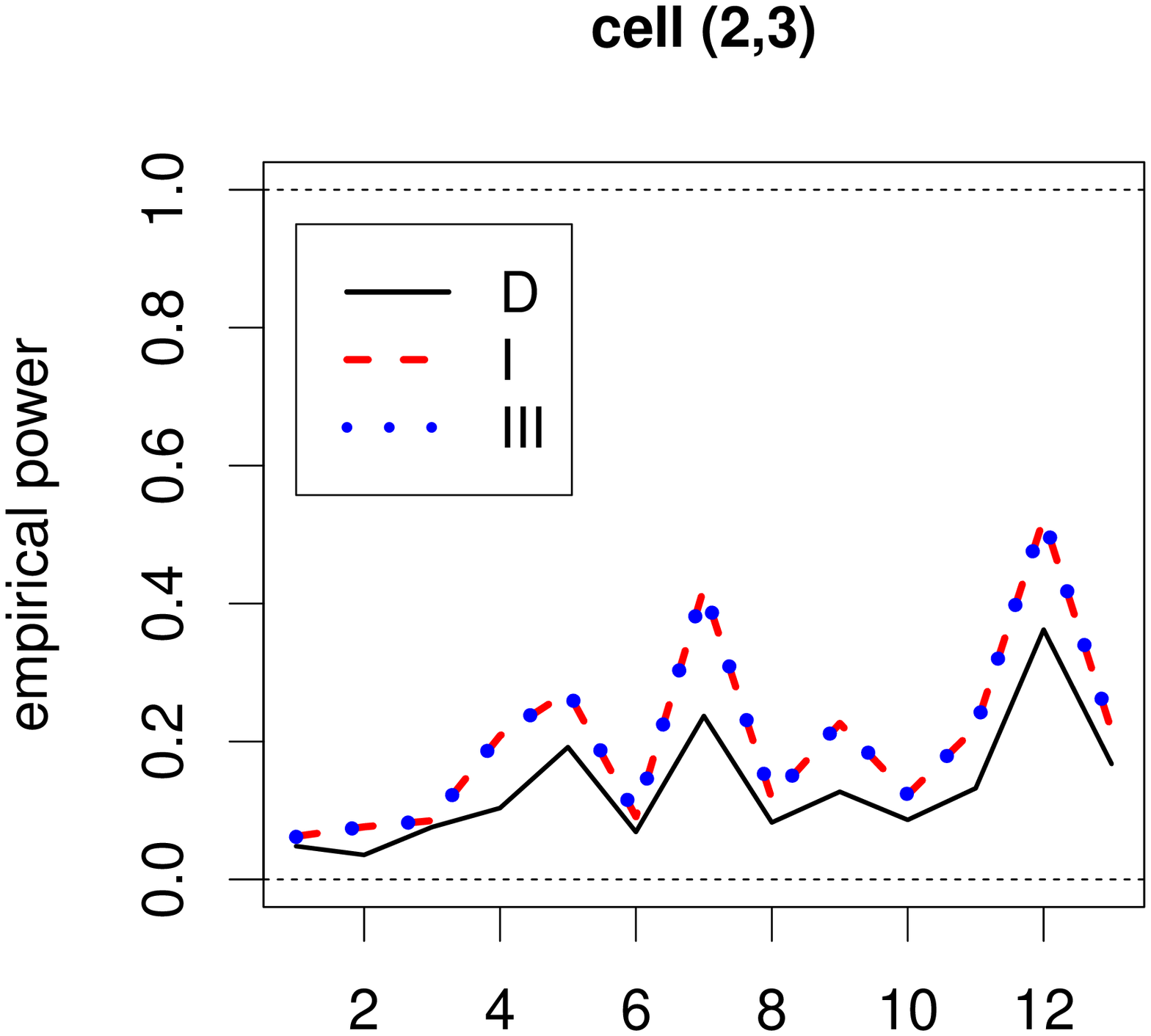} }}
Power Estimates under $H_{S_3}$\\
\rotatebox{-90}{ \resizebox{2.1 in}{!}{\includegraphics{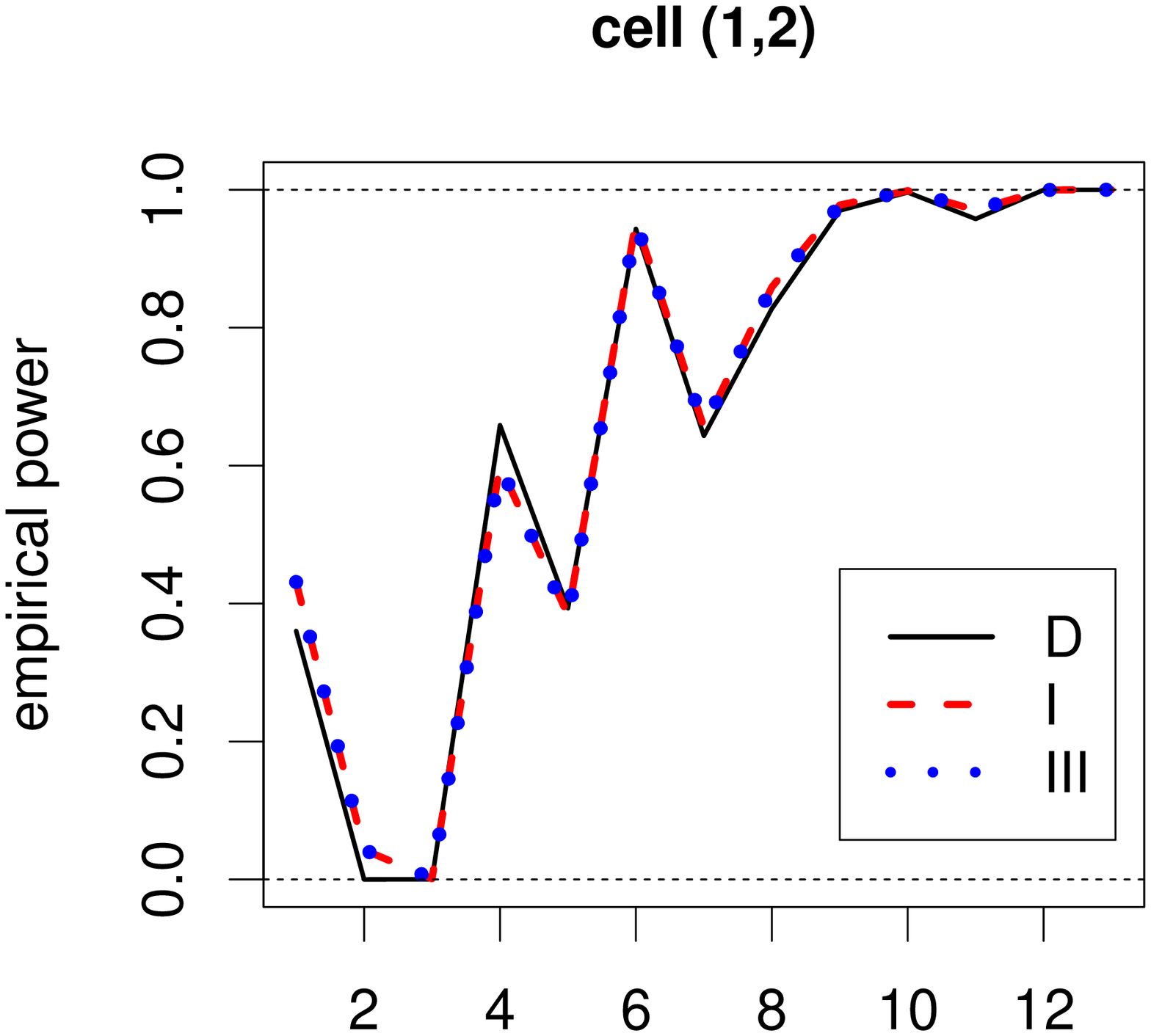} }}
\rotatebox{-90}{ \resizebox{2.1 in}{!}{\includegraphics{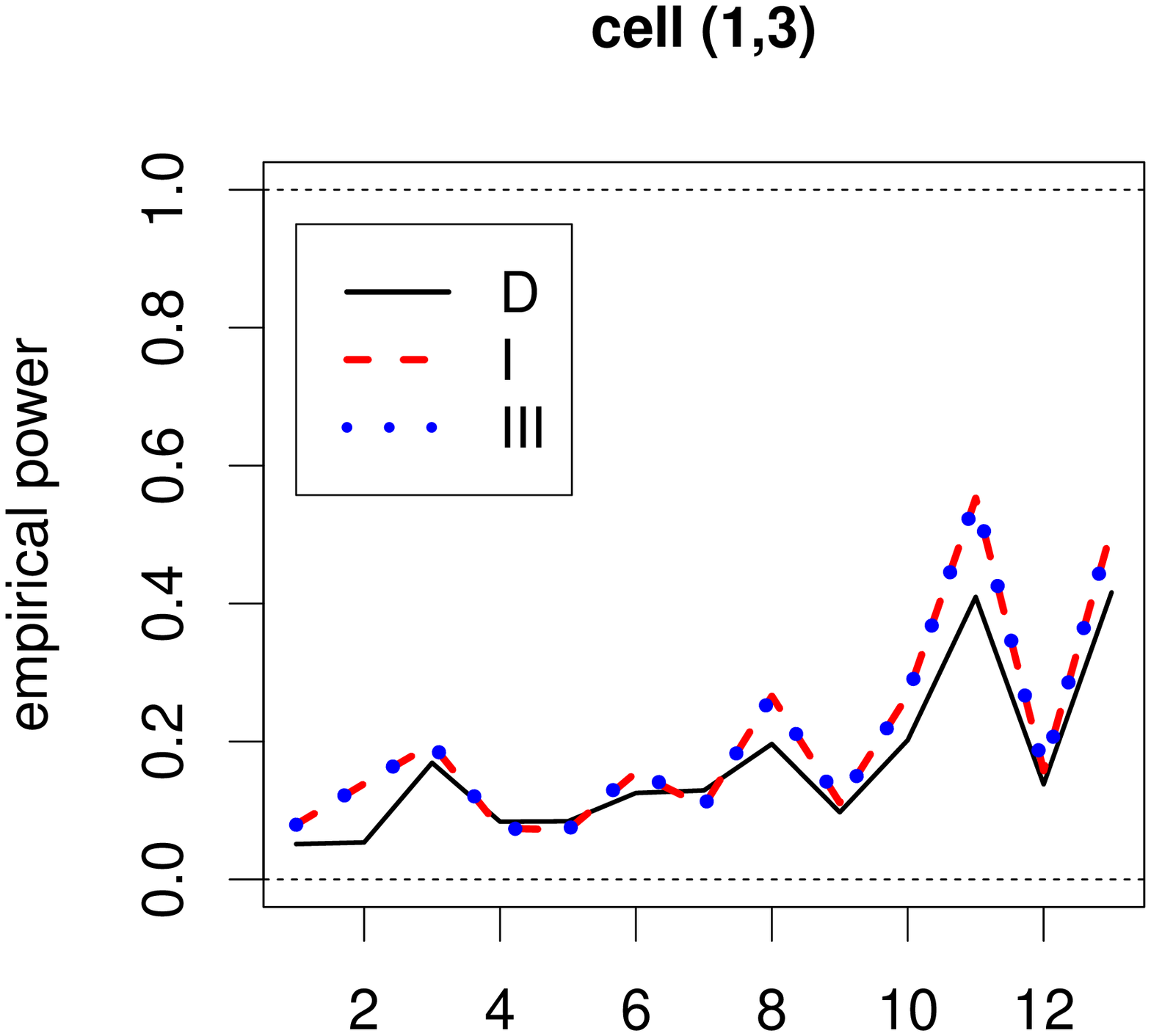} }}
\rotatebox{-90}{ \resizebox{2.1 in}{!}{\includegraphics{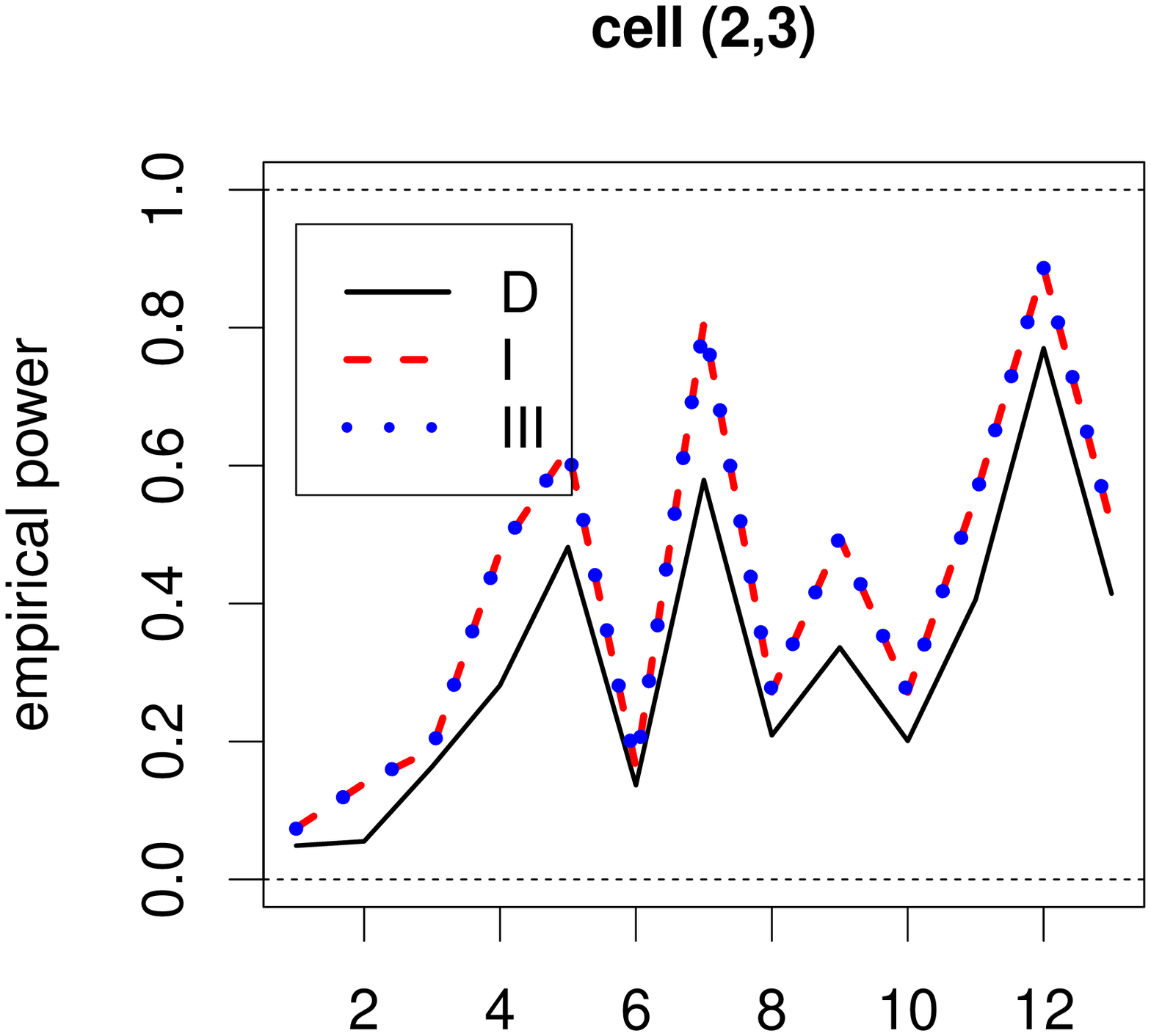} }}
\caption{
\label{fig:power-seg-cell-3cl-off-diag}
The empirical power estimates of the cell-specific tests for cells $(1,2)$, $(1,3)$, and $(2,3)$
under the segregation alternatives
$H_{S_1}$ (top), $H_{S_2}$ (middle), and $H_{S_3}$ (bottom) in the three-class case.
The legend labeling is as in Figure \ref{fig:emp-size-CSR-2cl}
and
horizontal axis labels are as in Figure \ref{fig:emp-size-CSR-cell-3cl}.
}
\end{figure}

\begin{figure} [hbp]
\centering
%\psfrag{Density}{ \Huge{\bf{Density}}}
Empirical Power Estimates of Overall Tests under $H_S$\\
\rotatebox{-90}{ \resizebox{2.1 in}{!}{\includegraphics{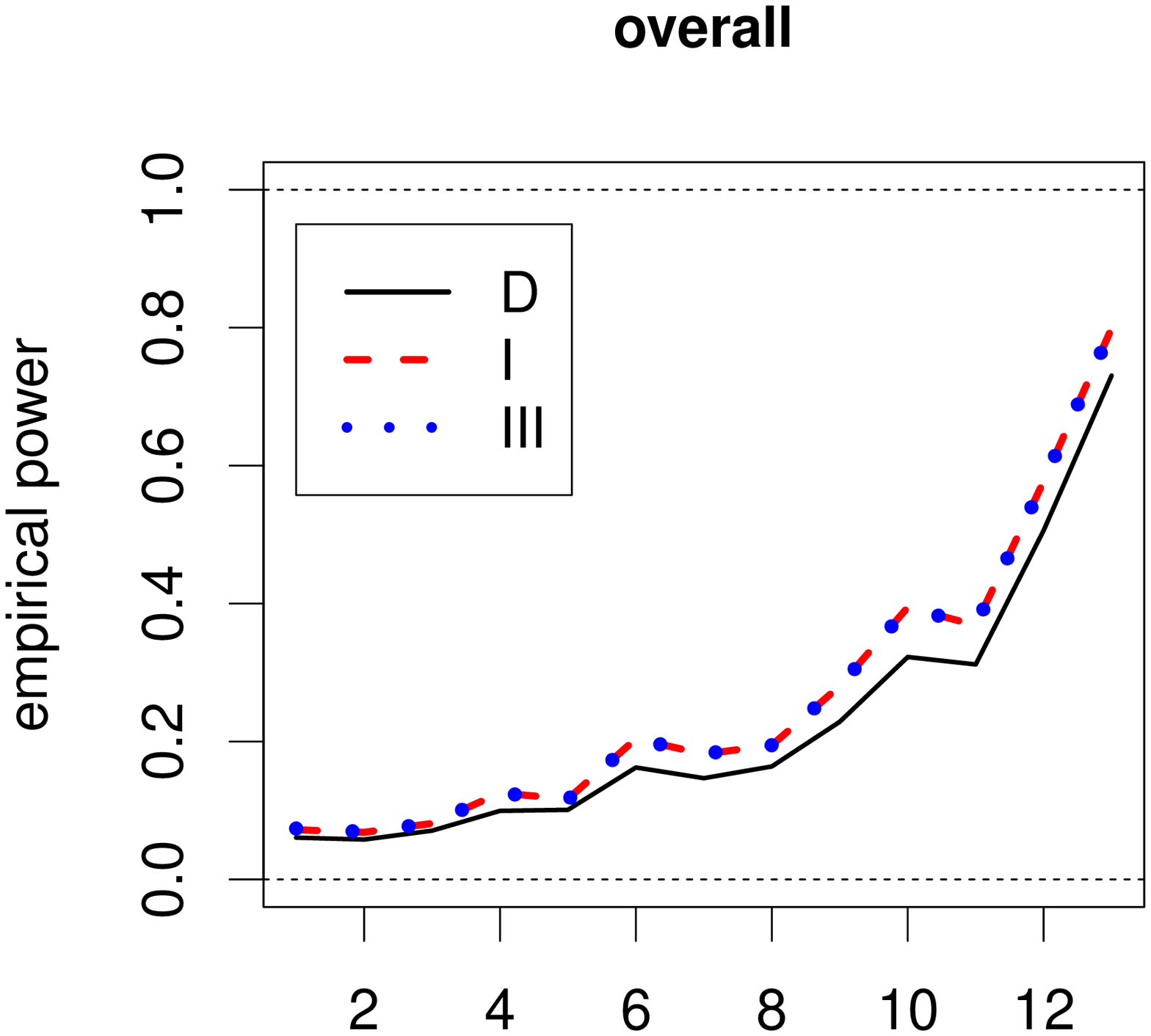} }}
\rotatebox{-90}{ \resizebox{2.1 in}{!}{\includegraphics{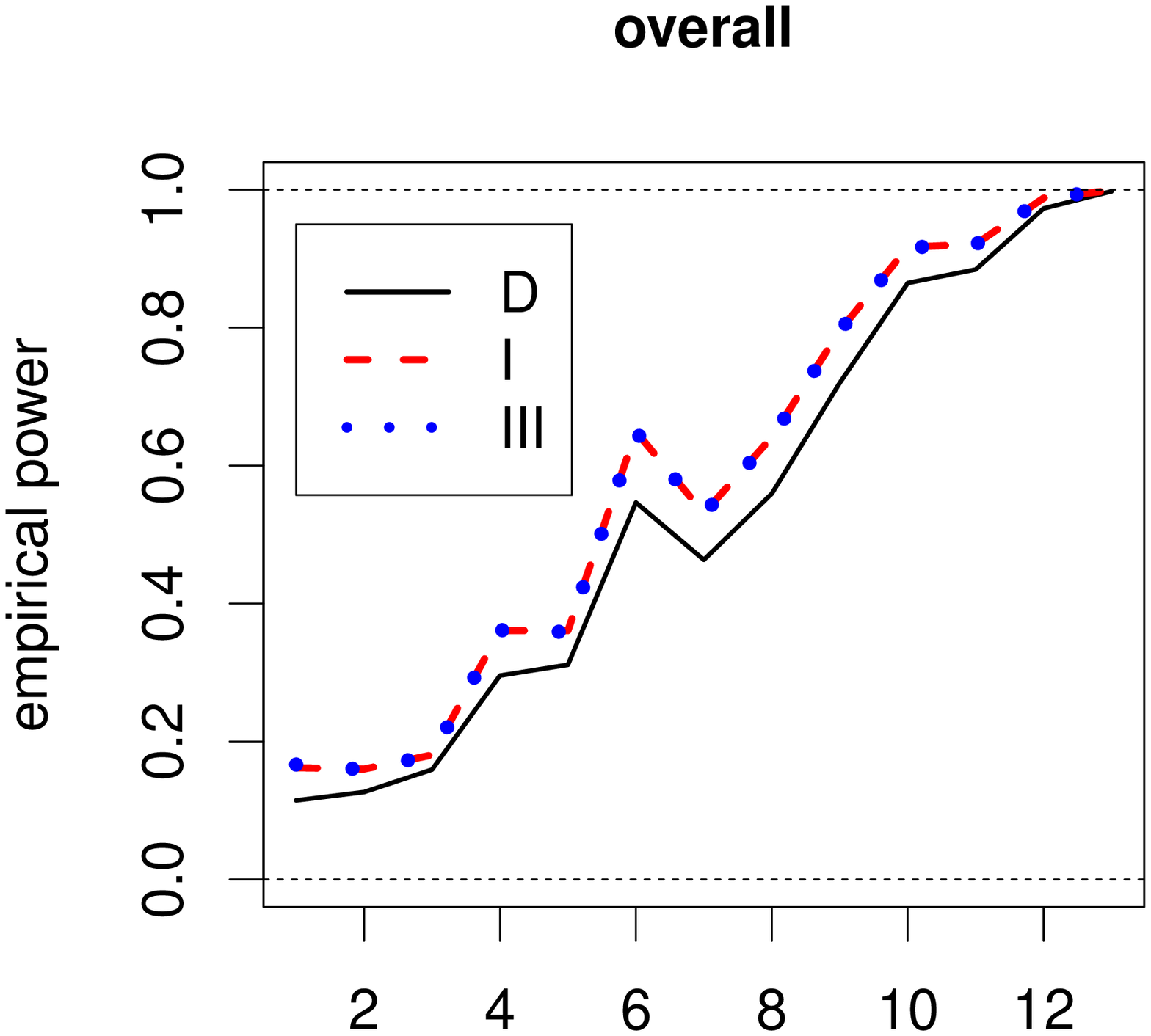} }}
\rotatebox{-90}{ \resizebox{2.1 in}{!}{\includegraphics{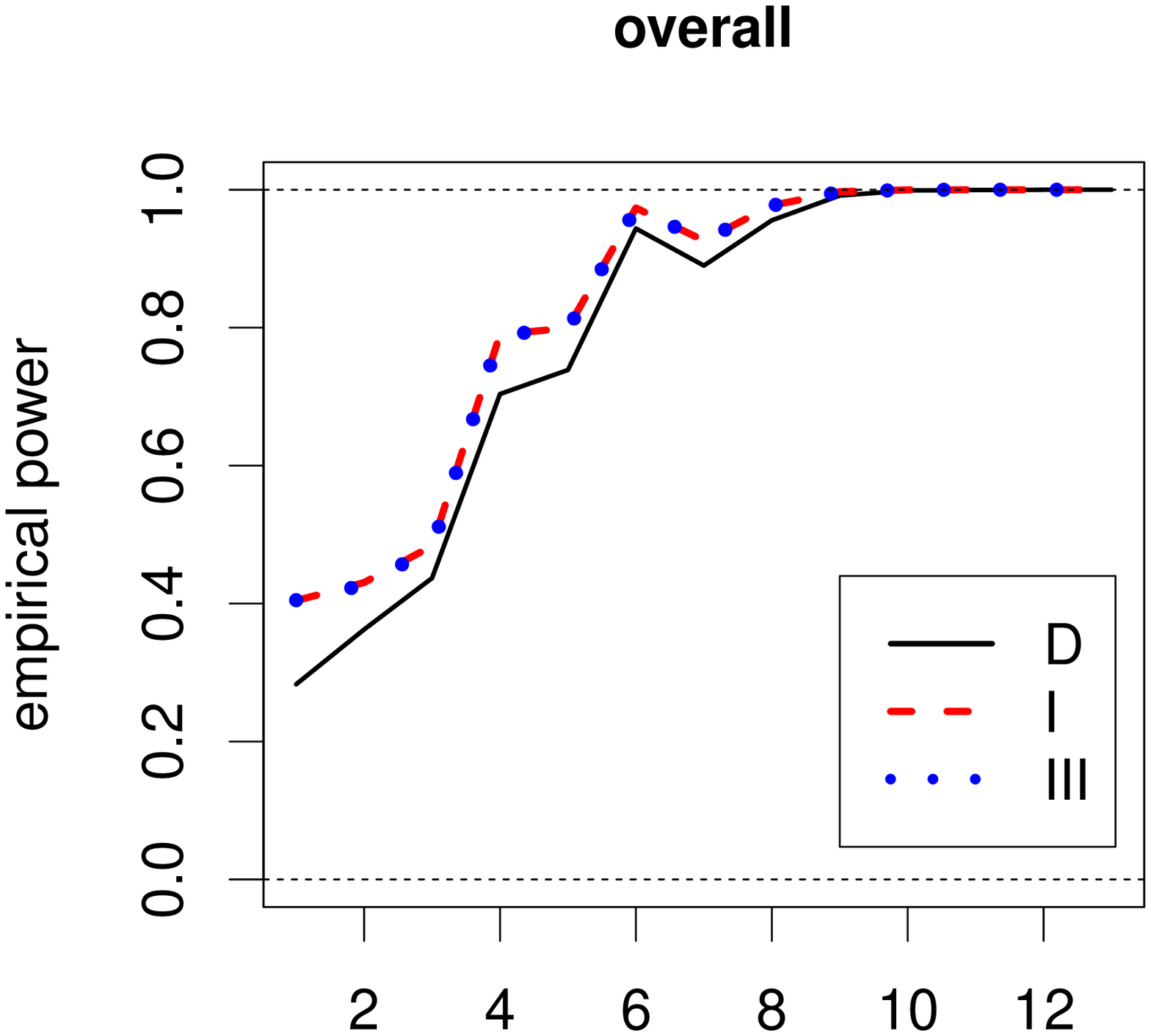} }}
\caption{
\label{fig:power-seg-overall-3cl}
The empirical power estimates of the overall tests under the segregation alternatives
$H_{S_1}$ (left), $H_{S_2}$ (middle), and $H_{S_3}$ (right) in the three-class case.
The legend labeling is as in Figure \ref{fig:emp-size-CSR-2cl}
and
horizontal axis labels are as in Figure \ref{fig:emp-size-CSR-cell-3cl}.
}
\end{figure}

Empirical power estimates for the two-sided alternatives for the diagonal cells $(1,1)$, $(2,2)$,
and $(3,3)$ under segregation alternatives
are plotted in Figure \ref{fig:power-seg-cell-3cl-diag}
and for the off-diagonal cells $(1,2)$, $(1,3)$, and $(2,3)$
are plotted in Figure \ref{fig:power-seg-cell-3cl-off-diag}.
For diagonal cells $(1,1)$ and $(2,2)$
type I and III tests have higher power,
while for diagonal cell $(3,3)$, all tests have similar power estimates.
For the off-diagonal cells $(1,2)$ and $(1,3)$
all tests have similar power estimates (although type I and III tests have slightly higher power),
while for cell $(2,3)$
type I and III tests have higher power.
In line with our simulation setup,
power estimates for cells $(1,1)$ and $(2,2)$ are higher compared to cell $(3,3)$,
as classes $X$ and $Y$ are more segregated compared to class $Z$.
For the same reason, power estimates for cell $(1,2)$ is higher compared to cells $(1,3)$ and $(2,3)$.

Empirical power estimates for the overall tests are presented in Figure \ref{fig:power-seg-overall-3cl}.
Type I and III tests have higher power compared to Dixon's test.

\subsection{Empirical Power Analysis under Association of Three Classes}
\label{sec:power-comp-assoc-3Cl}
Under the association alternatives, we also consider three cases.
We generate $X_i \stackrel{iid}{\sim} \U((0,1)\times(0,1))$ for $i=1,2,\ldots,n_1$.
Then we generate $Y_j$ and $Z_k$ for $j=1,2,\ldots,n_2$ and $k=1,2,\ldots,n_3$
as follows.
For each $j$, select an $i$ randomly, and set
$Y_j:=X_i+R^Y_j\,(\cos T_j, \sin T_j)'$
where
$R^Y_j \stackrel{iid}{\sim} \U(0,r_y)$ with $r_y \in (0,1)$
and
$T_j \stackrel{iid}{\sim} \U(0,2\,\pi)$.
Similarly,
for each $k$, select an $i'$ randomly,
and set
$Z_k:=X_{i'}+R^Z_k\,(\cos U_{\ell}, \sin U_{\ell})'$
where
$R^Z_k\stackrel{iid}{\sim} \U(0,r_z)$ with $r_z \in (0,1)$
and
$U_k\stackrel{iid}{\sim} \U(0,2\,\pi)$.
We consider the following association alternatives:
\begin{multline}
\label{eqn:assoc-alt-3Cl}
H_{A_1}: r_y=1/(2\sqrt{n_t}),\,r_z=1/(3\sqrt{n_t}),\;\;\;
H_{A_2}: r_y=1/(2\sqrt{n_t}),\,r_z=1/(4\sqrt{n_t}),\\
\text{ and }
H_{A_3}: r_y=1/(3\sqrt{n_t}),\,r_z=1/(4\sqrt{n_t})
\end{multline}
where $n_t=n_1+n_2+n_3$.
As $r_y$ and $r_z$ decrease, the level of association increases.
That is,
the association between $X$ and $Y$
gets stronger from $H_{A_1}$ to $H_{A_3}$;
and
the association between $X$ and $Z$
gets stronger from $H_{A_1}$ to $H_{A_2}$.
By construction,
classes $Y$ and $Z$ are associated with class $X$,
while classes $Y$ and $Z$ are not associated, but perhaps mildly segregated for small $r_y$ and $r_z$.
Furthermore, by construction,
classes $X$ and $Z$ are more associated compared to classes $X$ and $Y$
at each association alternative.

\begin{figure} [hbp]
\centering
%\psfrag{Density}{ \Huge{\bf{Density}}}
Empirical Power Estimates of Cell-Specific Tests under $H_A$\\
%\rotatebox{-90}{ \resizebox{2.1 in}{!}{\includegraphics{AssocI3Cl11.ps} }}
%\rotatebox{-90}{ \resizebox{2.1 in}{!}{\includegraphics{AssocII3Cl11.ps} }}
%\rotatebox{-90}{ \resizebox{2.1 in}{!}{\includegraphics{AssocIII3Cl11.ps} }}

\rotatebox{-90}{ \resizebox{2.1 in}{!}{\includegraphics{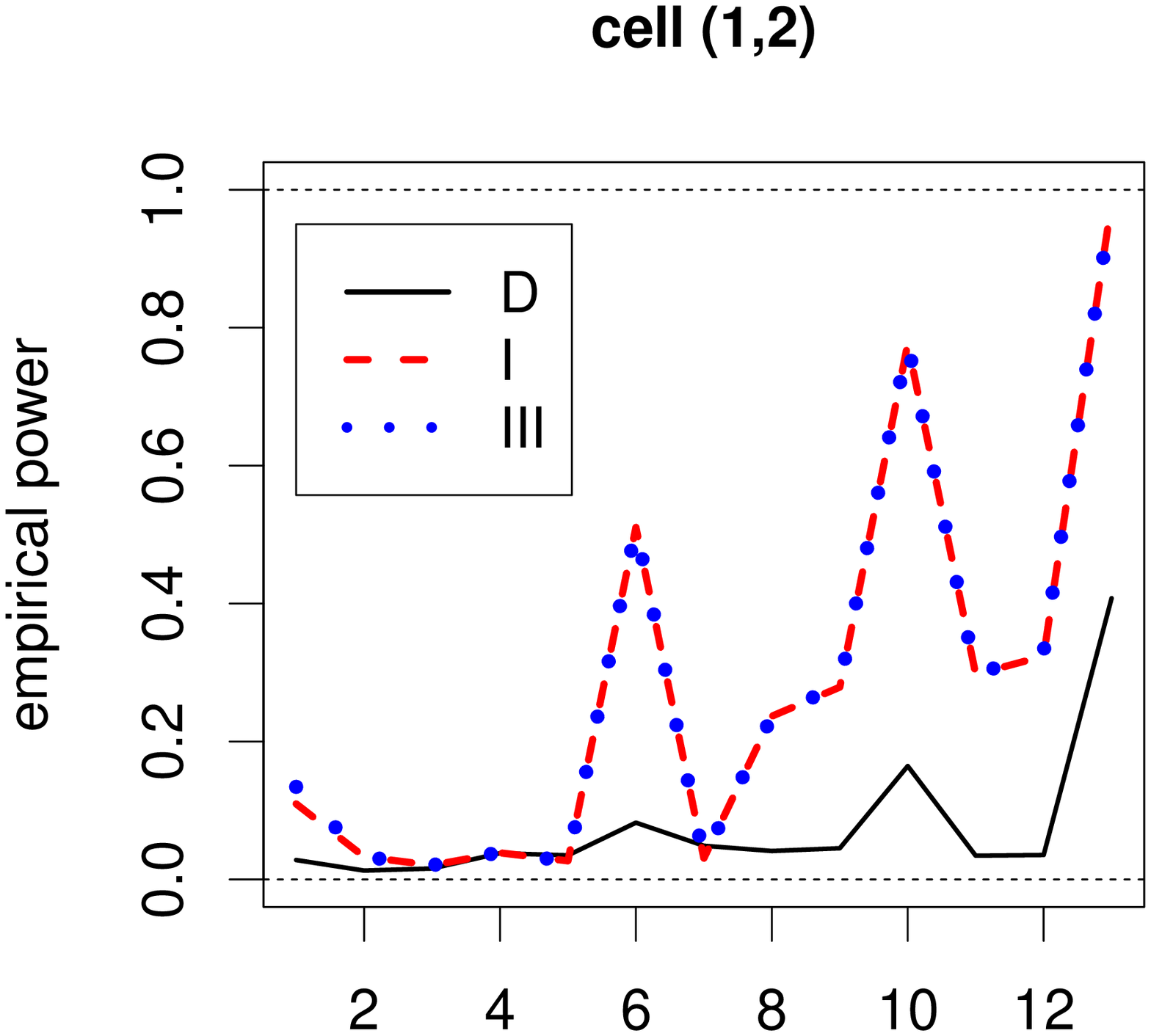} }}
\rotatebox{-90}{ \resizebox{2.1 in}{!}{\includegraphics{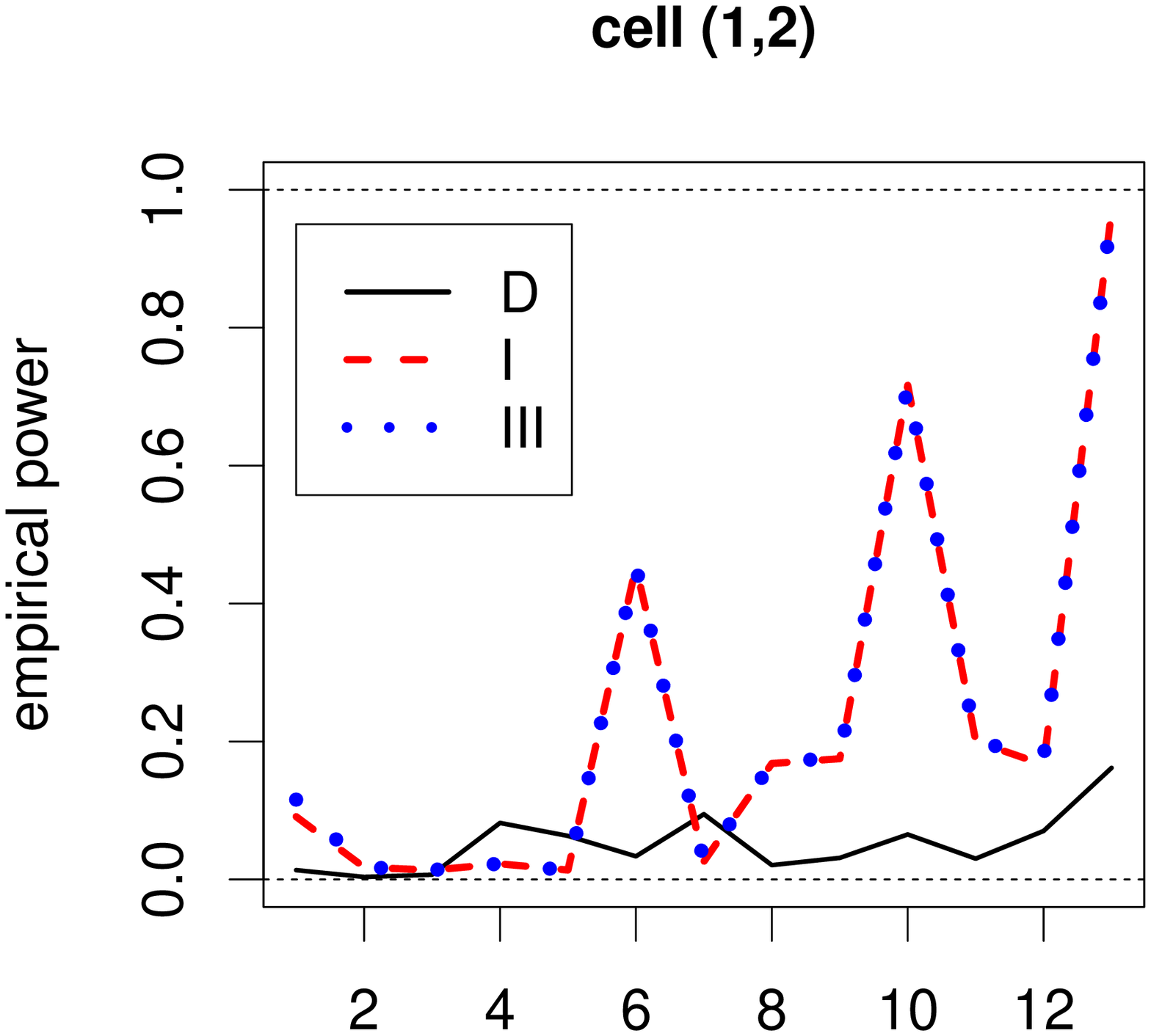} }}
\rotatebox{-90}{ \resizebox{2.1 in}{!}{\includegraphics{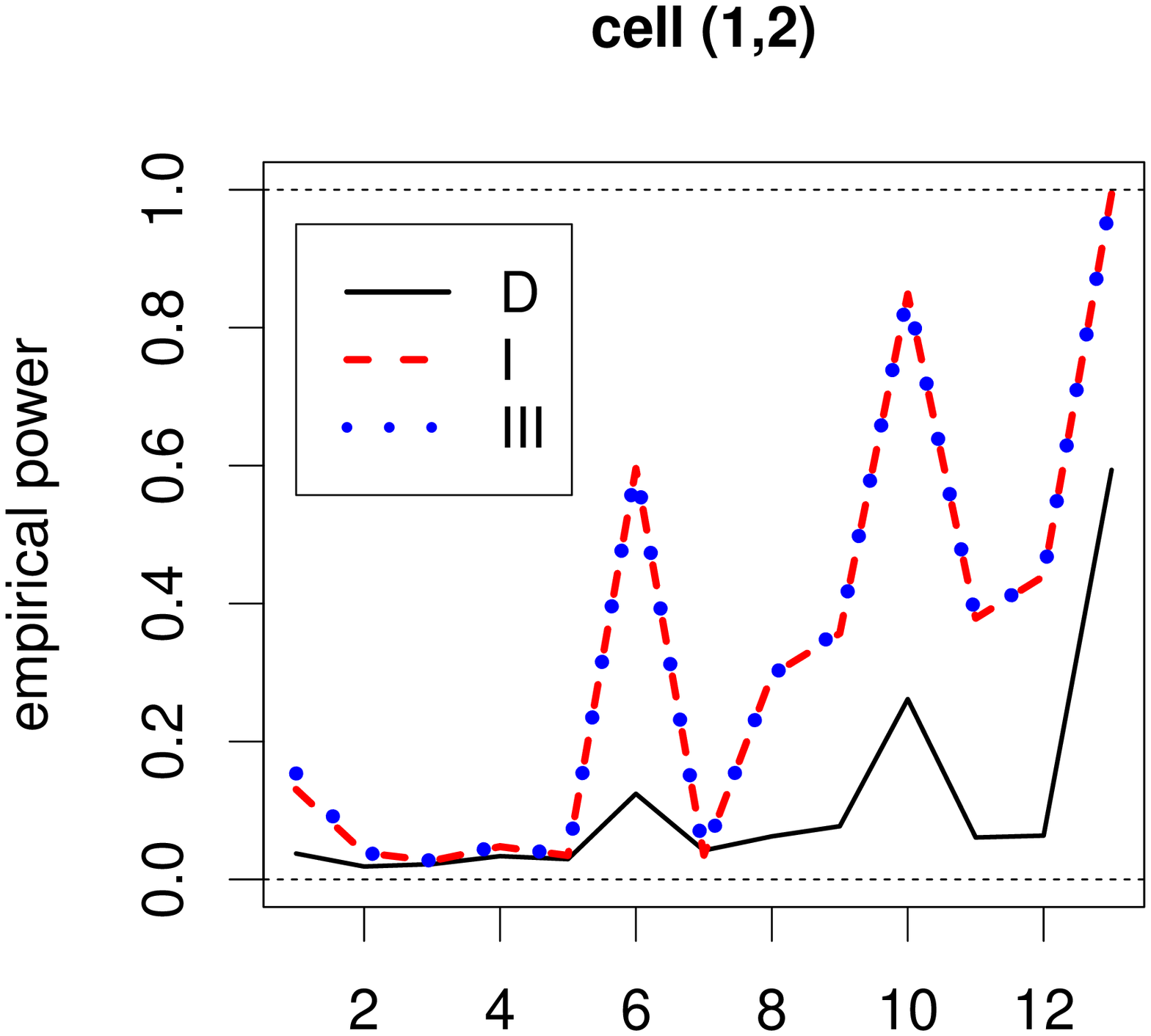} }}

\rotatebox{-90}{ \resizebox{2.1 in}{!}{\includegraphics{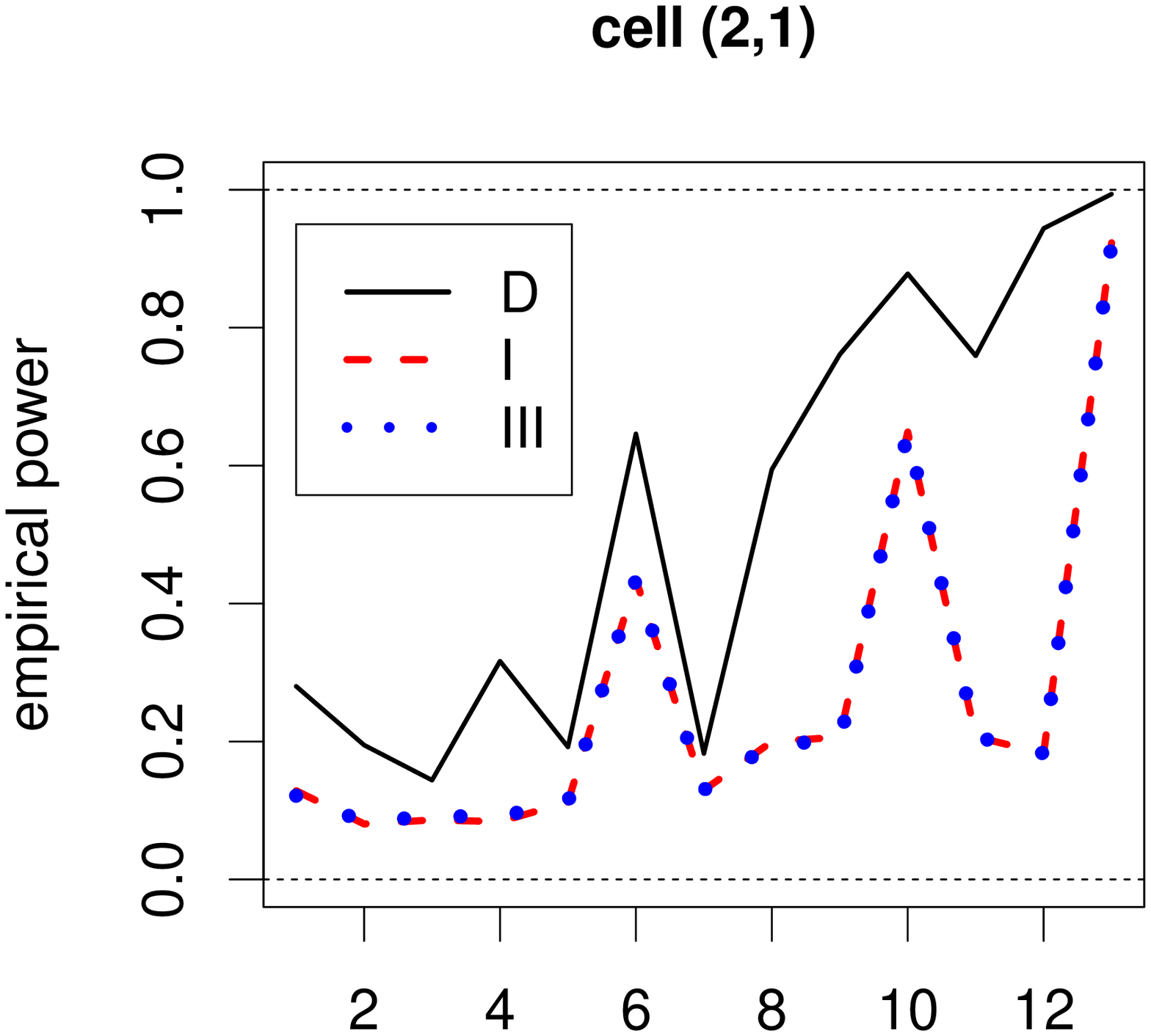} }}
\rotatebox{-90}{ \resizebox{2.1 in}{!}{\includegraphics{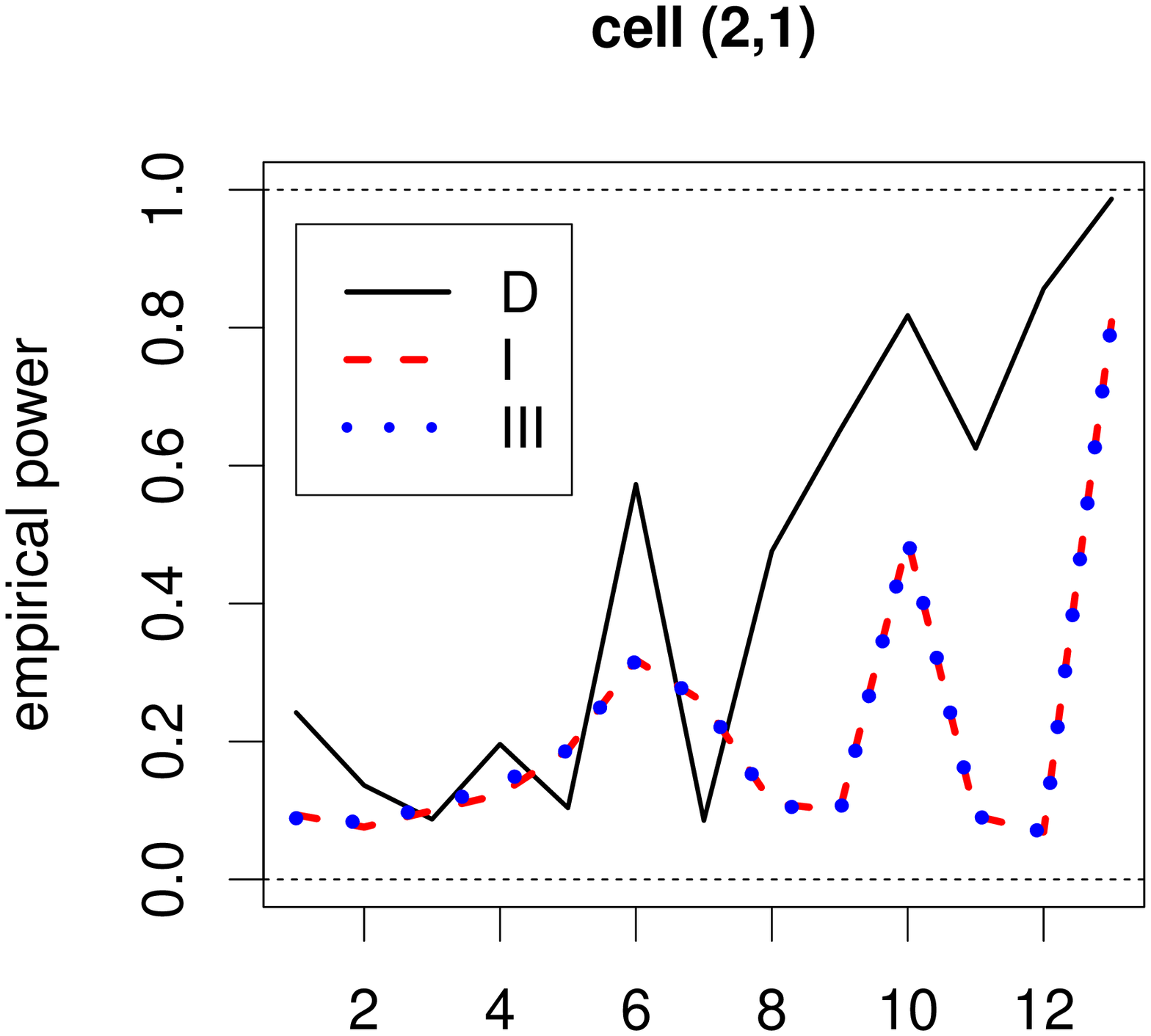} }}
\rotatebox{-90}{ \resizebox{2.1 in}{!}{\includegraphics{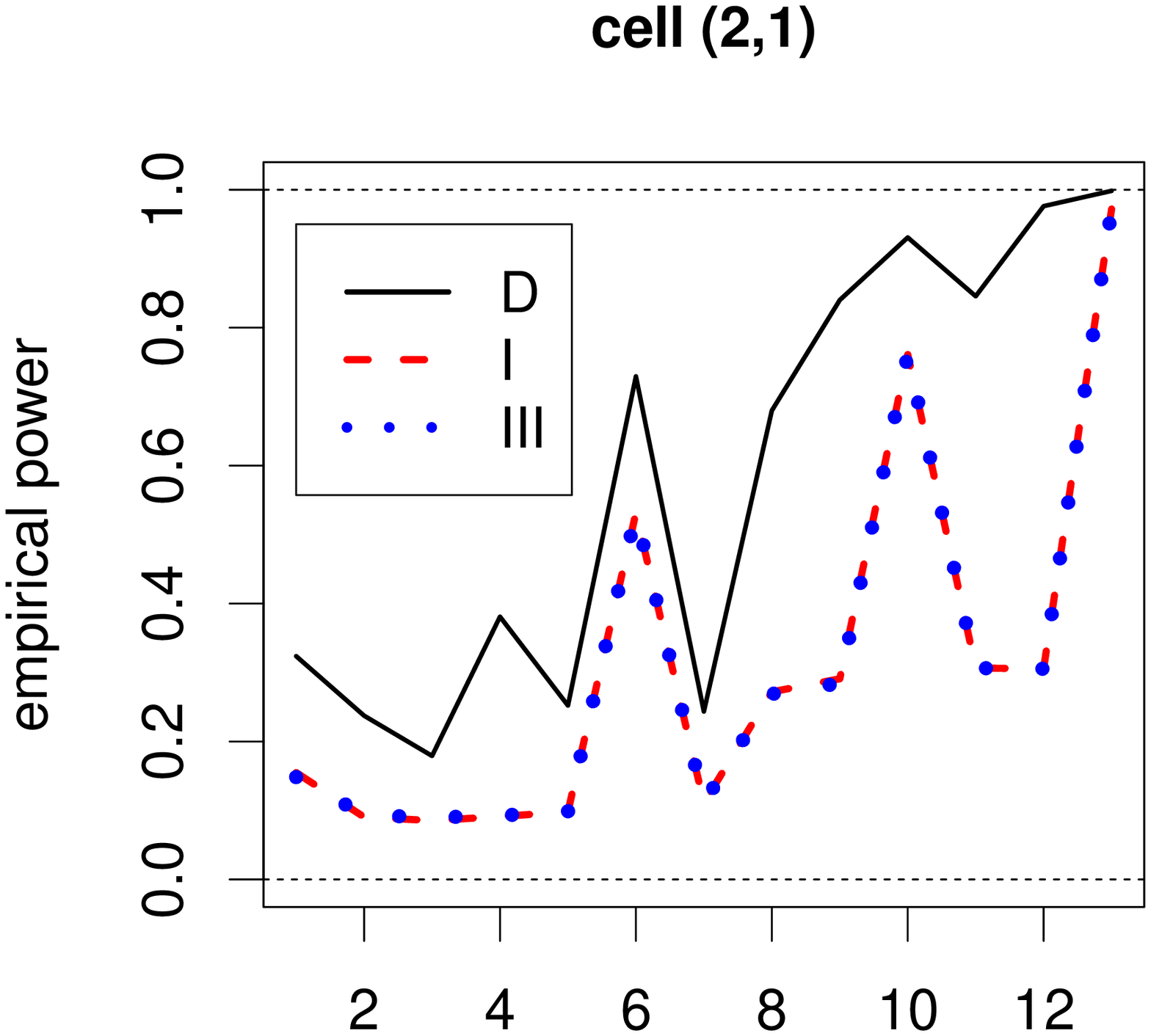} }}
 \caption{
\label{fig:power-assoc-cell-3cl-12}
The empirical power estimates of the
cell-specific tests for cells $(1,2)$ and $(2,1)$ under the association alternatives
$H_{A_1}$ (left), $H_{A_2}$ (middle), and $H_{A_3}$ (right) in the three-class case.
The legend labeling is as in Figure \ref{fig:emp-size-CSR-2cl}
and
horizontal axis labels are as in Figure \ref{fig:emp-size-CSR-cell-3cl}.
}
\end{figure}

\begin{figure} [hbp]
\centering
%\psfrag{Density}{ \Huge{\bf{Density}}}
Empirical Power Estimates of Cell-Specific Tests under $H_A$\\
\rotatebox{-90}{ \resizebox{2.1 in}{!}{\includegraphics{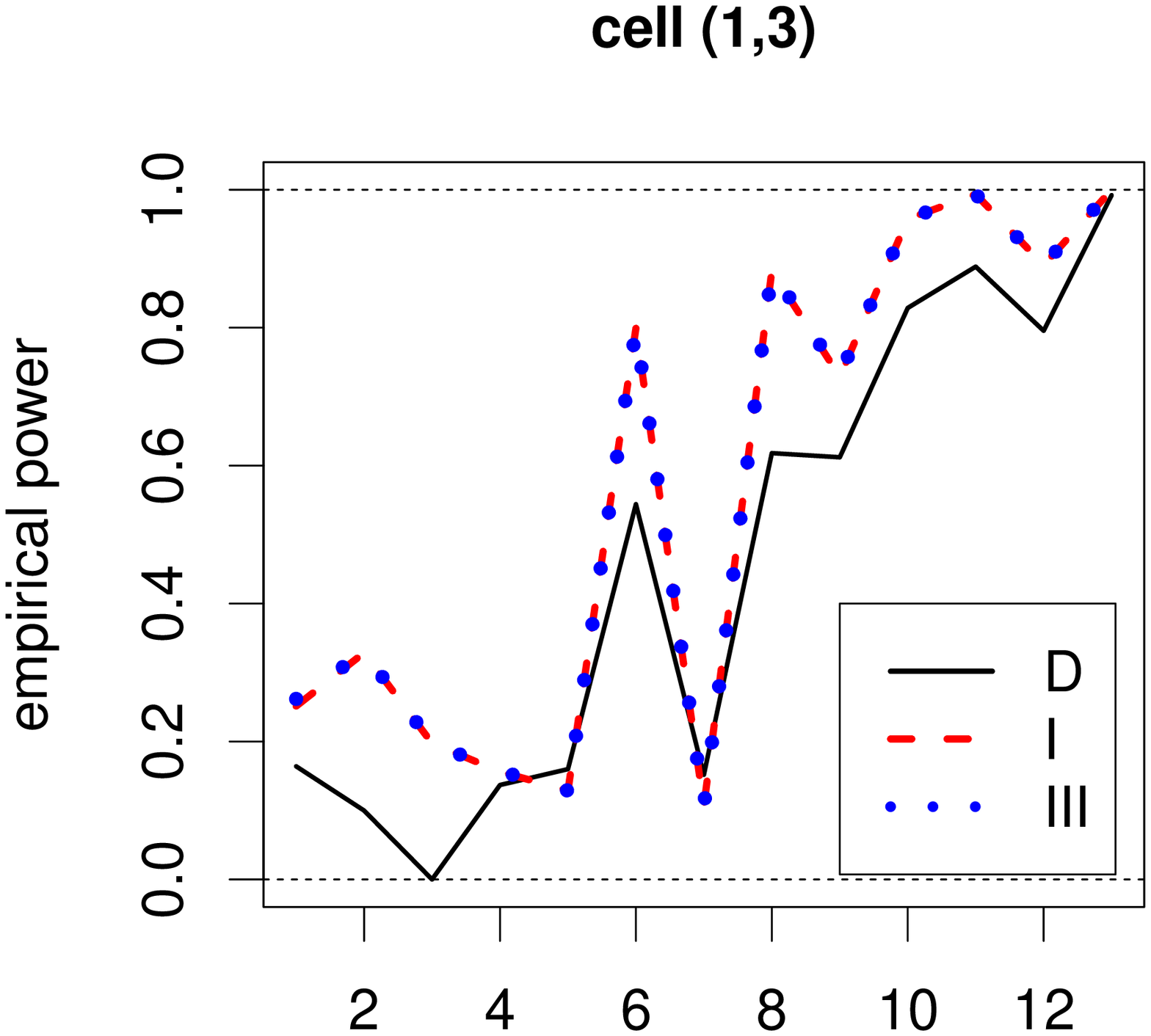} }}
\rotatebox{-90}{ \resizebox{2.1 in}{!}{\includegraphics{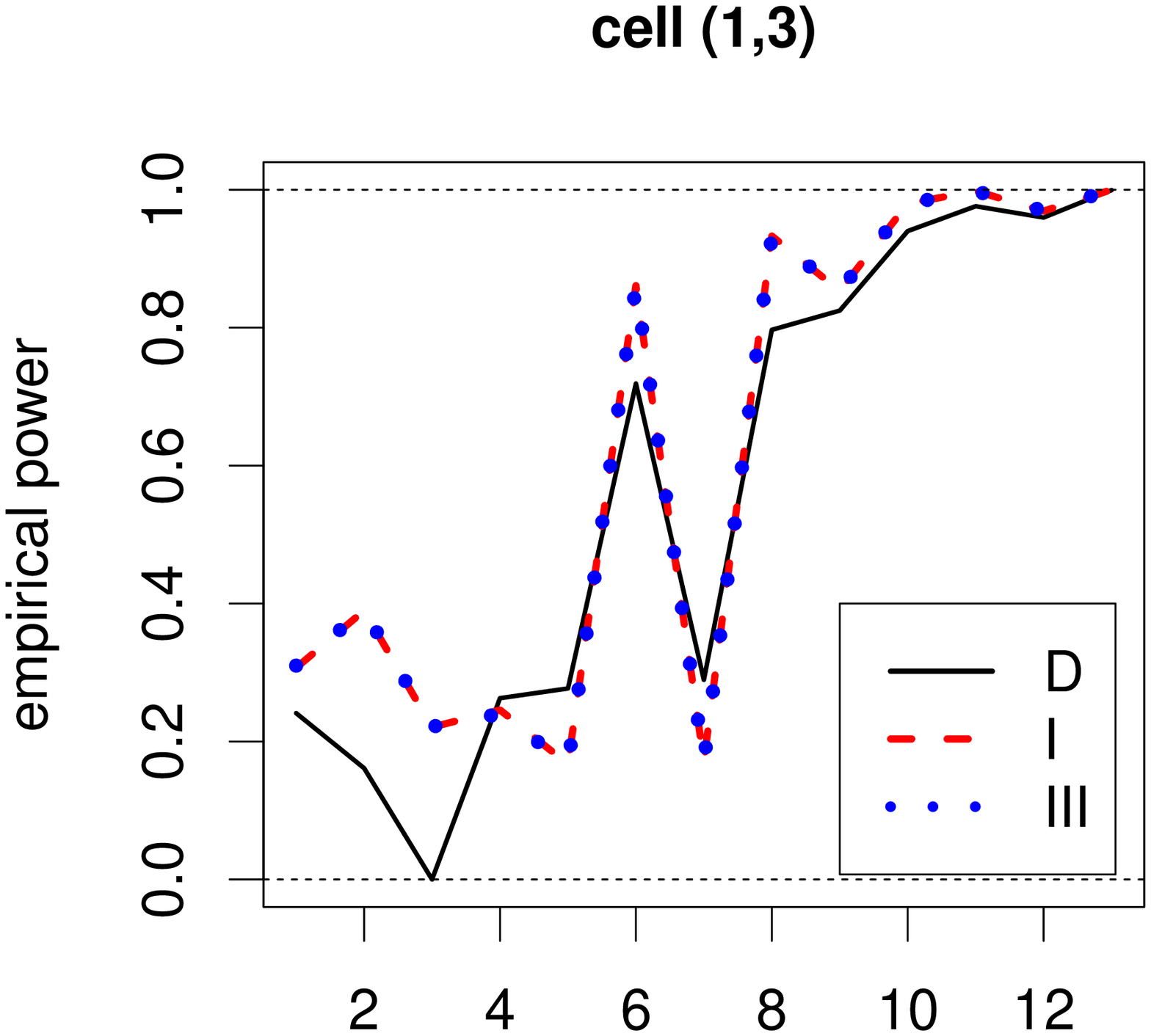} }}
\rotatebox{-90}{ \resizebox{2.1 in}{!}{\includegraphics{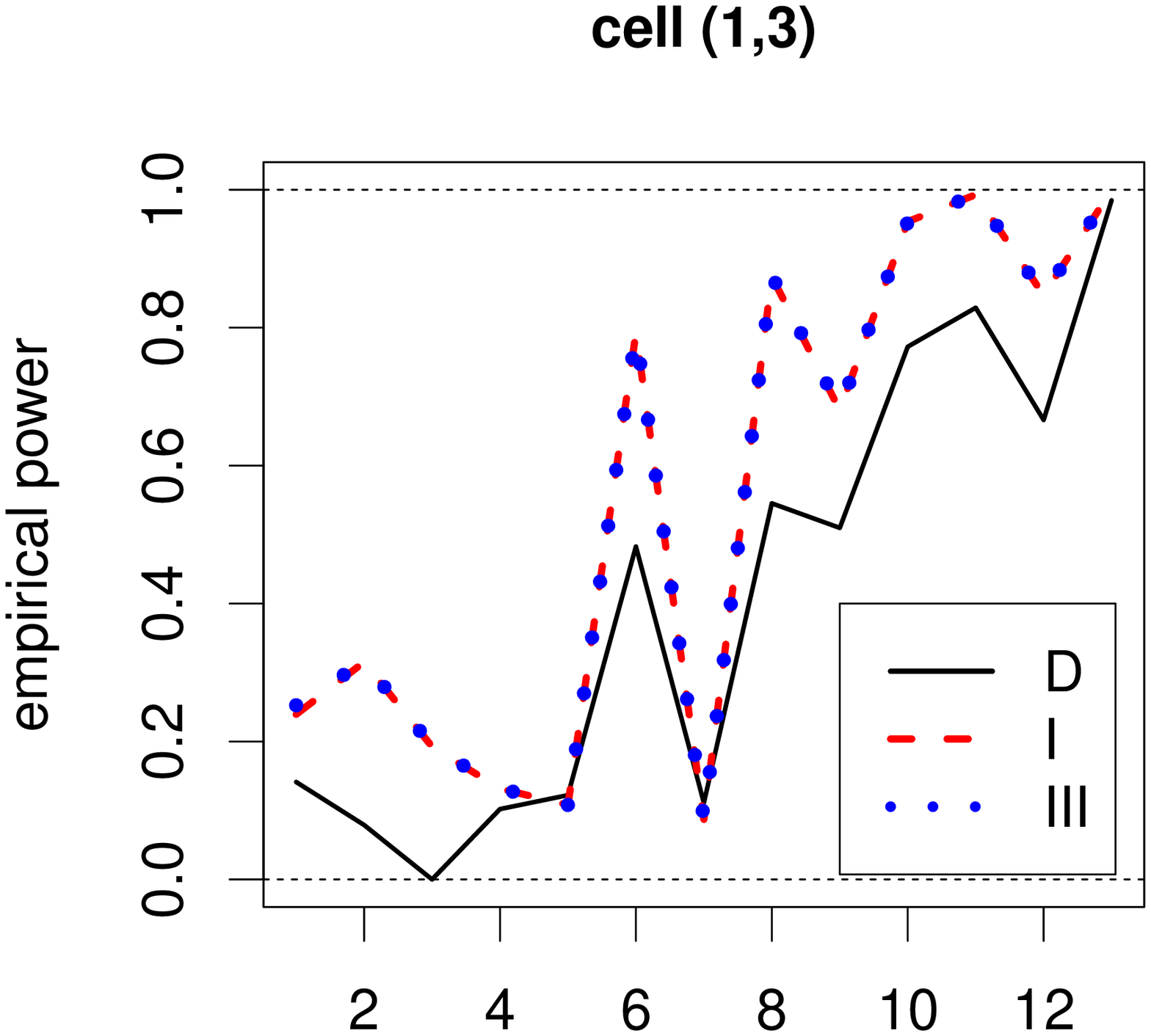} }}

\rotatebox{-90}{ \resizebox{2.1 in}{!}{\includegraphics{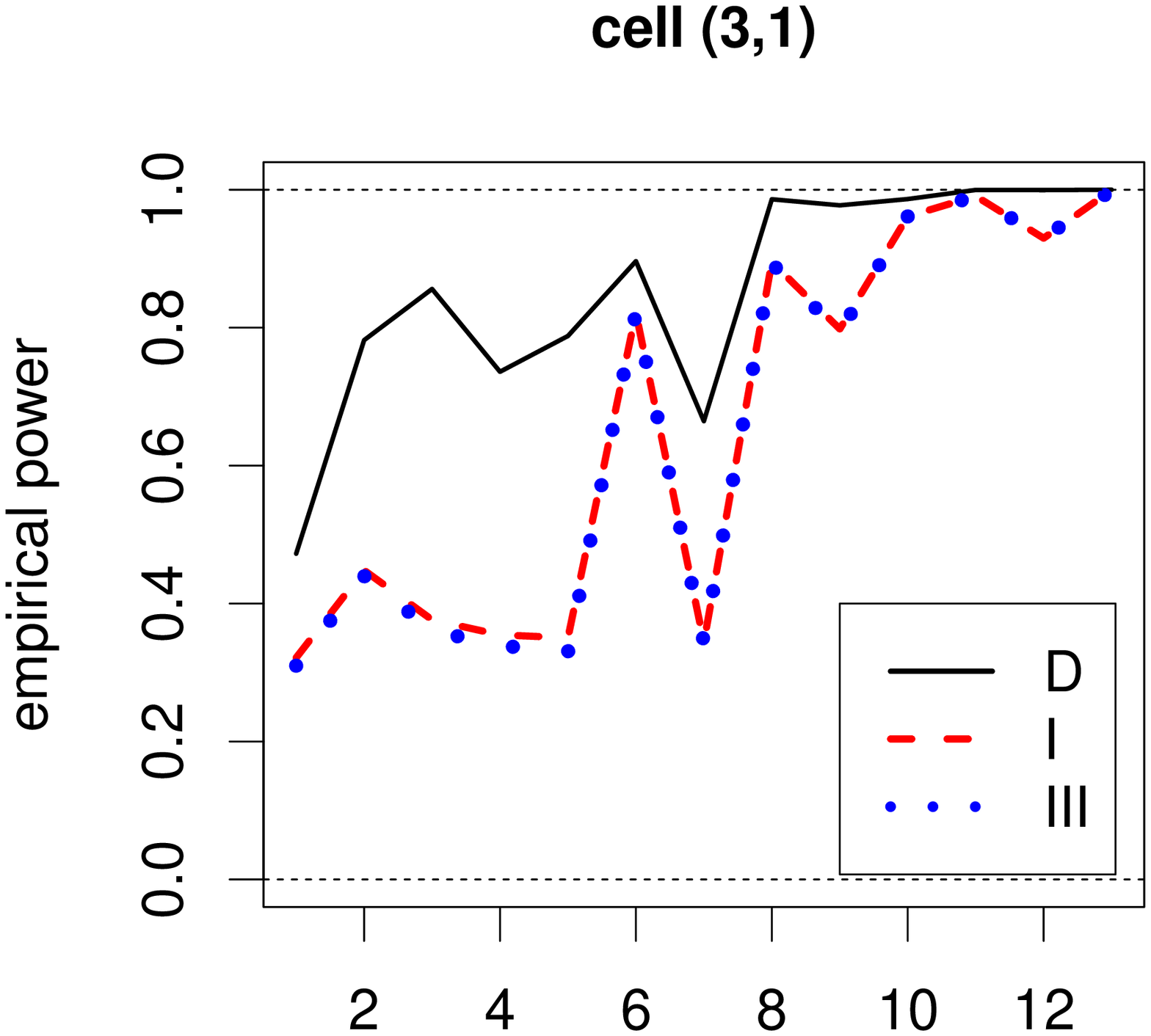} }}
\rotatebox{-90}{ \resizebox{2.1 in}{!}{\includegraphics{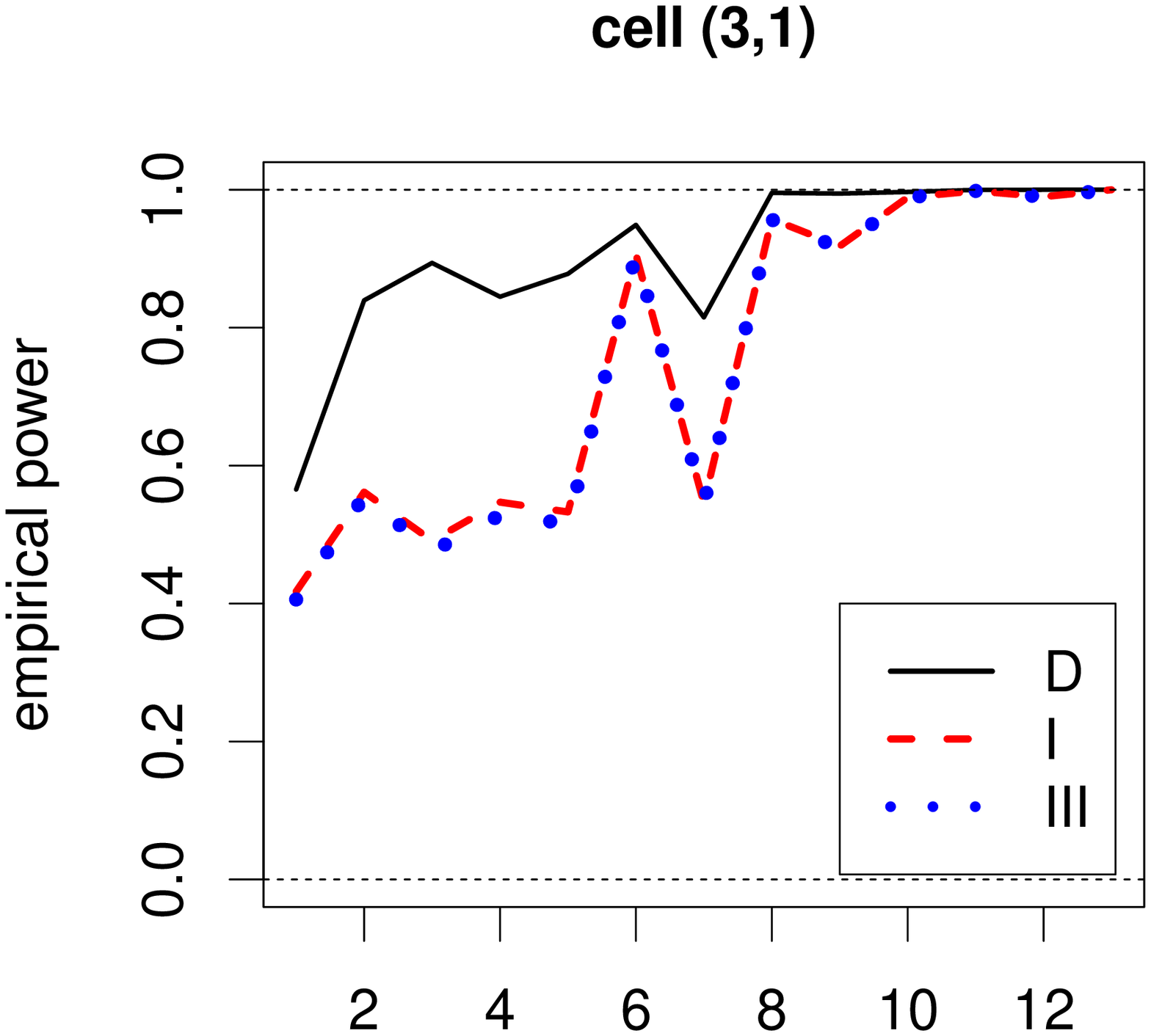} }}
\rotatebox{-90}{ \resizebox{2.1 in}{!}{\includegraphics{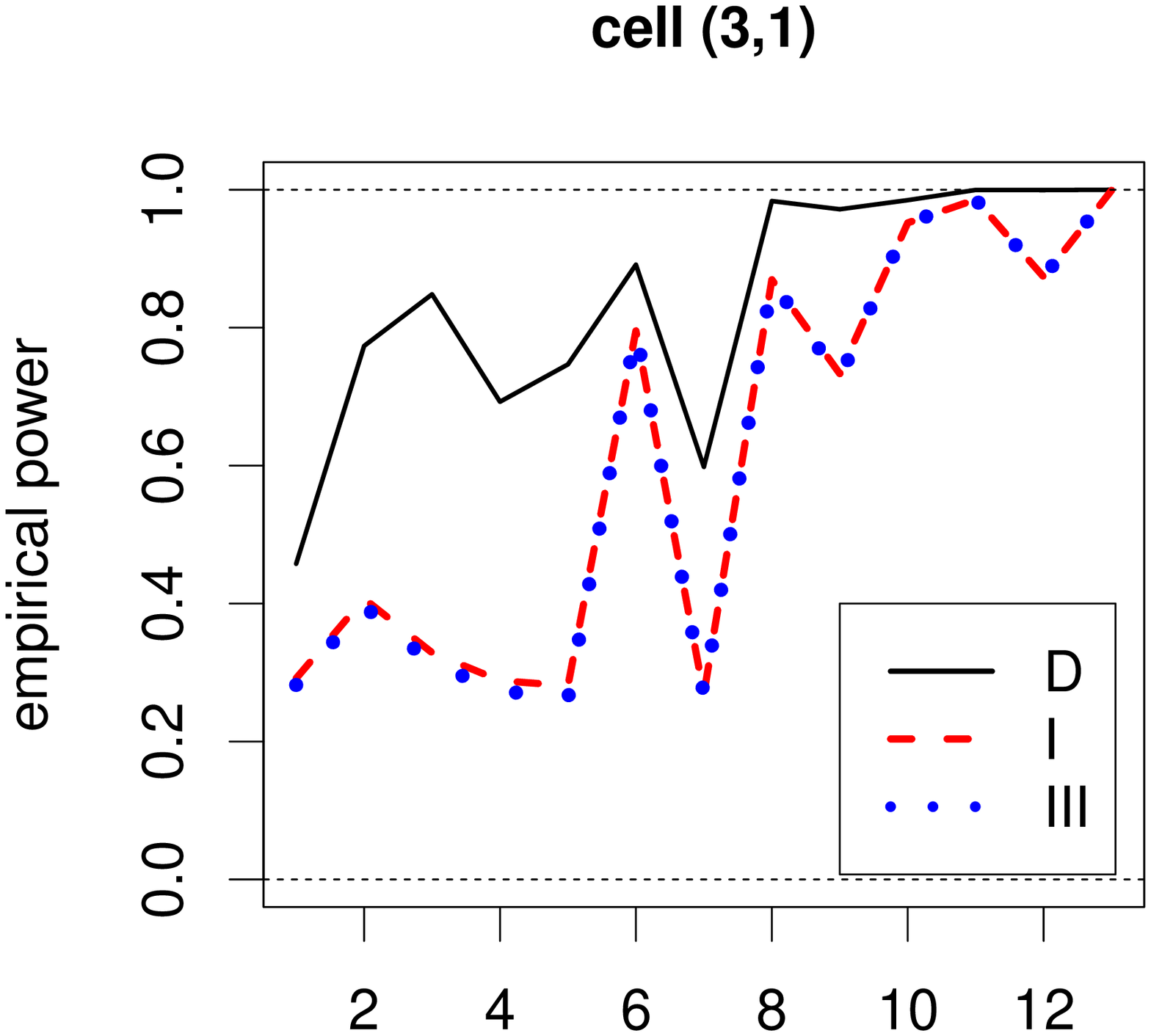} }}
 \caption{
\label{fig:power-assoc-cell-3cl-13}
The empirical power estimates of the
cell-specific tests for cells $(1,3)$ and $(3,1)$ under the association alternatives
$H_{A_1}$ (left), $H_{A_2}$ (middle), and $H_{A_3}$ (right) in the three-class case.
The legend labeling is as in Figure \ref{fig:emp-size-CSR-2cl}
and
horizontal axis labels are as in Figure \ref{fig:emp-size-CSR-cell-3cl}.
}
\end{figure}

\begin{figure} [hbp]
\centering
%\psfrag{Density}{ \Huge{\bf{Density}}}
Empirical Power Estimates of Overall Tests under $H_A$\\
\rotatebox{-90}{ \resizebox{2.1 in}{!}{\includegraphics{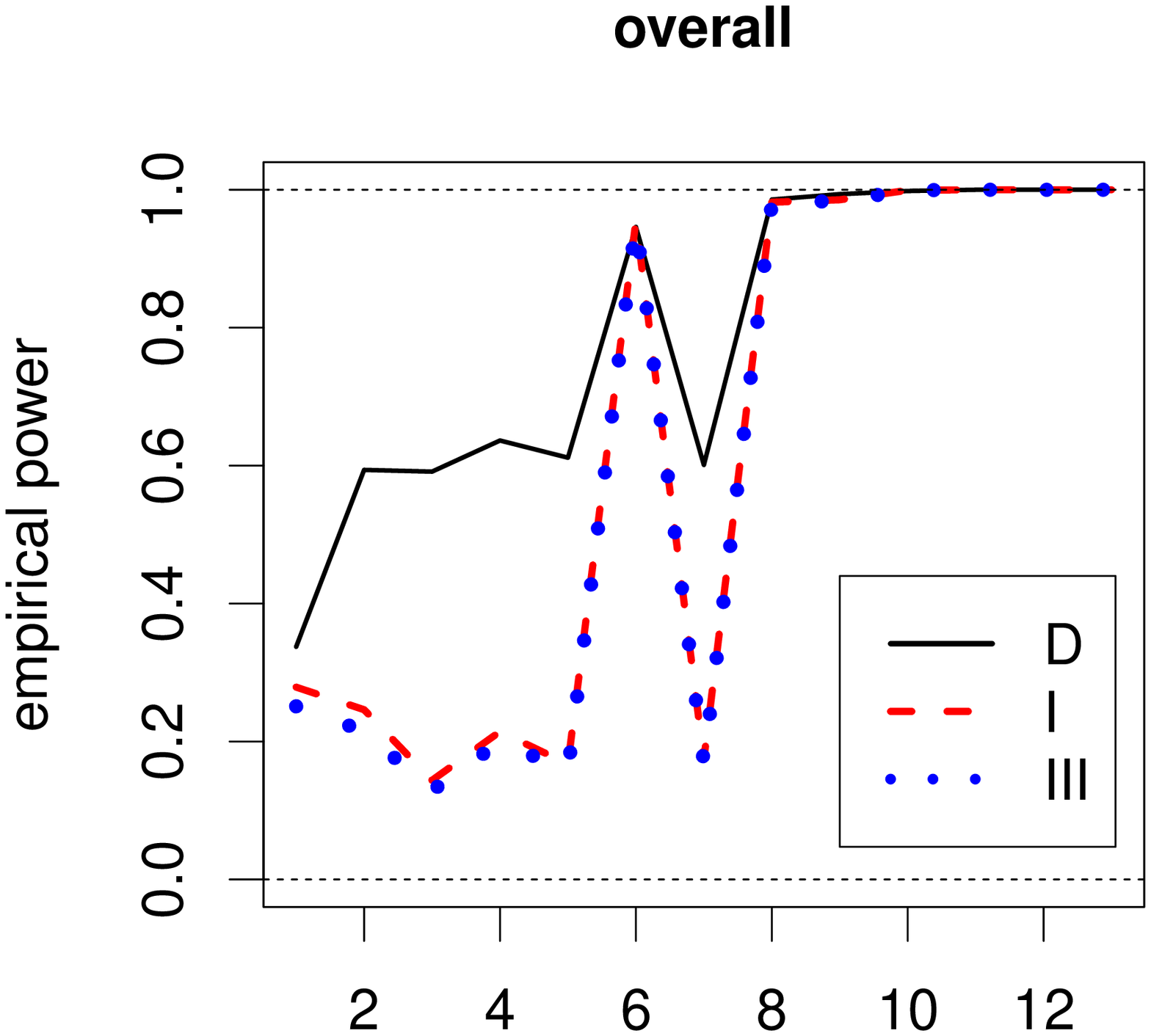} }}
\rotatebox{-90}{ \resizebox{2.1 in}{!}{\includegraphics{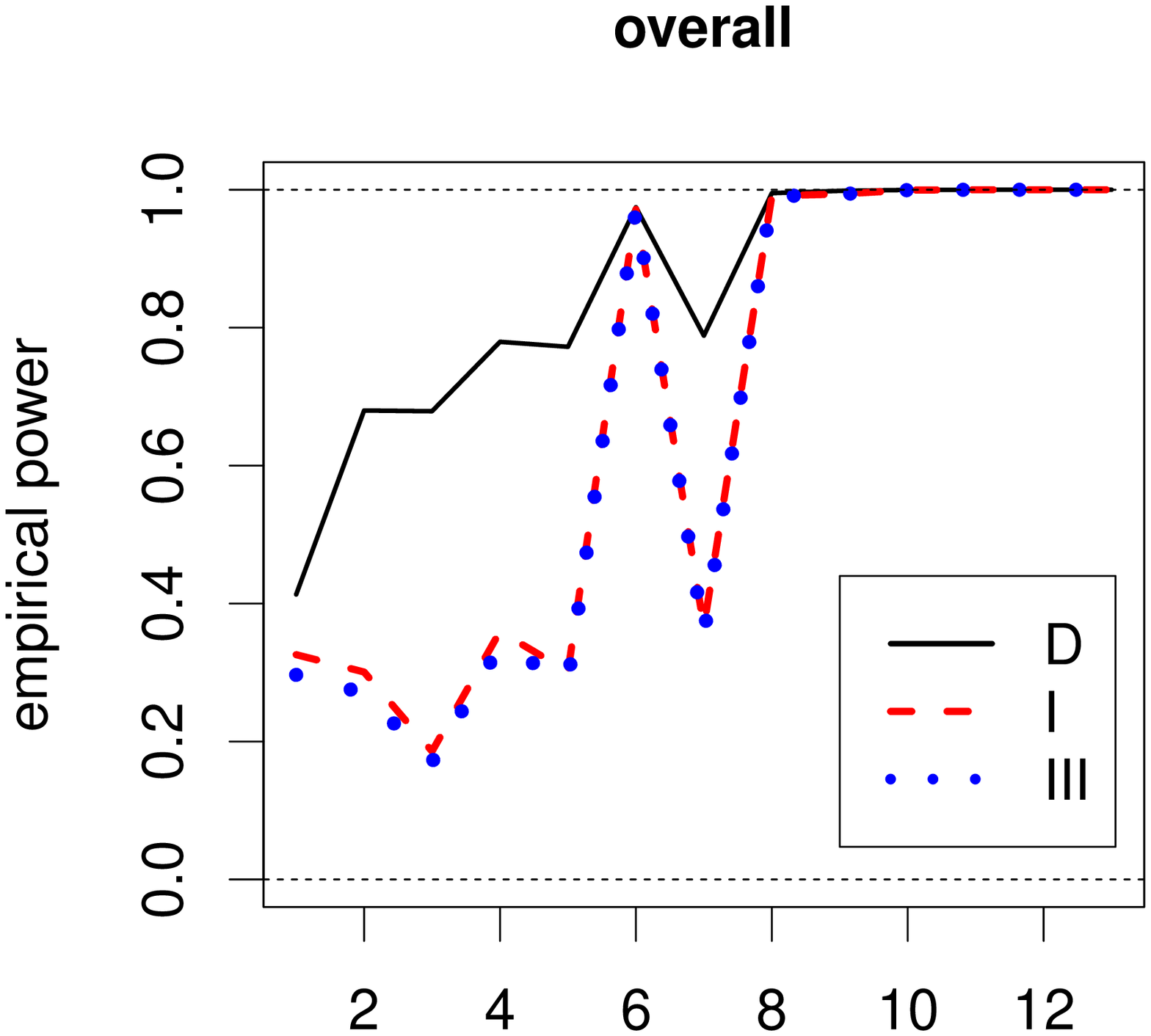} }}
\rotatebox{-90}{ \resizebox{2.1 in}{!}{\includegraphics{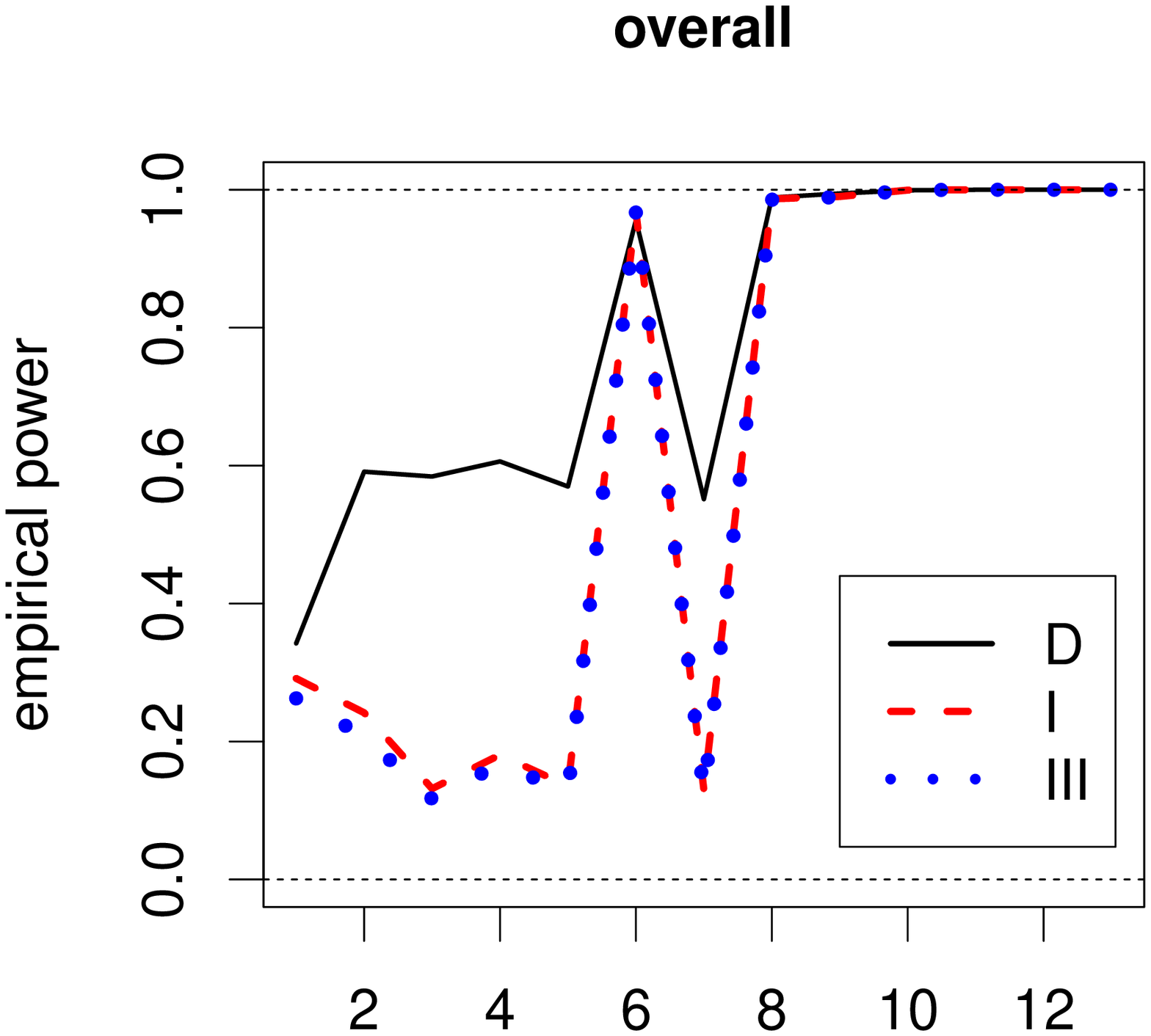} }}
\caption{
\label{fig:power-assoc-overall-3cl}
The empirical power estimates of the overall test under the association alternatives
$H_{A_1}$ (left), $H_{A_2}$ (middle), and $H_{A_3}$ (right) in the three-class case.
The legend labeling is as in Figure \ref{fig:emp-size-CSR-2cl}
and
horizontal axis labels are as in Figure \ref{fig:emp-size-CSR-cell-3cl}.
}
\end{figure}

The empirical power estimates for cells $(1,2)$ and $(2,1)$ are presented in Figure \ref{fig:power-assoc-cell-3cl-12},
and estimates for cells $(1,3)$ and $(3,1)$ are presented in Figure \ref{fig:power-assoc-cell-3cl-13}.
For cells $(1,2)$ and $(1,3)$,
type I and III cell-specific tests have higher power,
while for cells $(2,1)$ and $(3,1)$,
Dixon's cell-specific test has higher power.
The power estimates for the overall tests are presented in Figure \ref{fig:power-assoc-overall-3cl}.
For the overall tests,
Dixon's test has higher power estimates.

\section{Empirical Size and Power Analysis for the One-vs-Rest Type Tests in the Three Class Case}
\label{sec:one-vs-rest-size-power}
\begin{figure} [hbp]
\centering
%\psfrag{Density}{ \Huge{\bf{Density}}}
Empirical Size Estimates of Cell-Specific Tests for cell $(2,2)$ under CSR\\
\rotatebox{-90}{ \resizebox{2.1 in}{!}{\includegraphics{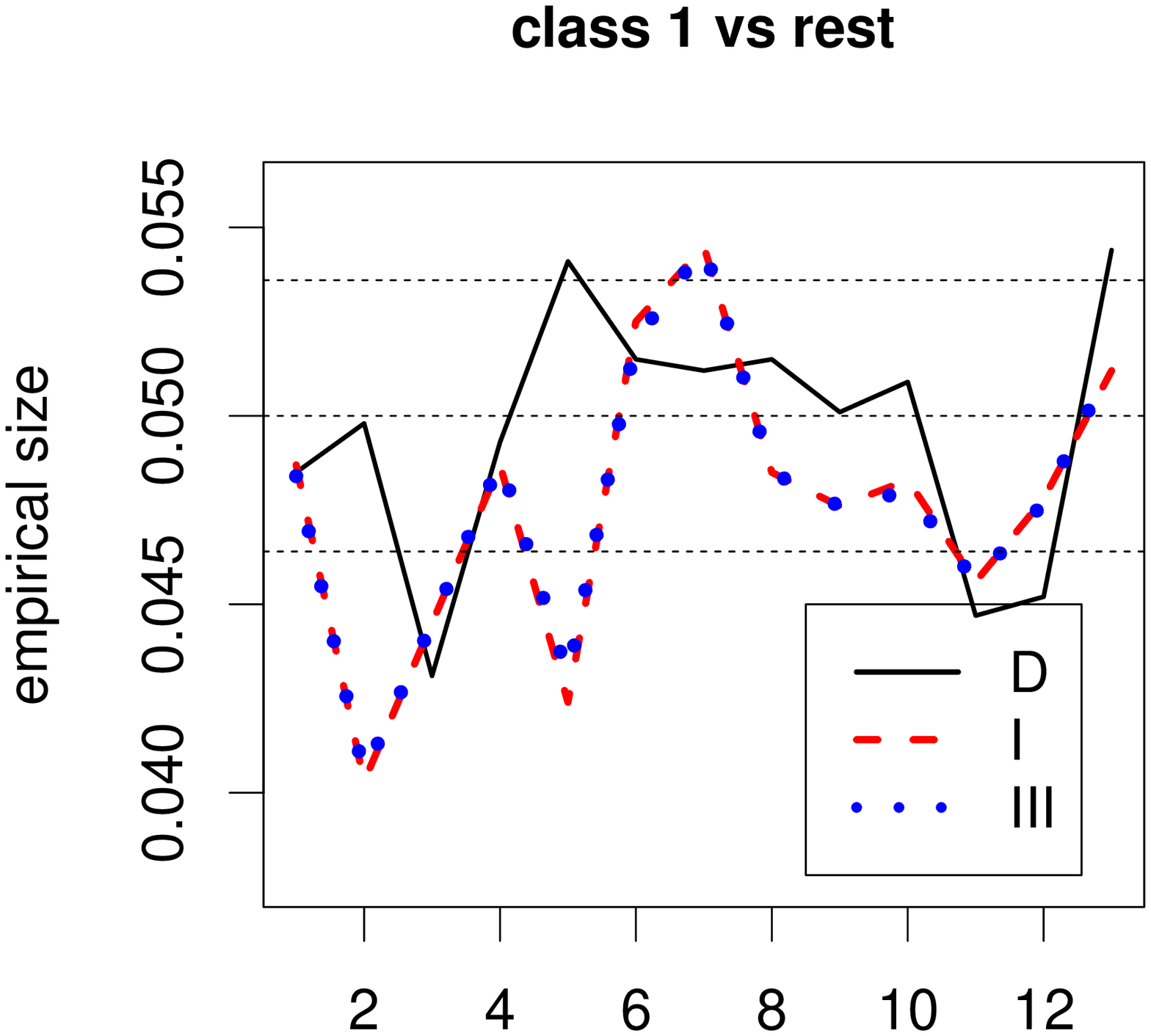} }}
\rotatebox{-90}{ \resizebox{2.1 in}{!}{\includegraphics{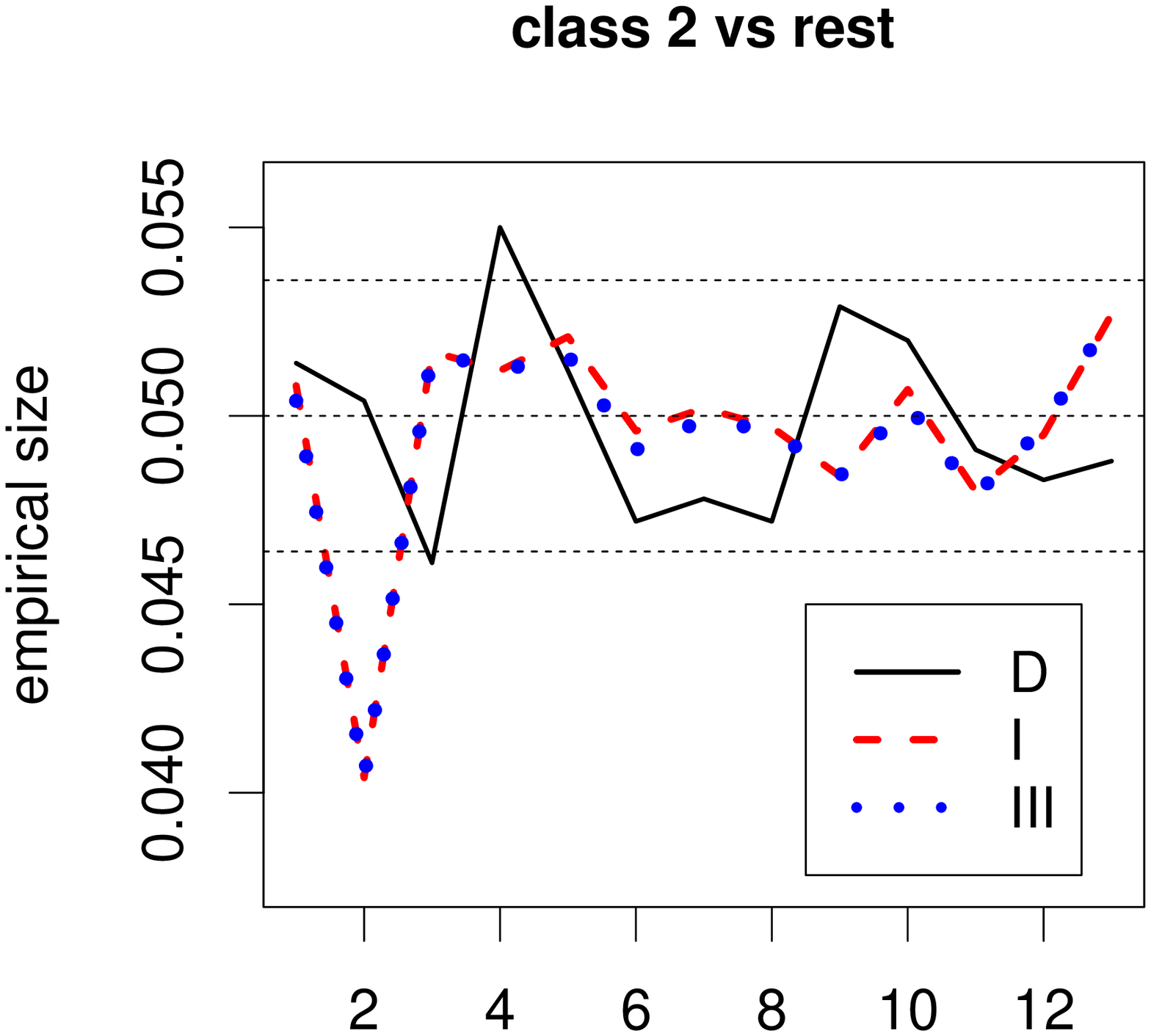} }}
\rotatebox{-90}{ \resizebox{2.1 in}{!}{\includegraphics{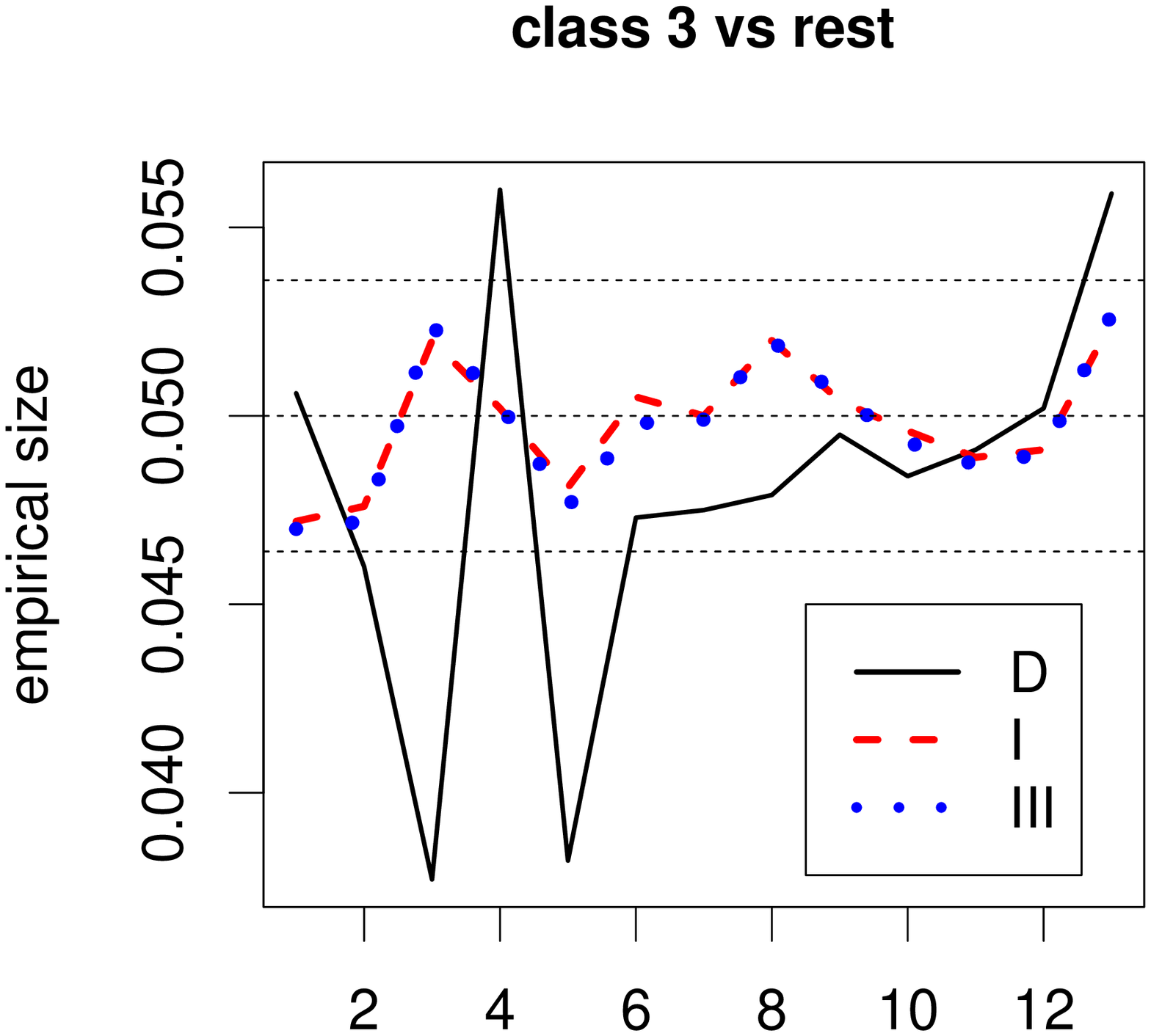} }}
Empirical Size Estimates of Overall Tests under CSR\\
\rotatebox{-90}{ \resizebox{2.1 in}{!}{\includegraphics{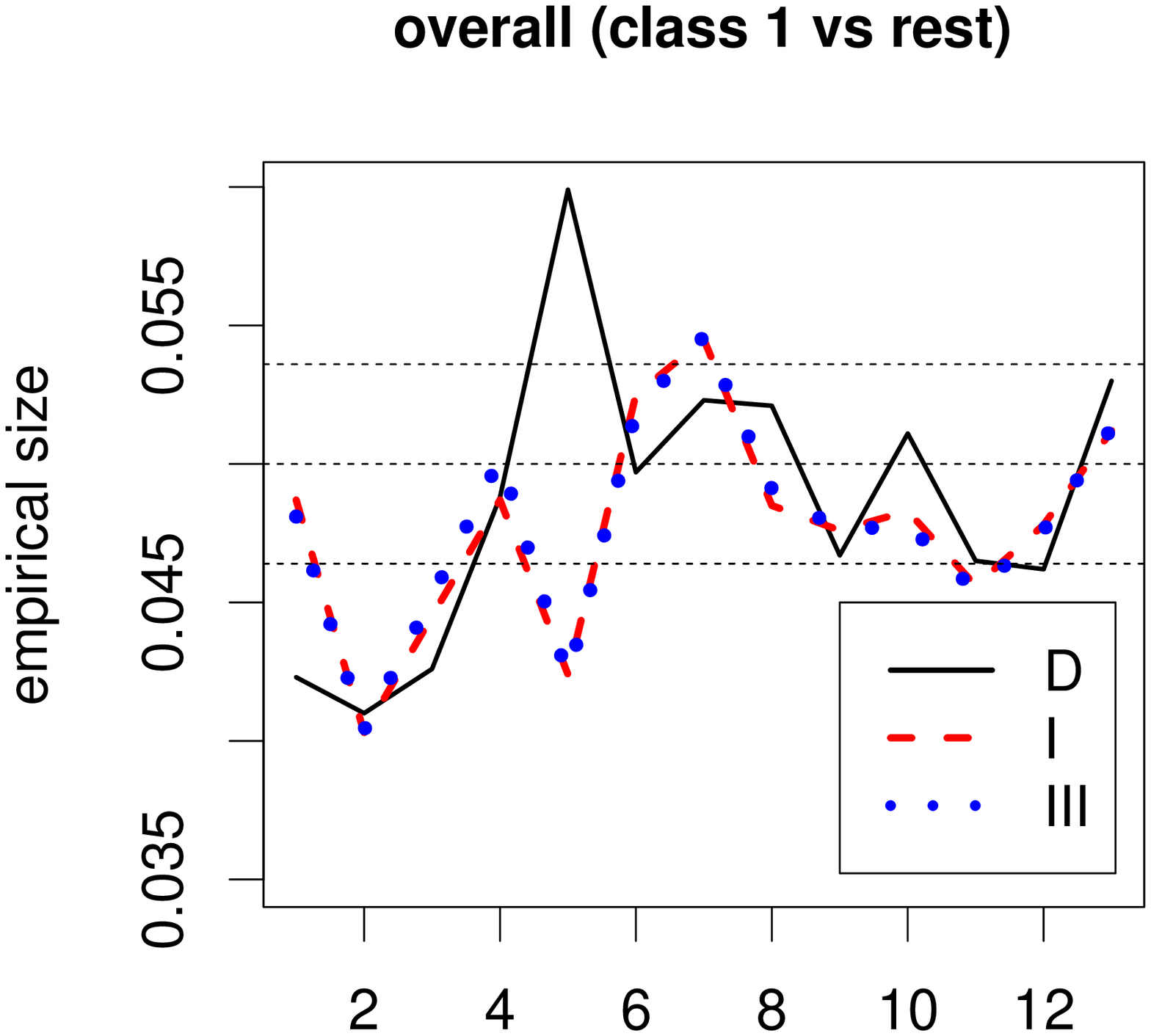} }}
\rotatebox{-90}{ \resizebox{2.1 in}{!}{\includegraphics{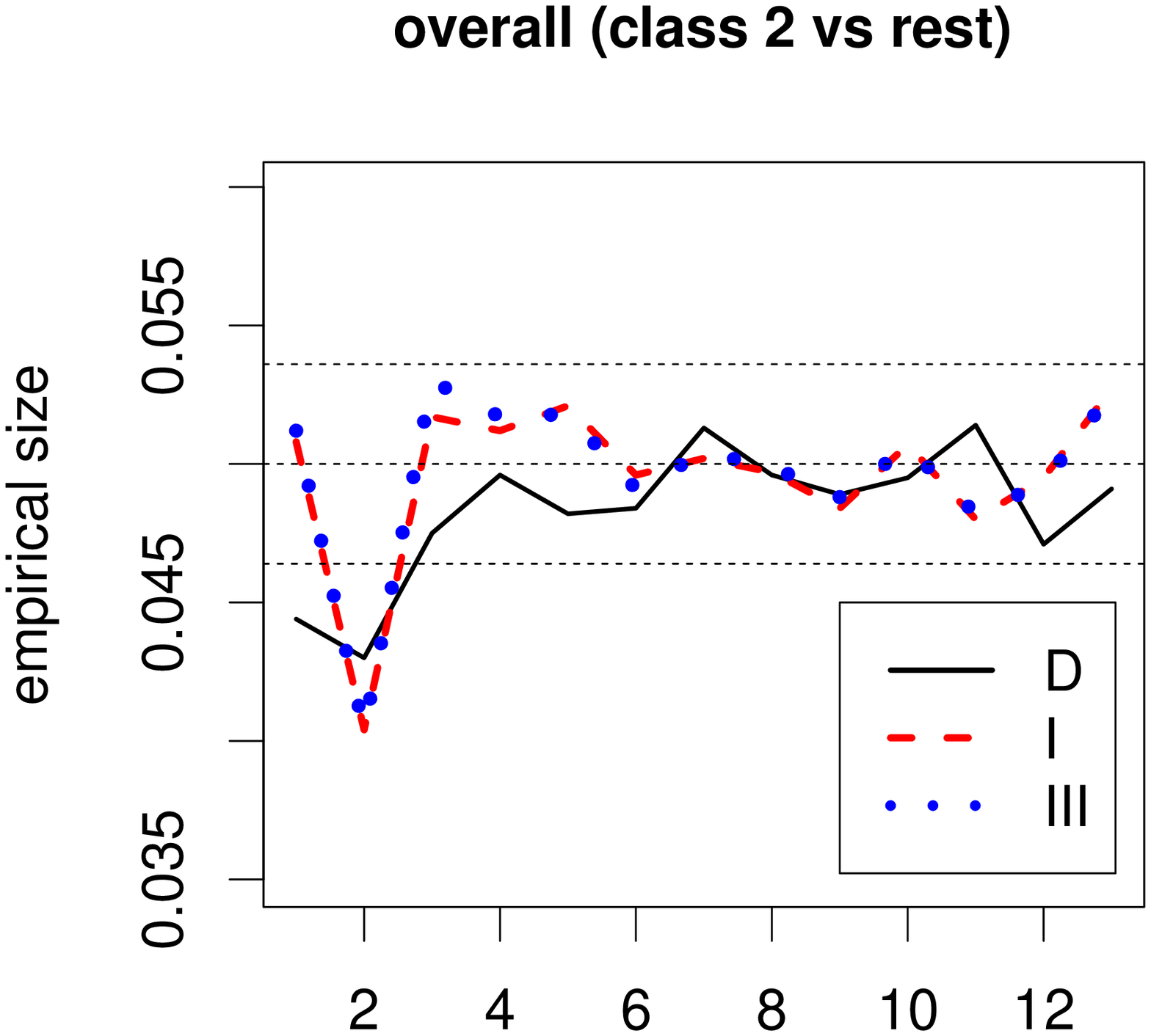} }}
\rotatebox{-90}{ \resizebox{2.1 in}{!}{\includegraphics{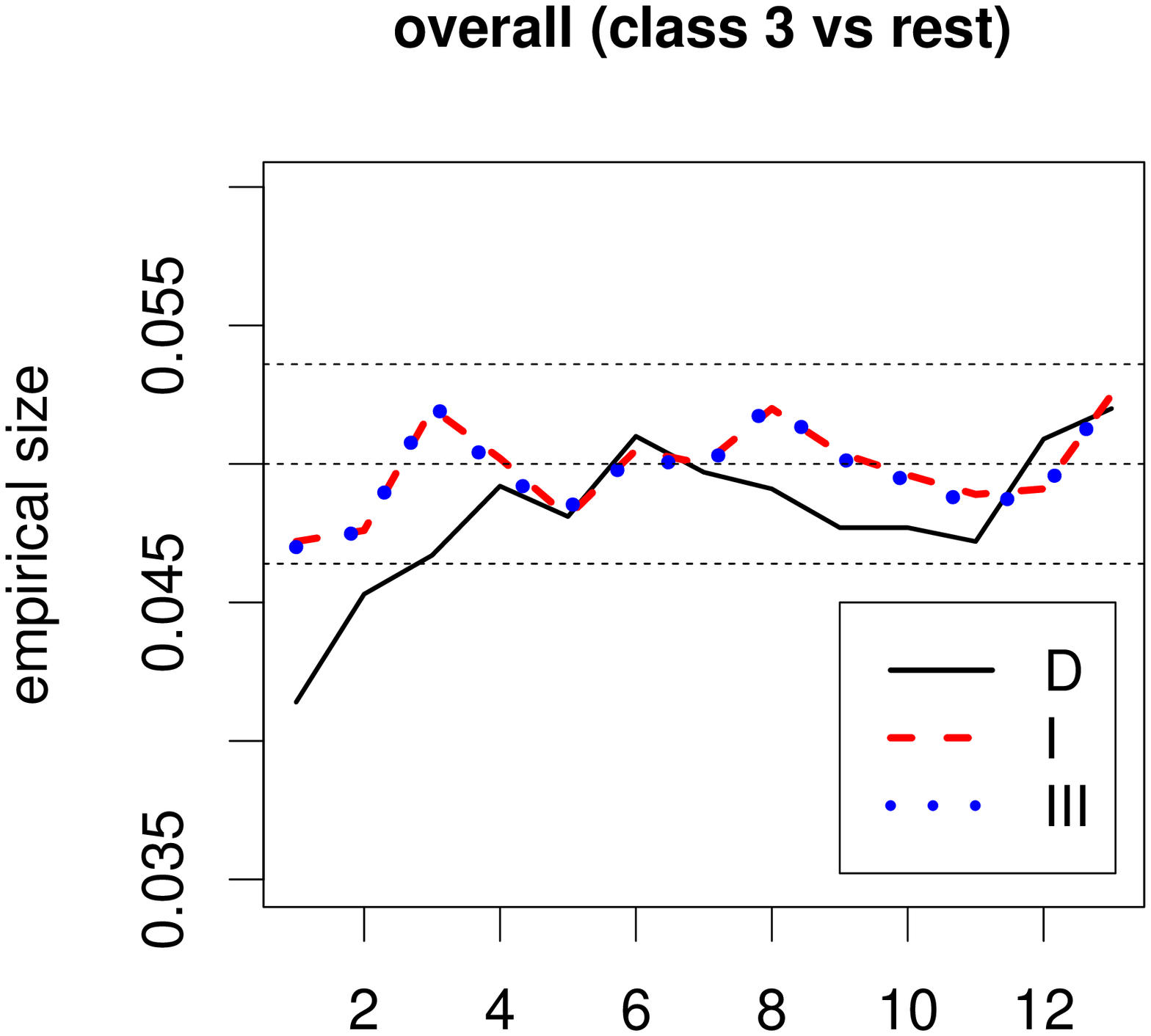} }}
\caption{
\label{fig:emp-size-CSR-1vsR}
The empirical size estimates of the cell-specific tests for cell $(2,2)$ and overall tests
under CSR independence with one-vs-rest type testing.
The legend labeling is as in Figure \ref{fig:emp-size-CSR-2cl}
and
horizontal axis labels are as in Figure \ref{fig:emp-size-CSR-cell-3cl}.
}
\end{figure}

In one-versus-rest type testing,
we implement Monte Carlo simulations as in Section \ref{sec:CSR-emp-sign-3Cl}
to assess the empirical size performance of these tests under CSR independence.
We present the empirical size estimates for various class size combinations
in Figure \ref{fig:emp-size-CSR-1vsR}
where only cell-specific tests for cell $(2,2)$
and the overall test are presented,
since the cell-specific test for cell $(1,1)$
is the same as in the $3\times 3$ NNCT analysis.
Among cell-specific tests,
types I and III tests perform better compared to Dixon's test,
since they are closer to the nominal level especially for large classes.
For the overall tests,
the tests are about the nominal level
with type I and III tests being slightly closer than Dixon's test.

\begin{figure} [hbp]
\centering
%\psfrag{Density}{ \Huge{\bf{Density}}}
Empirical Power Estimates of Cell-Specific Tests under $H_{S_1}$\\
\rotatebox{-90}{ \resizebox{2.1 in}{!}{\includegraphics{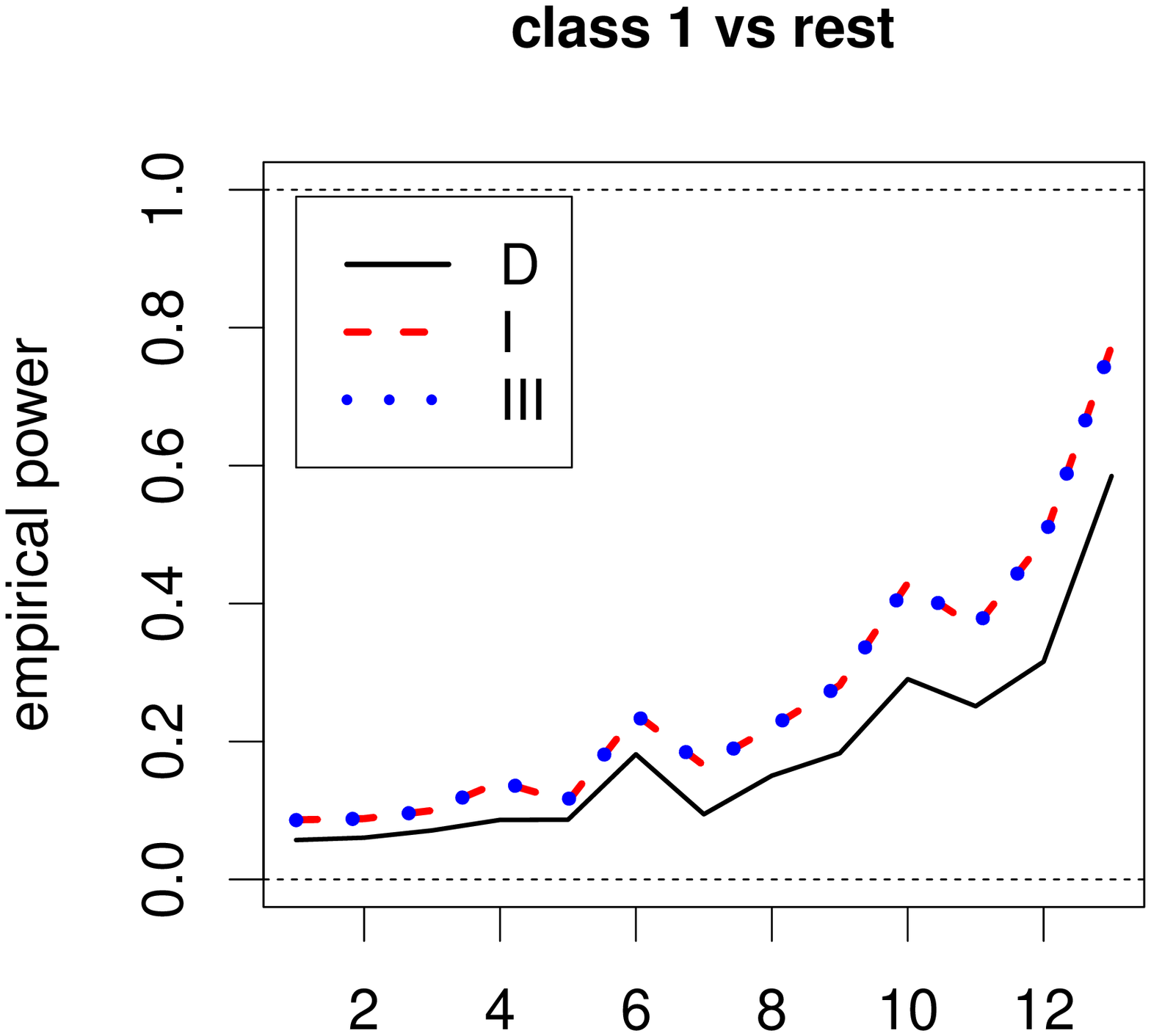} }}
\rotatebox{-90}{ \resizebox{2.1 in}{!}{\includegraphics{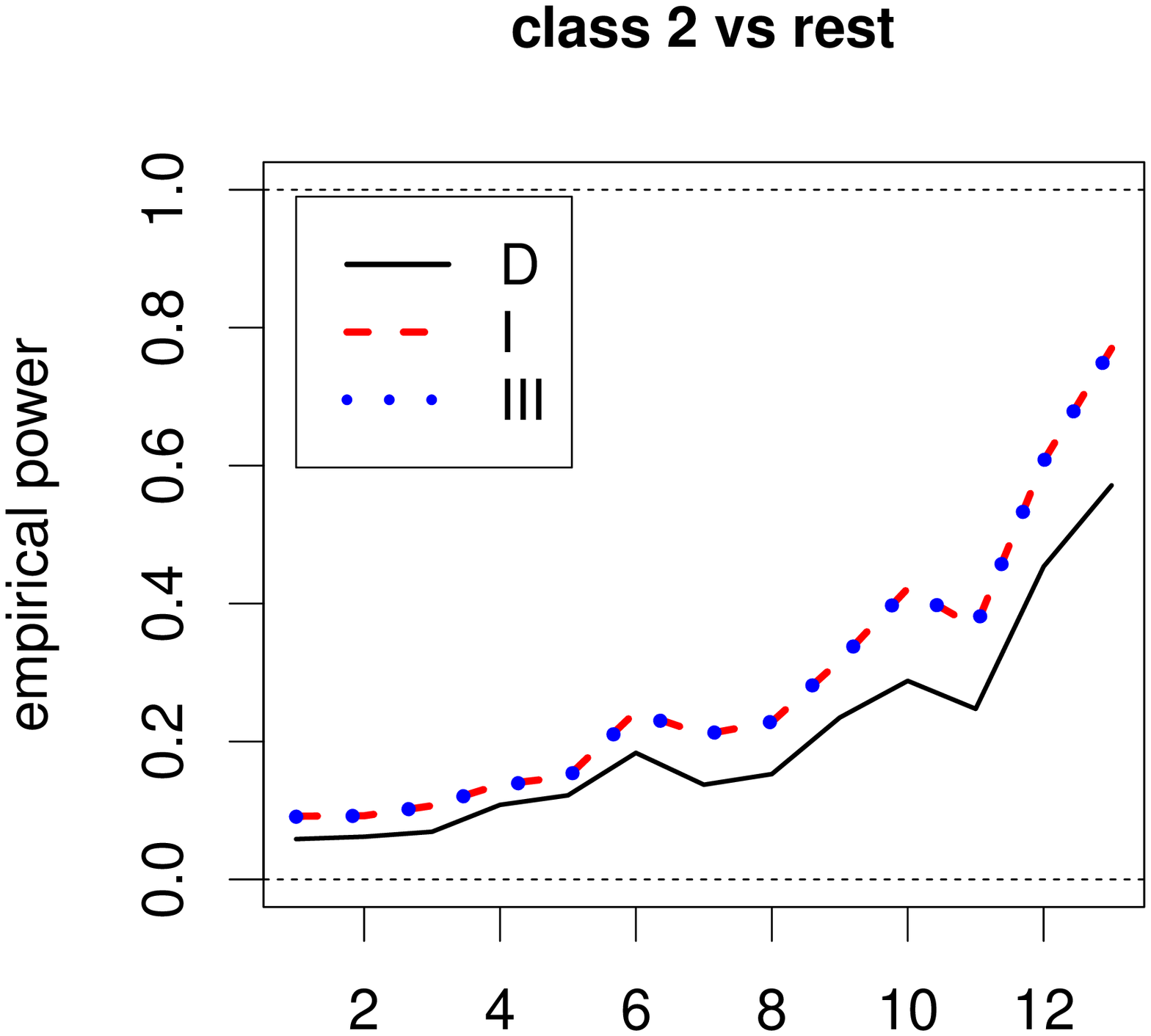} }}
\rotatebox{-90}{ \resizebox{2.1 in}{!}{\includegraphics{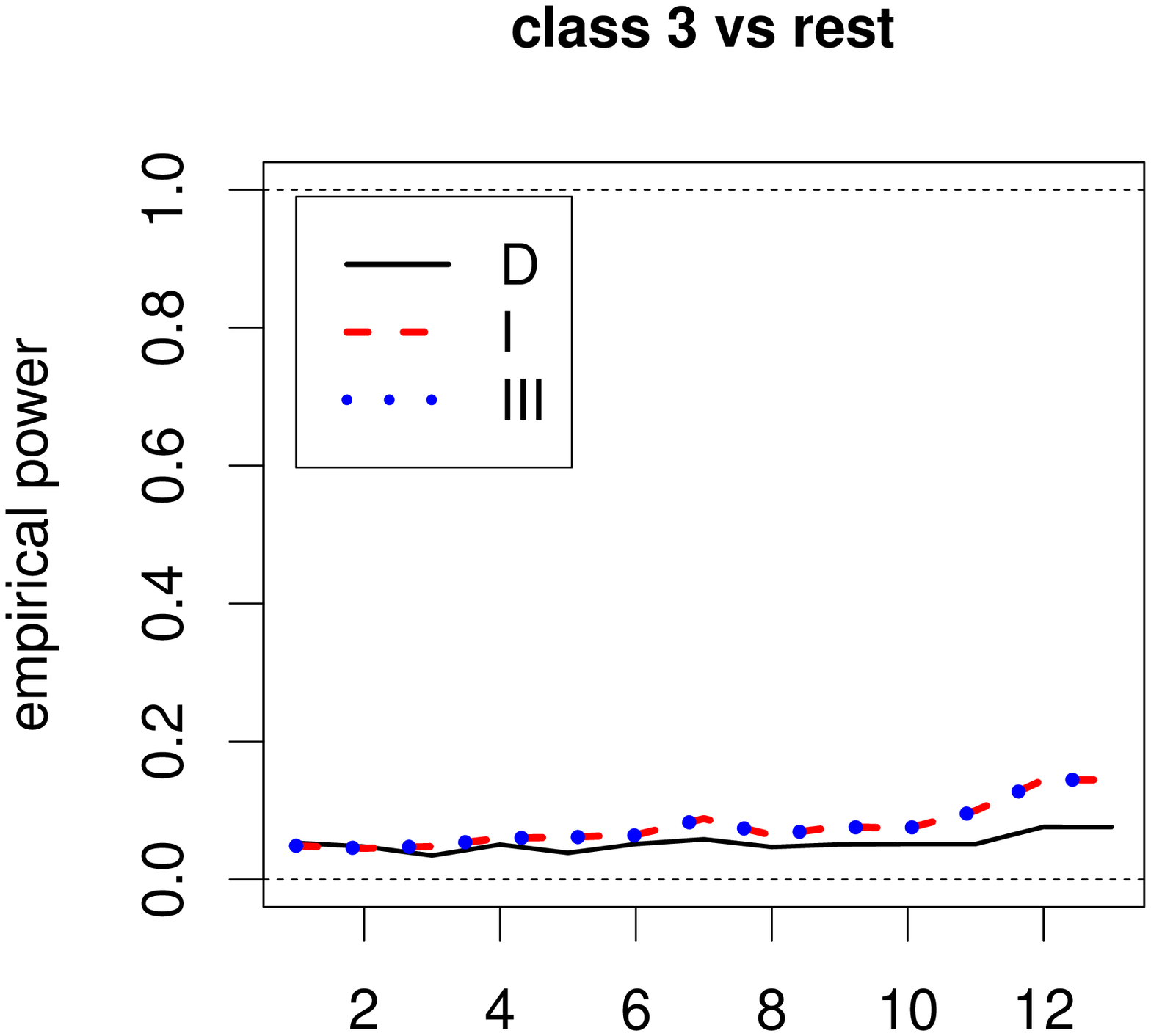} }}
Power Estimates under $H_{S_2}$\\
\rotatebox{-90}{ \resizebox{2.1 in}{!}{\includegraphics{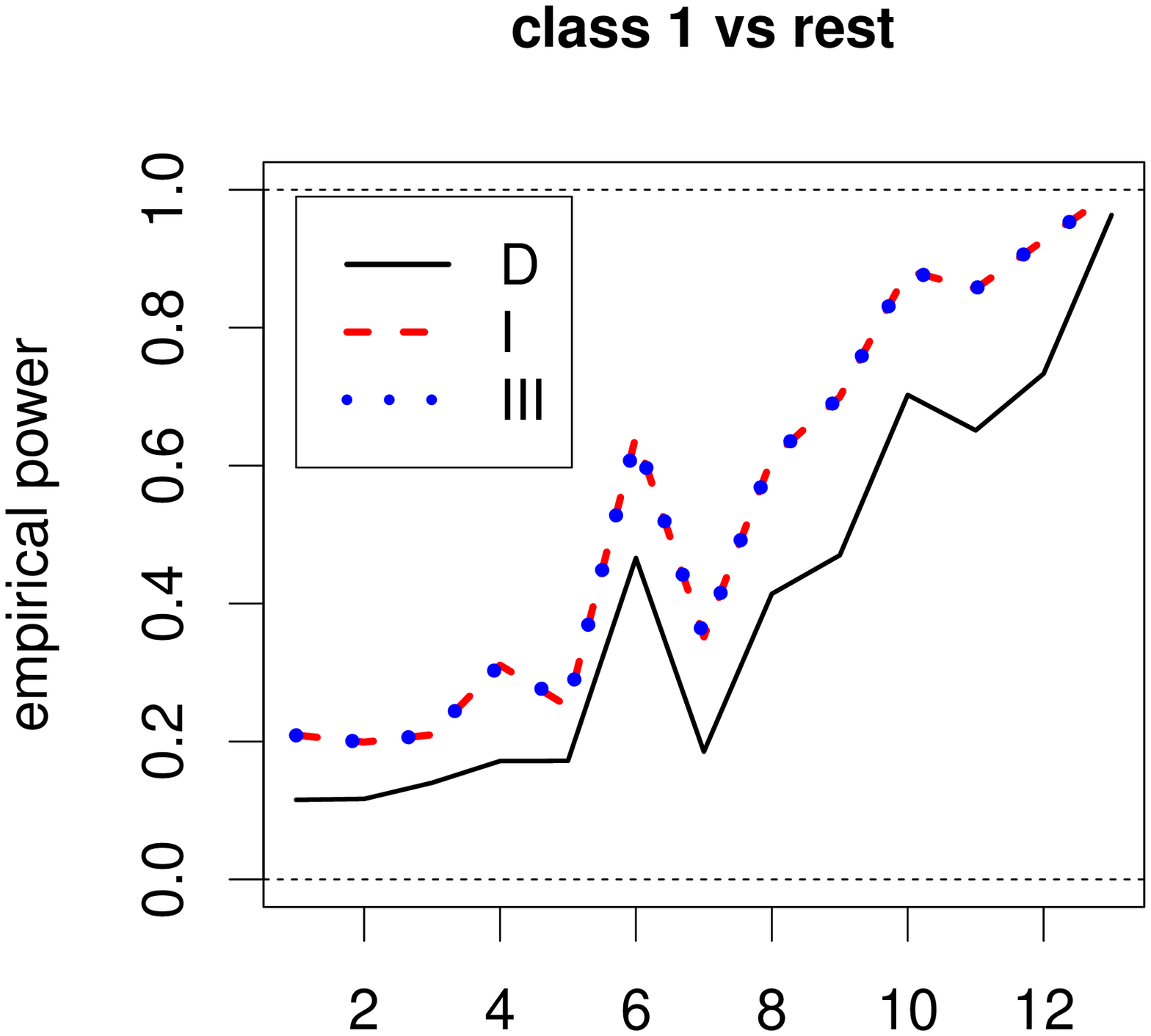} }}
\rotatebox{-90}{ \resizebox{2.1 in}{!}{\includegraphics{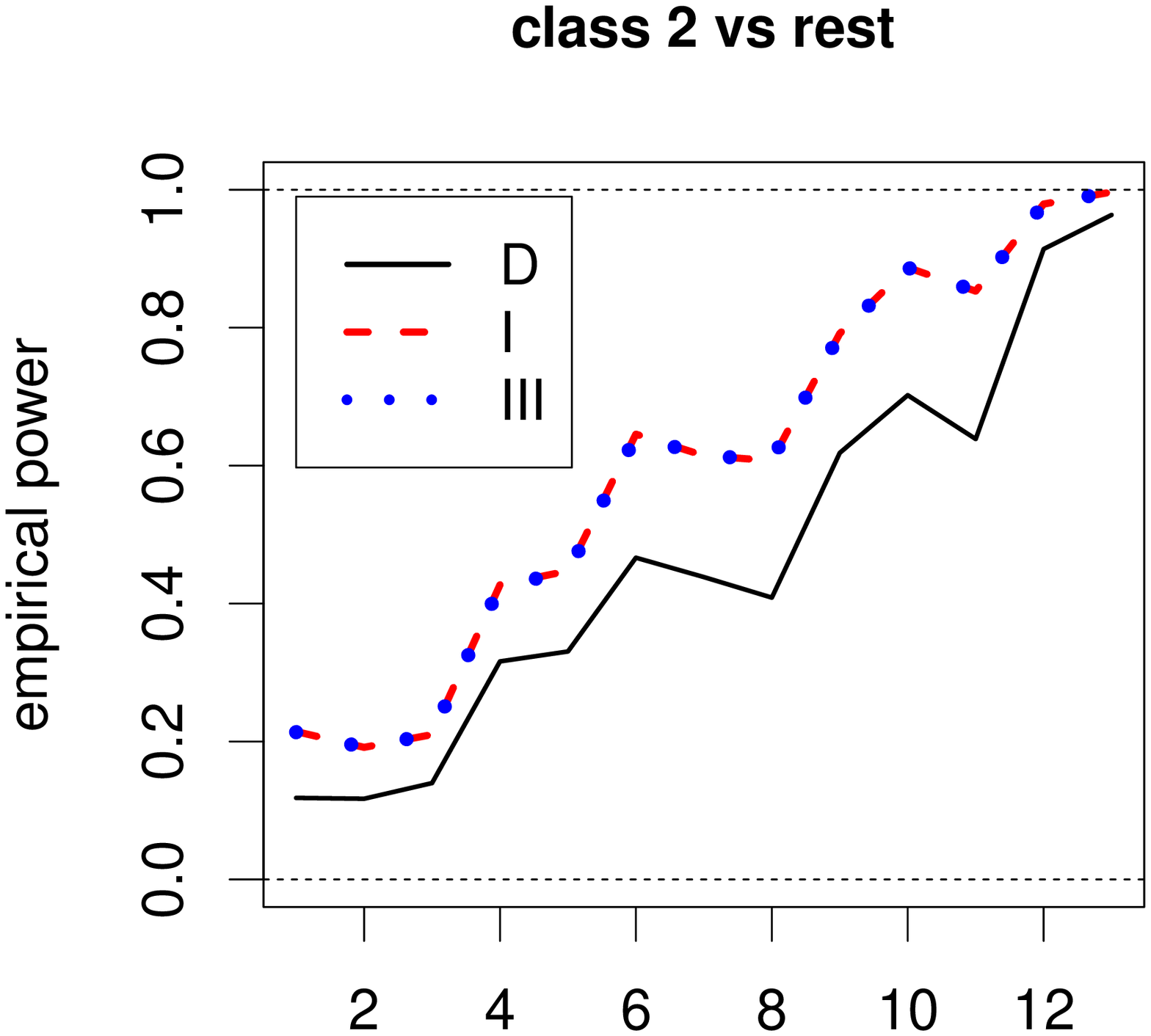} }}
\rotatebox{-90}{ \resizebox{2.1 in}{!}{\includegraphics{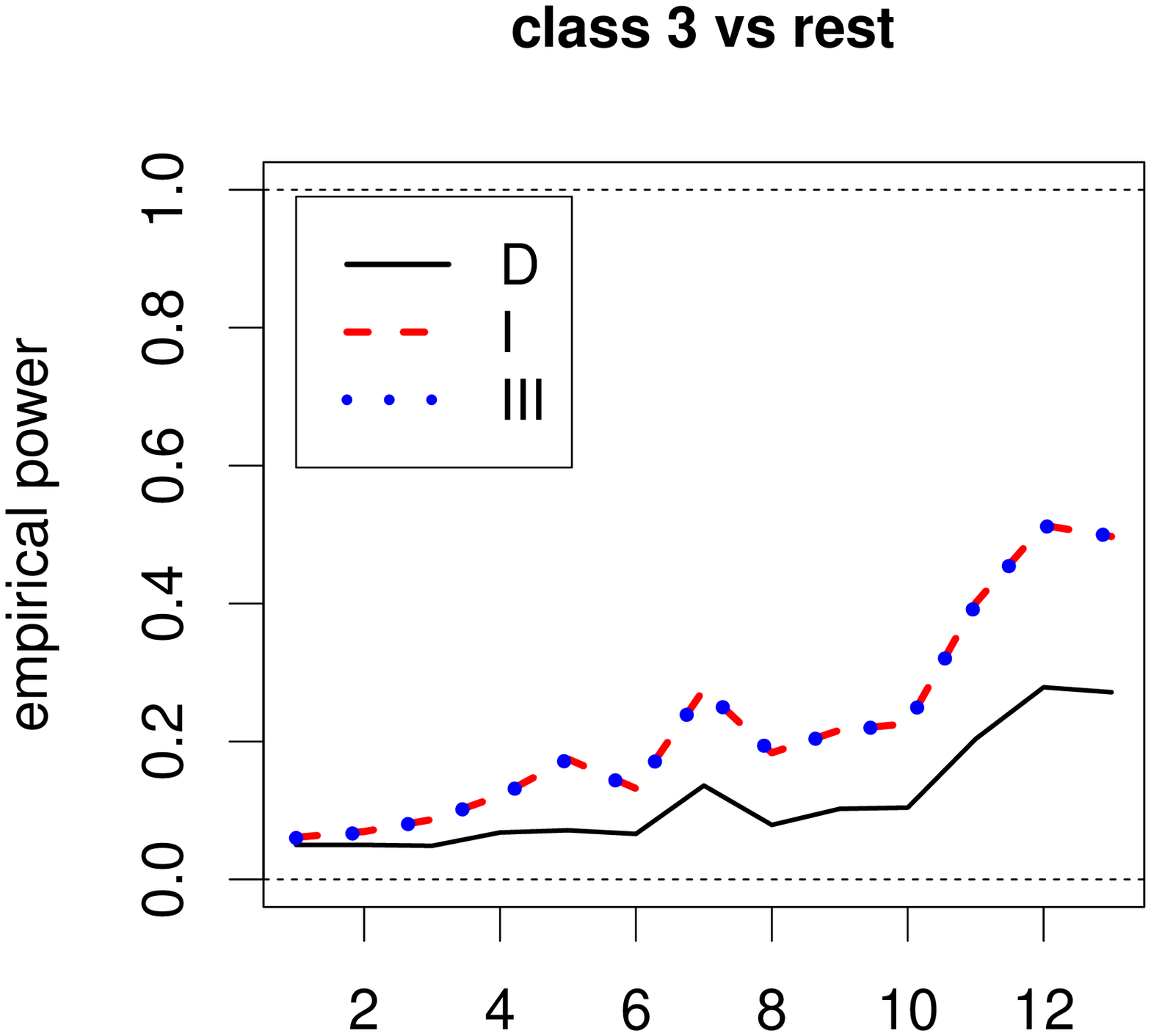} }}
Power Estimates under $H_{S_3}$\\
\rotatebox{-90}{ \resizebox{2.1 in}{!}{\includegraphics{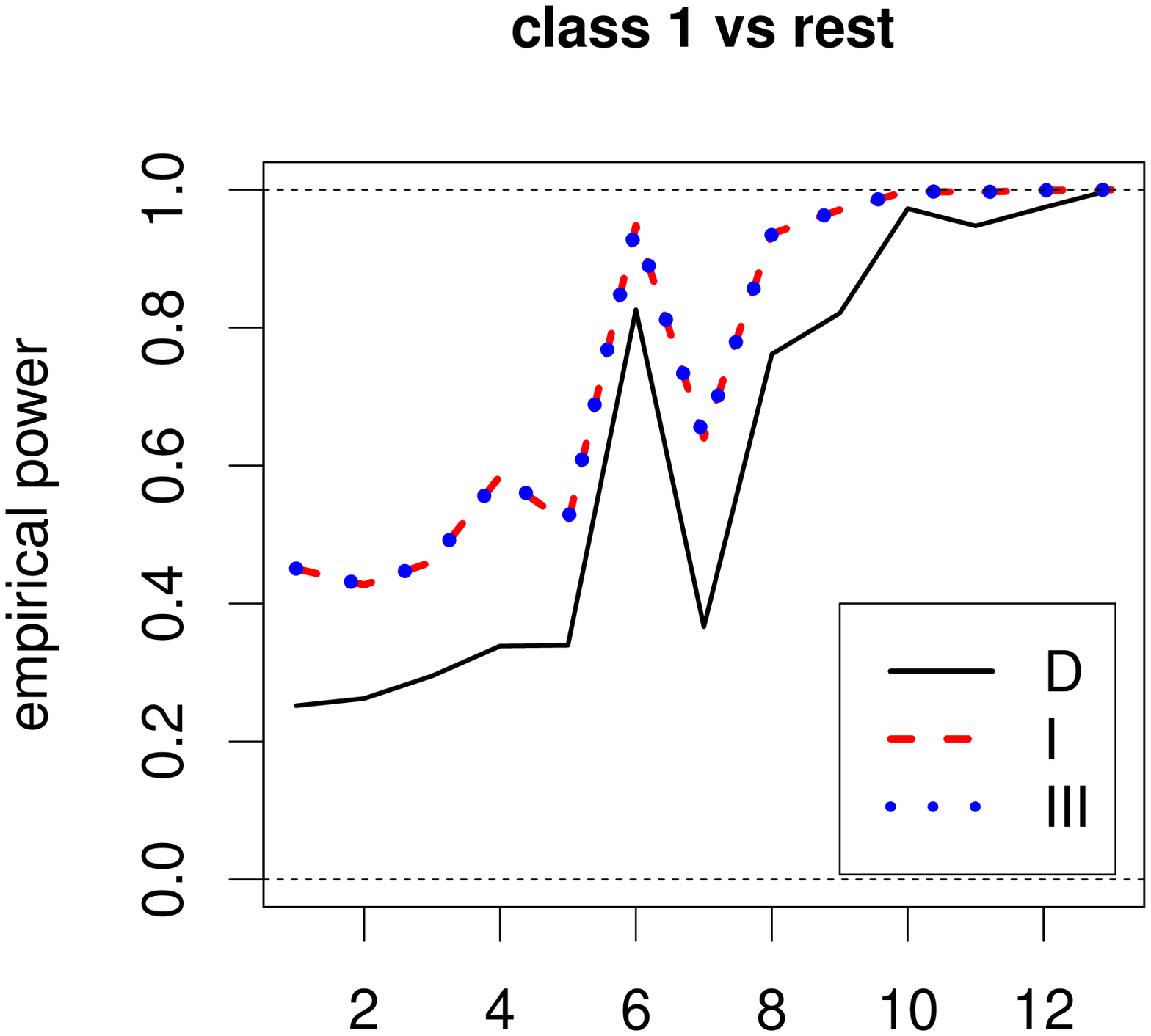} }}
\rotatebox{-90}{ \resizebox{2.1 in}{!}{\includegraphics{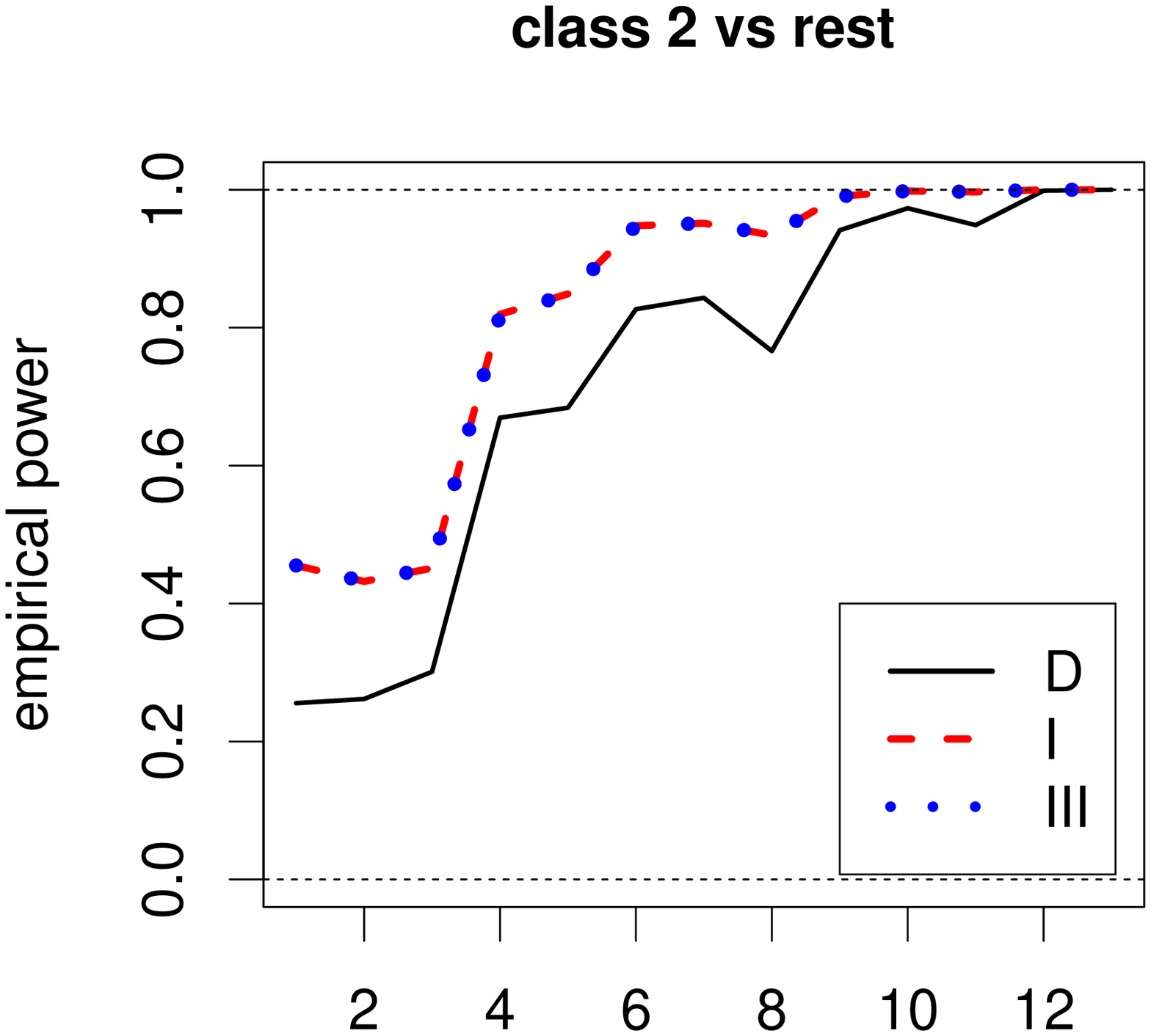} }}
\rotatebox{-90}{ \resizebox{2.1 in}{!}{\includegraphics{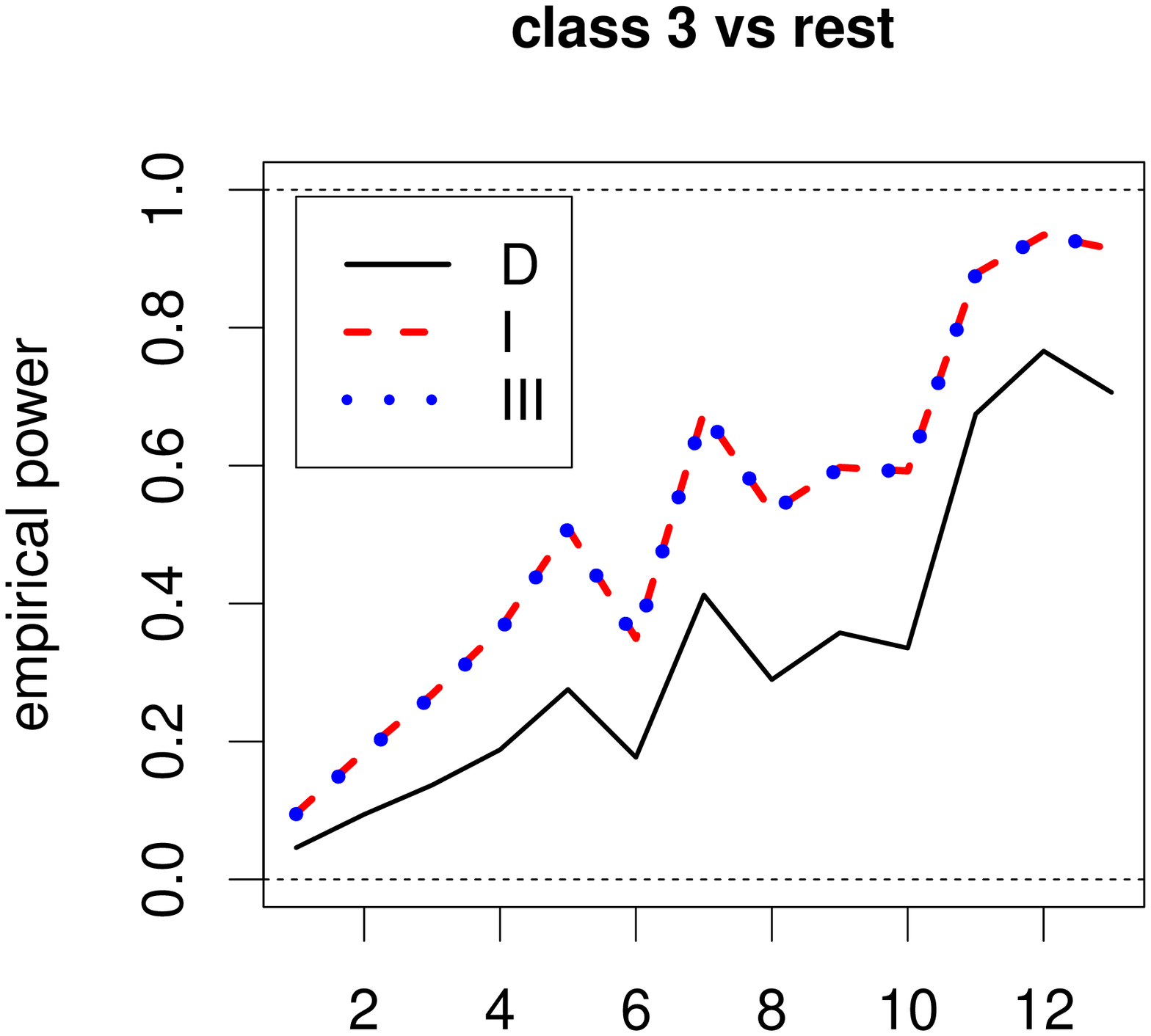} }}
\caption{
\label{fig:power-seg-cell-3cl-1vsR}
The empirical power estimates of the cell-specific tests for cell $(2,2)$,
under the segregation alternatives
$H_{S_1}$ (top), $H_{S_2}$ (middle), and $H_{S_3}$ (bottom) in the three-class case
with the one-vs-rest type testing.
The legend labeling is as in Figure \ref{fig:emp-size-CSR-2cl}
and
horizontal axis labels are as in Figure \ref{fig:emp-size-CSR-cell-3cl}.
}
\end{figure}

\begin{figure} [hbp]
\centering
%\psfrag{Density}{ \Huge{\bf{Density}}}
Empirical Power Estimates of Overall Tests under $H_{S_1}$\\
\rotatebox{-90}{ \resizebox{2.1 in}{!}{\includegraphics{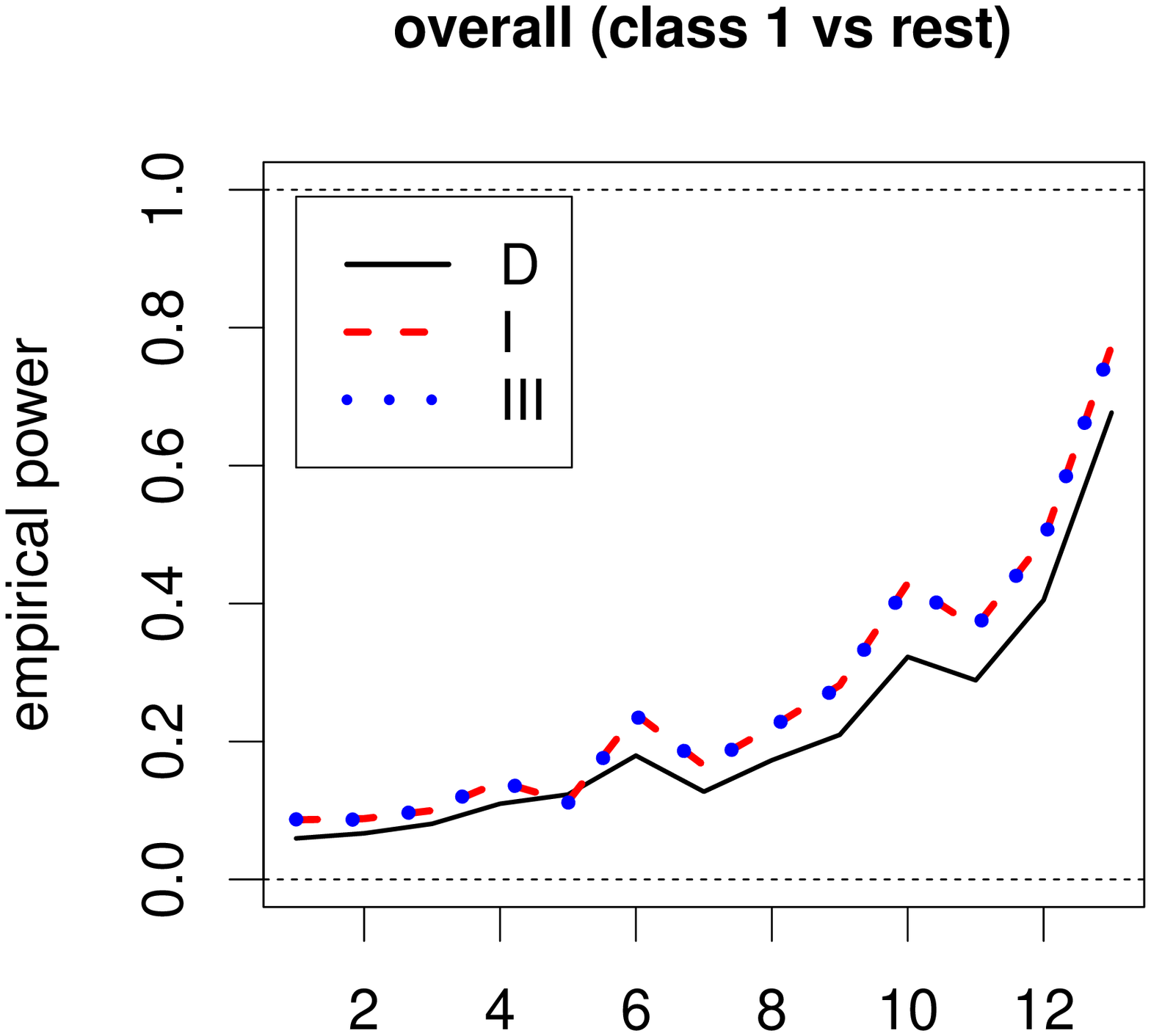} }}
\rotatebox{-90}{ \resizebox{2.1 in}{!}{\includegraphics{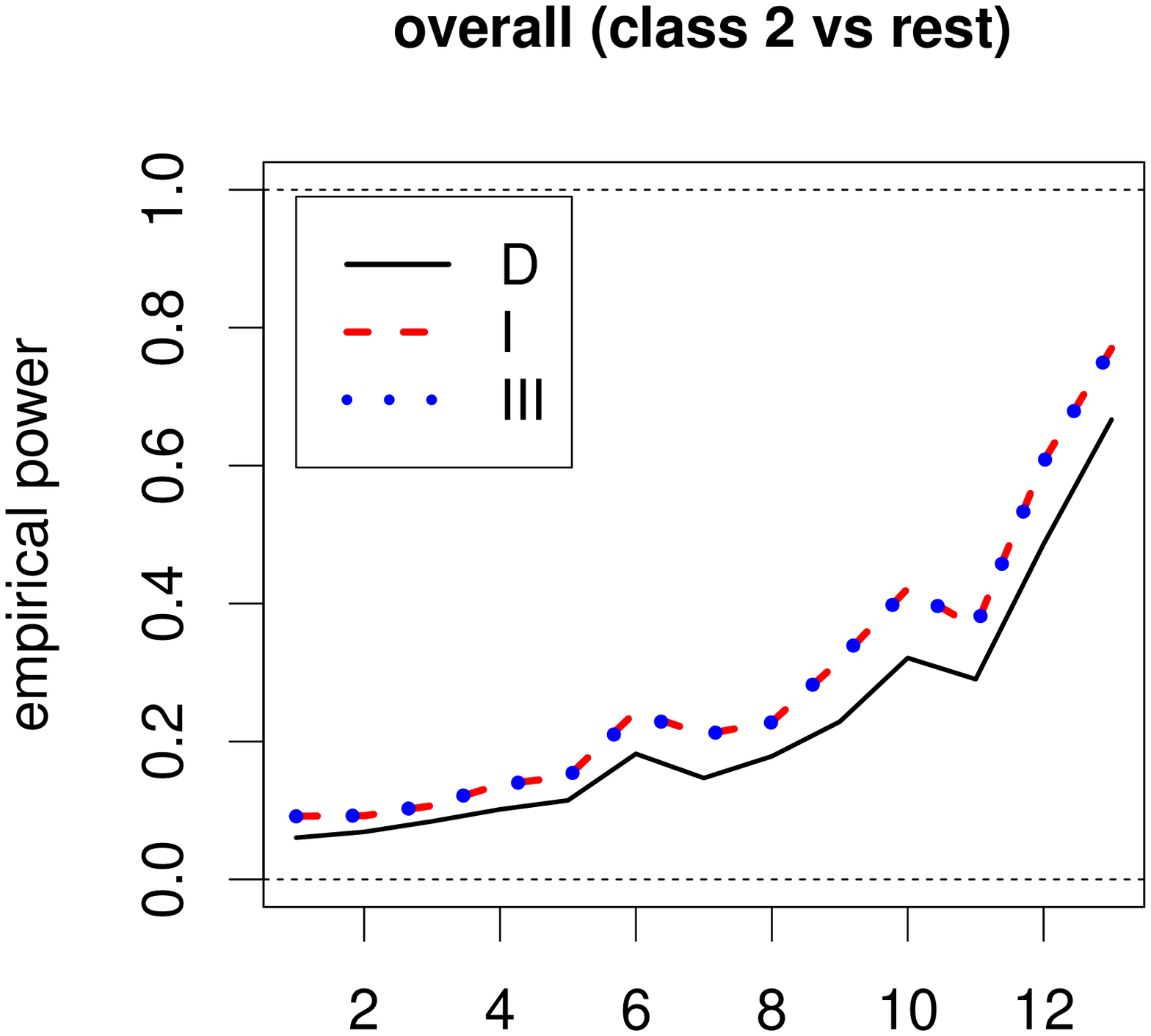} }}
\rotatebox{-90}{ \resizebox{2.1 in}{!}{\includegraphics{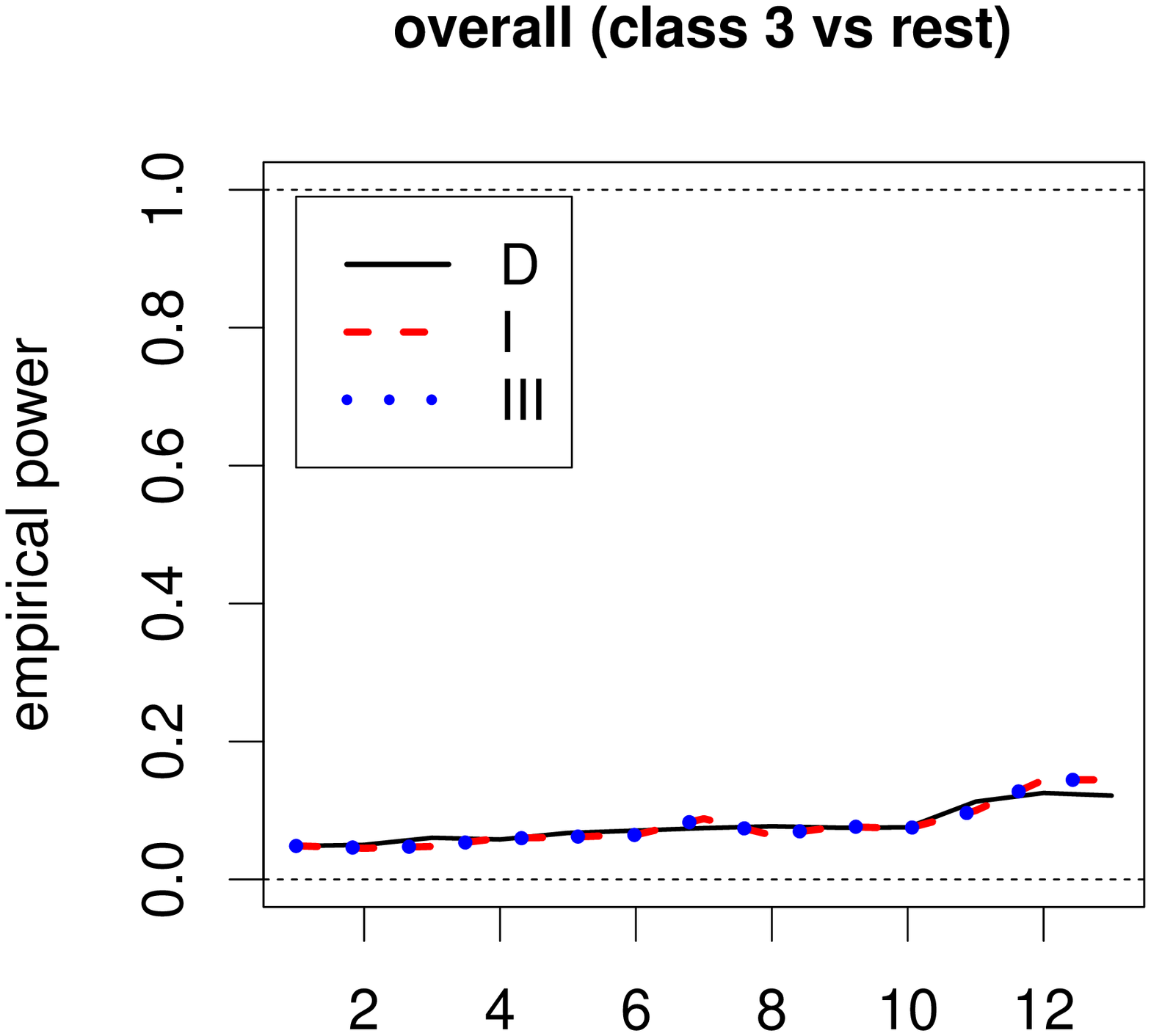} }}
Power Estimates under $H_{S_2}$\\
\rotatebox{-90}{ \resizebox{2.1 in}{!}{\includegraphics{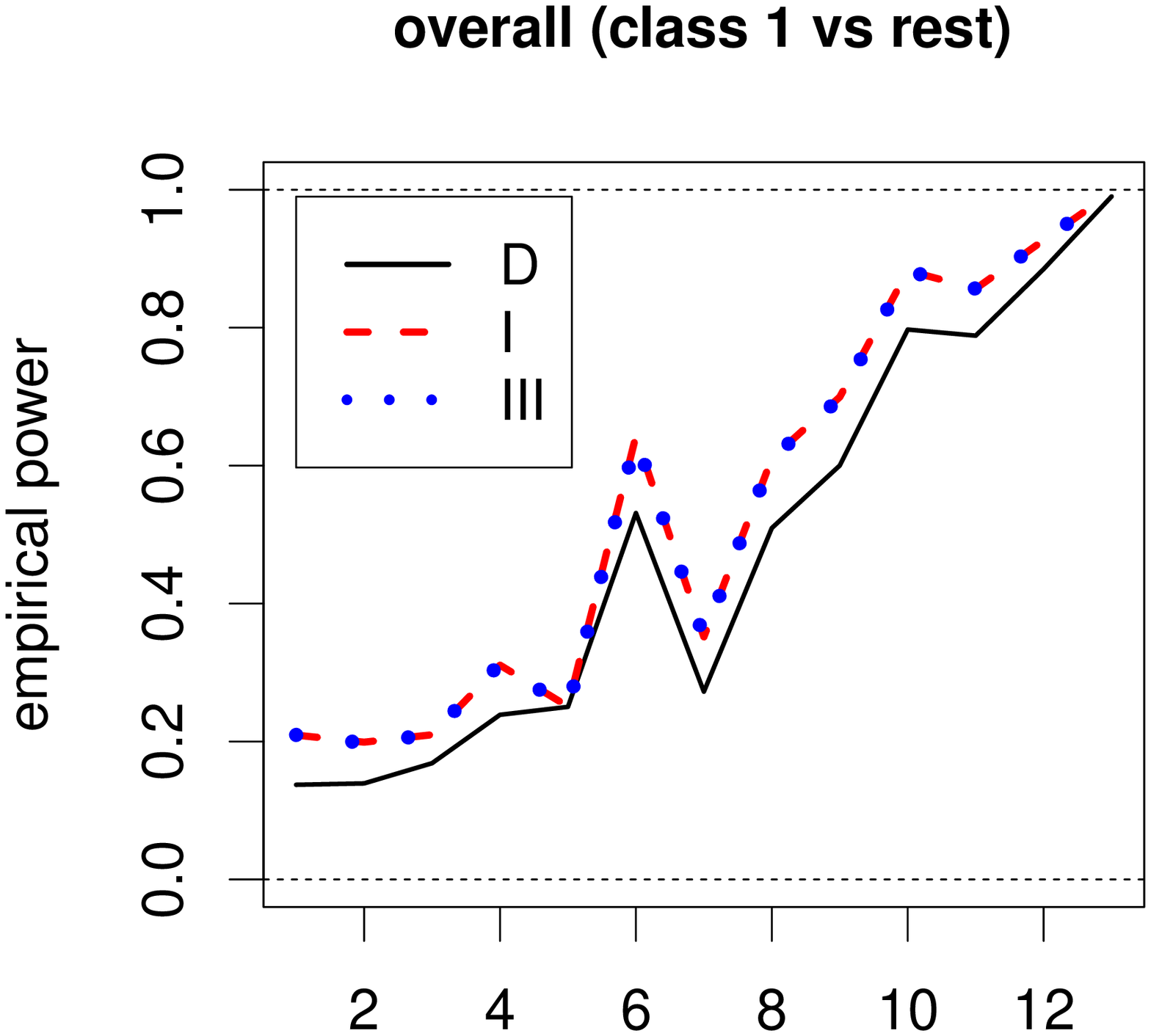} }}
\rotatebox{-90}{ \resizebox{2.1 in}{!}{\includegraphics{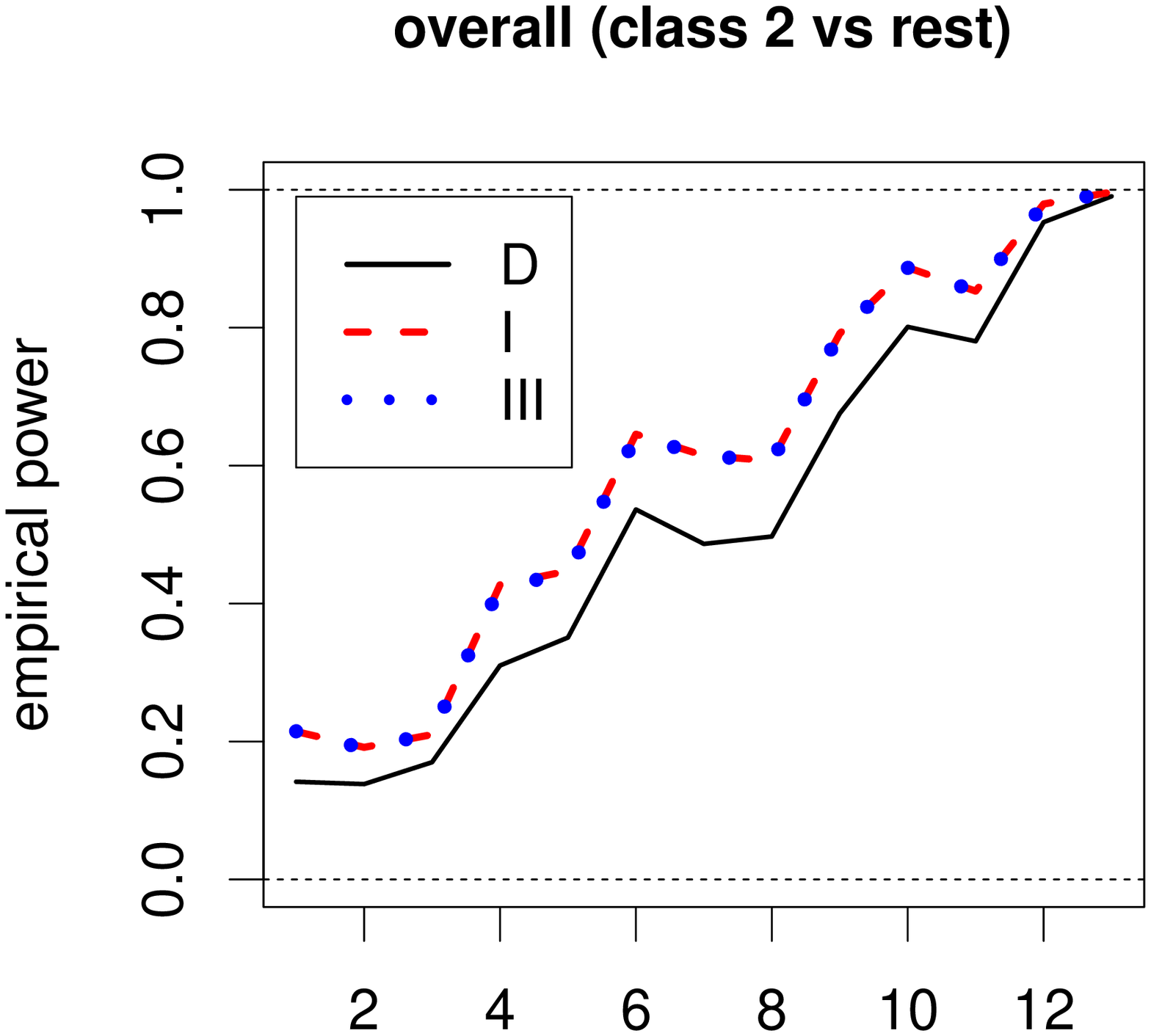} }}
\rotatebox{-90}{ \resizebox{2.1 in}{!}{\includegraphics{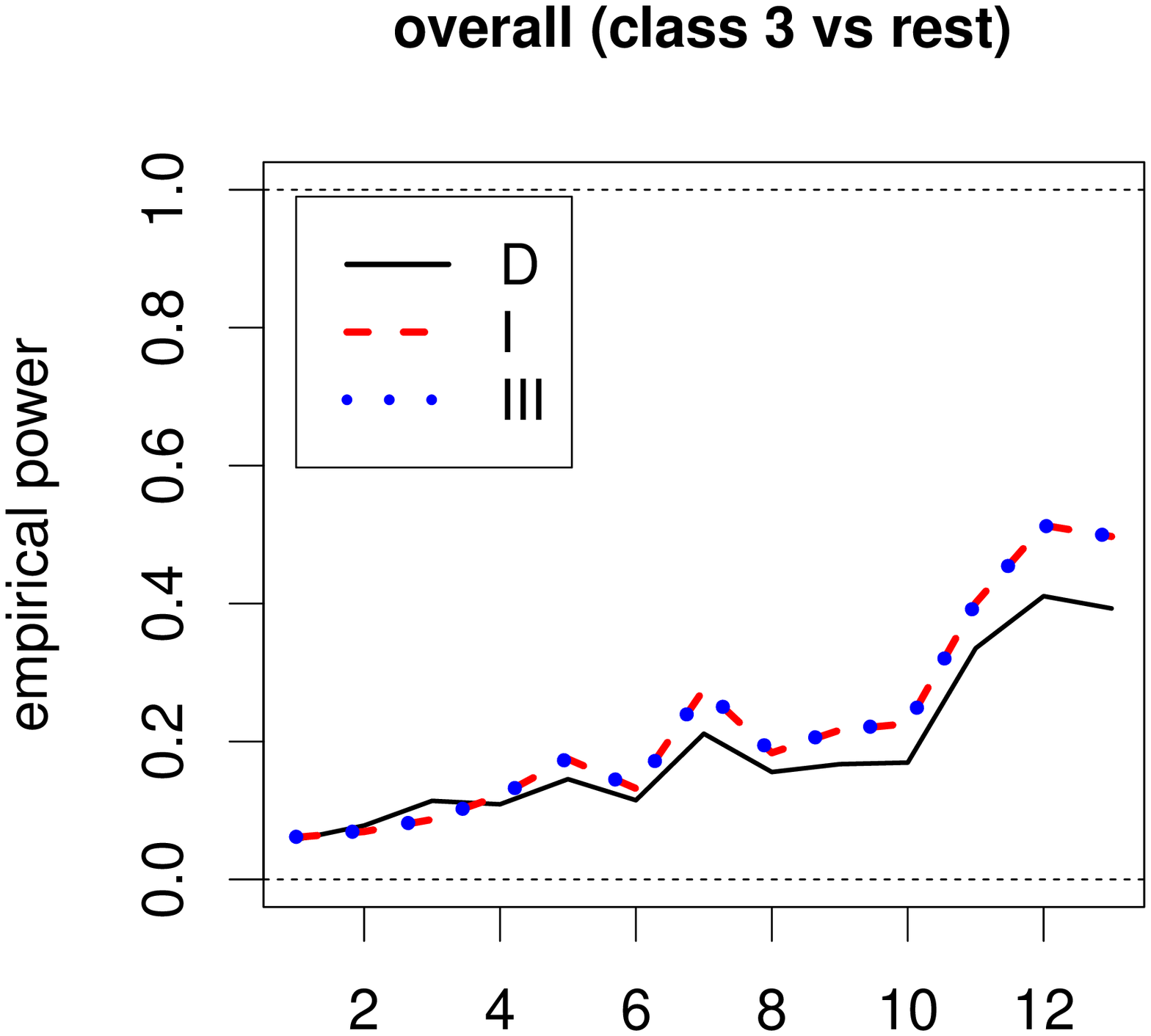} }}
Power Estimates under $H_{S_3}$\\
\rotatebox{-90}{ \resizebox{2.1 in}{!}{\includegraphics{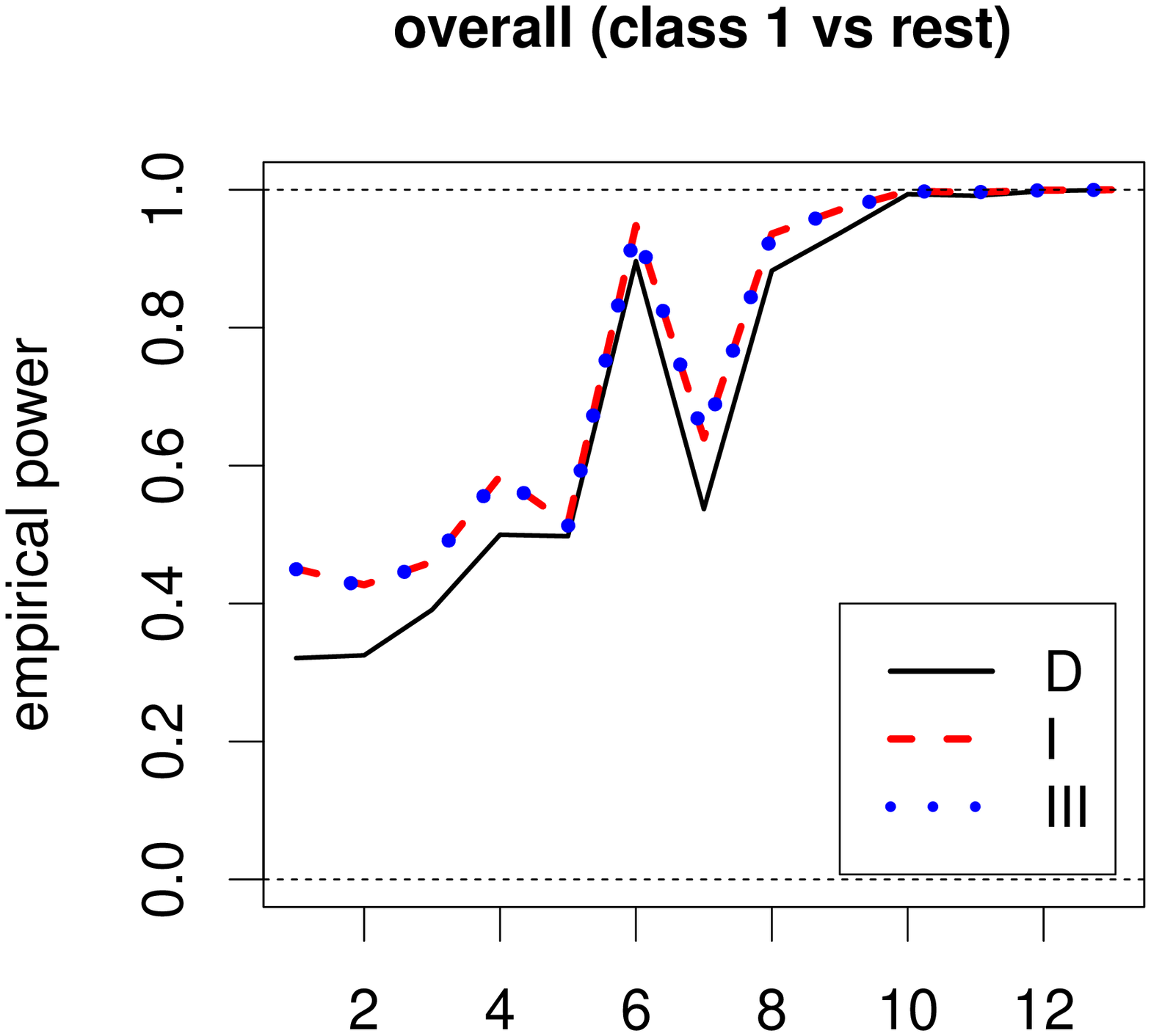} }}
\rotatebox{-90}{ \resizebox{2.1 in}{!}{\includegraphics{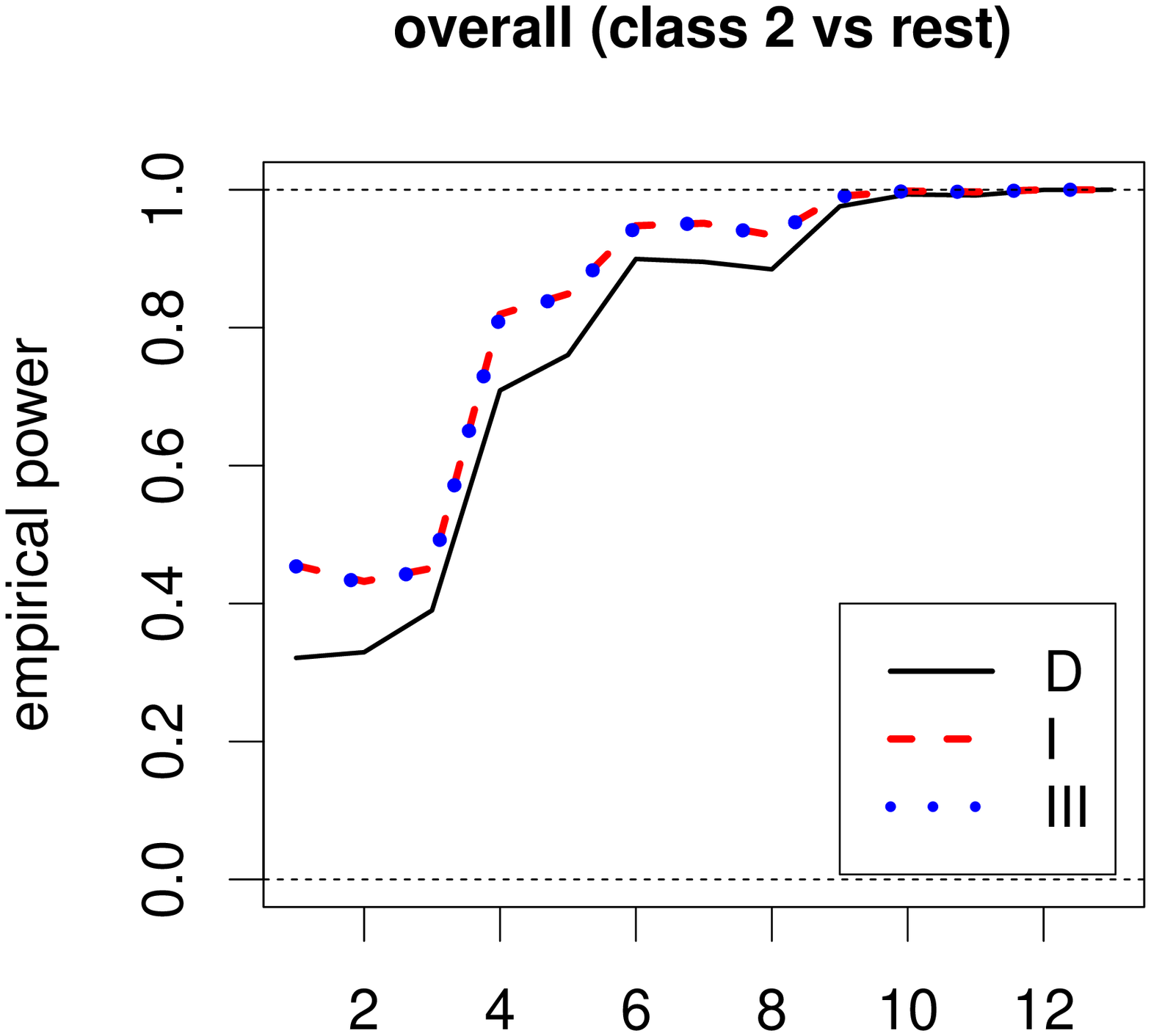} }}
\rotatebox{-90}{ \resizebox{2.1 in}{!}{\includegraphics{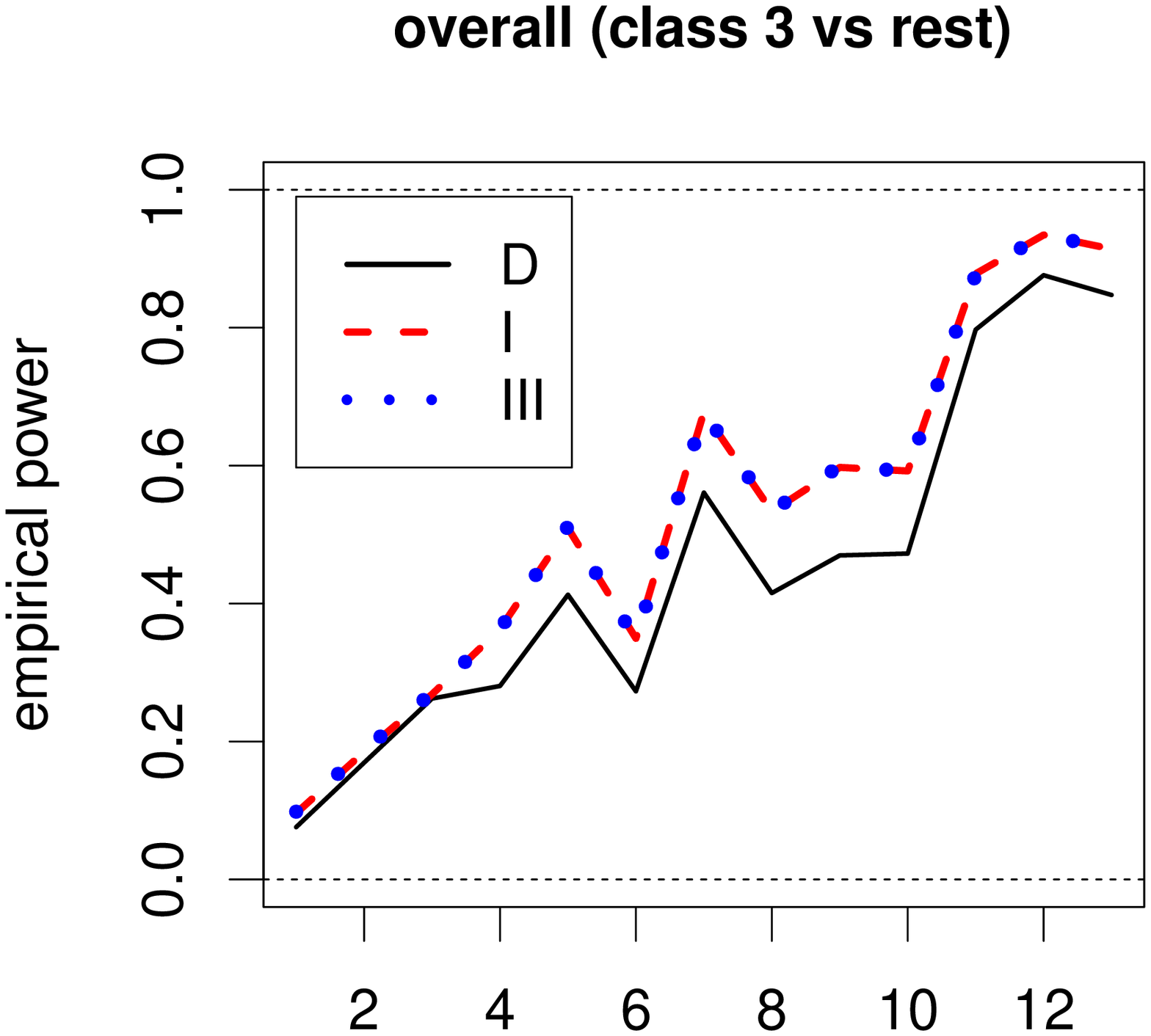} }}
\caption{
\label{fig:power-seg-overall-3cl-1vsR}
The empirical power estimates of the overall tests
under the segregation alternatives
$H_{S_1}$ (top), $H_{S_2}$ (middle), and $H_{S_3}$ (bottom) in the three-class case
with the one-vs-rest type testing.
The legend labeling is as in Figure \ref{fig:emp-size-CSR-2cl}
and
horizontal axis labels are as in Figure \ref{fig:emp-size-CSR-cell-3cl}.
}
\end{figure}

To evaluate the power performance of these tests,
we perform simulations under segregation alternatives as in Section \ref{sec:power-comp-seg-3Cl}.
The empirical power estimates under the three segregation alternatives are
presented in Figures \ref{fig:power-seg-cell-3cl-1vsR}
and \ref{fig:power-seg-overall-3cl-1vsR}.
Among the tests,
type I and III tests have higher power estimates compared to Dixon's test.
One class-vs-rest tests for classes 1 and 2 have higher power estimates
compared to that of class 3.
This occurs, since by construction,
classes 1 and 2 are equally segregated from other classes,
and these classes are more segregated compared to class 3.

\begin{figure} [hbp]
\centering
%\psfrag{Density}{ \Huge{\bf{Density}}}
Empirical Power Estimates of Cell-Specific Tests under $H_{A_1}$\\
\rotatebox{-90}{ \resizebox{2.1 in}{!}{\includegraphics{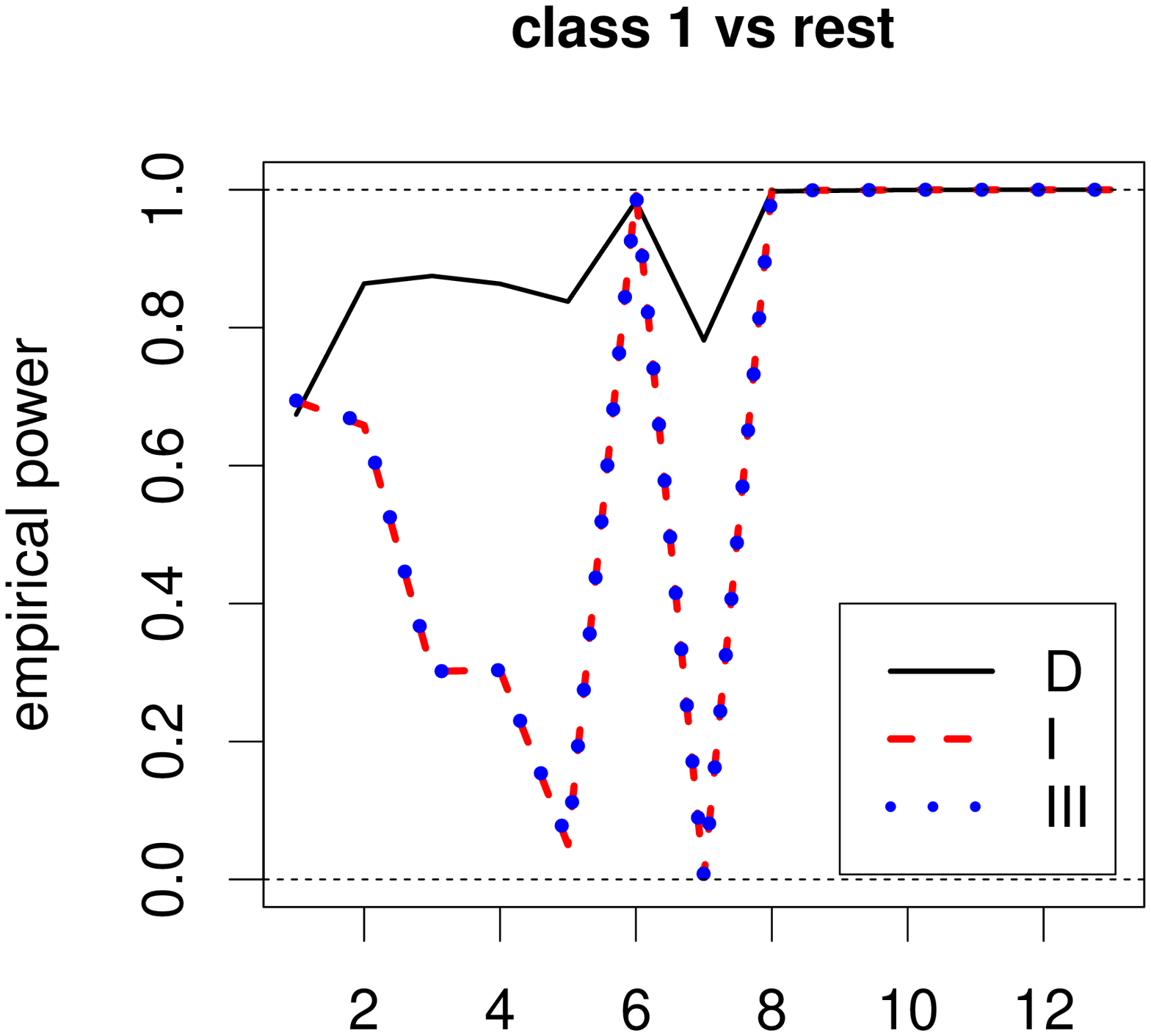} }}
\rotatebox{-90}{ \resizebox{2.1 in}{!}{\includegraphics{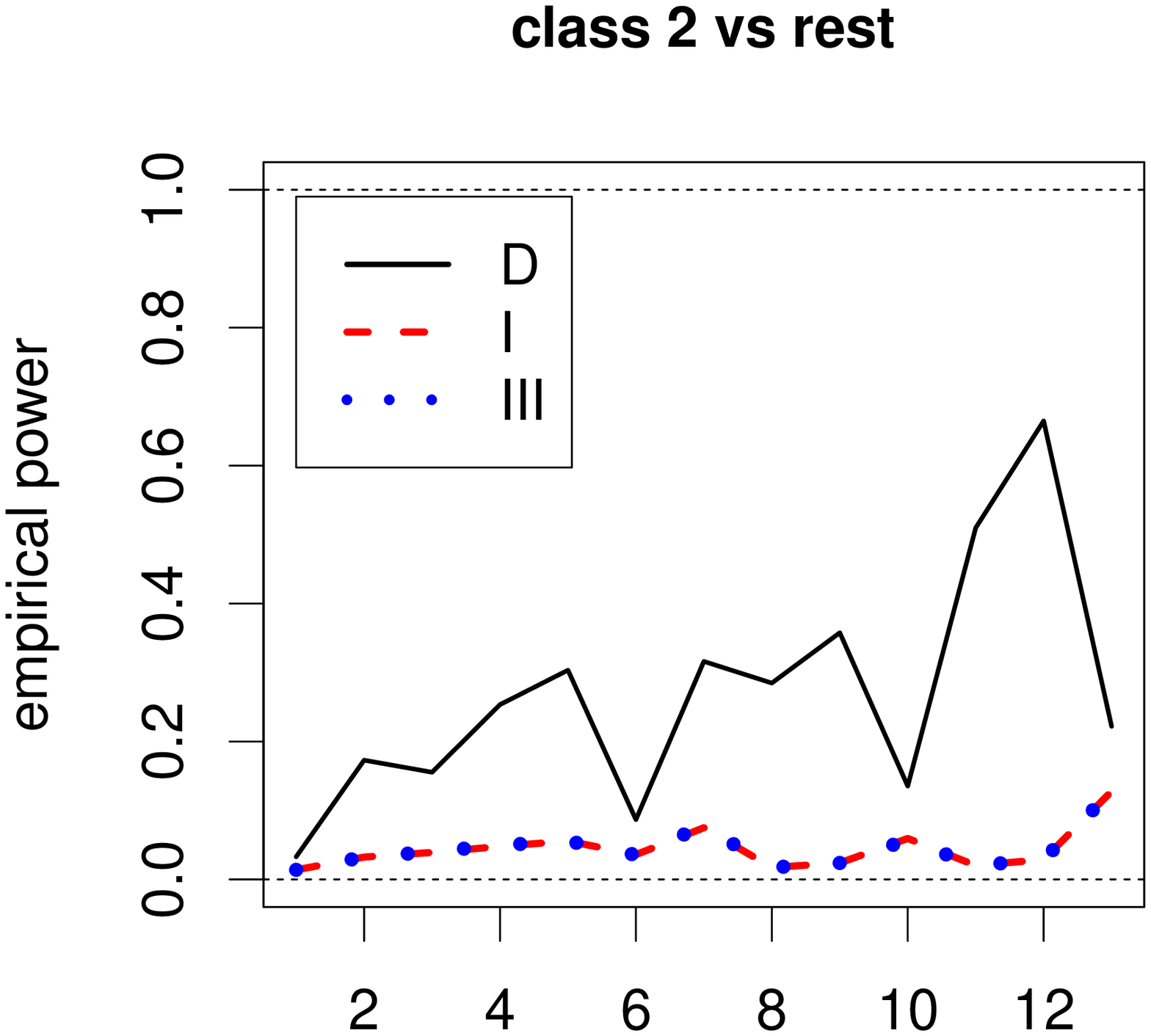} }}
\rotatebox{-90}{ \resizebox{2.1 in}{!}{\includegraphics{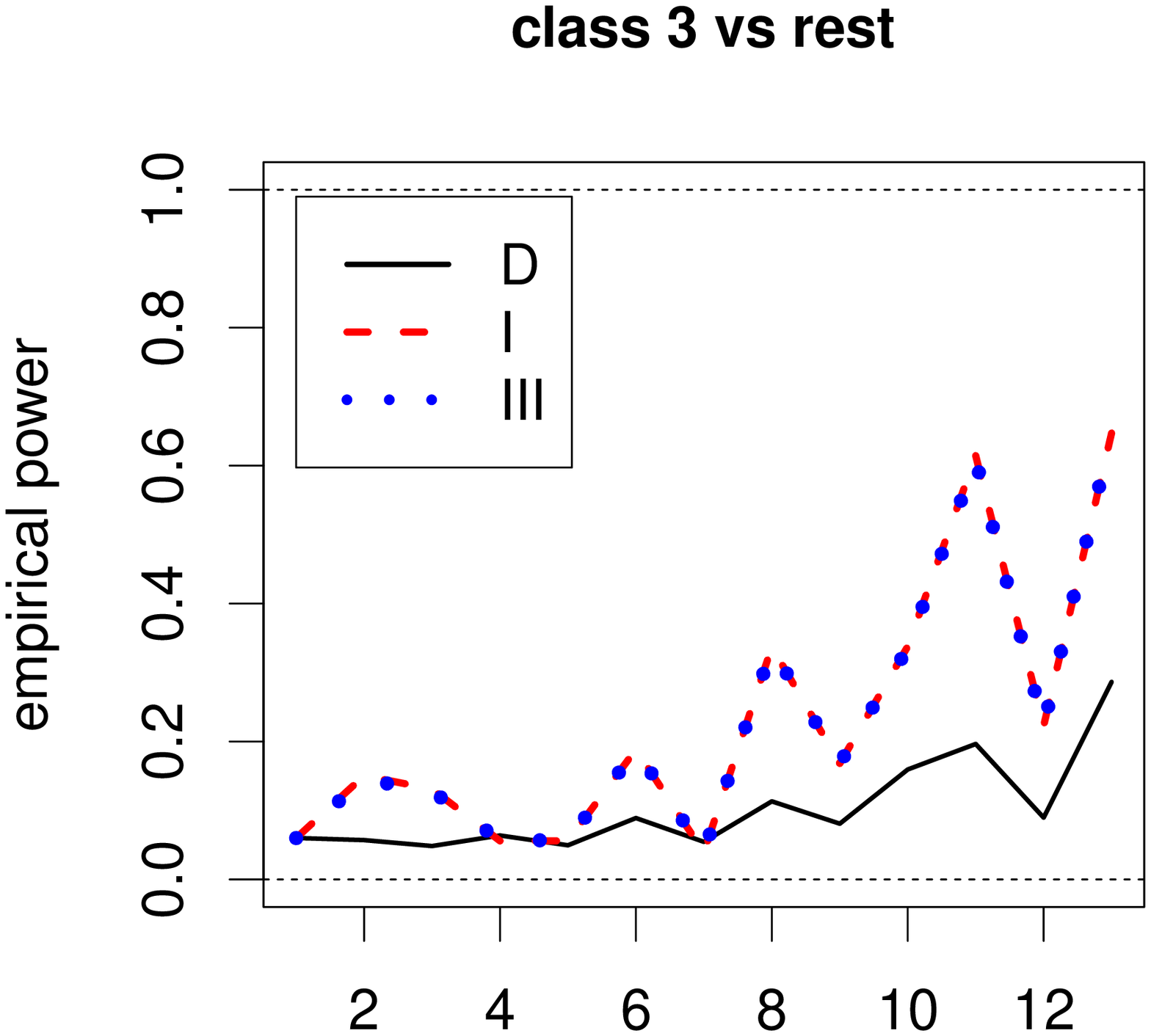} }}
Power Estimates under $H_{A_2}$\\
\rotatebox{-90}{ \resizebox{2.1 in}{!}{\includegraphics{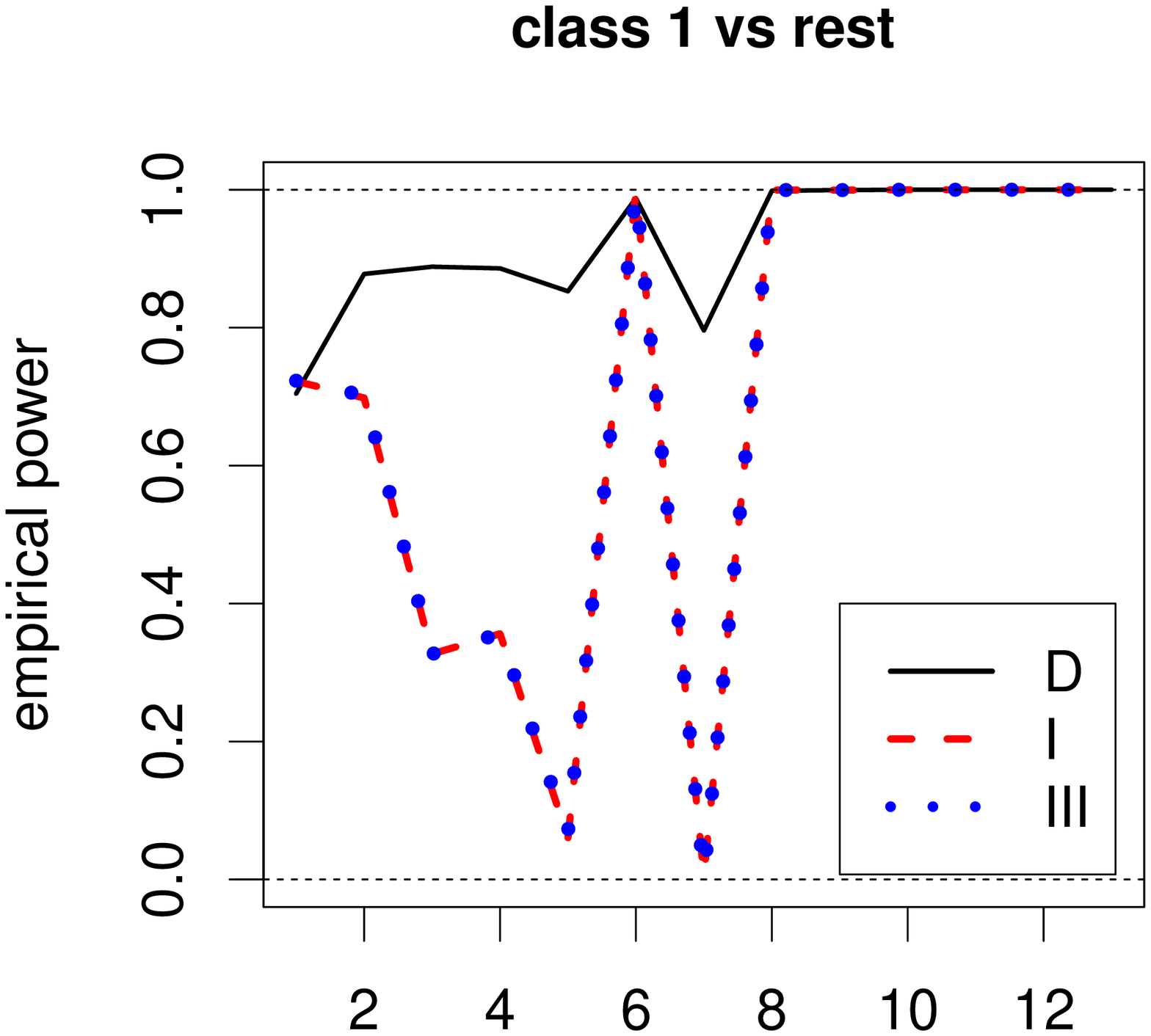} }}
\rotatebox{-90}{ \resizebox{2.1 in}{!}{\includegraphics{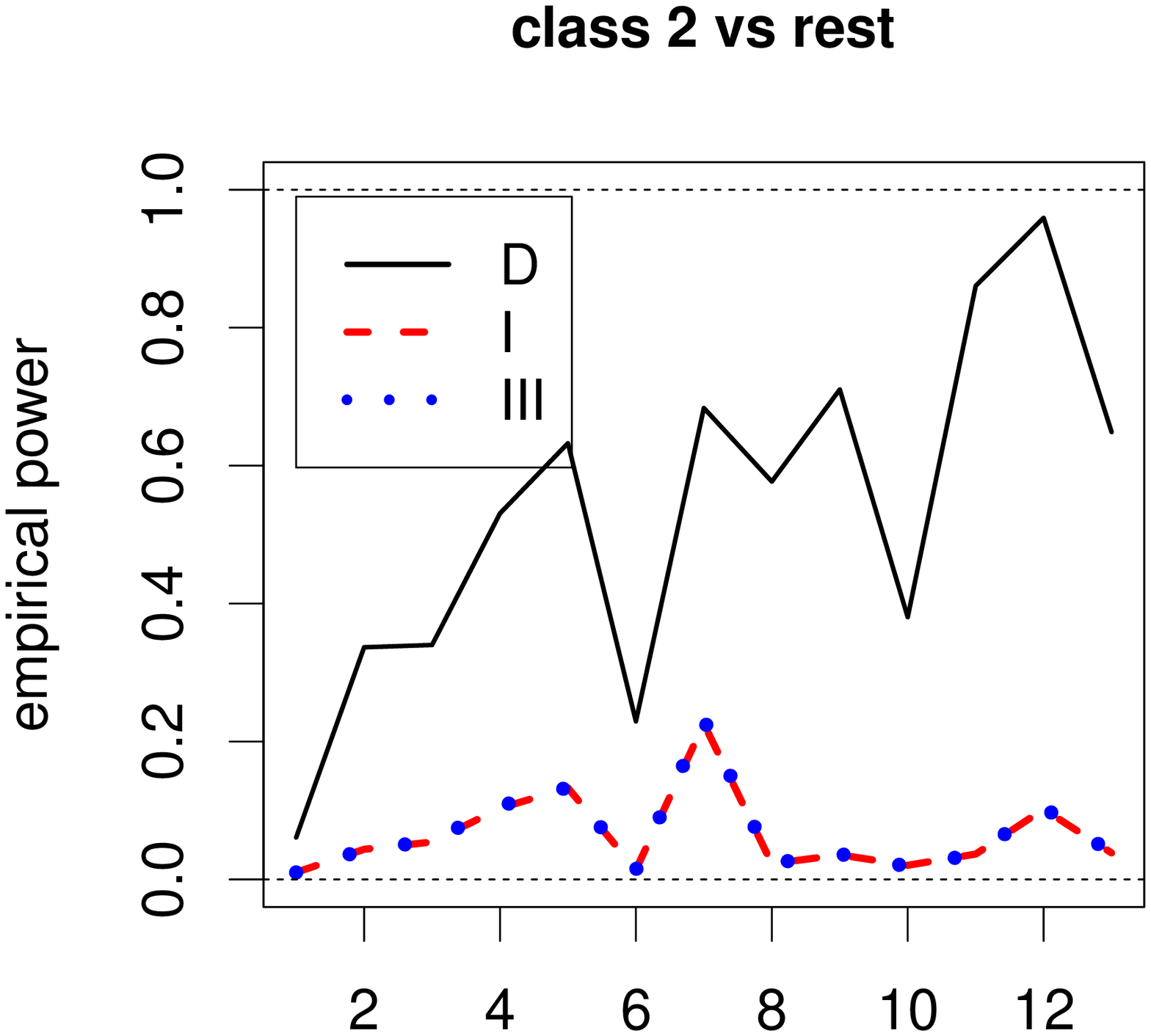} }}
\rotatebox{-90}{ \resizebox{2.1 in}{!}{\includegraphics{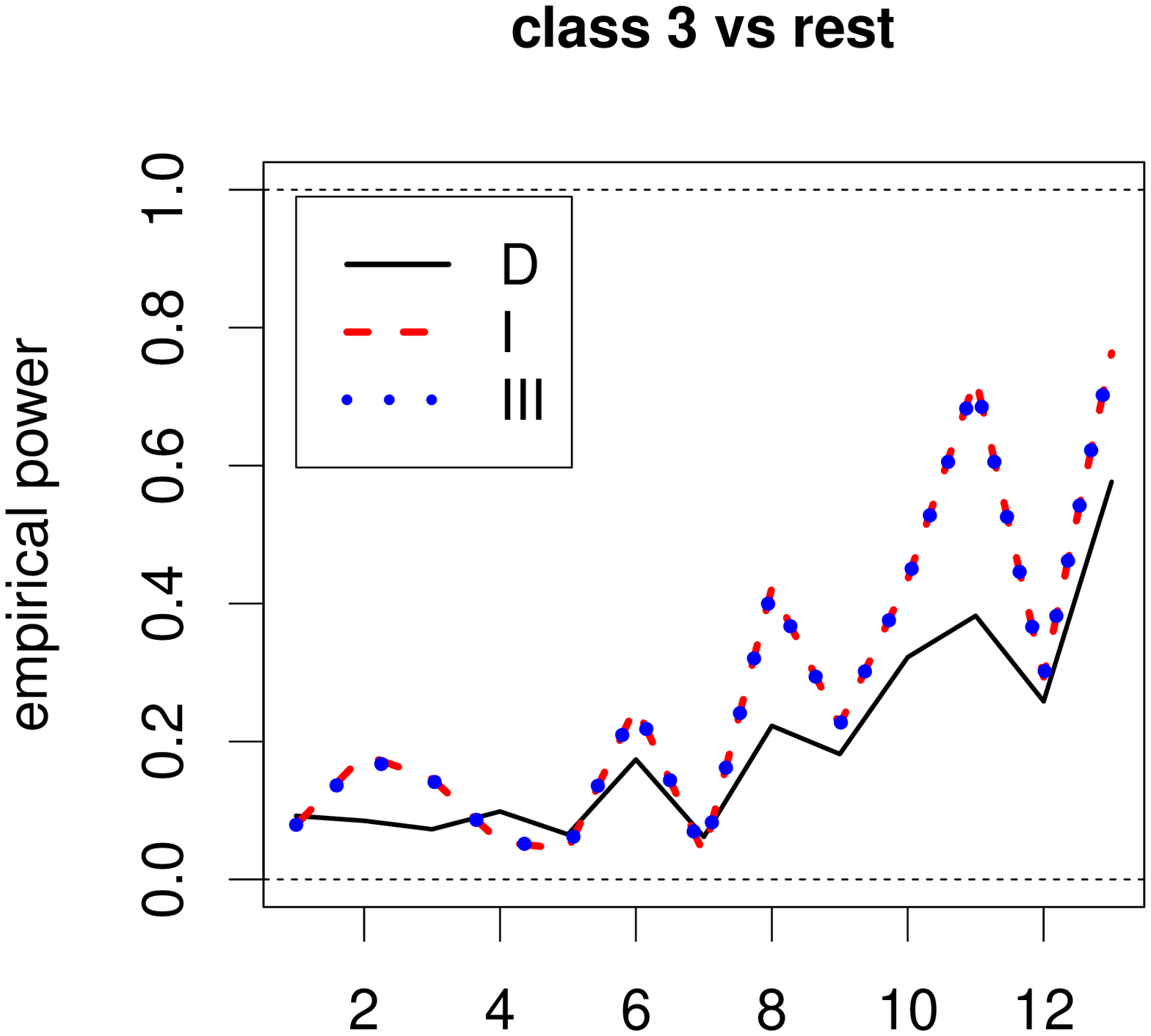} }}
Power Estimates under $H_{A_3}$\\
\rotatebox{-90}{ \resizebox{2.1 in}{!}{\includegraphics{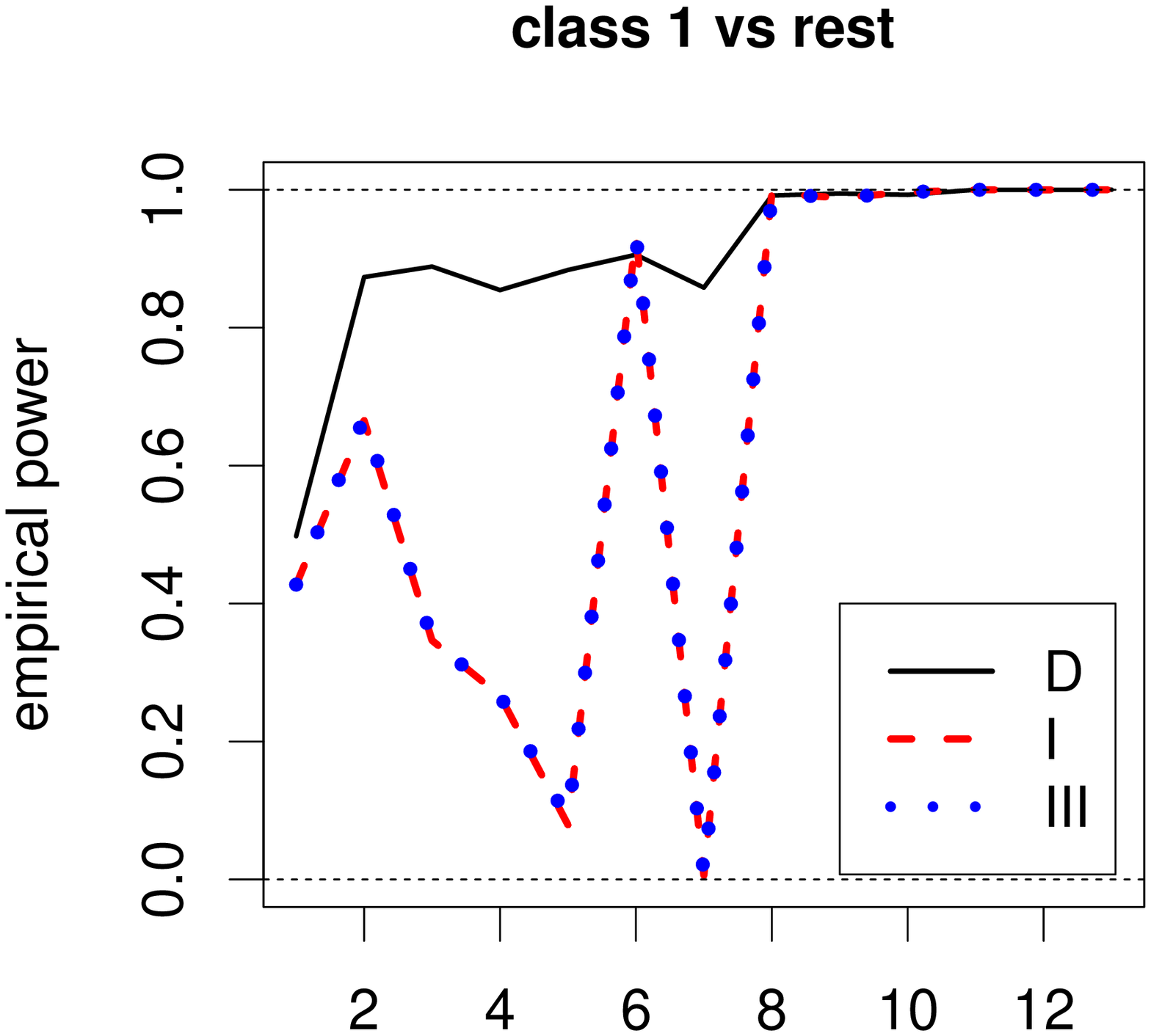} }}
\rotatebox{-90}{ \resizebox{2.1 in}{!}{\includegraphics{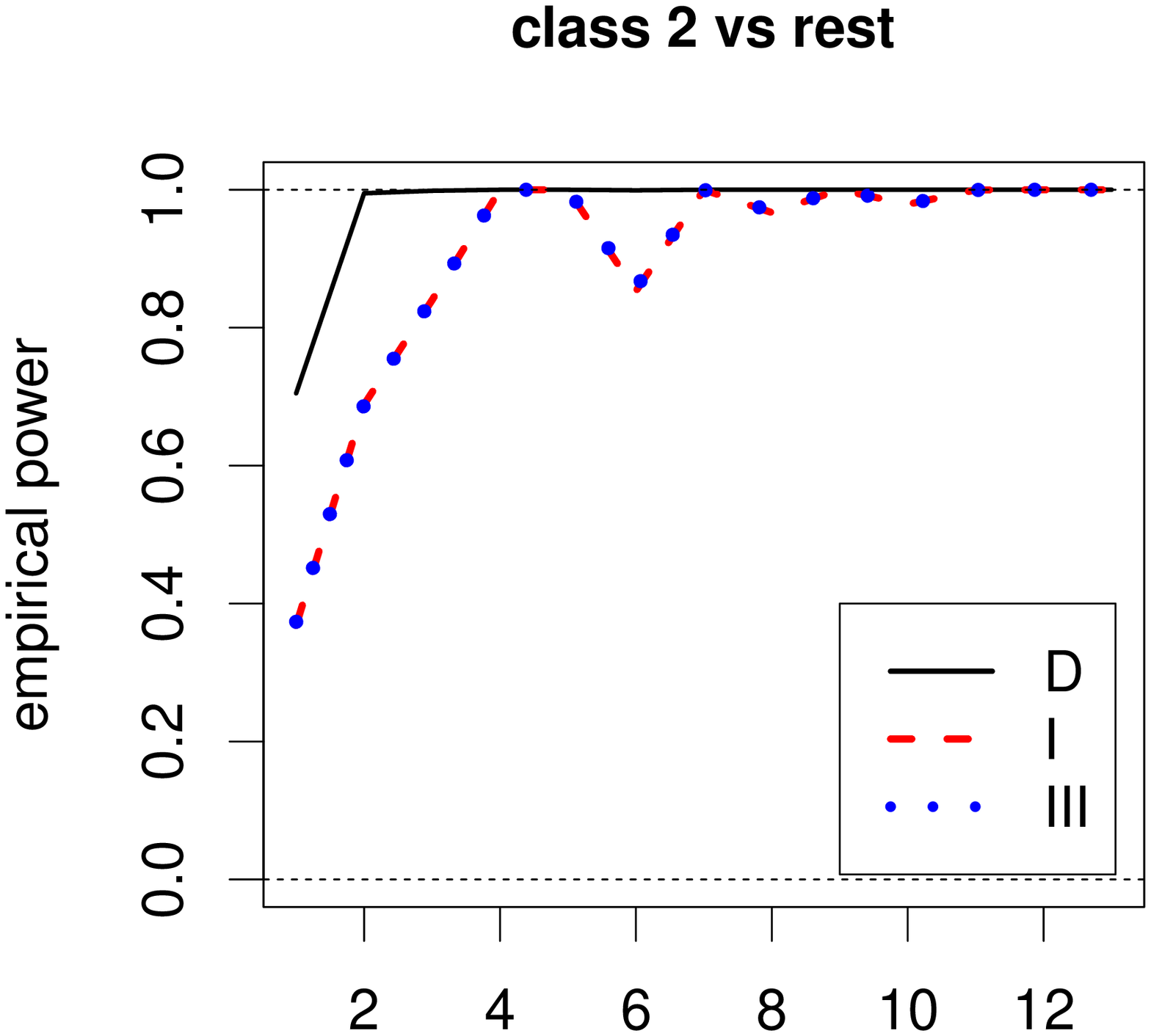} }}
\rotatebox{-90}{ \resizebox{2.1 in}{!}{\includegraphics{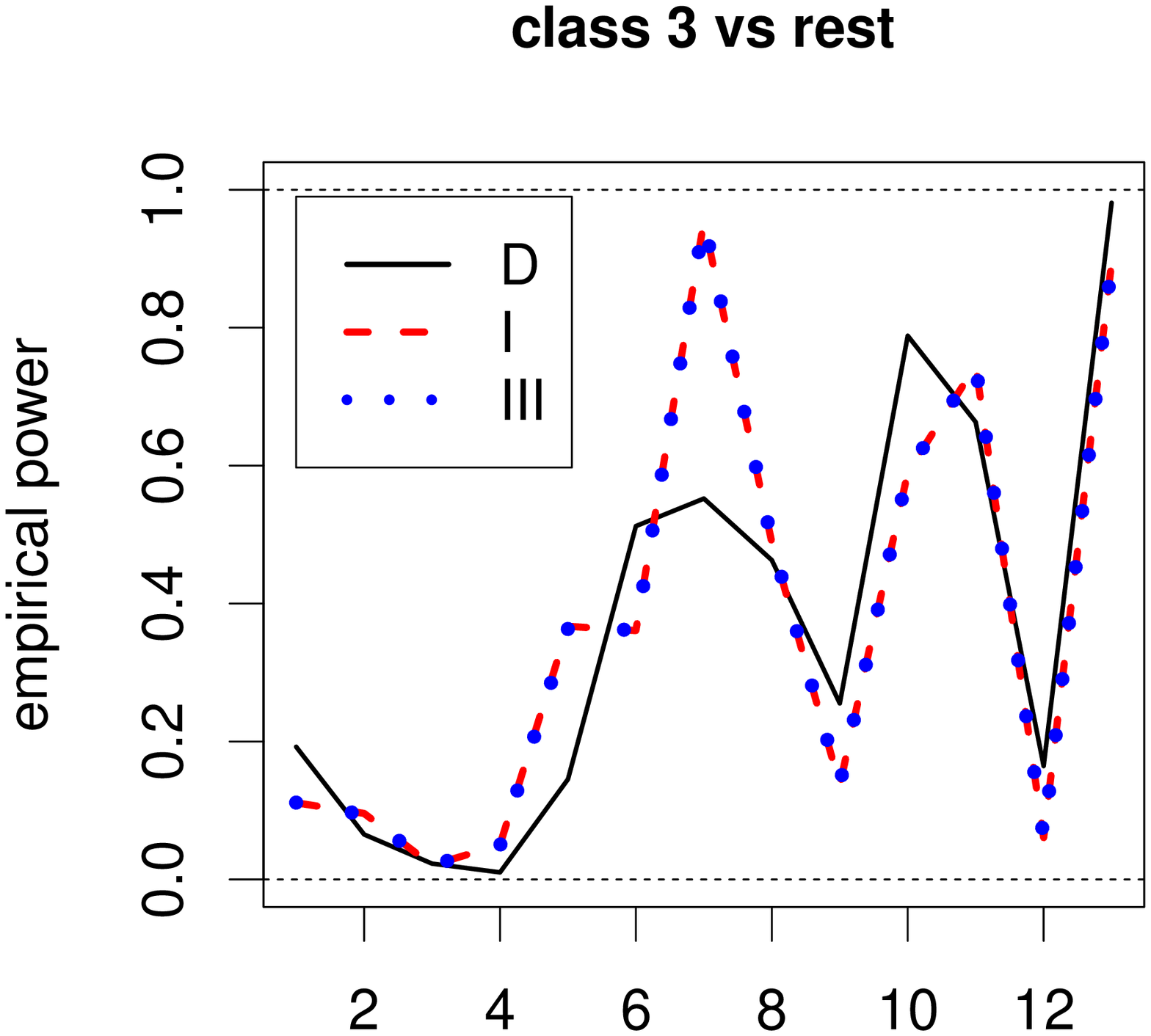} }}
\caption{
\label{fig:power-assoc-cell-3cl-1vsR}
The empirical power estimates of the cell-specific tests for cell $(2,2)$,
under the association alternatives
$H_{A_1}$ (top), $H_{A_2}$ (middle), and $H_{A_3}$ (bottom) in the three-class case
with the one-versus-rest type testing.
The legend labeling is as in Figure \ref{fig:emp-size-CSR-2cl}
and
horizontal axis labels are as in Figure \ref{fig:emp-size-CSR-cell-3cl}.
}
\end{figure}

\begin{figure} [hbp]
\centering
%\psfrag{Density}{ \Huge{\bf{Density}}}
Empirical Power Estimates of Overall Tests under $H_{A_1}$\\
\rotatebox{-90}{ \resizebox{2.1 in}{!}{\includegraphics{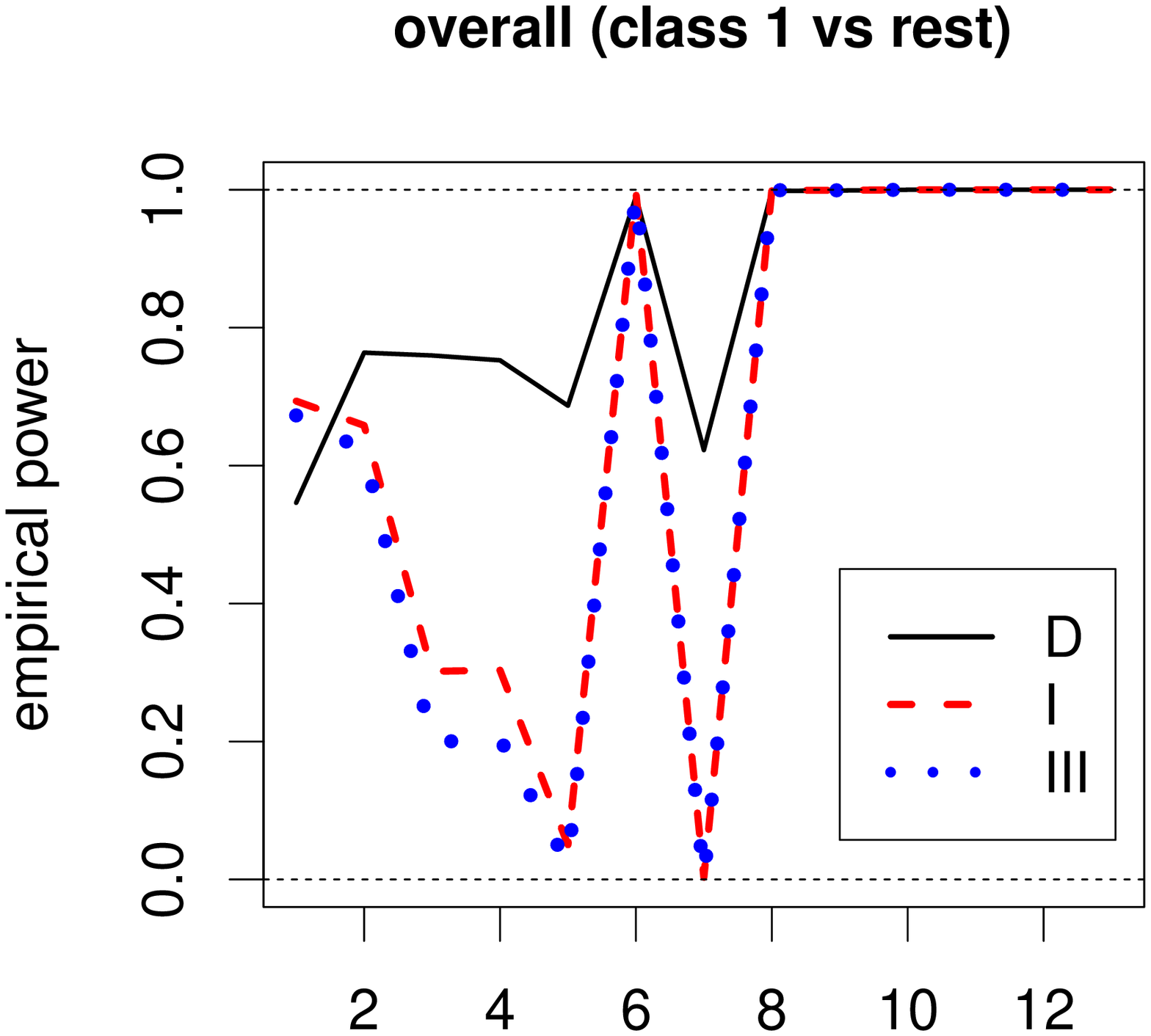} }}
\rotatebox{-90}{ \resizebox{2.1 in}{!}{\includegraphics{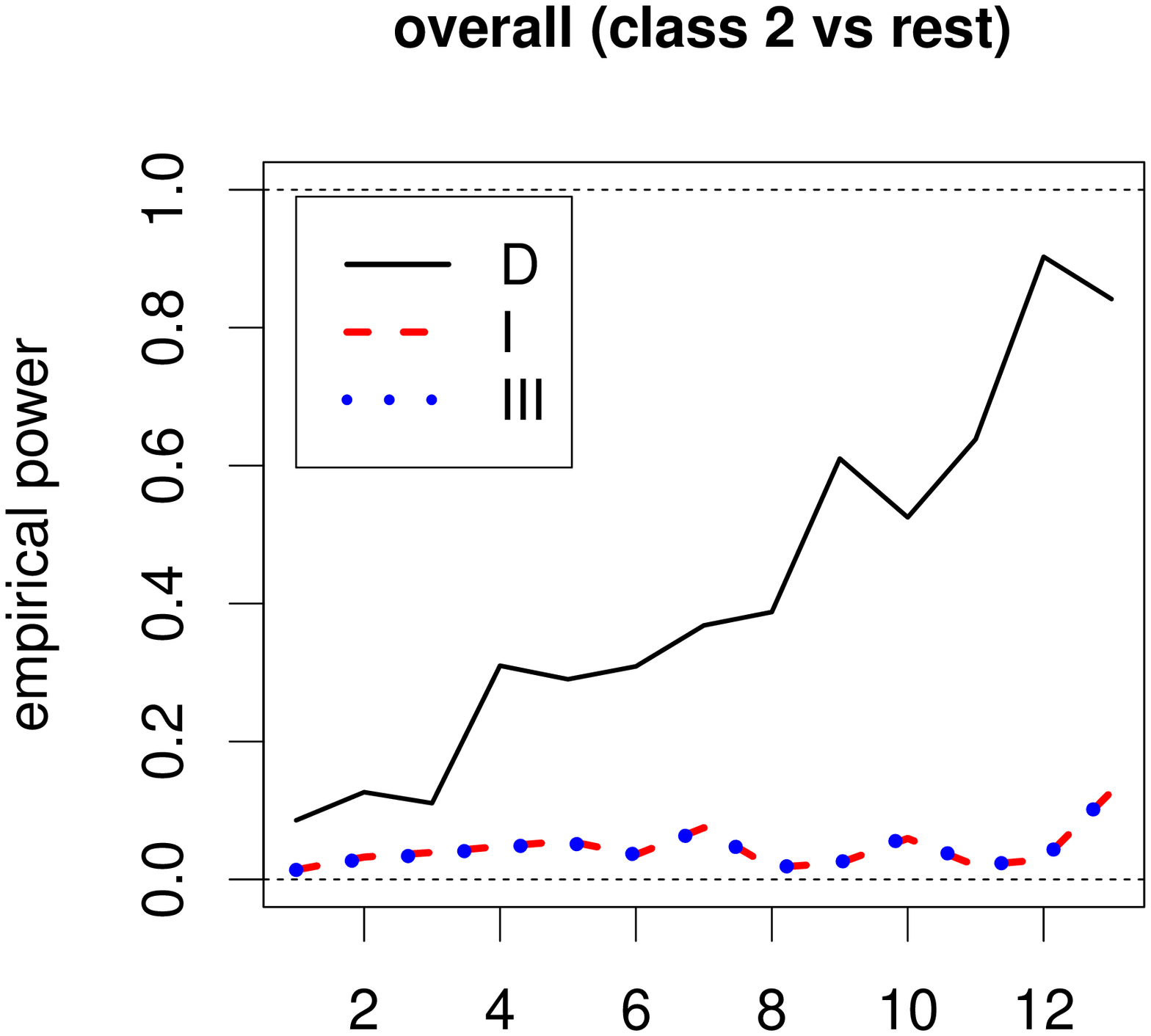} }}
\rotatebox{-90}{ \resizebox{2.1 in}{!}{\includegraphics{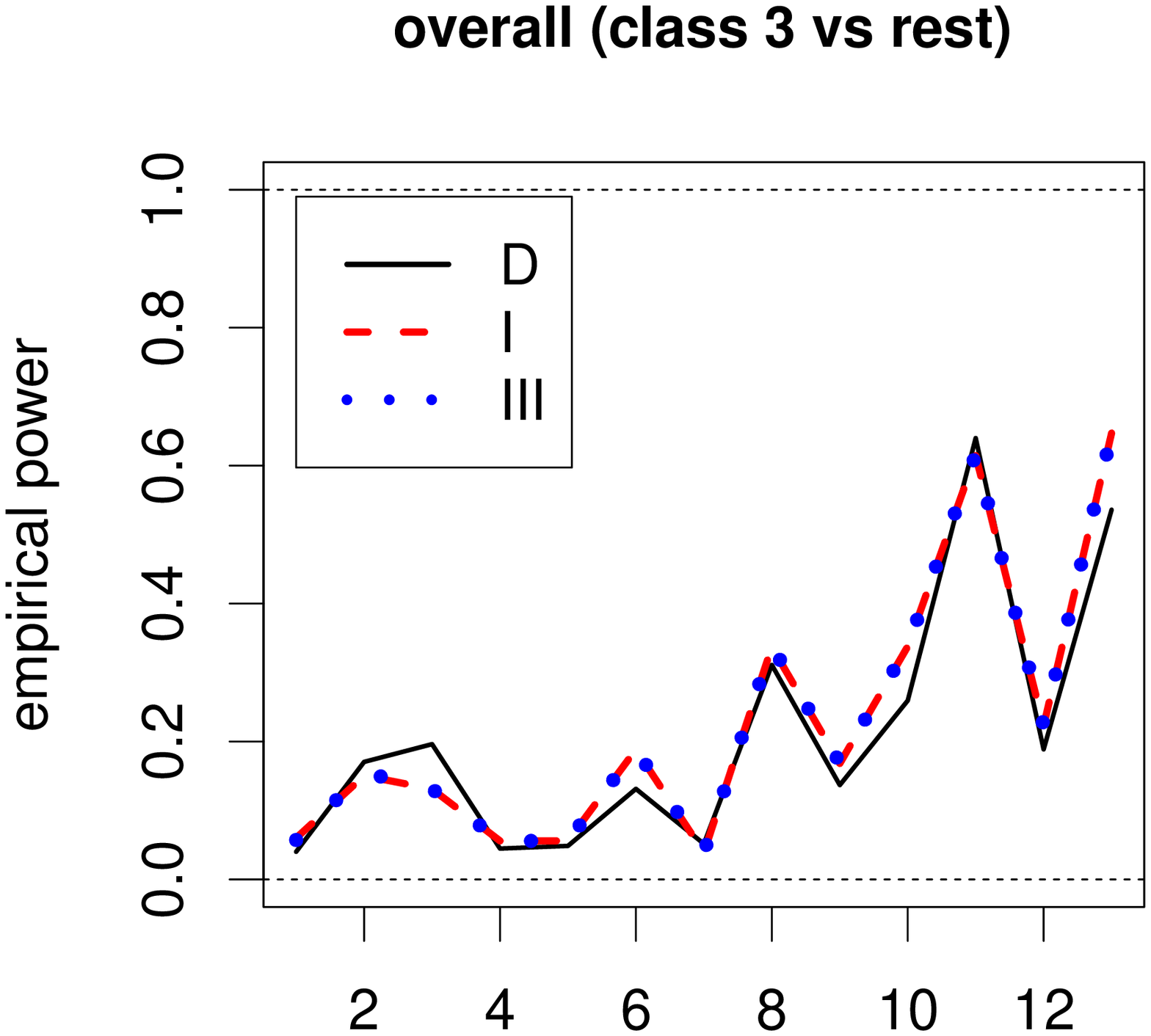} }}
Power Estimates under $H_{A_2}$\\
\rotatebox{-90}{ \resizebox{2.1 in}{!}{\includegraphics{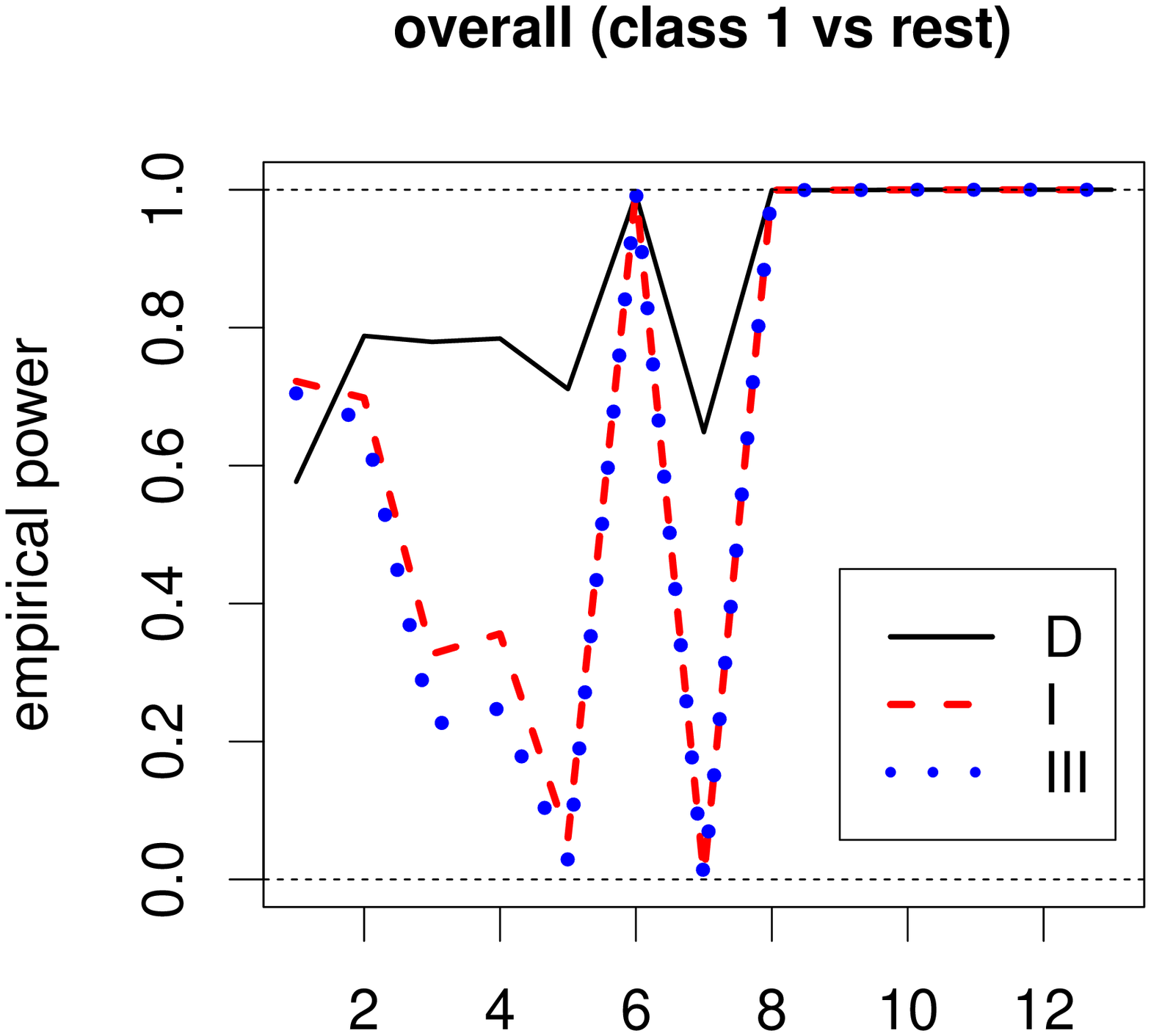} }}
\rotatebox{-90}{ \resizebox{2.1 in}{!}{\includegraphics{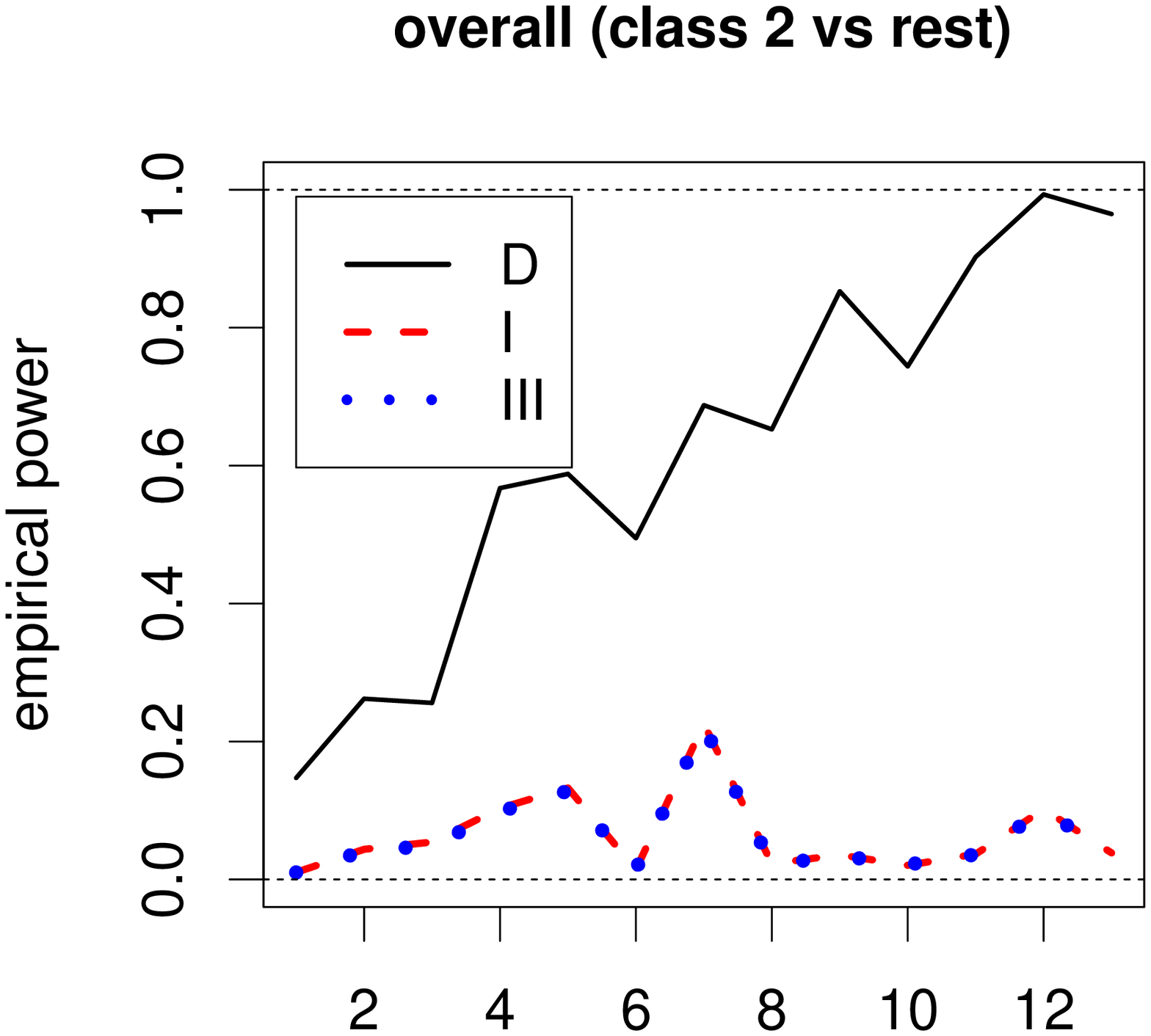} }}
\rotatebox{-90}{ \resizebox{2.1 in}{!}{\includegraphics{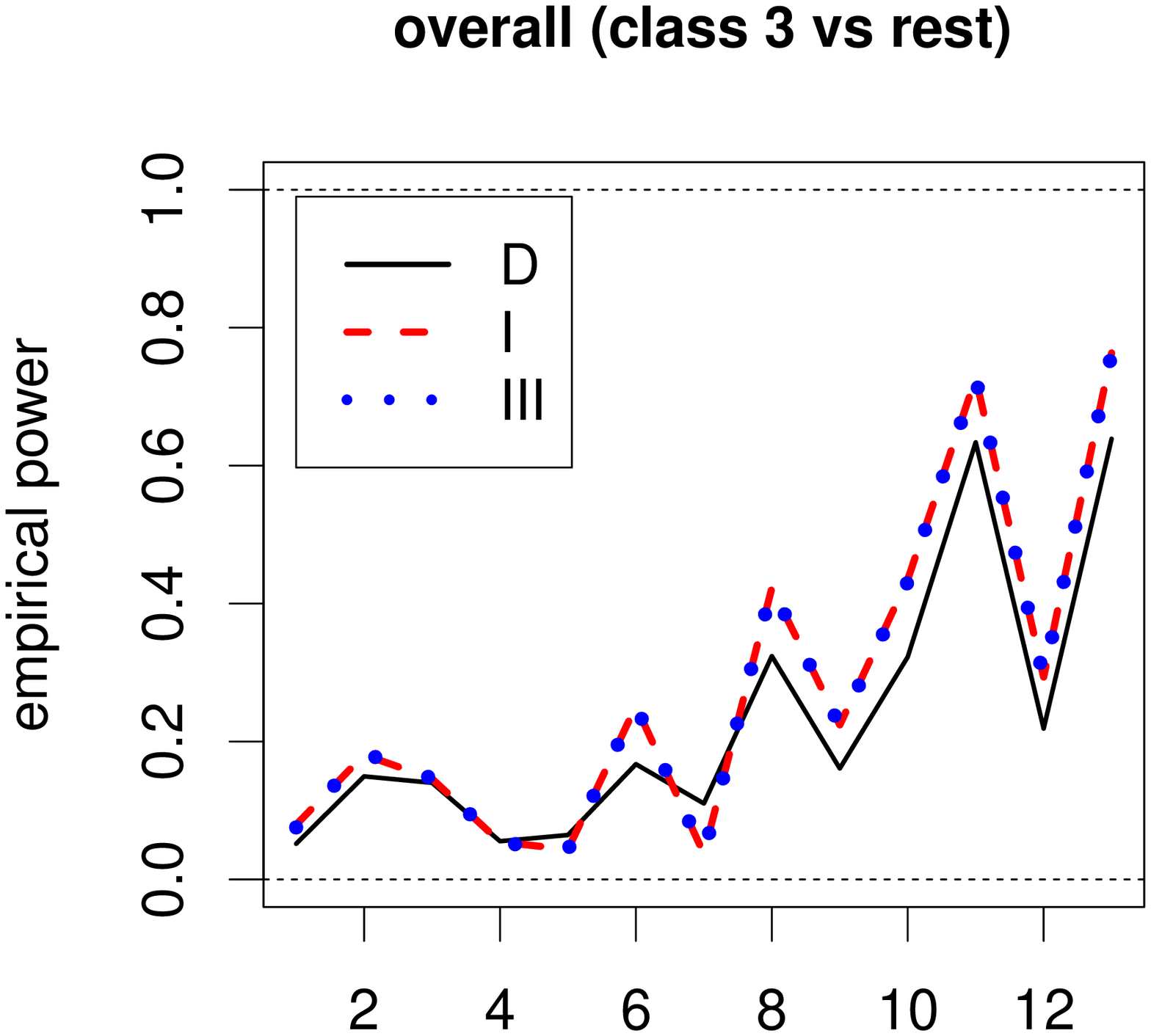} }}
Power Estimates under $H_{A_3}$\\
\rotatebox{-90}{ \resizebox{2.1 in}{!}{\includegraphics{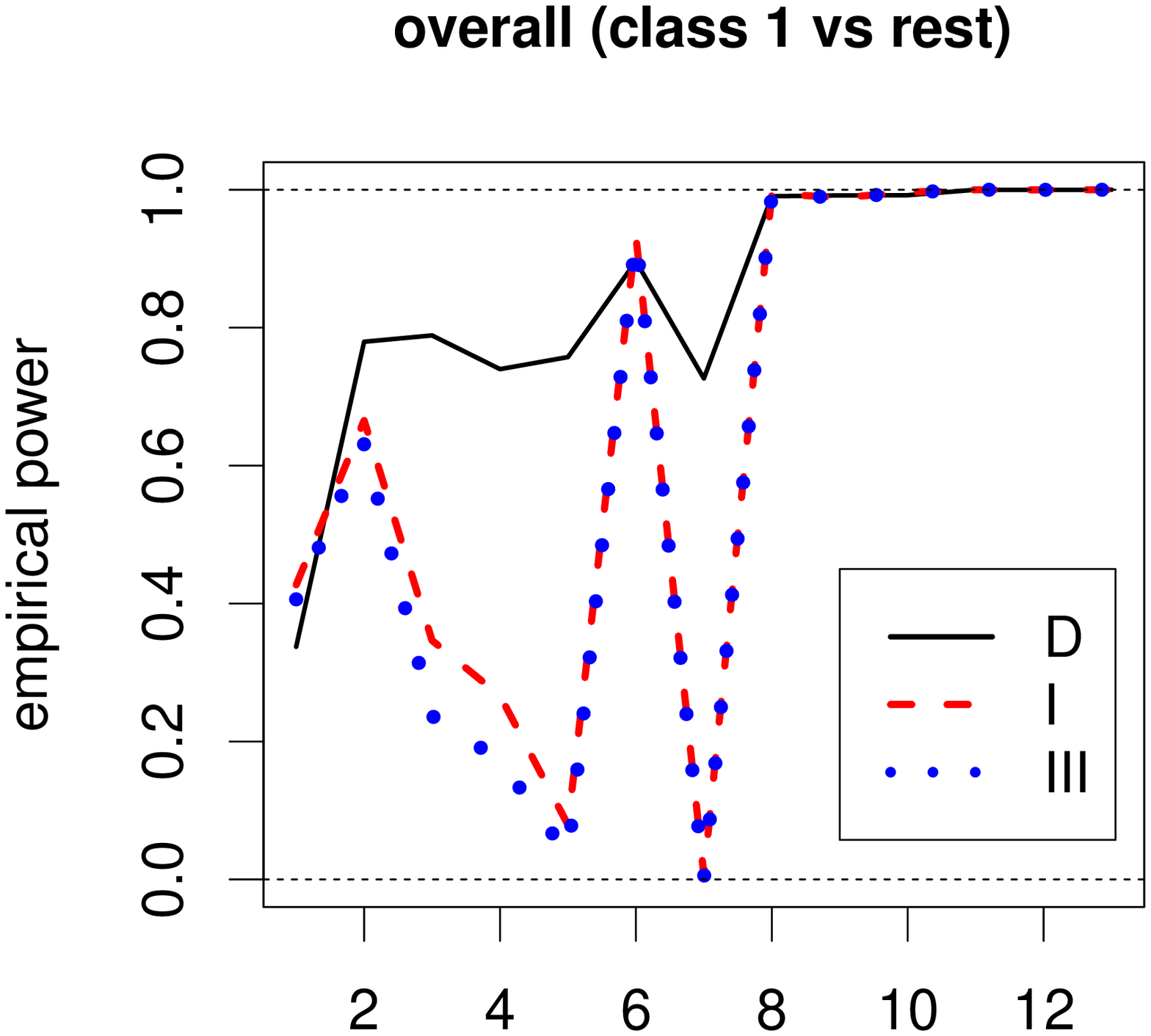} }}
\rotatebox{-90}{ \resizebox{2.1 in}{!}{\includegraphics{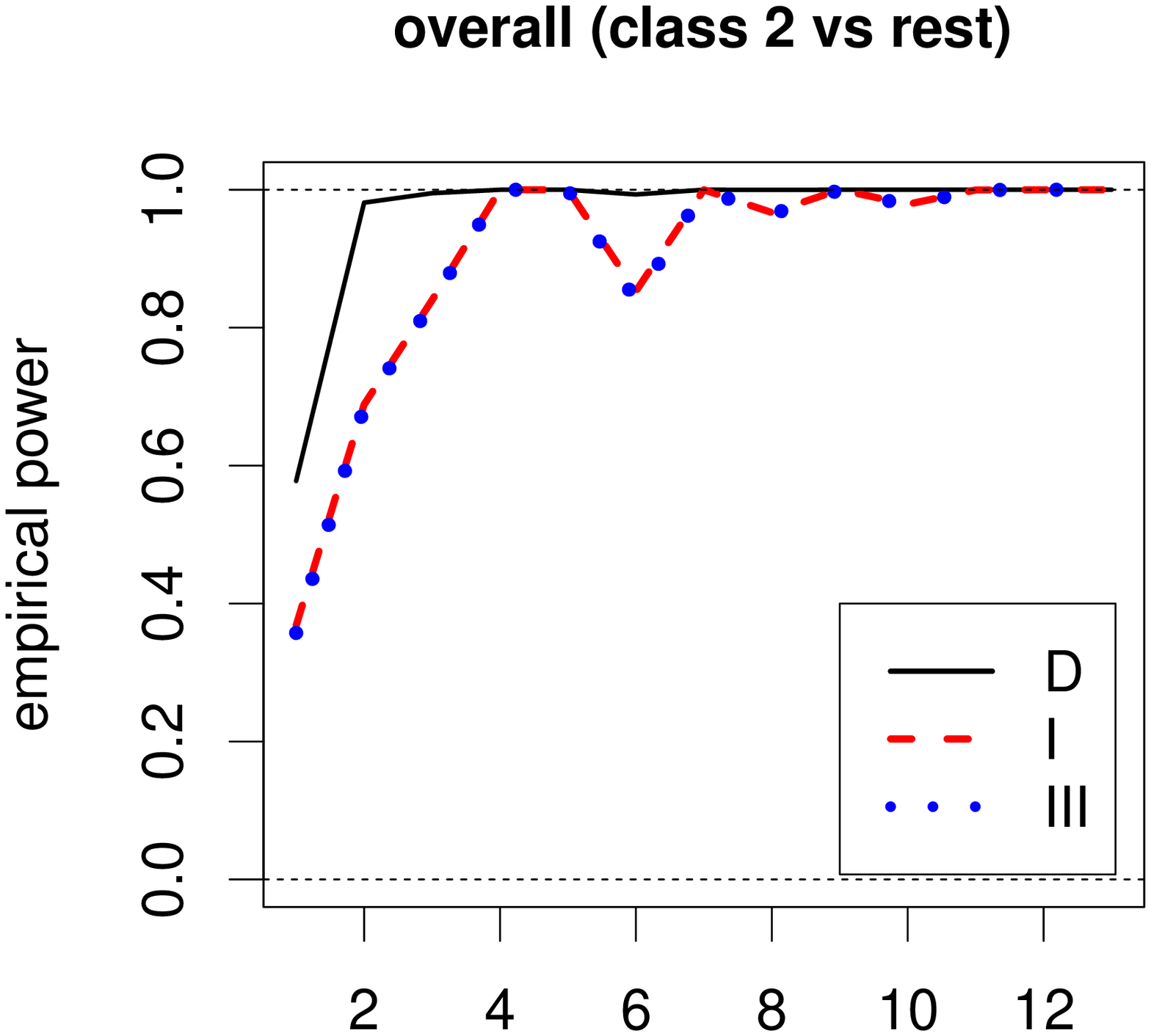} }}
\rotatebox{-90}{ \resizebox{2.1 in}{!}{\includegraphics{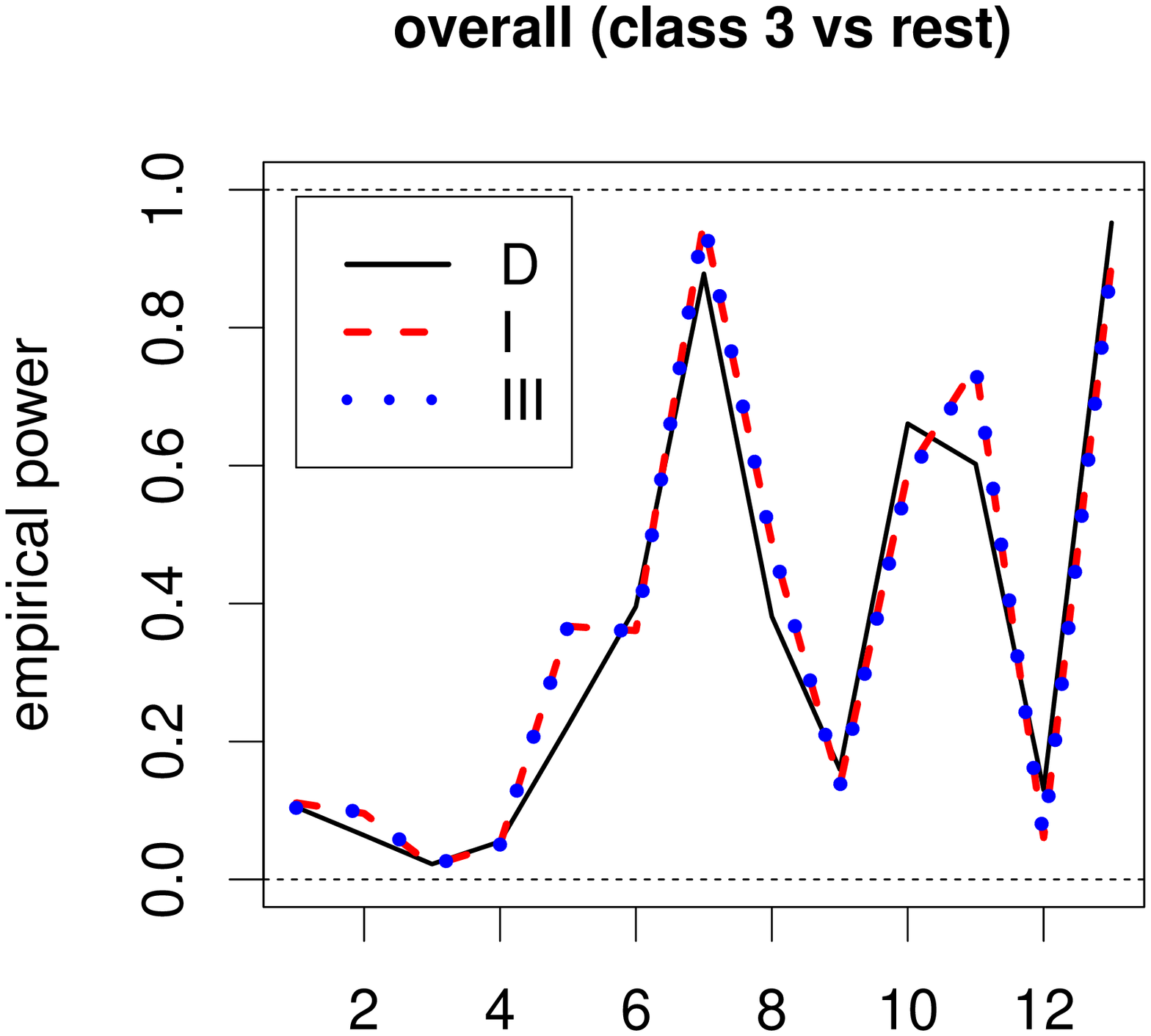} }}
\caption{
\label{fig:power-assoc-overall-3cl-1vsR}
The empirical power estimates of the overall tests
under the association alternatives
$H_{A_1}$ (top), $H_{A_2}$ (middle), and $H_{A_3}$ (bottom) in the three-class case
with the one-vs-rest type testing.
The legend labeling is as in Figure \ref{fig:emp-size-CSR-2cl}
and
horizontal axis labels are as in Figure \ref{fig:emp-size-CSR-cell-3cl}.
}
\end{figure}

For the association alternatives,
we perform the simulations as in Section \ref{sec:power-comp-assoc-3Cl}.
The corresponding power estimates under the three association alternatives are presented
in Figures \ref{fig:power-assoc-cell-3cl-1vsR}
and \ref{fig:power-assoc-overall-3cl-1vsR}.
For the one-vs-rest cell-specific tests,
Dixon's test has higher power for class 1-vs-rest and 2-vs-rest tests,
and type I and III have higher power for class 3-vs-rest test.
For the overall one-vs-rest tests,
Dixon's test has higher power for classes 1 and 2,
and for class 3,
all tests have similar power estimates.

\section{Example Data: Swamp Tree Data}
\label{sec:example}
\begin{figure} [hbp]
\centering
%\psfrag{Density}{ \Huge{\bf{Density}}}
\rotatebox{-90}{ \resizebox{3.25 in}{!}{\includegraphics{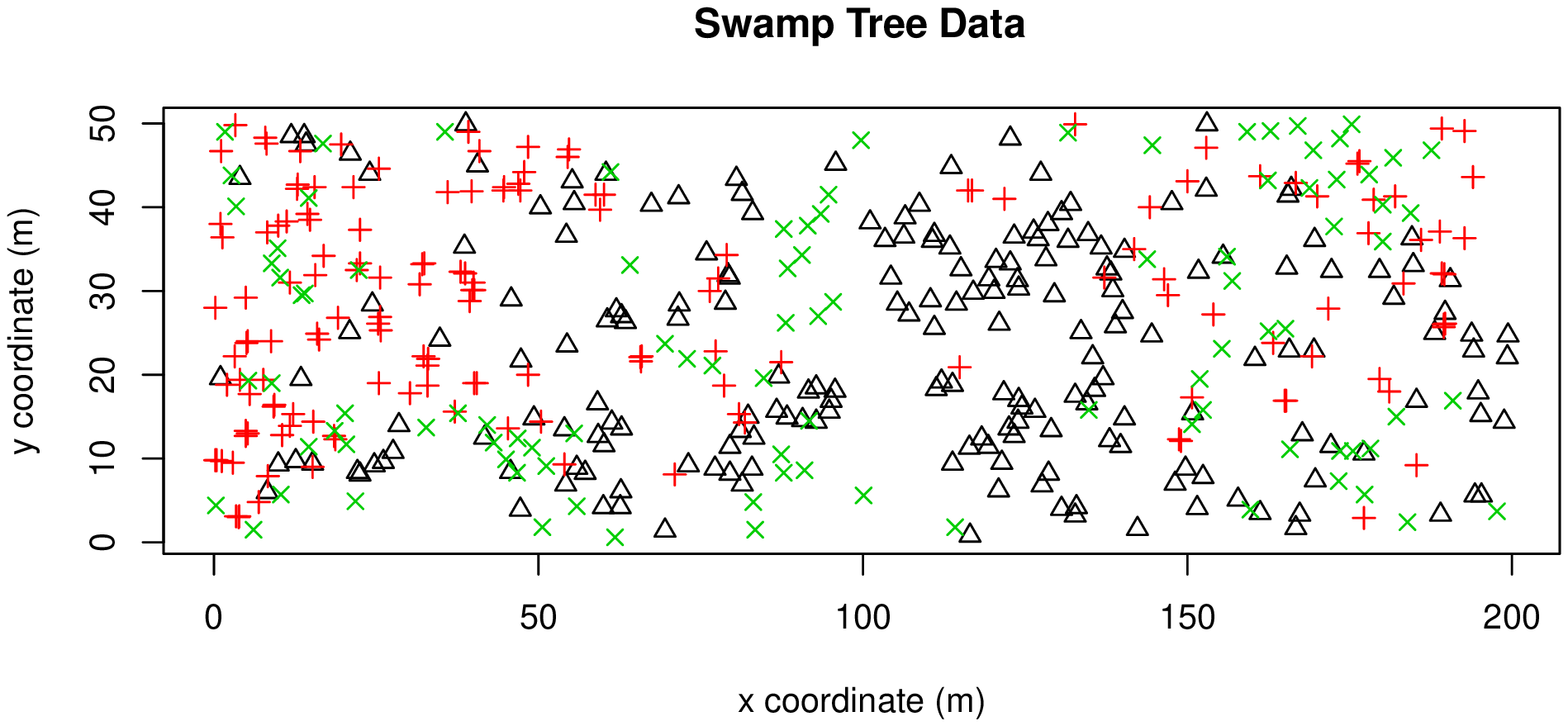} }}
 \caption{
\label{fig:SwampTrees}
The scatter plot of the locations of black gum trees (triangles {\footnotesize $\triangle$}),
Carolina ashes (pluses $+$), bald cypress trees (crosses $\times$).
}
\end{figure}

The NNCT methodology is illustrated on an ecological data set:
the swamp tree data of \cite{good:1982} which was also analyzed by \cite{dixon:1994, dixon:NNCTEco2002}.
The data set is described in detail in \cite{ceyhan:corrected}.
Briefly,
the plot contains 13 different tree species,
of which four species account for over 90 \% of the 734 tree stems.
In our analysis, we only consider
black gums (\emph{Nyssa sylvatica}),
Carolina ashes (\emph{Fraxinus caroliniana}),
and
bald cypresses (\emph{Taxodium distichum})
as if only these three tree species exist in the area,
so we are ignoring the possible effects of other species
on the spatial interaction between these species for illustrative purposes.
Thus, we perform a $3 \times 3$ NNCT-analysis on this data set.
See Figure \ref{fig:SwampTrees} for the location of the trees in this plot
and Table \ref{tab:NNCT-swamp} for the associated
$3 \times 3$ NNCT together with cell percentages based on the base class sizes, and
marginal percentages based on the grand sum, $n$.
When, e.g.,
black gum is the base species and Carolina ash is the NN species,
the cell count is 40 which is 20 \% of the black gums (and Carolina ashes are 34 \% of all trees).
The percentages in Table \ref{tab:NNCT-swamp} and the Figure \ref{fig:SwampTrees}
suggest that each tree species is segregated from the other trees
as the observed percentages of species in the diagonal cells are much larger
than the row percentages (or species percentages).

\begin{table}
\centering
\begin{tabular}{cc|ccc|c}
\multicolumn{2}{c}{}& \multicolumn{3}{c}{NN}& \\
\multicolumn{2}{c}{}& B.G. & C.A. &  B.C. &   sum  \\
\hline
& B.G. &    142  (69 \%, .31) &  40 (20 \%, .09) &  23 (11 \%, .05)  &  205 (45 \%) \\
& C.A. &    34  (22 \%, .07) &   97 (62 \%, .21) &  25 (16 \%, .05) &   156 (34 \%) \\
\raisebox{2.5ex}[0pt]{base}
& B.C. &    38  (39 \%, .08) &   32 (33 \%, .07) &  28 (29 \%, .06) &   98  (21 \%) \\
\hline
&sum   &   214  (47 \%) &  169 (37 \%) & 76 (17 \%) &   459 (100 \%)\\
\end{tabular}
\caption{ \label{tab:NNCT-swamp}
The NNCT for swamp tree data
and the corresponding percentages and $\widehat \pi_{ij}=N_{ij}/n$ values (in parentheses),
where the cell percentages are with respect to the size of the base species (i.e., row sums),
and marginal percentages are with respect to the total size, $n$.
B.G. = black gums, C.A. = Carolina ashes, and B.C. = bald cypresses.
}
\end{table}

\begin{table} [ht]
\centering
\begin{tabular}{|c|c|c|c|}
\multicolumn{4}{c}{Overall tests} \\
\hline
& $\X_D$ & $\X_I$ & $\X_{III}$ \\
\hline
 & 75.78  &  65.35 & 65.39 \\
\hline
$\pasy$  & $<.0001$ & $<.0001$  & $<.0001$  \\
\hline
$\pmc$   & $<.0001$ & $<.0001$  & $<.0001$  \\
\hline
$\prand$ & $<.0001$ & $<.0001$  & $<.0001$  \\
\hline
\end{tabular}
\caption{ \label{tab:pval-overall-swamp-tree}
Test statistics and $p$-values for the overall tests
and the corresponding $p$-values.
$\pasy$, $\pmc$, and $\prand$ stand for the $p$-values based on the asymptotic
approximation, Monte Carlo simulation, and randomization of the tests, respectively.
$\X_D$ stands for Dixon's overall test,
$\X_I$ and $\X_{III}$ are for types I and III overall tests, respectively.}
\end{table}

\begin{table} [ht]
\centering
\begin{tabular}{|c|c|c|c|}
\multicolumn{4}{c}{Dixon's cell-specific tests} \\
\hline
& B.G. & C.A. &  B.C. \\
\hline
B.G. & 6.57 ($<.0001$, $<.0001$, $<.0001$,)     &  -4.46 ($<.0001$, $<.0001$, $<.0001$,) & -3.74 (.0002, $<.0001$, .0003)   \\
\hline
C.A. & -5.65 ($<.0001$, $<.0001$, $<.0001$) & 6.60 ($<.0001$, $<.0001$, $<.0001$)   &  -1.70 (.0893, .0918, .1032)  \\
\hline
B.C. & -1.18 (.2395, .2470, .2596)    & -0.30 (.7672, .7796, .8140)   &  1.51 (.1320, .1345, .1445)   \\
\hline
\multicolumn{4}{c}{Type I cell-specific tests} \\
\hline
 & B.G. & C.A. &  B.C. \\
\hline
B.G. &   6.91 ($<.0001$, $<.0001$, $<.0001$)     &  -6.29 ($<.0001$, $<.0001$, $<.0001$) & -2.37 (.0177, .0170, .0176)   \\
\hline
C.A. &   -6.86 ($<.0001$, $<.0001$, $<.0001$)     & 6.49 ($<.0001$, $<.0001$, $<.0001$)   &  -0.21 (.8352, .8439, .8408)  \\
\hline
B.C. &   -1.67 (.0944, .0954, .0900)    & -0.96 (.3382, .3407, .3433)   &  2.61 (.0091, .0087, .0081)   \\
\hline
\multicolumn{4}{c}{Type III cell-specific tests} \\
\hline
 & B.G. & C.A. &  B.C. \\
\hline
B.G. &   6.91 ($<.0001$, $<.0001$, $<.0001$)     &  -6.29 ($<.0001$, $<.0001$, $<.0001$) & -2.37 (.0180, .0172, .0179)   \\
\hline
C.A. &   -6.86 ($<.0001$, $<.0001$, $<.0001$)  & 6.49 ($<.0001$, $<.0001$, $<.0001$)   &  -0.20 (.8381, .8455, .8436)  \\
\hline
B.C. &   -1.67 (.0943, .0953, .0898)    & -0.96 (.3375, .3401, .3426)   &  2.60 (.0094, .0088, .0084)   \\
\hline
\end{tabular}
\caption{ \label{tab:pval-cell-swamp-tree}
Test statistics and $p$-values for the cell-specific tests
and the corresponding $p$-values (in parentheses).
The $p$-values are given in the order of $\pasy$, $\pmc$, and $\prand$,
whose labeling is as in Table \ref{tab:pval-overall-swamp-tree}.
B.G. = black gums, C.A. = Carolina ashes, and
B.C. = bald cypresses.
}
\end{table}

The null model in a NNCT analysis depends on the particular ecological context.
\cite{goreaud:2003} state that under CSR independence,
the two classes are \emph{a priori}
the result of different processes (e.g., individuals of different species or age cohorts).
On the other hand,
under RL,
some processes affect \emph{a posteriori}
the individuals of a single population
(e.g., diseased vs. non-diseased individuals of a single species).
Hence, in the swamp tree data,
the locations of the tree species can be viewed a priori resulting from different processes,
so the more appropriate null hypothesis is the CSR independence pattern.
We compute
$Q=282$ and $R=288$ for this data set
and our inference will be conditional on these values.
Dixon's and the new overall segregation tests and the associated $p$-values are presented
in Table \ref{tab:pval-overall-swamp-tree},
where $\pasy$ stands for the $p$-value based on the asymptotic approximation,
$\pmc$ is the $p$-value based on $10000$ Monte Carlo replication of the CSR independence
pattern in the same plot and $\prand$ is based on Monte Carlo
randomization of the labels on the given locations of the trees 10000 times.
Notice that $\pasy$, $\pmc$, and $\prand$ are all significant.
The cell-specific test statistics and the associated $p$-values are presented
in Table \ref{tab:pval-cell-swamp-tree},
where $p$-values are calculated as in Table \ref{tab:pval-overall-swamp-tree}.
Again, all three $p$-values in Table \ref{tab:pval-cell-swamp-tree} are similar for each cell-specific test.

The overall segregation tests are all highly significant
which implies that there is
significant deviation from the CSR independence pattern
for at least one of the tree species.
To determine which species exhibit segregation or association,
we perform the cell-specific tests as a post-hoc analysis.
At 0.05 level,
Dixon's and the new cell-specific tests
agree for all cells in term of significance
except for (B.C.,B.C.) cell,
at which Dixon's test is not significant but types I and III are significant.
At 0.10 level tests agree for cells
except (B.C.,B.G) and (C.A.,B.C.),
at cell (B.C.,B.G) Dixon's test is not significant but types I and III are significant,
while at cell (C.A.,B.C.) Dixon's test is significant but types I and III are not.
At 0.01 level tests agree at cells except for cell (B.G.,B.C.)
at which Dixon's test is significant while types I and III are not.
The test statistics are all positive (negative) for the diagonal (off-diagonal) cells
which also support the segregation of species.

%\begin{table}
%\centering
%\begin{tabular}{cc|ccc}
%\multicolumn{2}{c}{}& \multicolumn{3}{c}{NN} \\
%\multicolumn{2}{c}{}& B.G. & C.A. &  B.C.\\
%\hline
%& B.G. &  .69 &  .26 &  .23 \\
%& C.A. &  .17 &  .62 &  .26 \\
%\raisebox{2.5ex}[0pt]{base}
%& B.C. &  .19 &  .21 &  .29 \\
%\hline
%\end{tabular}
%\caption{ \label{tab:prob-NNCT-swamp}
%The observed probabilities of the NN structure for swamp tree data
%B.G. = black gums, C.A. = Carolina ashes, and B.C. = bald cypresses.
%}
%\end{table}

For a given class $i$,
we estimate probabilities $\pi_{ij}$ of Section \ref{sec:part-seg} as
$\widehat \pi_{ij}=N_{ij}/n$.
The estimated probabilities are presented in parentheses as decimals in Table \ref{tab:NNCT-swamp}.
For example, for (B.G.,C.A.) cell,
$\widehat \pi_{12}=N_{12}/n=40/459 \approx 0.09$.
For black gums,
we have $\widehat \pi_{11}= 0.31> \widehat \pi_{12}+\widehat \pi_{13}=0.09+0.05=0.14$,
so black gums exhibit total segregation from the other two tree species.
Similarly,
for California ashes,
we have $\widehat \pi_{22}= 0.21 > \widehat \pi_{21}+\widehat \pi_{23}=0.07+0.05=0.12$,
so Carolina ashes exhibit total segregation from the other two tree species.
However,
bald cypresses exhibit neither strong nor total segregation,
since $\widehat \pi_{33}= 0.06 < \widehat \pi_{31}=0.08$ and $\widehat \pi_{33}< \widehat \pi_{32}=0.07$.
Furthermore,
black gums seem to be strongly associated with bald cypresses
as $\widehat \pi_{31}=0.08 > \widehat \pi_{32}=0.07$ and $\widehat \pi_{31}>\widehat \pi_{33}=0.06$.

\begin{table} [ht]
\centering
\begin{tabular}{|c|c|c|c|}
\multicolumn{4}{c}{One-vs-rest Cell-specific Tests} \\
\hline
& $Z^D_{22}$ & $Z^I_{22}$ & $Z^{III}_{22}$ \\
\hline
B.G.-vs-rest & 5.09 ($<.0001$)  &  6.91 ($<.0001$) & 6.91 ($<.0001$)  \\
\hline
C.A.-vs-rest & 3.86 ($.0001$)  &  6.49 ($<.0001$) & 6.49 ($<.0001$)  \\
\hline
B.C.-vs-rest & 4.12 ($<.0001$)  &  2.61 ($.0046$) & 2.61 ($.0045$)  \\
\hline
\end{tabular}
\begin{tabular}{|c|c|c|c|}
\multicolumn{4}{c}{One-vs-rest Overall Tests} \\
\hline
& $\X_D$ & $\X_I$ & $\X_{III}$ \\
\hline
B.G.-vs-rest  & 48.86 ($<.0001$)  &  47.70 ($<.0001$) & 47.72 ($<.0001$)  \\
\hline
C.A.-vs-rest  & 44.79 ($<.0001$)  &  42.11 ($<.0001$) & 42.15 ($<.0001$) \\
\hline
B.C.-vs-rest  & 16.96 ($.0002$) &  6.79 ($.0091$) & 6.75 ($.0094$) \\
\hline
\end{tabular}
\caption{ \label{tab:one-vs-rest-swamp-tree}
Test statistics and $p$-values for one-vs-rest cell-specific tests for cell $(2,2)$
and one-vs-rest overall tests.
The corresponding $p$-values are presented in parentheses.
$Z^D_{22}$ stands for Dixon's cell-specific test,
and
$Z^I_{22}$ and $Z^{III}_{22}$ stand for type I and III cell-specific tests.
$\X_D$ stands for Dixon's overall test and
$\X_I$ and $\X_{III}$ are for type I and III overall tests.}
\end{table}

We also present the one-vs-rest cell-specific and overall tests (see Table \ref{tab:one-vs-rest-swamp-tree}).
For each species,
we observe that the other species combined tend to be segregated from the species in consideration,
but to a lesser extent for bald cypresses.

%Based on the Monte Carlo simulation analysis,
%the new tests are more reliable to attach significance to these situations.
The spatial interaction is significant for each species,
but at different levels.
In particular,
black gums exhibit significant segregation from other species
(they are significantly segregated from both Carolina ashes and bald cypresses),
Carolina ashes exhibit significant segregation from other species
(they are significantly segregated from black gums but not from bald cypresses),
and
Bald cypresses exhibit significant segregation from other species
(they are moderately segregated from black gums only but when the two
species of black gums and Carolina ashes are considered together,
the (B.C.,B.C.) cell is significant).

\begin{figure} [hbp]
\centering
%\psfrag{Density}{ \Huge{\bf{Density}}}
%\rotatebox{-90}{ \resizebox{2 in}{!}{\includegraphics{SwTrPCFall.ps} }}
\rotatebox{-90}{ \resizebox{2 in}{!}{\includegraphics{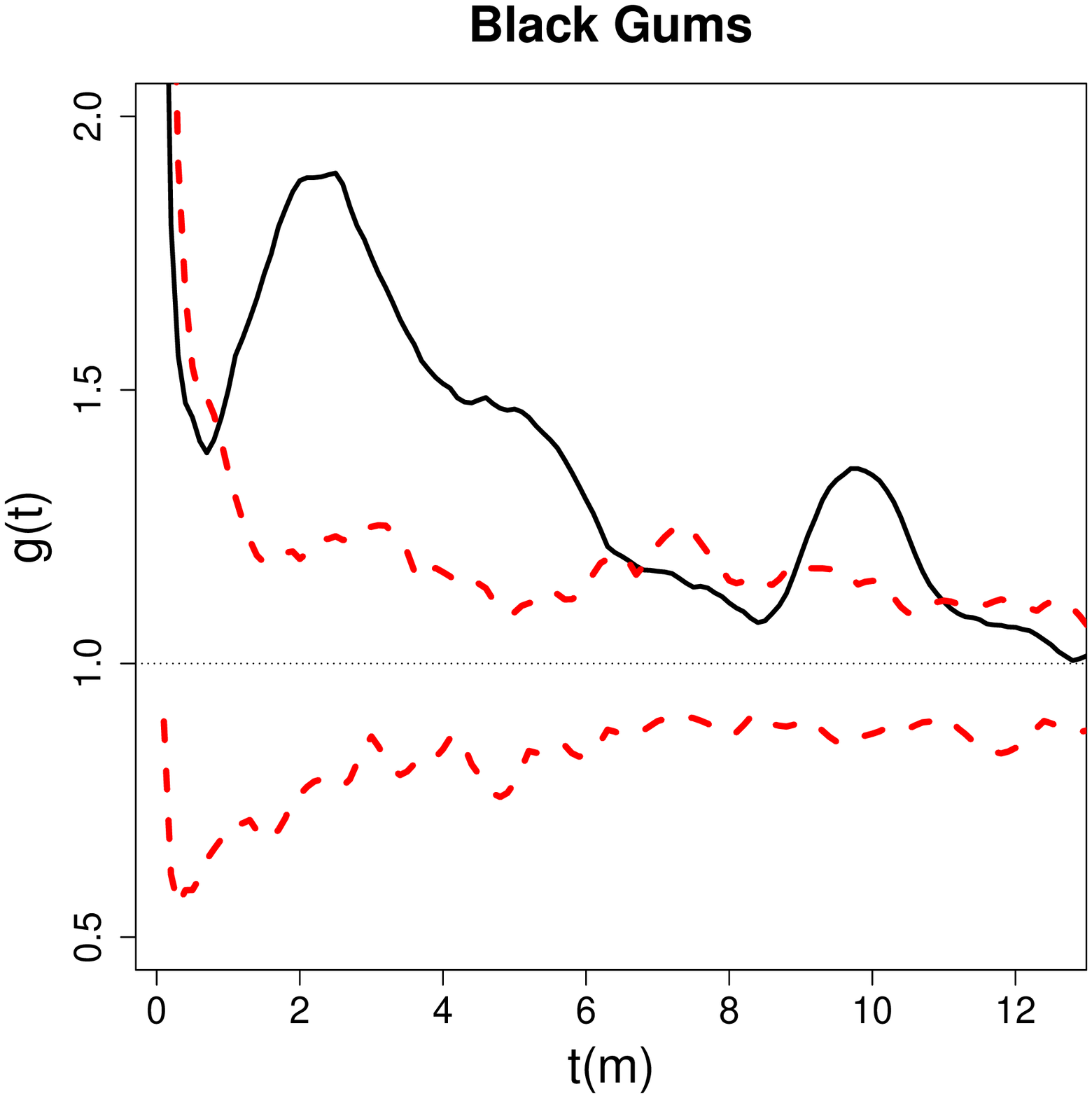} }}
\rotatebox{-90}{ \resizebox{2 in}{!}{\includegraphics{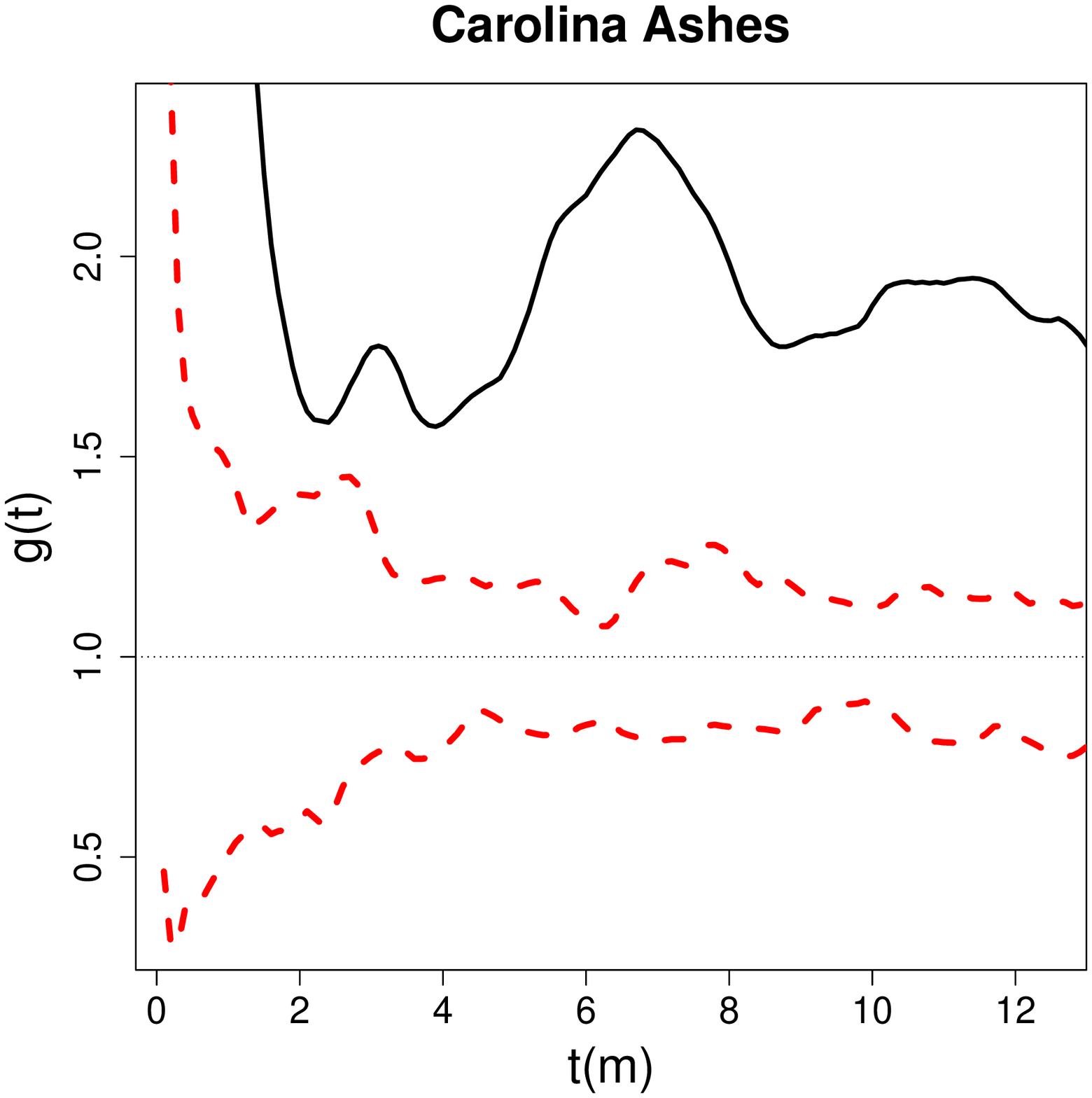} }}
\rotatebox{-90}{ \resizebox{2 in}{!}{\includegraphics{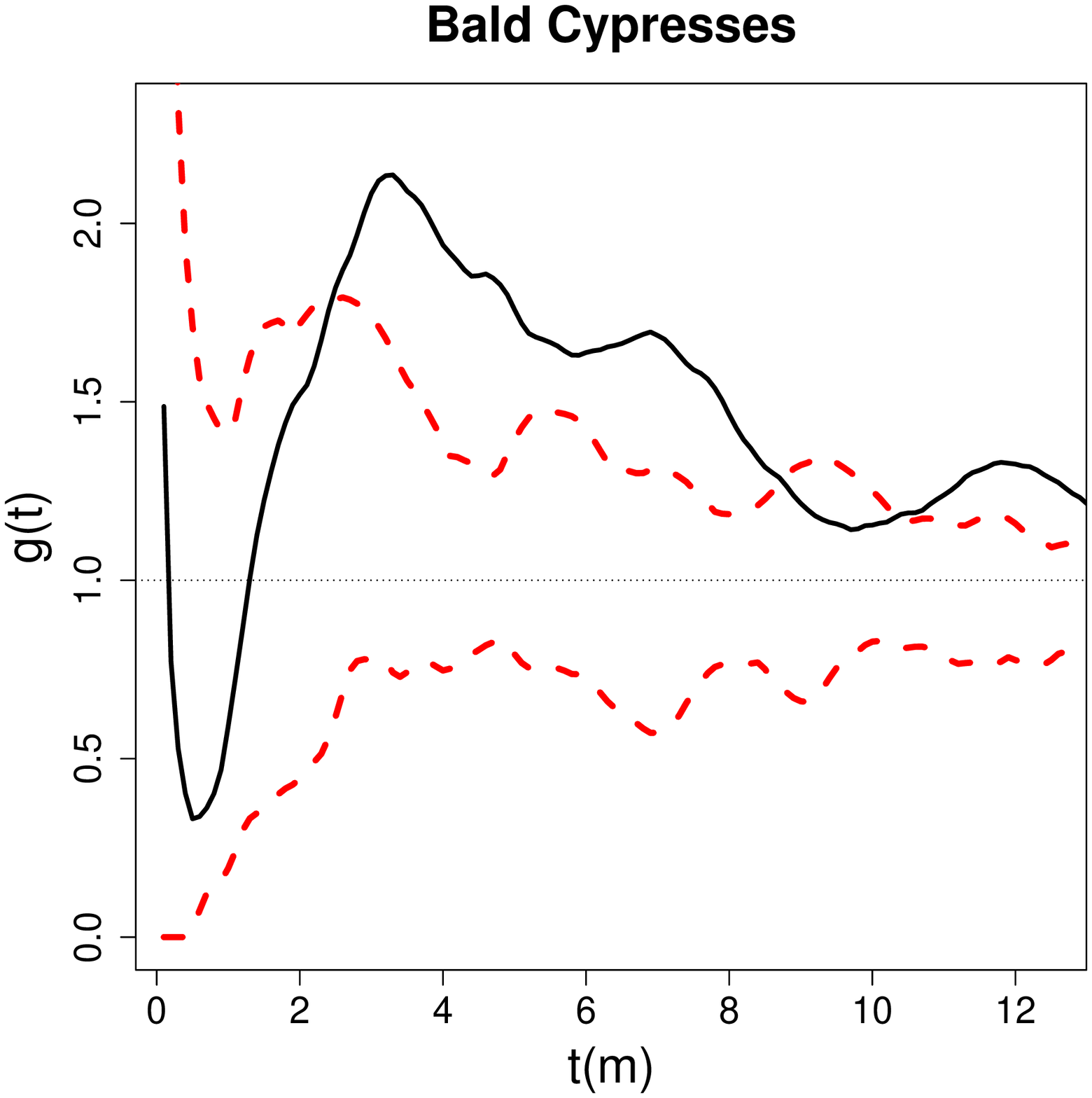} }}
%\rotatebox{-90}{ \resizebox{2 in}{!}{\includegraphics{SwampTreesPCF44.ps} }}
%\rotatebox{-90}{ \resizebox{2 in}{!}{\includegraphics{SwampTreesPCF55.ps} }}
\caption{
\label{fig:swamp-PCFii}
Pair correlation functions for each species in the swamp tree data.
Wide dashed lines around 1 (which is the theoretical value)
are the upper and lower (pointwise) 95 \% confidence bounds for the
pair correlation functions based on Monte Carlo simulation under the CSR independence pattern.}
\end{figure}

However, these results pertain to interaction at about the average NN distances.
For the swamp tree data average NN distance ($\pm$ standard deviation)
is about 2.1 ($\pm$ 1.35) meters.
We might also be interested in the possible causes of the segregation
and the type and level of interaction between the tree species
at different distances between the trees.
Along this line,
we also present the second-order analysis of the swamp tree data
by the pair correlation function $g(t)$ (\cite{stoyan:1994}).
The pair correlation function of a (univariate)
stationary point process is defined as
$g(t) = \frac{K'(t)}{2\,\pi\,t}$
where $K'(t)$ is the derivative of Ripley's $K(t)$ function.
For a univariate stationary Poisson process, $g(t)=1$;
values of $g(t) > 1$ suggest clustering (or aggregation)
and
the values of $g(t) < 1$ suggest inhibition (or regularity) between points.
The pair correlation functions for each species are plotted
in Figure \ref{fig:swamp-PCFii}.
Black gums are aggregated for distance values of about 1-6 and 9-11 m;
Carolina ashes are aggregated for all the range of the plotted distances;
and
bald cypresses are aggregated for distance values of about 2-8 and around 11 m.
These distance ranges at which species are aggregated include the mean NN distance for our data,
hence this aggregation could be the reason of the significant segregation between the species.

\begin{figure} [hbp]
\centering
%\psfrag{Density}{ \Huge{\bf{Density}}}
\rotatebox{-90}{ \resizebox{2 in}{!}{\includegraphics{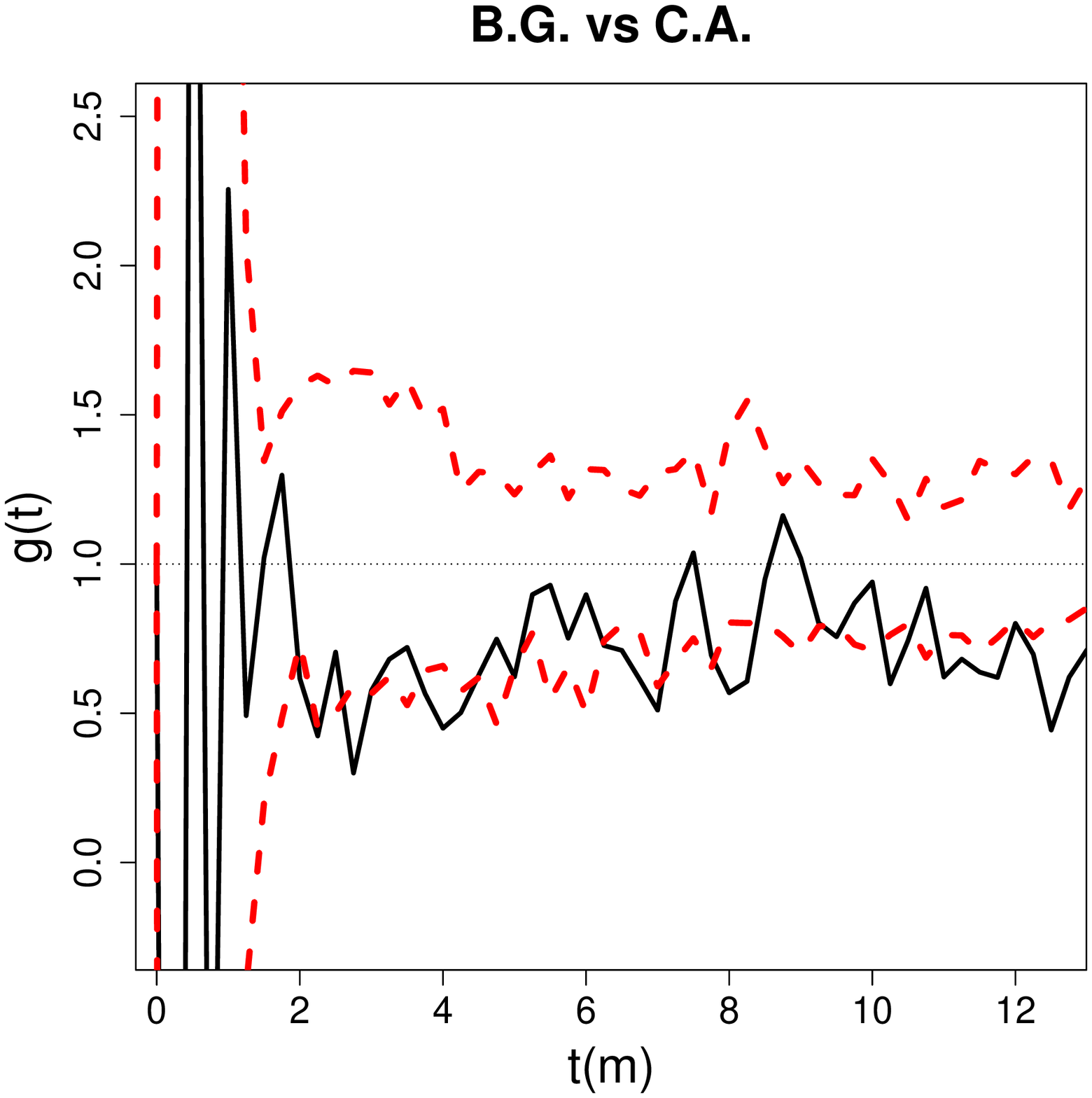} }}
\rotatebox{-90}{ \resizebox{2 in}{!}{\includegraphics{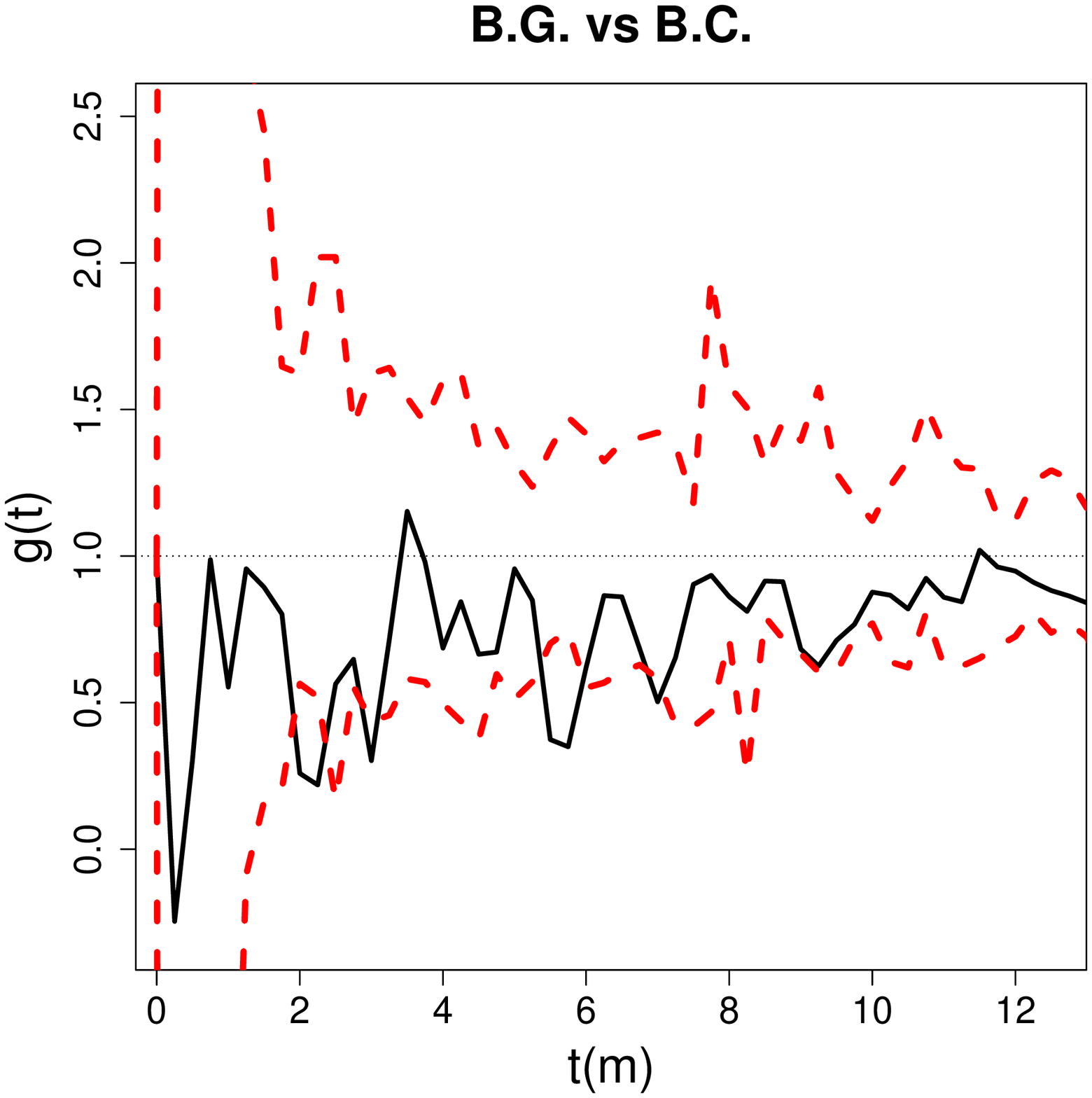} }}
\rotatebox{-90}{ \resizebox{2 in}{!}{\includegraphics{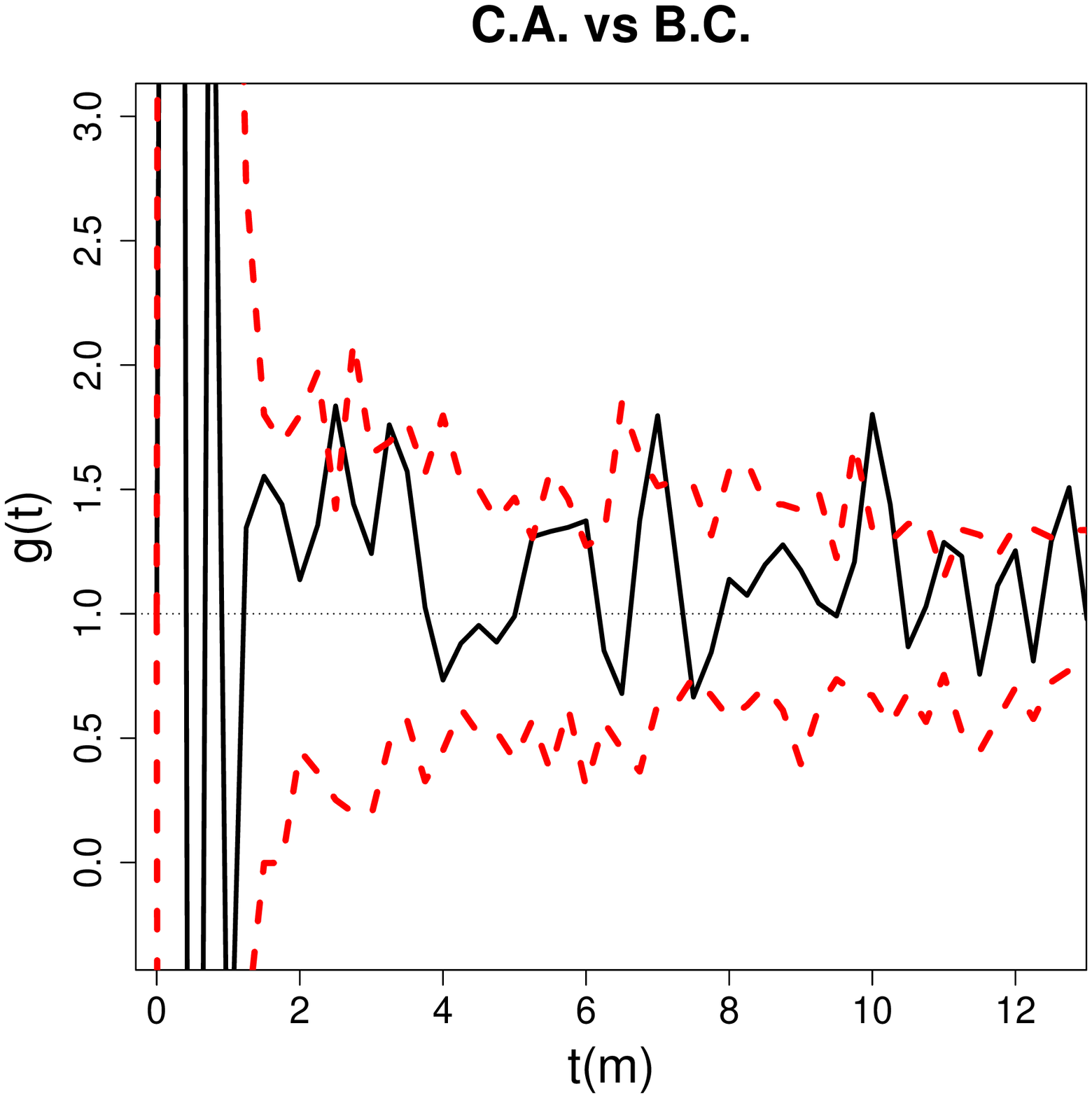} }}
\caption{
\label{fig:swamp-PCFij}
Pair correlation functions for each pair of species in the swamp tree data.
Wide dashed lines around 1 (which is the theoretical value)
are the upper and lower (pointwise) 95 \% confidence bounds for the
pair correlation functions based on Monte Carlo simulations under the CSR independence pattern.
B.G. = black gums, C.A. = Carolina ashes,
and B.C.  = bald cypresses.}
\end{figure}

The same definition of the pair correlation function
can be applied to Ripley's bivariate (i.e., two-class) $K$ or $L$-functions.
Under CSR independence,
we have $g(t) = 1$;
$g(t) > 1$ suggests association of the classes;
and
$g(t) < 1$ suggests segregation of the classes.
The bivariate pair correlation functions for the
species in swamp tree data are plotted in Figure \ref{fig:swamp-PCFij}.
Black gums and Carolina ashes are segregated for about 2-2.5, 3.5-4.5, 7.5-8.5, and 10.5-12 m;
black gums and bald cypresses are segregated for about 2.5, 3, and 6 m;
and
Carolina ashes and bald cypresses are associated for 7 and 9 m.

The pair correlation function estimates have considerably high variability
for small $t$ if $g(t)>0$,
hence not so reliable for small distances (\cite{stoyan:1996}).
See for example Figures \ref{fig:swamp-PCFii} and \ref{fig:swamp-PCFij}
where the confidence bands for small $t$ values are much wider compared
to those for larger $t$ values.
So pair correlation function analysis is more reliable for larger distances,
%and it is safer to use $g(t)$ for distances larger than
say,
larger than about the average NN distance in the data set.
While the pair correlation function
provides information on the univariate and bivariate patterns
at all distances,
NNCT-tests summarize the spatial interaction
for distances about the average NN distance in the data set.

\section{Discussion and Conclusions}
\label{sec:disc-conc}
We introduce new cell-specific and overall segregation tests
based on nearest neighbor contingency tables (NNCTs).
NNCT-tests are used in testing randomness in the nearest neighbor (NN)
structure between two or more classes with NN probabilities being proportional to the class frequencies.
The overall test is used for testing any deviation
from the null pattern in all the NNCT cells combined;
cell-specific test for cell $(i,j)$ is used for testing any deviation from the null case in cell $(i,j)$,
i.e.,
the probability of a (base,NN) in which base class is $i$ and NN class is $j$
is proportional to the product of frequencies of classes $i$ and $j$.
This statistic tests the segregation or lack of it, if $i=j$;
the association or lack of it between classes $i$ and $j$, if $i \neq j$.
Among many possible patterns,
the null pattern is implied by the RL or CSR independence patterns.
We demonstrate that under the CSR independence pattern,
NNCT-tests are conditional on $Q$ and $R$,
while under the RL pattern, these tests are unconditional.

Although we consider five types of cell-specific and overall tests,
we demonstrate that actually, these tests yield three distinct types of cell-specific or overall tests.
More specifically,
Dixon's tests and type II tests are identical,
and so are type III and type IV tests.
Hence in our empirical size and power analysis
(as well as in the example data),
we only use and present Dixon's, type I and III test statistics.
In the two-class case, cell-specific tests are essentially different only for at most two cells,
since cell $(1,1)$ and $(1,2)$ yield the same test statistic in absolute value for Dixon's
cell-specific test, likewise for cells $(2,1)$ and $(2,2)$.
Similarly, cell $(1,1)$ and $(2,1)$ yield the same test statistic in absolute value for
the type III cell-specific test, likewise for cells $(1,2)$ and $(2,2)$.
For type I cell-specific test cells $(1,1)$ and $(2,2)$ yield the same test statistic,
and the off-diagonal cells give the negative of this value.

We demonstrate that the cell-specific tests tend to standard normal distribution,
as the sample size gets larger.
On the other hand,
the overall tests tend to chi-square distribution with the corresponding degrees of freedom
with the increasing sample size.
In terms of the asymptotic distribution of the overall tests,
we have two groups of tests.
For $m$ classes,
Dixon's overall test has $\chi^2$ distribution with $m(m-1)$ df,
while type I and III tests have $\chi^2$ distribution with $(m-1)^2$ df.
Two major types of asymptotic structures for spatial data exist in literature:
infill asymptotics and increasing domain asymptotics (\cite{lahiri:1996}).
In ``infill asymptotics"
the region of interest is a fixed bounded region and the number of observed points gets larger in this region.
Hence the minimum distance between data points tends to zero
as the sample size tends to infinity.
In ``increasing domain asymptotics",
any two observations are required to be at least a fixed distance apart,
hence as the number of observations increase, the region on which the process
is observed eventually becomes unbounded (\cite{cressie:1993}).
The sampling structure in our asymptotic sampling distribution
could be either one of these asymptotic structures.
Because we only consider the class sizes and hence the total sample size
tending to infinity regardless of the size of the study region.

Based on our Monte Carlo simulations,
we observe that the asymptotic approximation for the cell-specific-tests
is appropriate only when the corresponding cell count in the NNCT is larger than 10;
and for the overall tests when all cell counts are at least 5.
For NNCTs with smaller cell counts,
we recommend the Monte Carlo randomization of the tests.
In the two-class case,
types I and III cell-specific tests have better empirical size performance
for the cell corresponding to the smaller class,
while Dixon's cell-specific test has better size performance for the cell corresponding to the larger class.
For the overall test, the performance of the tests are similar for Dixon's
and types I and III tests.
In the three class case,
types I and III cell-specific tests have better size performance,
and
overall tests have similar size estimates.
We also observe that
types I and III cell-specific tests
and
type III overall test are more robust to the differences in class sizes
(i.e., differences in relative abundance).
Under the \emph{segregation alternatives},
in the two-class case,
types I and III cell-specific tests have similar power estimates which are larger than those of Dixon's,
and the same holds for the overall tests as well.
In the three class case,
types I and III and Dixon's cell-specific tests have similar power estimates,
with type I and III being slightly higher.
The same holds for the overall tests as well.
Under the \emph{association alternatives},
in the two-class case,
types I and III cell-specific and overall tests tend to
have higher power estimates for most of the class size combinations.
The only exception is when the classes are highly unbalanced and
the cell-specific test is for the diagonal cell for the larger class.
In this case,
Dixon's tests have higher power.
In the three class case,
types I and III cell-specific tests have higher power estimates for cell $(i,j)$
if $n_i$ is less than $n_j$,
while
Dixon's cell-specific tests have higher power estimates
if $n_i$ is larger than $n_j$.
For the overall tests,
Dixon's overall test has the highest power estimates.
When empirical size and power performances are considered together,
among cell-specific tests,
types I and III cell-specific tests are recommended against the segregation alternatives,
while types I, III, and Dixon's cell-specific test are recommended against the association alternatives
depending on the class sizes in the off-diagonal cells.
Among overall tests,
type I and III overall tests are recommended against the segregation alternatives,
while Dixon's overall test is recommended against the association alternatives.
We extend this recommendation to one-vs-rest type tests as well.
Furthermore,
for one-vs-rest type tests,
all the tests have similar size performance,
but type I and III are more robust to differences in relative abundances.

NNCT-tests summarize the pattern in the data set for small scales
%more specifically, they provide information on the pattern
around the average NN distance between all points.
On the other hand, pair correlation function $g(t)$
and Ripley's classical $K$ or $L$-functions and other variants (\cite{baddeley:2000b}) provide
information on the pattern at various scales
(i.e., around other distance values).
Hence NNCT-tests and pair correlation or $K$-functions are not comparable but
provide complimentary information about the pattern in question.
However,
an advantage of overall NNCT-tests is that
they provide the interaction in a multi-class setting
in the presence of all classes,
while the second order analysis with $K$ or $g$ functions allow a comparison
of pairs of classes (one at a time).
Furthermore,
when an overall NNCT-test is significant,
it offers various post-hoc tests to follow up the specifics of the interaction:
(i) cell-specific tests,
(ii) one-class-vs-rest type tests,
and
(iii) class-specific tests.
In the cell-specific tests for cell $(i,j)$,
the interaction between classes $i$ and $j$ are examined in the presence of all other classes,
and in the class $i$-vs-rest testing,
the interaction of all the classes other than class $i$ with class $i$ is investigated.
The pair correlation function and $K$-functions
can also be adapted for one-vs-rest type analysis,
as classes $i$ and the rest of the classes can be treated as the two classes in our analysis.
On the other hand,
the bivariate pair correlation function or $K$-functions are also applicable for classes $i$ and $j$,
however, this analysis is restricted to the classes $i$ and $j$ only
in the sense that it ignores the influence of the other classes present in the region.
To the author's knowledge,
the class-specific tests has no counterpart among the $K$-function type second order methods.

The course of action we recommend depends on which null hypothesis is more appropriate.
If CSR independence is the reasonable null pattern,
we recommend the overall segregation test
to detect the spatial interaction at small scales at about the mean NN distance.
If it yields a significant result, then to determine which pairs
of classes have significant spatial interaction,
the cell-specific tests or one-vs-rest type tests can be performed
(we recommend both versions as they provide information on different aspects
of the spatial interaction).
To detect spatial interaction at larger distances,
pair correlation function is recommended (\cite{stoyan:2000}),
due to the cumulative nature of Ripley's $K$- or $L$-functions for larger distances.
On the other hand, if the RL pattern is the reasonable null pattern,
we recommend the NNCT-tests to detect the interaction at about the mean NN distance,
and Diggle's $D$-function (\cite{diggle:2003}) or
modified version of Ripley's $K$ function (\cite{baddeley:2000b}) to detect the interaction at higher distances.

\section*{Acknowledgments}
%I would like to thank an anonymous associate editor and two referees,
%whose constructive comments and suggestions greatly improved the presentation
%and flow of the paper.
Most of the Monte Carlo simulations presented in this article
were executed at Ko\c{c} University High Performance Computing Laboratory.
This research was supported by the research agency TUBITAK via Project \# 111T767
and the European Commission under the Marie Curie International Outgoing Fellowship Programme
via Project \# 329370 titled PRinHDD.

%\bibliography{References}
%\bibliographystyle{apalike}
%%\bibliographystyle{plain}

\end{document}